\def\doctype{0}
\def\flagreview{1}

\ifcase\doctype
  \documentclass{article}
  \usepackage{arxiv}
\or
  \documentclass[times,sageh]{sagej_fixed}
\or
  \documentclass[linenum]{SSA-BSSA}
\or
  \documentclass[linenum]{SSA-SRL}
\or
  \documentclass[linenum]{SSA-TSR}
\else
  \errmessage{Invalid value for \doctype}
\fi

\ifnum\doctype<2
    \usepackage[utf8]{inputenc} 
    \usepackage[T1]{fontenc}    
    \usepackage{geometry}       
    \usepackage{hyperref}       
    \usepackage{nameref}        
    \usepackage{url}            
    \usepackage{microtype}      
    \usepackage{xcolor}         
    \usepackage{nicefrac}       
    \usepackage{amssymb}
    \usepackage{booktabs}       
    \usepackage{amsfonts}       
    \usepackage{amsmath}
    \usepackage{amssymb}
    \usepackage{natbib}         
    \usepackage{doi}
    \usepackage{cleveref}
    \usepackage{graphicx} 
    \usepackage{multirow}
    \usepackage{float}
    \usepackage{caption}
    \usepackage{subcaption}
\fi
\ifnum\doctype>1
    \usepackage{caption}
    \usepackage{subcaption}
\fi

\if\doctype0
    \usepackage{lineno}
    \usepackage{pdflscape} 
    \usepackage{amsmath}
    \usepackage{graphicx} 
    \usepackage{multirow}
    \usepackage{float}
    \usepackage{geometry}
\fi

\usepackage{forest}          
\usepackage{lipsum}
\usepackage[percent]{overpic}
\usepackage{subcaption}
\usepackage{todonotes}
\usepackage{longtable}

\DeclareMathOperator{\E}{\mathbb{E}}

\if\doctype0 
\fi
\if\doctype1 
    \newcommand\BibTeX{{\rmfamily B\kern-.05em \textsc{i\kern-.025em b}\kern-.08em
    T\kern-.1667em\lower.7ex\hbox{E}\kern-.125emX}}
\fi

\title{Principal Surface Fault Rupture Hazard: Moving Beyond Aggregate Displacement Metrics}
\if\doctype0
    \renewcommand{\shorttitle}{Manuscript Short Title} 
    \newcommand{\headeright}{}
    \newcommand{\undertitle}{Submitted for Consideration to Journal}
    \hypersetup{
        pdftitle={PDF Title},
        pdfsubject={physics.geo-ph, stat.AP},
        pdfauthor={Author1, Author2, Author3},
        pdfkeywords={key1, key2, key3},
    }
\fi
\if\doctype1 
    \runninghead{Marmarchinia and Lavrentiadis}
\fi
\ifnum\doctype>1
  \ifnum\doctype<5
    \citethisauthor{Saba Marmarchinia, Grigorios Lavrentiadis}
    \doi{00.0000/0000000000}
    \recdate{00 Month 0000}
  \fi
\fi

\if\doctype0
    \author{
        \href{https://orcid.org/0009-0000-4221-5200}{\includegraphics[scale=0.06]{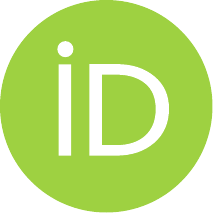}\hspace{1mm}Saba Marmarchinia} \\
        Department of Civil, Structural and Environmental Engineering\\
        State University of New York at Buffalo\\
        Buffalo, New York, U.S.A. \\
    	\texttt{sabamarm@buffalo.edu} \\
        \And
        \href{https://orcid.org/0000-0001-6546-1340}{\includegraphics[scale=0.06]{orcid.pdf}\hspace{1mm}Grigorios Lavrentiadis} \\
        Department of Civil, Structural and Environmental Engineering\\
        State University of New York at Buffalo\\
        Buffalo, New York, U.S.A. \\
    	\texttt{glavrent@buffalo.edu} 
    } 
\fi
\if\doctype1
    \author{Saba Marmarchinia \affilnum{1}, meeri,  Grigorios Lavrentiadis \affilnum{1}, meeri}
    \affiliation{
        \affilnum{1}Department of Civil, Structural and Environmental Engineering, State University of New York at Buffalo, Buffalo, New York, U.S.A.  \\
    }
    \corrauth{Grigorios Lavrentiadis, Department of Civil, Structural and Environmental Engineering, State University of New York at Buffalo, Buffalo, New York, U.S.A.}
    \email{glavrent@buffalo.edu}
\fi
\ifnum\doctype>1
  \ifnum\doctype<5
    \author[*1,2\orc{0000-0000-0000-0001}]{Author 1}%
    \author[1\orc{0000-0000-0000-0002}]{Author 2}
    \author[3\orc{0000-0000-0000-0003}]{Author 2}
    
    \affil[1]{First Affiliation, Institute, City, Country}{\auorc{0000-0000-0000-0001}{(AAF)}\auorc{0000-0000-0000-0002}{(ABS)}\auorc{0000-0000-0000-0003}{(ACT)}\auorc{0000-0000-0000-0004}{(AF)}}
    \affil[2]{Second Affiliation, Institute, City, Country}{\auorc{0000-0000-0000-0005}{(ADF)}\auorc{0000-0000-0000-0006}{(AS)}}
    \affil[3]{Third Affiliation, Institute, City, Country}{}
    \corau{*Corresponding author: cauthor@email.edu}
  \fi
\fi

\ifnum\flagreview>0
    \usepackage{blindtext}
    \usepackage{setspace}
    \usepackage{lineno}
\fi

\begin{document}

\if\doctype0
    \maketitle
\fi
\begin{abstract} 
    Structures crossing active faults are subjected to permanent ground deformation that can impose severe demands on infrastructure. Although probabilistic fault displacement hazard analysis (PFDHA) has advanced considerably in recent years, existing principal fault displacement models are formulated in terms of aggregate displacement, implicitly assuming that all overlapping rupture strands contribute simultaneously to the displacement demand at a site. This assumption is appropriate for regional hazard characterization but does not adequately represent the finite dimensions of engineering structures, which may intersect only a subset of the ruptures within the deformation zone. This study presents a new probabilistic framework for principal surface fault rupture hazard that reformulates aggregate displacement into a rupture-level representation suitable for infrastructure-specific hazard assessment. Using the Fault Displacement Hazard Initiative (FDHI) database of 75 historical surface-rupturing earthquakes, individual principal rupture strands are reconstructed from mapped rupture geometries and displacement observations through a preprocessing procedure based on event and segment coordinate systems. The proposed framework models three key quantities governing rupture-level hazard: (1) the number of overlapping principal ruptures, (2) the partitioning of aggregate displacement among individual rupture strands while preserving the aggregate displacement budget, and (3) the spatial distribution of rupture strands relative to the primary fault trace. These components are integrated into a generalized PFDHA formulation that explicitly accounts for earthquake magnitude, style of faulting, rupture curvature, structural footprint, and uncertainty in fault-trace location. Application of the methodology to benchmark fault-crossing scenarios demonstrates its ability to estimate the probability that individual rupture strands intersect a structure and to quantify the corresponding displacement demand. The proposed framework provides a physically consistent extension of existing aggregate-based fault displacement models and establishes the foundation for infrastructure-specific probabilistic fault displacement hazard analyses that explicitly account for rupture geometry and finite crossing dimensions.
\end{abstract}

\ifnum\doctype>0
   \maketitle
\fi

\section{Keywords}

Fault displacement models, seismic hazard, principal fault displacement, surface fault rupture, fault displacement hazard analysis (PFDHA).

\ifnum\flagreview>0
    \large
    \doublespacing
\fi

\section{Introduction}

While Probabilistic Seismic Hazard Analysis (PSHA) is well established for quantifying design-level ground shaking and is broadly supported by modern design codes, Probabilistic Fault Displacement Hazard Analysis (PFDHA) remains comparatively less mature despite its importance for infrastructure crossing or located near active faults. For systems such as pipelines, bridges, and tunnels, fault avoidance is often impractical, and permanent tectonic displacement can impose large differential movements and strains that may lead to severe damage or rupture. As a result, the hazard associated with permanent ground deformation must be explicitly considered in the design and assessment of such systems.

This requires characterization of both the amplitude of discrete displacement and the spatial extent of the associated deformation. In practice, these parameters are often estimated using a combination of empirical models and engineering judgment. While such approaches have proven useful, their reliance on subjective interpretation can limit the transparency, reproducibility, and consistency of hazard characterization.

Over the past three decades, numerous studies have focused on developing probabilistic models of surface fault displacement. \cite{wells1994new} compiled a global database of historical surface-rupturing earthquakes and developed empirical relationships between earthquake magnitude and rupture characteristics, including rupture length, rupture area, rupture width, and both average and maximum surface displacement. These relationships established the basis for characterizing different aspects of surface rupture.

Building on, \cite{yucca2001probabilistic} and \cite{youngs2003methodology} introduced a probabilistic framework for fault displacement hazard analysis for normal faulting by adapting the PSHA framework. In this framework, fault displacement models (FDMs) are used to estimate conditional displacement exceedance probabilities, which are combined with earthquake occurrence rates and the probability of surface rupture to quantify displacement hazard. The methodology distinguished between principal faulting, associated with slip on the primary seismogenic fault, and distributed faulting, associated with secondary ruptures occurring within the broader deformation zone.

Subsequent studies extended the applicability of PFDHA to additional faulting environments. \cite{moss2011probabilistic} developed models for reverse-faulting earthquakes using Weibull and Gamma distributions to represent normalized displacement profiles, while \cite{petersen2011fault} developed probabilistic models for principal and distributed surface displacement associated with strike-slip earthquakes using the rupture compilation of \cite{wesnousky2008displacement}. Their framework adopted lognormal displacement distributions and incorporated uncertainty in the location of principal rupture relative to the mapped fault trace.

More recently, \cite{lavrentiadis2019generation} proposed a kinematic-rupture-inspired framework for fault displacement modeling that represents displacement as a continuous spatial field along the fault trace. Unlike previous formulations based on normalized rupture coordinates, the proposed framework incorporates physics-informed scaling relationships and explicitly models the spatial correlation structure of fault displacement.

A recent advancement in PFDHA came through the Fault Displacement Hazard Initiative (FDHI) project. As part of this effort, \cite{sarmiento2025database} assembled a standardized, quality-controlled, and georeferenced database of surface fault displacement observations from 75 historical earthquakes spanning a broad range of magnitudes and faulting styles. Previously published surface rupture databases (e.g. \cite{baize2020worldwide} and \cite{nurminen2022sure}), were also reviewed and re-evaluated during the database development process. The database established consistent definitions of displacement metrics and provided the foundation for a new generation of fault displacement models. These include CEA25 \citep{chiou2025fault}, KEA24 \citep{kuehn2025fault}, LA23 \citep{lavrentiadis2025fault}, and MEA24 \citep{moss2025new}, which introduced updated statistical formulations incorporating nonlinear magnitude scaling, random effects, and non-Gaussian representations of aleatory variability, resulting in improved agreement with observed surface displacement data. In parallel, \cite{lavrentiadis2025seismologically} developed a surface rupture length model for use within the FDHI framework that explicitly distinguishes between width-limited and unbounded ruptures. A comprehensive comparison of the resulting fault displacement models was subsequently presented by \cite{sarmiento2025comparisons}, who documented systematic differences in model predictions, including larger mean displacements for large-magnitude events and reduced upper-tail variability.


A comprehensive overview of the evolution of PFDHA over the past 25 years is provided by \cite{valentini2025twenty}.

Despite these developments, current fault displacement models remain focused primarily on aggregate displacement metrics. For example, CEA25 predicts aggregate principal displacement, KEA24 and MEA24 characterize aggregate displacement resulting from the contribution of both principal and distributed faulting, while LA23 provides models for both aggregate principal displacement and aggregate total displacement. Although these quantities are appropriate for many hazard applications, they do not explicitly characterize the individual rupture components that contribute to the aggregate displacement.

This limitation becomes important when evaluating the performance of fault-crossing infrastructure. Structures may intersect only a subset of the ruptures contributing to the aggregate displacement, particularly when the dimensions of the structure are small relative to the width of the deformation zone. In addition, engineering analyses often require information regarding the location and displacement of individual rupture strands to estimate localized deformation demands and strain concentrations. As a result, aggregate displacement models provide only limited information for applications requiring characterization of discrete rupture geometry and displacement partitioning.

This study develops a probabilistic framework for decomposing aggregate principal displacement into its constituent discrete rupture components. The proposed methodology models three quantities: (1) the number of overlapping principal ruptures, (2) the partitioning of aggregate displacement among individual ruptures, and (3) the strike-normal location of each rupture relative to the primary fault trace. Uncertainty in the location of the primary fault trace is incorporated through alternative fault-trace definitions. The framework is formulated such that the aggregate principal displacement is conserved as a displacement budget, thereby maintaining consistency with existing aggregate-based fault displacement models.

The remainder of this paper is organized as follows. First, the FDHI database and the preprocessing procedures used to identify and characterize overlapping principal ruptures are described. Next, a probabilistic framework is developed to reformulate aggregate principal displacement hazard in terms of individual rupture hazard. Regression models are then presented for the number of overlapping ruptures, displacement partitioning, and rupture location. The proposed methodology is subsequently applied to the IAEA benchmark cases \citep{valentini2021probabilistic,valentini2025iaea}, and the resulting hazard estimates are presented and discussed. Finally, the notation adopted throughout this study is summarized.

\section{Dataset and Data Processing}

\subsection{FDHI Database}

The Fault Displacement Hazard Initiative (FDHI) database \citep{sarmiento2025database} was used in this study. The database was developed through the compilation, review, and homogenization of fault displacement measurements and surface rupture maps from 75 published surface-rupturing earthquakes that occurred between 1872 and 2019. The primary objective of this effort was to provide a standardized dataset for the development and evaluation of fault displacement models.

Only earthquakes with sufficiently complete surface rupture maps, displacement measurements, and two-dimensional geospatial control of both rupture traces and displacement observations were included. The majority of the events correspond to shallow crustal earthquakes in North America, while nine events are from the stable continental region of Australia. Although differences in rupture characteristics may exist between these tectonic environments, both datasets were combined to maximize the empirical dataset available for model development.

\begin{figure}[htbp]
    \centering
    \includegraphics[width=0.8\linewidth]{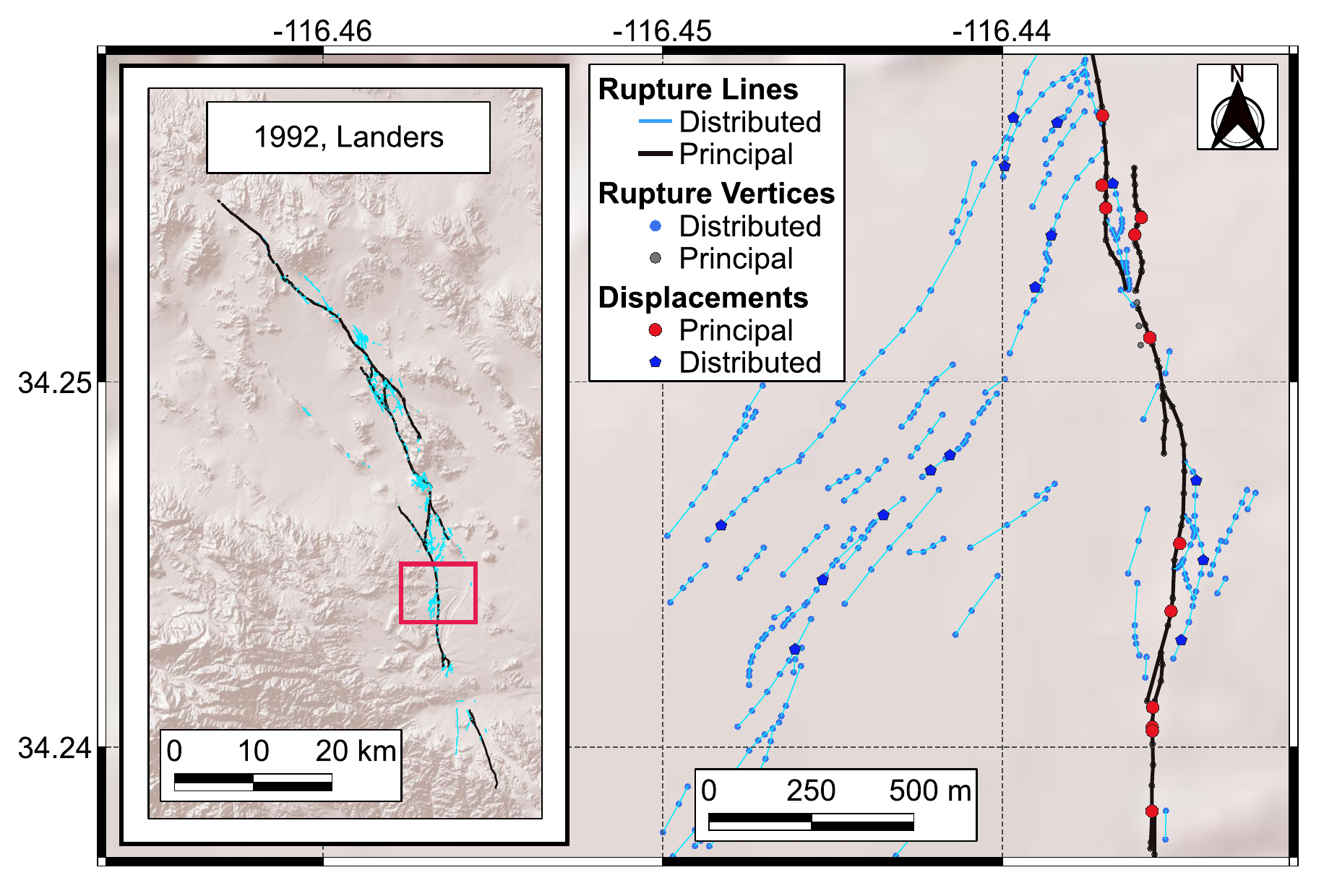}
    \caption{Classification of principal and distributed rupture vertices and displacement observations for the 1992 Landers earthquake ($M_w=7.28$).}
    \label{fig:Landers(rupvertices)}
\end{figure}

The final database contains more than 40,000 individual fault displacement measurements and approximately 28,000 measurement sites spanning all styles of faulting, together with 91,051 rupture lines derived from 763,691 rupture vertices. As part of the review and homogenization process, consistent definitions of principal and distributed faulting were applied throughout the database to both displacement observations and mapped rupture geometries. Principal faulting corresponds to surface slip occurring on the primary fault, whereas distributed faulting corresponds to deformation occurring on secondary faults, splays, pre-existing fractures, or surrounding shear zones. Figure~\ref{fig:Landers(rupvertices)} illustrates this classification for the 1992 $M~ 7.28$ Landers earthquake.

\begin{figure}[htbp]
    \centering

    \begin{subfigure}[c]{0.49\linewidth}
        \centering
        \caption{}
        \includegraphics[width=\linewidth]{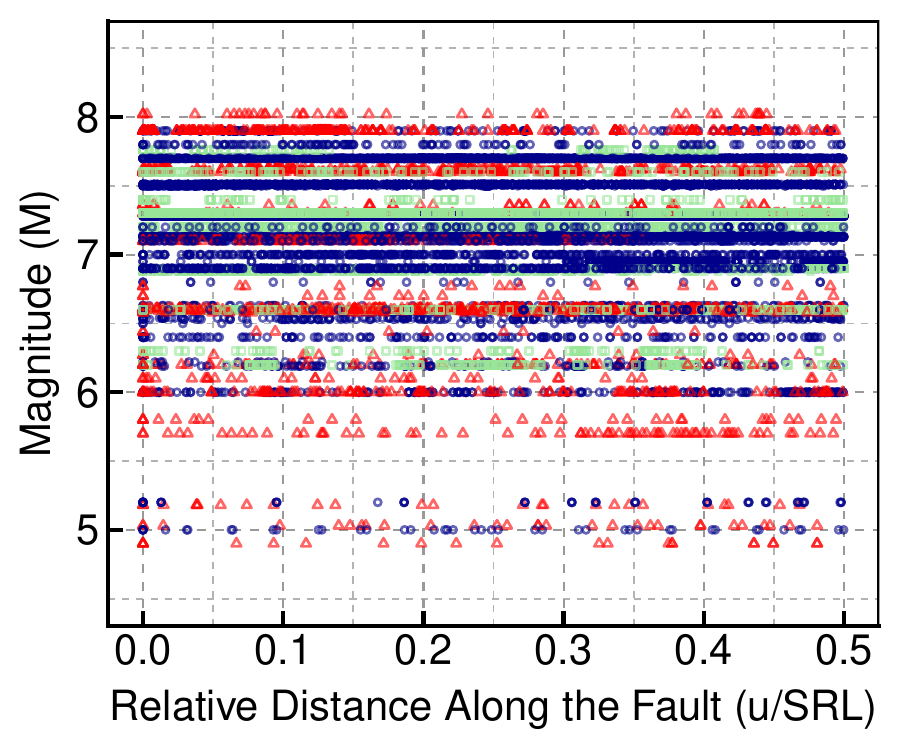}
        \label{fig:dataset_mag_scs_u2_norm}
    \end{subfigure}
    \hfill
    \begin{subfigure}[c]{0.49\linewidth}
        \centering
        \caption{}
        \includegraphics[width=\linewidth]{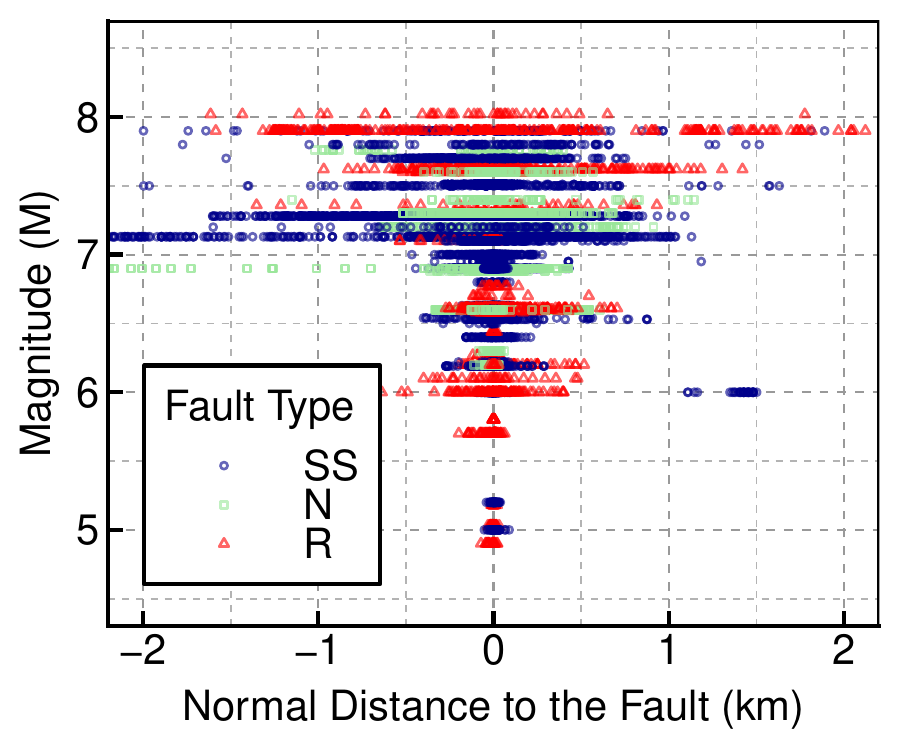}
        \label{fig:dataset_mag_scs_t_rup}
    \end{subfigure}

    \caption{Summary of principal displacement observations in the FDHI database: (a) earthquake magnitude versus normalized along-strike position and (b) earthquake magnitude versus strike-normal distance. Different markers correspond to different styles of faulting.}
    \label{fig:dataset_details}
\end{figure}

Figure~\ref{fig:dataset_details} summarizes the spatial distribution of principal displacement observations in the FDHI database. Most observations correspond to moderate-to-large magnitude earthquakes ($M$ 6--8) and span the full along-strike extent of the rupture (Figure~\ref{fig:dataset_mag_scs_u2_norm}). Figure~\ref{fig:dataset_mag_scs_t_rup} further shows that the strike-normal extent of principal faulting generally increases with earthquake magnitude, with large-magnitude events producing principal ruptures at distances of a few kilometers from the primary rupture trace.

\subsection{Reconstruction of Individual Principal Ruptures}

The objective of this study is to characterize the individual rupture components that contribute to the aggregate displacement. This requires identification of all principal rupture strands present at a common along-strike location together with their associated displacement amplitudes and strike-normal locations.

Although the FDHI database provides the geometry of mapped rupture traces and the locations of individual displacement observations, it does not directly provide the displacement associated with all principal rupture strands at a common along-strike location. This is because displacement measurements are typically collected at natural or anthropogenic features that provide suitable markers for estimating offset and are therefore distributed irregularly in space. Consequently, observations are generally unavailable at the same along-strike location across multiple parallel rupture strands. To address this limitation, an additional processing procedure was developed to reconstruct the displacement amplitudes and locations of individual principal rupture components from the original geospatial dataset.

\begin{figure}[htbp]
    \centering
    \includegraphics[width=0.8\linewidth]{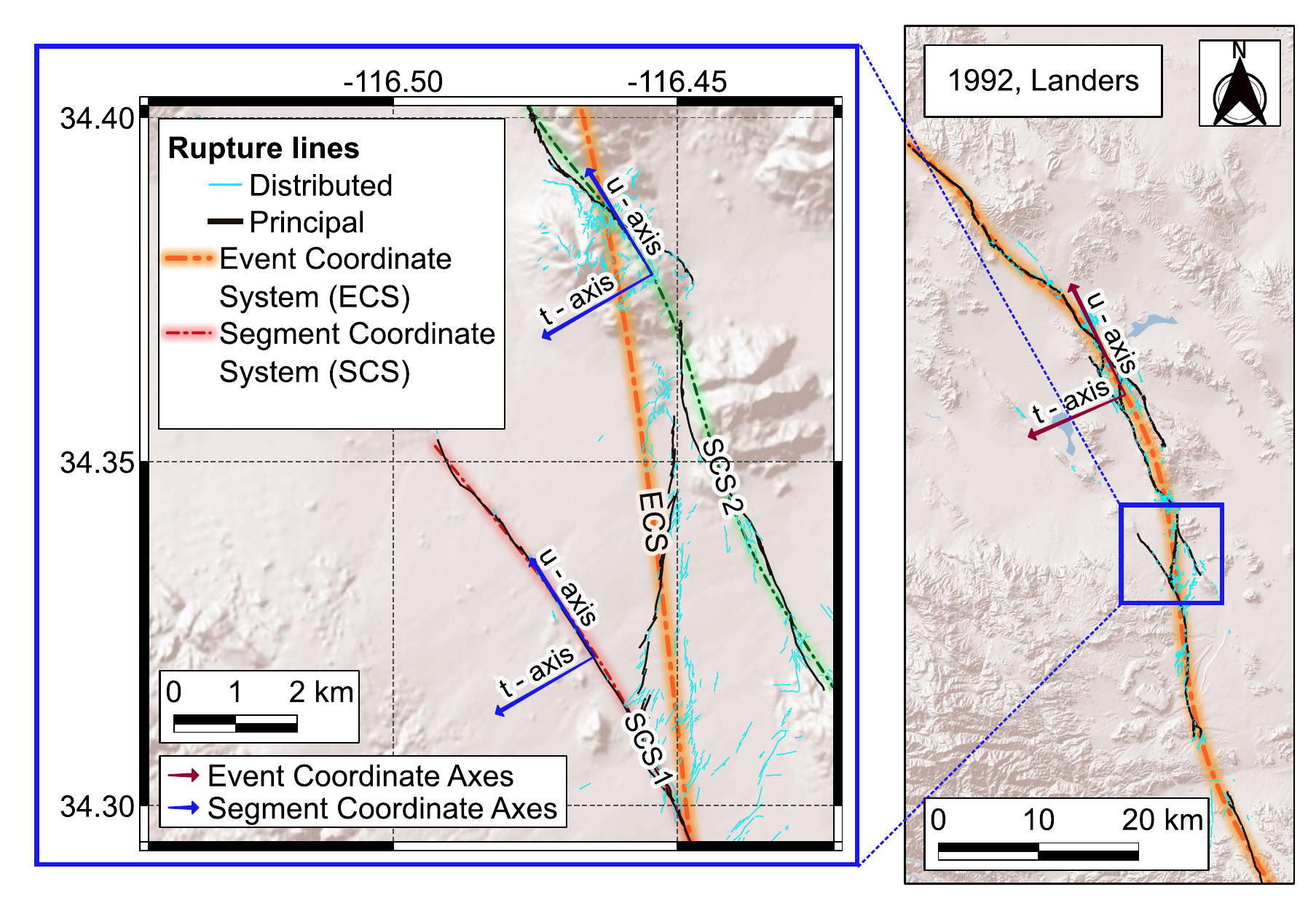}
    \caption{Event and segment rupture traces for the 1992 $M~ 7.28$ Landers earthquake together with the corresponding ECS and SCS coordinate systems.}
    \label{fig:landers_gc2_axes}
\end{figure}

The reconstruction procedure is performed using the Event Coordinate System (ECS) and the Segment Coordinate System (SCS). Both coordinate systems define a nominal rupture trace based on the spatial distribution of principal displacement observations \citep{lavrentiadis2019new,lavrentiadis2024event}. The ECS is defined for the entire rupture event, whereas the SCS is defined independently for each rupture segment, following the segmentation established in \citep{lavrentiadis2025fault}. Once the nominal rupture traces are established, along-strike ($u$) and strike-normal ($t$) coordinates are computed using the Generalized Coordinate System 2 (GC2) framework of \cite{Spudich2015}. Figure~\ref{fig:landers_gc2_axes} illustrates the resulting event and segment coordinate systems and local coordinate axes for the 1992 $M~ 7.28$ Landers earthquake.

For each rupture segment, all principal rupture strands intersecting a given along-strike location $u$ are identified. A rupture strand is considered to overlap location $u$ if at least one pair of consecutive rupture vertices defining the strand spans that location:

\begin{equation}
    \exists \, (u_{k,i},u_{k,j}) \in \mathcal{V}_k
    \quad \text{such that} \quad
    (u_{k,i}-u)(u_{k,j}-u) \le 0
\end{equation}

where $\mathcal{V}_k$ denotes the set of vertices defining the geometry of the $k^{\mathrm{th}}$ rupture strand. The above condition indicates that the along-strike coordinate $u$ lies between vertices $i$ and $j$, and therefore that the rupture strand intersects the specified location. If no such vertex pair exists, the rupture strand does not overlap location $u$.

\begin{figure}[htbp]
    \centering
    \includegraphics[width=0.8\linewidth]{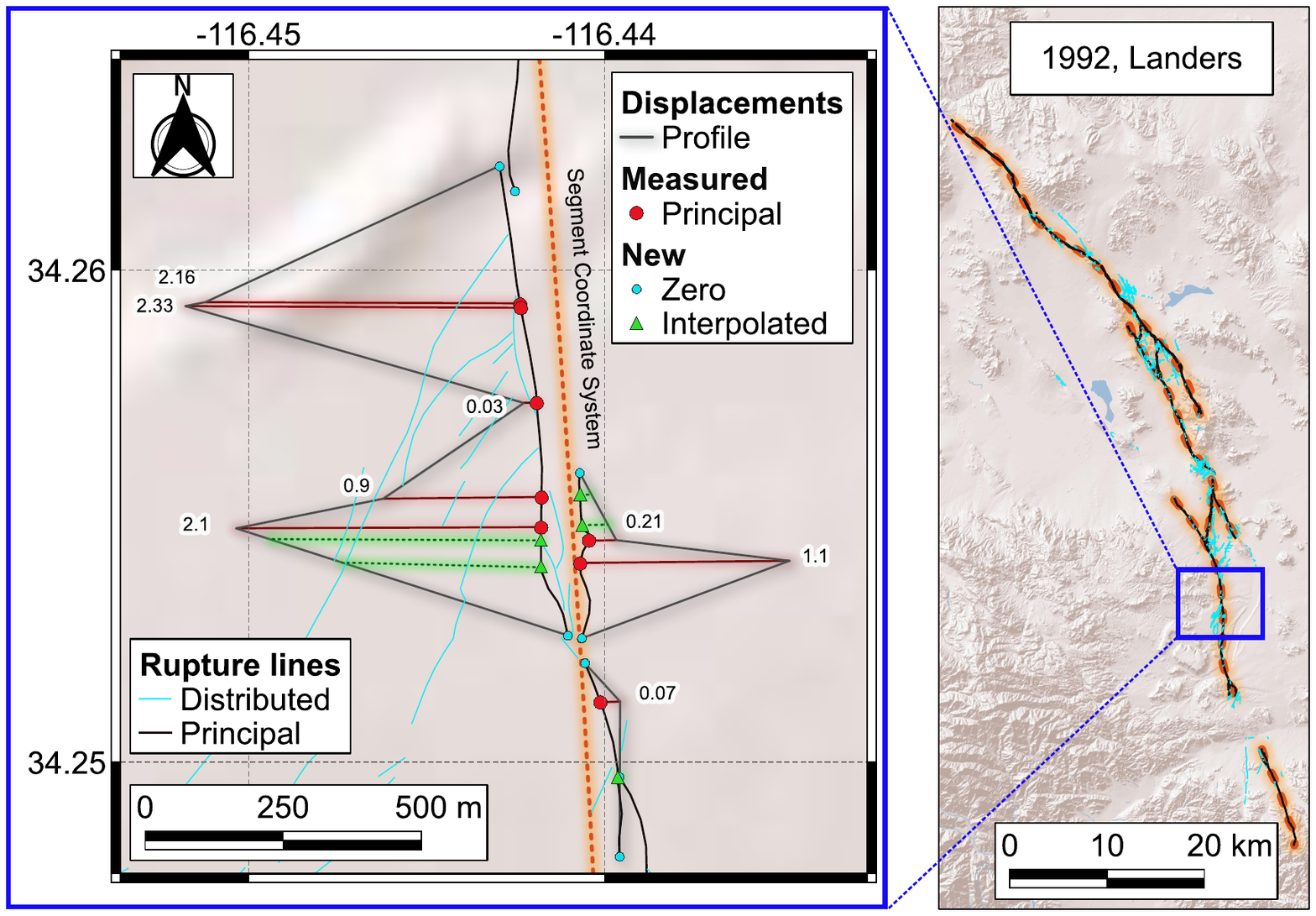}
    \caption{Illustration of the displacement interpolation procedure used to estimate displacement values along overlapping rupture strands for the 1992 Landers earthquake.}
    \label{fig:landers_disp_map}
\end{figure}

In many cases, a rupture strand overlaps a given along-strike location but no displacement measurement is available at that position. For such cases, displacement is estimated through interpolation using neighboring observations located along the same rupture strand. Specifically, displacement is assumed to reduce to zero at both rupture endpoints and to vary linearly between consecutive displacement observations. Figure~\ref{fig:landers_disp_map} illustrates the interpolation procedure adopted for the 1992 Landers earthquake.

\begin{figure}[htbp]
    \centering
    \includegraphics[width=0.8\linewidth]{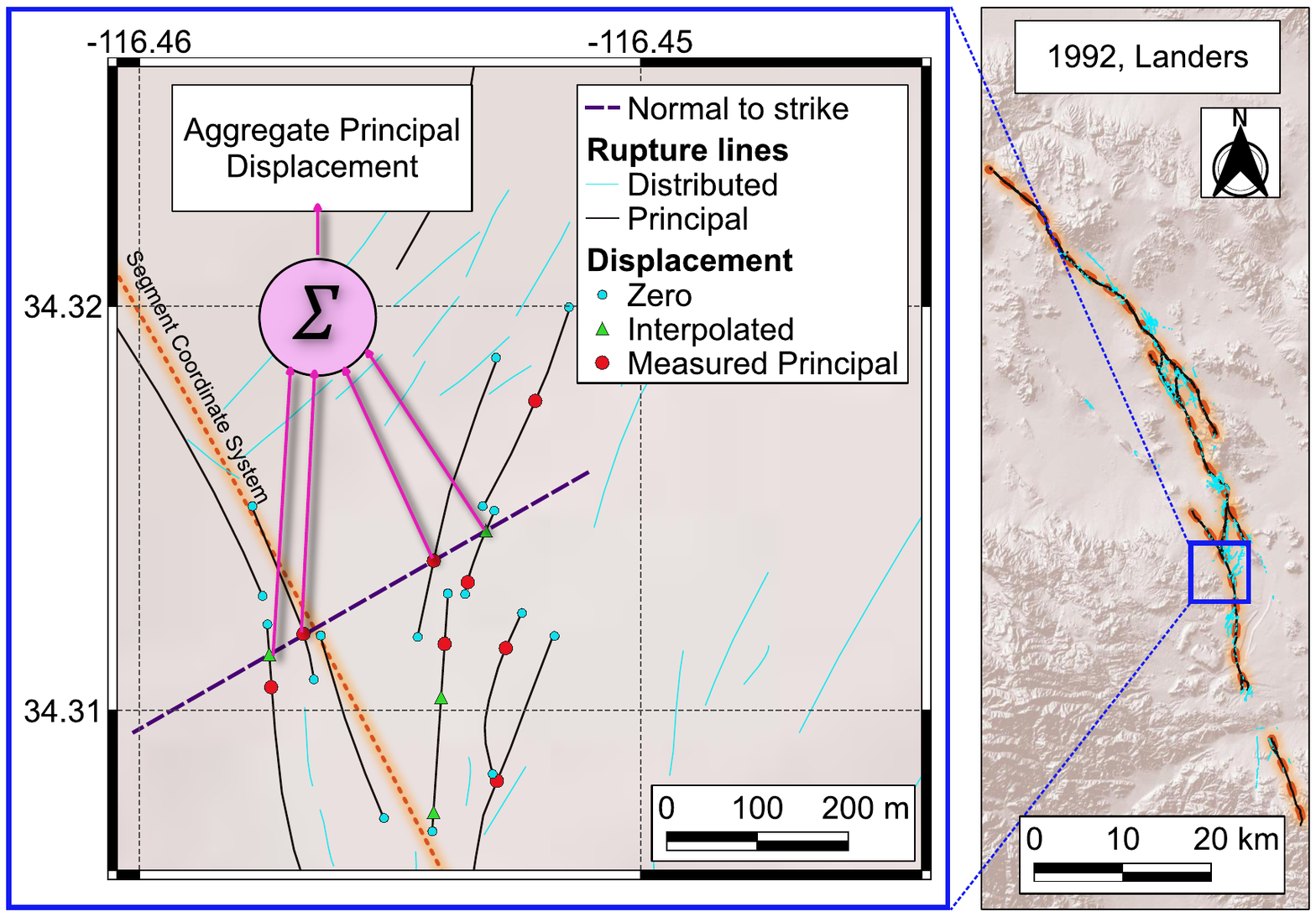}
    \caption{Schematic illustration of principal and aggregate principal displacement at a given along-strike location for the 1992 Landers earthquake.}
    \label{fig:1992_Landers_aggvsprncp}
\end{figure}

\begin{figure}[htbp]
    \centering
    \includegraphics[width=0.8\linewidth]{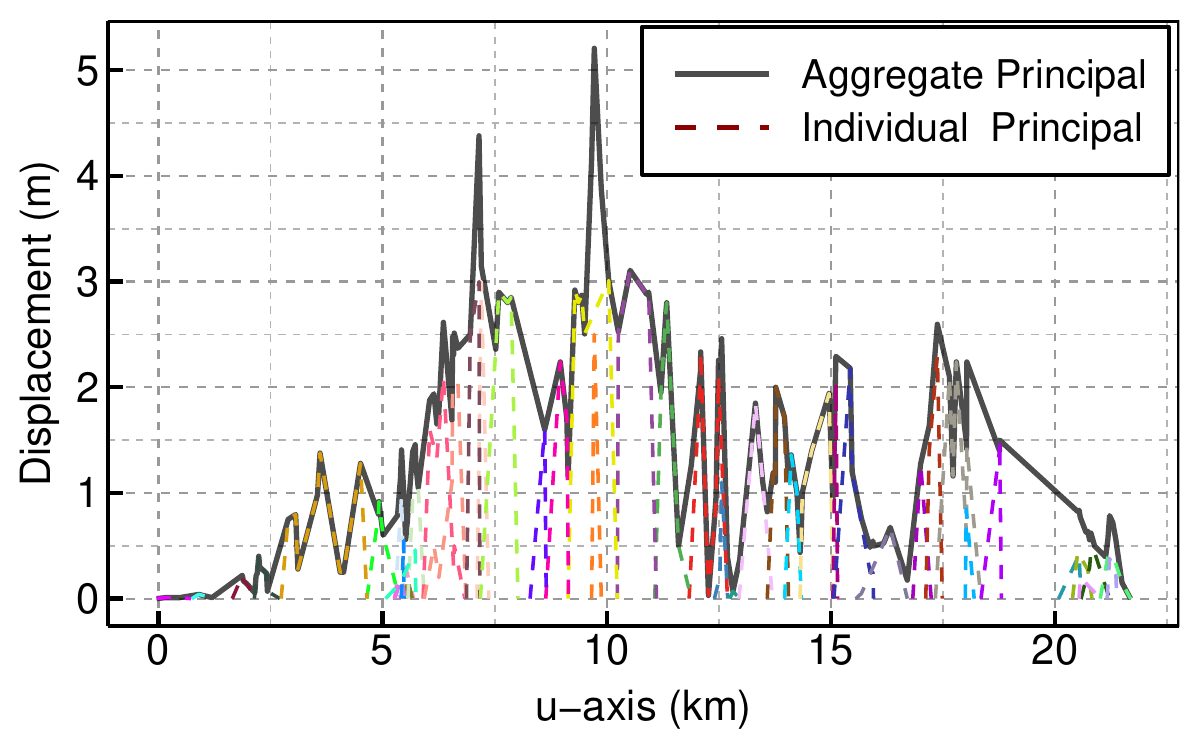}
    \caption{Individual principal displacement profiles and corresponding aggregate principal displacement along a segment of the 1992 Landers rupture.}
    \label{fig:landers_disp_prof}
\end{figure}

As described in Figure \ref{fig:1992_Landers_aggvsprncp} the displacement associated with an individual principal rupture strand is referred to as the principal displacement, $D_{p,~k}$. At a given along-strike location $u$, the aggregate principal displacement, $D_{p,~\mathrm{agg}}$, is defined as the sum of the displacements associated with all overlapping principal rupture strands,

\begin{equation}
D_{p,~\mathrm{agg}}(u)
=
\sum_{k=1}^{N_{p,~\mathrm{tot}}(u)}
D_{p,~k}(u)
\end{equation}

where $N_{p,~\mathrm{tot}}(u)$ denotes the number of overlapping principal rupture strands at location $u$. The resulting principal displacement profiles for one segment of the 1992 Landers rupture are shown in Figure~\ref{fig:landers_disp_prof}. Each rupture strand is represented by a separate displacement profile derived from the mapped rupture geometry and interpolated displacement observations. The figure illustrates that multiple rupture strands may be present at a common along-strike location and therefore contribute to the aggregate displacement field.

Given the aggregate principal displacement at each analyzed along-strike location, a principal displacement ratio is computed for every overlapping rupture strand:

\begin{equation}
r_{p,~k}(u)
=
\frac{D_{p,~k}(u)}
     {D_{p,~\mathrm{agg}}(u)}
\end{equation}

where $D_{p,~k}(u)$ is the displacement associated with rupture strand $k$ and $D_{p,~\mathrm{agg}}(u)$ is the aggregate principal displacement at the same location.

At the conclusion of this processing stage, a dataset is obtained containing, for each analyzed location, the number of overlapping principal ruptures, aggregate principal displacement, displacement ratio of each rupture, corresponding strike-parallel and strike-normal coordinates, and a rupture identifier. This dataset forms the basis for development of the probabilistic models presented in the following sections.

\section{Existing and Proposed FDHA Framework} \label{sec:pfdha_framework}

\subsection{Existing PFDHA Framework} \label{sec:pfdha_framework_existing}

The existing probabilistic framework for PFDHA is formulated analogously to PSHA \citep{youngs2003methodology}, in which ground-motion intensity measures are replaced by permanent ground-surface displacement. However, unlike seismic ground shaking, permanent tectonic fault displacement does not necessarily reach the ground surface for every earthquake, and even when surface rupture occurs, principal displacement may locally reduce to zero between rupture segments or within portions of a rupture where deformation is accommodated through distributed slip. Consequently, all of these effects must be explicitly incorporated into the hazard formulation. In \cite{lavrentiadis2025fault}, which forms the basis of the present study, the aggregate principal displacement hazard is formulated as:

\begin{equation}\label{eq:Previous_hazard_framework}
\begin{aligned}
Haz\,(D_{p,~\mathrm{agg}} \ge z)
&= \sum_{i=1}^{N_{eq}} \nu
     \times \, P(SR \mid M)\\
   & \quad \times \, (1 - P\!\left( \text{GAP} \mid M, u/SRL , SOF \right) ) \\
&\quad \times \, (1 - P\!\left( D_{p,~\mathrm{agg}} = 0 \mid M, u/SRL , SOF \right) ) \\
   & \quad \times \, P\!\left( D_{p,~\mathrm{agg}} \ge z \mid M, u/SRL , SOF  \right)
\end{aligned}
\end{equation} 
\noindent
where the terms equivalent to those in PSHA are $\nu$, representing the annual frequency of occurrence of the $i^{\mathrm{th}}$ earthquake scenario, and 
$P(D_{p,~\mathrm{agg}} \ge z \mid M, u/SRL, SOF)$ representing the conditional probability that the aggregate principal displacement exceeds a threshold $z$, analogous to the exceedance probability obtained from a ground-motion model in PSHA. 
The remaining terms are unique to PFDHA. Specifically, 
$P(SR \mid M)$ is the probability that the rupture reaches the ground surface, accounting for the portion of earthquakes of a given magnitude that produce surface rupture; 
$P(\mathrm{GAP} \mid M, u/SRL, SOF)$ is the probability that the site lies within a gap between rupture segments where the principal displacement is zero; and 
$P(D_{p,~\mathrm{agg}} = 0 \mid M, u/SRL, SOF)$ is the conditional probability that the principal displacement locally reduces to zero within an active rupture segment due to distributed deformation.

However, using aggregate displacement in hazard calculations implicitly assumes a structure of infinite length that intersects all overlapping rupture strands, while also obscures the localized deformation demands that a structure may experience over a finite crossing length.

\subsection{Proposed PFDHA Framework} \label{sec:pfdha_framework_proposed}

To address this limitation in the current practice, a new framework is proposed in this study which accounts for the spatial distribution of the principal rupture strands and the extent of the structure. In this formulation, the aggregate principal displacement is decomposed into discrete principal ruptures by explicitly modeling (i) the number of rupture strands at a given location along strike, (ii) the partitioning of the aggregate displacement among the principal displacement strands, and (iii) the distances of these strands from the primary fault trace. The resulting hazard formulation, developed to account for the aforementioned effects, is expressed as follows:

\begin{equation}\label{eq:Proposed_hazard_framework}
\begin{aligned}
Haz\,(D_{p,~\mathrm{cross}} \ge z)
&= \sum_{i=1}^{N_{eq}} \nu
     \times \, P(SR \mid M)\\
   & \quad \times \, (1 - P\!\left( \text{GAP} \mid M, u/SRL , SOF \right) ) \\
&\quad \times \, (1 - P\!\left( D_{p,~\mathrm{agg}} = 0 \mid M, u/SRL , SOF \right) ) \\
   & \quad \times \, P\!\left( D_{p,~\mathrm{cross}} \ge z \mid M, u/SRL , SOF , t_{1}, t_{2}, \kappa \right)
\end{aligned}
\end{equation}
\noindent 

The key component of Equation~(\ref{eq:Proposed_hazard_framework}) is the conditional probability of exceedance for principal displacement,
$P(D_{p,~\mathrm{cross}} \ge z \mid M, u/SRL,SOF,t_1,t_2,\kappa)$.
This probability accounts for all possible configurations of overlapping principal rupture strands, including the number of strands present, the number intersecting the structure, the distribution of aggregate principal displacement among those strands, and their spatial occurrence within the structure span. It is formulated as:

\begin{equation}
\label{eq:Probability_principal_displacement_exceedance}
\begin{aligned}
P\!\left( D_{p,~\mathrm{cross}} \ge z \mid M, u/SRL, SOF, t_1, t_2, \kappa \right)
&= \sum_{N_{p,~\mathrm{tot}}=1}^{\infty} P(N_{p,~\mathrm{tot}}  \mid M, \kappa, SOF) \\[6pt]
&\quad \times \sum_{N_{p,~\mathrm{cross}}=1}^{N_{p,~\mathrm{tot}}} \int_{0}^{\infty} \int_{0}^{1}
f\!\left( r_{p,~\mathrm{cross}} \mid M, SOF, N_{p,~\mathrm{tot}}, N_{p,~\mathrm{cross}} \right) \\
&\quad \times 
P\!\left( t \in [t_1, t_2] \mid M, SOF, 
\hat{D}_p, N_{p,~\mathrm{cross}}\right) \\
&\quad \times 
\, f\!\left( D_{p,~\mathrm{agg}} \mid M, SOF, u/SRL \right) \\
&\quad \times I(D_{p,~\mathrm{agg}}~r_{p,~\mathrm{cross}} - z)\, dr \, dD_{p,~\mathrm{agg}}
\end{aligned}
\end{equation}
\noindent 

\noindent

The formulation sums over all possible configurations of overlapping principal rupture strands. The term $P(N_{p,~\mathrm{tot}} \mid M,\kappa,SOF)$ represents the probability that $N_{p,\mathrm{tot}}$ principal rupture strands are present at the considered along-strike location. The outer summation accounts for all admissible numbers of principal rupture strands, while the inner summation considers all possible numbers of strands intersecting the structure, $N_{p,~\mathrm{cross}}$, conditioned on the total number present.

For a given rupture configuration, the fraction of aggregate principal displacement associated with the rupture strands intersecting the structure is described by the probability density function $f(r_{p,~\mathrm{cross}} \mid M,SOF,N_{p,~\mathrm{tot}},N_{p,~\mathrm{cross}})$, where $r_{p,~\mathrm{cross}}$ denotes the ratio of displacement carried by the intersecting rupture strands relative to the aggregate principal displacement. The spatial occurrence of the intersecting rupture strands within the structure span is represented by $P(t \in [t_1,t_2] \mid M,SOF,\hat{D}_p, N_{p,~\mathrm{cross}})$, where $t_1$ and $t_2$ define the limits of the structure normal to the fault trace.

The aggregate principal displacement is characterized by the conditional probability density function $f(D_{p,~\mathrm{agg}} \mid M,SOF,u/SRL)$, obtained from the aggregate principal displacement model of \cite{lavrentiadis2025fault}. This term describes the distribution of aggregate principal displacement conditioned on earthquake magnitude, style of faulting, and normalized along-strike location.

The contribution of each rupture configuration to the exceedance probability of a principal displacement threshold $z$ is evaluated through the indicator function $I(D_{p,~\mathrm{agg}}~r_{p,~\mathrm{cross}}-z)$. The indicator function equals unity when the total principal displacement associated with the rupture strands intersecting the structure ($D_{p,~\mathrm{cross}} = D_{p,~\mathrm{agg}}~r_{p,~\mathrm{cross}}$) exceeds $z$, and zero otherwise.

Throughout the formulation, $M$ denotes earthquake magnitude, $\kappa$ denotes fault curvature, $u/SRL$ denotes the normalized along-strike location relative to the event surface rupture length ($SRL$), and $SOF$ denotes the style of faulting.

\section{Methodology} \label{sec:Methodology}

As described in \hyperref[sec:pfdha_framework_proposed]{Section: Proposed PFDHA Framework}, the decomposition of aggregate principal displacement into individual rupture components requires characterization of three quantities: (i) the number of overlapping principal rupture strands, which determines the number of individual displacement contributions present at a given location; (ii) the partitioning of aggregate principal displacement among those rupture strands, represented through displacement ratios; and (iii) the strike-normal location of each rupture strand relative to a nominal fault trace. The development of probabilistic models for each of these components is described in the following sections.

Due to the relatively limited number of events associated with normal and reverse faulting in the FDHI database, multiple fault-type groupings were considered during model development to ensure statistically stable parameter estimates. Specifically, five model variants were developed for each component: a model using all fault types combined (All), and separate models for strike-slip (SS), normal (N), reverse (R), and combined normal-reverse (NR) faulting. This approach seeks to balance the representation of fault-type-dependent behavior with the limitations imposed by the available dataset.

The reduced quantity of data available for normal and reverse faulting introduces additional epistemic uncertainty into the corresponding regressions. Accordingly, when applying the proposed framework to PFDHA, it is recommended that predictions from the All, NR, and the corresponding fault-type-specific models be considered to better capture the range of epistemic uncertainty associated with these events.

\subsection{Number of Overlapping Ruptures}

The number of overlapping ruptures denotes the number of principal rupture strands present at a given along-strike location and determines the number of individual contributions comprising the aggregate principal displacement. The distribution of overlapping rupture strands for different faulting styles is presented in Figure S1 of the electronic supplement.

\begin{figure}[htbp]
    \centering
    \includegraphics[width=.8\linewidth]{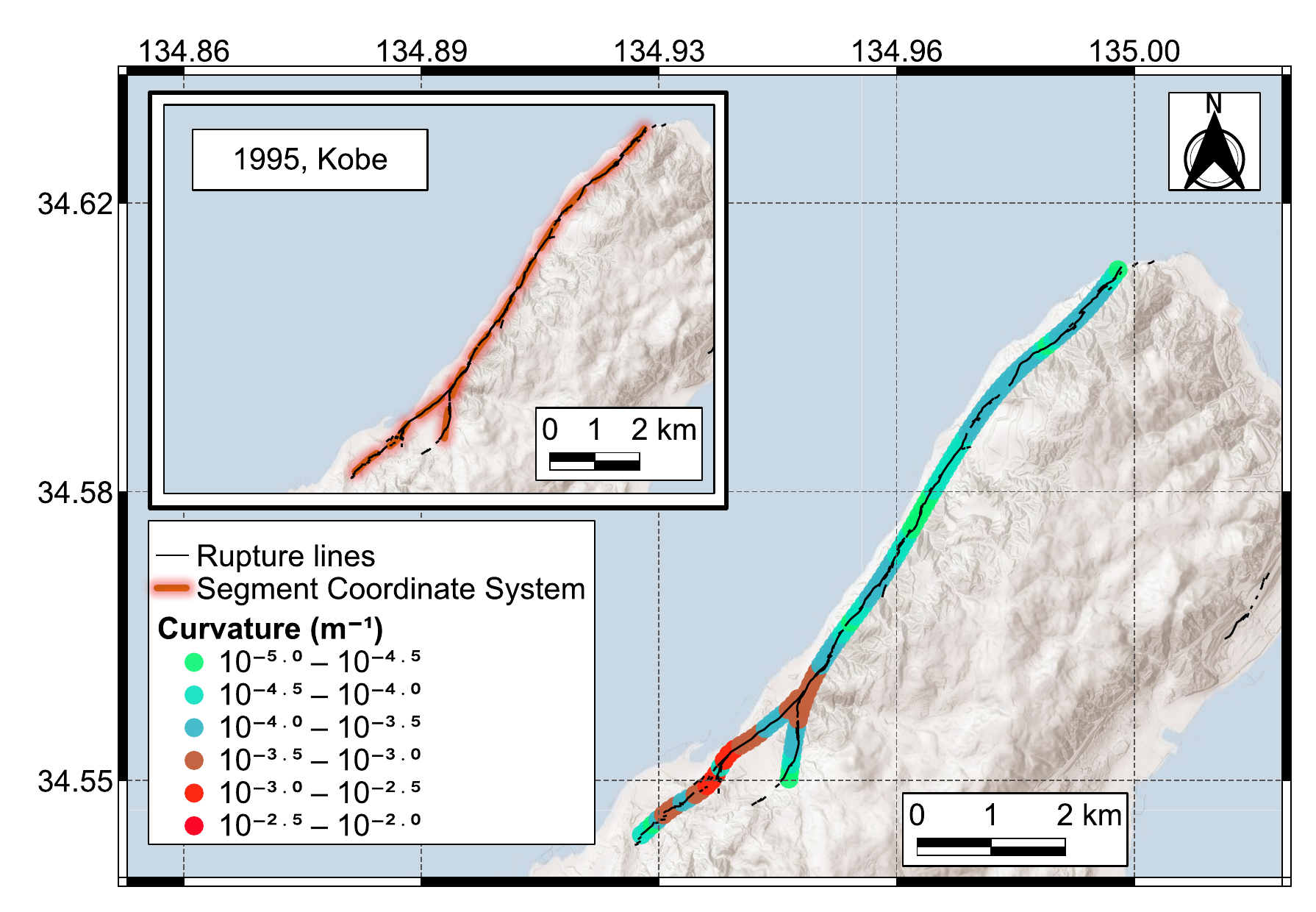}
    \caption{Variation of fault curvature along the segment reference coordinate system of the 1995 Kobe, Japan earthquake ($M_w = 6.9$). Colors denote ranges of computed curvature values, where higher curvature corresponds to greater changes in fault-trace direction.}
    \label{fig:curvature}
\end{figure}

The influence of earthquake magnitude and rupture geometry on the number of overlapping principal ruptures was evaluated empirically and assessed for physical consistency. Rupture geometry is characterized through fault curvature, which quantifies the degree of bending or directional change along the fault trace. The fault curvature is defined as:

\begin{equation}
\label{eq:fault_curvature}
\kappa(u) =
\sqrt{
\left(\frac{\partial^2x}{\partial u^2}\right)^2 +
\left(\frac{\partial^2y}{\partial u^2}\right)^2
},
\end{equation}

where $\partial^2x/\partial u^2$ and $\partial^2y/\partial u^2$ denote the second patial derivatives of the fault-trace coordinates with respect to $u$. The derivatives are estimated numerically using central finite-difference approximations. Larger curvature values correspond to greater bending of the fault trace. Figure~\ref{fig:curvature} illustrates the variation of curvature along the segment trace of the 1995 $M_w$ 6.9 Kobe, Japan earthquake.

\begin{figure}[htbp]
    \centering

    \begin{subfigure}[b]{0.48\linewidth}
        \centering
        \caption{}
        \includegraphics[width=\linewidth]{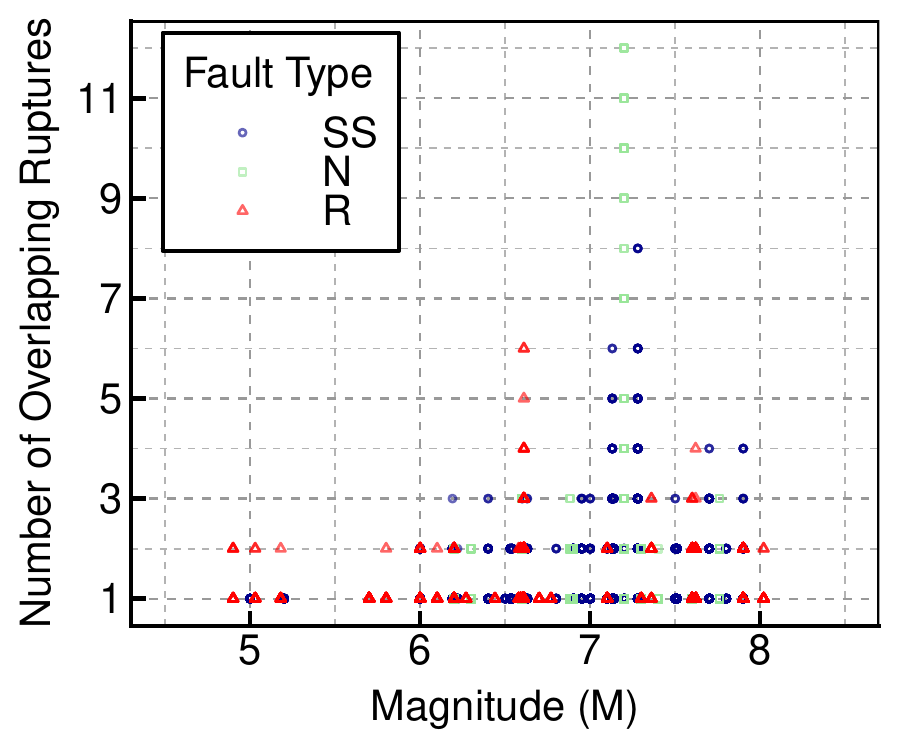}
        \label{fig:nprup_magnitude}
    \end{subfigure}
    \hfill
    \begin{subfigure}[b]{0.48\linewidth}
        \centering
        \caption{}
        \includegraphics[width=\linewidth]{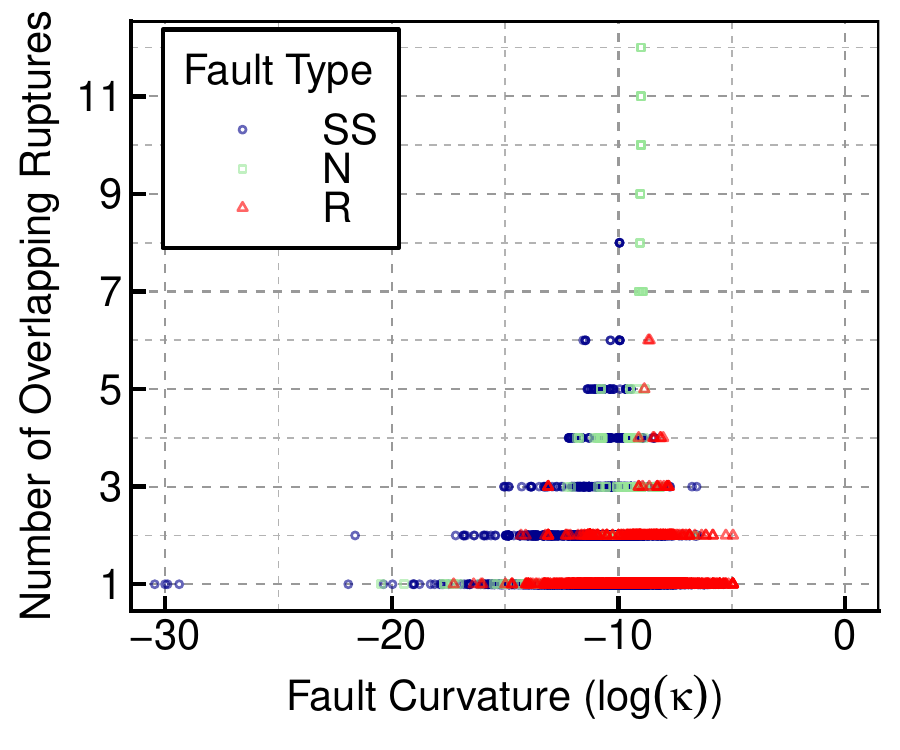}
        \label{fig:nprup_curvature}
    \end{subfigure}

    \caption{Influence of earthquake magnitude and fault curvature on the number of overlapping principal ruptures for strike-slip (SS), normal (N), and reverse (R) faulting styles. (a) illustrates the variation in the number of overlapping principal ruptures with earthquake magnitude ($M$), and (b) shows the corresponding variation with the logarithm of fault curvature ($\log(\kappa)$).}
    \label{fig:nprup_combined_mag_and_curv}
\end{figure}

Figure~\ref{fig:nprup_magnitude} shows that the number of overlapping rupture strands generally increases with earthquake magnitude. This trend is consistent with the expectation that larger earthquakes produce broader rupture zones, increasing the probability of multiple overlapping rupture strands at a given location.

A similar trend is observed with fault curvature. Figure~\ref{fig:nprup_curvature} indicates that locations with higher curvature tend to exhibit a larger number of overlapping rupture strands. This observation is consistent with the interpretation of curvature as a proxy for fault complexity, where geometrically complex ruptures are more likely to distribute deformation across multiple rupture strands rather than localizing slip on a single continuous fault trace.

Additional explanatory variables, including normalized location along the fault and perpendicular distance from the fault trace, were also evaluated. The corresponding relationships are presented in Figure S2 of the electronic supplement. Compared with magnitude and fault curvature, these variables exhibited substantially weaker explanatory power and were therefore excluded from the final model formulation based on both visual assessment and preliminary statistical analyses.

Following the visual distribution of the data, which characterized by a large proportion of single-rupture cases alongside a non-negligible frequency of multiple overlapping ruptures, a Hurdle model was adopted. In this formulation, a Bernoulli distribution is used to model the probability of observing a single rupture strand versus multiple overlapping rupture strands, while a Poisson distribution models the number of rupture strands conditional on the occurrence of a multi-rupture case, as defined in Equations~(\ref{eq:Hurdle_distribution(1)})--(\ref{eq:Hurdle_distribution(2)}). Earthquake magnitude ($M$) and fault curvature ($\kappa$) are incorporated as predictors in both components of the model, consistent with the empirical trends with the formulation given in Equation~(\ref{eq:Hurdle_distribution(3)}). To ensure physically meaningful behavior, the scaling coefficients associated with both magnitude and fault curvature were constrained to be positive.

\begin{equation} \label{eq:Hurdle_distribution(1)}
P(N_{p,~\mathrm{tot}} = j) =
\begin{cases}
P_1,                            & N_{p,~\mathrm{tot}} = 1, \\[6pt]
(1 - P_1)\, f_{+}(N_{p~\mathrm{tot}}),  & N_{p,~\mathrm{tot}} > 1
\end{cases}
\end{equation}

\begin{equation} \label{eq:Hurdle_distribution(2)}
\begin{aligned}
P_1 &= \frac{e^{\theta}}{1 + e^{\theta}} = \frac{1}{1 + e^{-\theta}}
    \qquad \text{(Bernoulli model)} \\[6pt]
f_{+}(N_{p,~\mathrm{tot}}) &= \frac{\lambda^{N_{p,~\mathrm{tot}}} e^{-\lambda}}
               {N_{p,~\mathrm{tot}}! \bigl(1 - P(N_{p,~\mathrm{tot}} \le 1)\bigr)}
    \qquad \text{(Poisson)}
\end{aligned}
\end{equation}

\begin{equation} \label{eq:Hurdle_distribution(3)}
\begin{aligned}
\lambda &= \exp\!\left(a_0 + a_1 M + a_2 \log({\kappa})\right) \\[6pt]
\theta  &= b_0 + b_1 M + b_2 \log({\kappa}) \\[6pt]
\text{where} \;& a_1 > 0,\; a_2 > 0,\qquad 
                b_1 < 0,\; b_2 < 0
\end{aligned}
\end{equation}

The preferred model was estimated based on the by maximizing log-likelihood with respect to the parameters, given the imposed scaling constraints.
To reduce the risk of overfitting, the coefficients associated with magnitude and curvature ($a_1$, $a_2$, $b_1$, and $b_2$) are assumed to be invariant across faulting styles. In addition, an event-level weighting factor, defined in Equation~(\ref{eq:Weight_factor}), is included to balance the contribution of individual earthquakes as a function of the number of observations ($N_i$). This weighting scheme was necessary because the level of detail of rupture mapping varies substantially among events. The event-level weighting factor prevents densely sampled earthquakes from disproportionately influencing the final model.

\begin{equation} \label{eq:Weight_factor}
w_{e,~i} = \frac{1}{N_{e}}
\end{equation}

The calibrated coefficients of the hurdle model for the number of overlapping rupture strands are summarized in Table~\ref{tab:hurdle_coefficients}.

\begin{table}[htbp]
\centering
\caption{Estimated coefficients for the model governing the number of overlapping principal ruptures.}
\label{tab:hurdle_coefficients}

\begin{tabular}{c|cccccc}
\hline
SOF & $a_0$ & $a_1$ & $a_2$ & $b_0$ & $b_1$ & $b_2$ \\
\hline

All & $-4.59$
    & \multirow{5}{*}{$0.77$}
    & \multirow{5}{*}{$0.16$}
    & $2.74$
    & \multirow{5}{*}{$-0.27$}
    & \multirow{5}{*}{$-0.11$} \\

SS  & $-4.13$
    & & & $2.13$
    & & \\

NR  & $-5.42$
    & & & $3.09$
    & & \\

N   & $-5.30$
    & & & $3.03$
    & & \\

R   & $-12.14$
    & & & $3.47$
    & & \\

\hline
\end{tabular}

\vspace{2mm}
\small SOF: style of faulting; SS: strike-slip; NR: normal-reverse; N: normal; R: reverse.

\end{table}

Standard errors are commonly used to quantify the uncertainty associated with parameter estimates. However, several parameters in the model ($a_1$, $a_2$, $b_1$, and $b_2$) were estimated under explicit constraints, which may violate the assumptions underlying Hessian-based (Wald) standard errors and confidence intervals. Therefore, parameter uncertainty was evaluated using profile likelihood confidence intervals based on Wilks' likelihood-ratio theorem \citep{wilks1938large}.

For each parameter $\theta_i$, a sequence of fixed values around the optimum estimate was considered while all remaining parameters were re-estimated. The resulting profile negative log-likelihood, $NLL_P(\theta_i)$, was then used to quantify the increase in negative log-likelihood relative to the optimum solution, $NLL_{\min}$, as defined below:

\begin{equation} \label{eq:negative_fixed_NLL}
NLL_P(\theta_i)
=
\min_{\theta_{-i}}
\left[-\log L(\theta_i,\theta_{-i})\right]
\end{equation}

\begin{equation} \label{eq:deltaNLL}
\Delta NLL(\theta_i) = NLL_P(\theta_i) - NLL_{min} 
\end{equation}

According to Wilks' likelihood-ratio theorem \citep{wilks1938large}, the likelihood-ratio statistic asymptotically follows a Chi-square distribution with one degree of freedom when a single parameter is fixed and all remaining parameters are re-estimated:

\begin{equation} \label{eq:chisquare}
2\,\Delta NLL(\theta_i)
\sim
\chi^2_1
\end{equation}

The 95\% confidence interval is therefore defined as:

\begin{equation} \label{eq:95percent_interval}
\Delta NLL(\theta_i)
\le
\frac{\chi^2_{1,0.95}}{2}
=
1.92
\end{equation}

This procedure was repeated for all model parameters, and the resulting lower and upper confidence bounds are summarized in Table \ref{tab:hurdle_profilelikelihood}.

\begin{table}[htbp]
\centering
\caption{Lower and upper bounds of the 95\% profile-likelihood confidence intervals for the hurdle-model parameters describing the number of overlapping principal ruptures for different faulting styles.}
\label{tab:hurdle_profilelikelihood}

\begin{tabular}{c|cccccc}
\hline
SOF & $a_0$ & $a_1$ & $a_2$ & $b_0$ & $b_1$ & $b_2$ \\
\hline

All & $[-17.39,\;\phantom{-}0.79]$
    & \multirow{5}{*}{$[0,\;2.20]$}
    & \multirow{5}{*}{$[0,\;0.79]$}
    & $[-0.31,\;6.47]$
    & \multirow{5}{*}{$[-0.90,\;0]$}
    & \multirow{5}{*}{$[-0.36,\;0]$} \\

SS  & $[-17.31,\;\phantom{-}0.71]$
    & & & $[-1.28,\;6.07]$
    & & \\

NR  & $[-16.03,\;\phantom{-}0.82]$
    & & & $[0.03\phantom{-},\;6.79]$
    & & \\

N   & $[-16.21,\;\phantom{-}0.48]$
    & & & $[-0.09,\;6.79]$
    & & \\

R   & $[-21.21,\;-0.12]$
    & & & $[-0.25,\;8.45]$
    & & \\

\hline
\end{tabular}

\vspace{2mm}
\small SOF: style of faulting; SS: strike-slip; NR: normal-reverse; N: normal; R: reverse.

\end{table}

\begin{figure}[htbp]
    \centering

    \begin{subfigure}[b]{0.47\linewidth}
        \centering
        \caption{}
        \includegraphics[width=\linewidth]{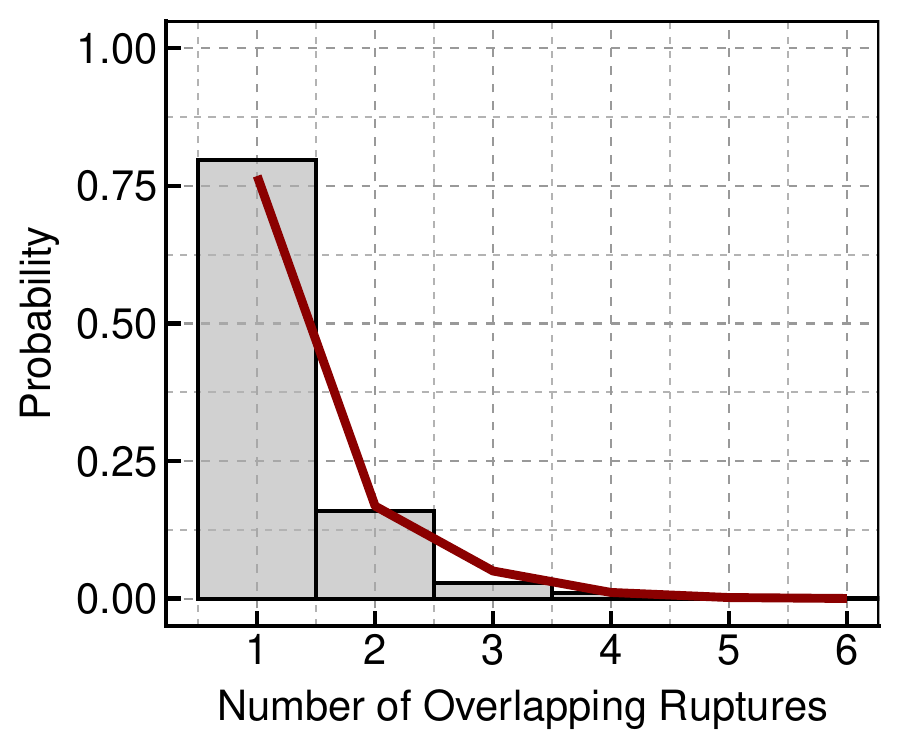}
        \label{fig:Fitted_Hurdle_SS-M2-C2}
    \end{subfigure}
    \hfill
    \begin{subfigure}[b]{0.47\linewidth}
        \centering
        \caption{}
        \includegraphics[width=\linewidth]{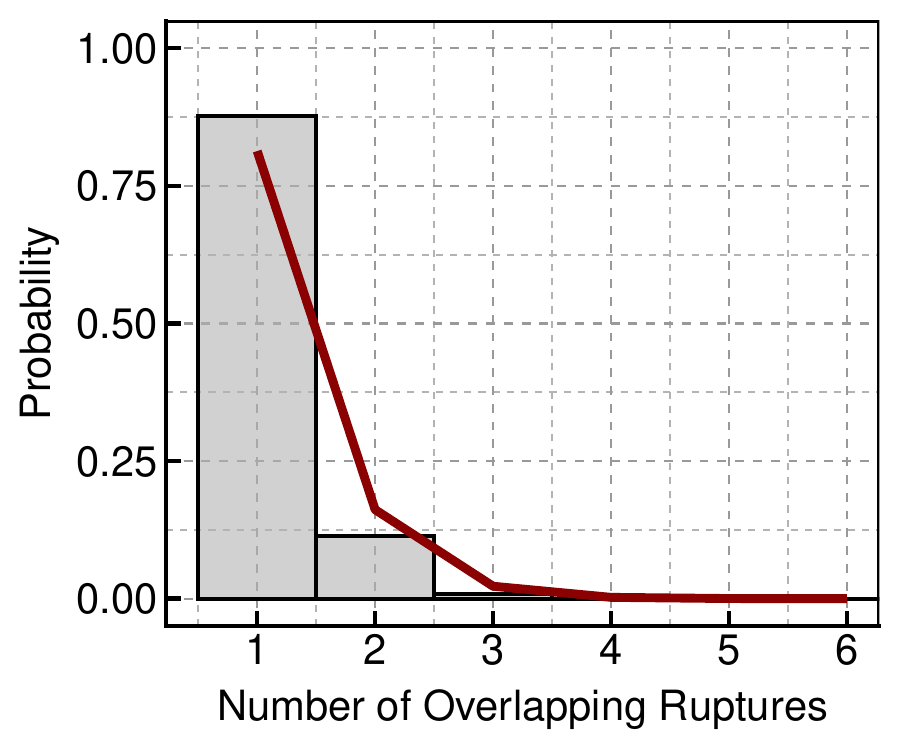}
        \label{fig:Fitted_Hurdle_SS-M1-C2}
    \end{subfigure}
    \begin{subfigure}[b]{0.47\linewidth}
        \centering
        \caption{}
        \includegraphics[width=\linewidth]{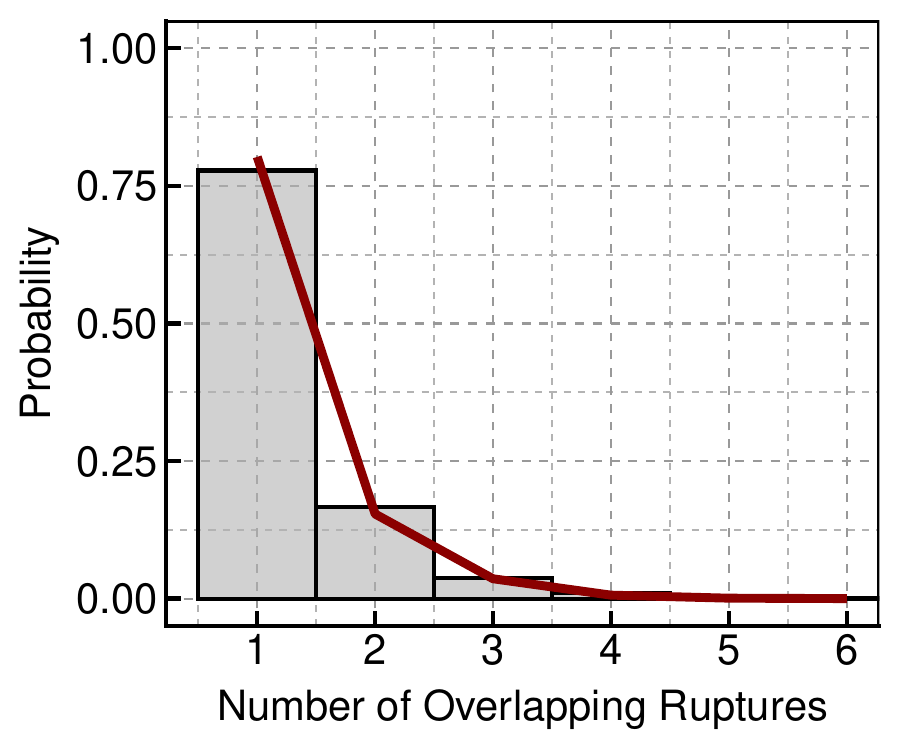}
        \label{fig:Fitted_Hurdle_SS-M2-C1}
    \end{subfigure}
    \begin{subfigure}[b]{0.47\linewidth}
        \centering
        \caption{}
        \includegraphics[width=\linewidth]{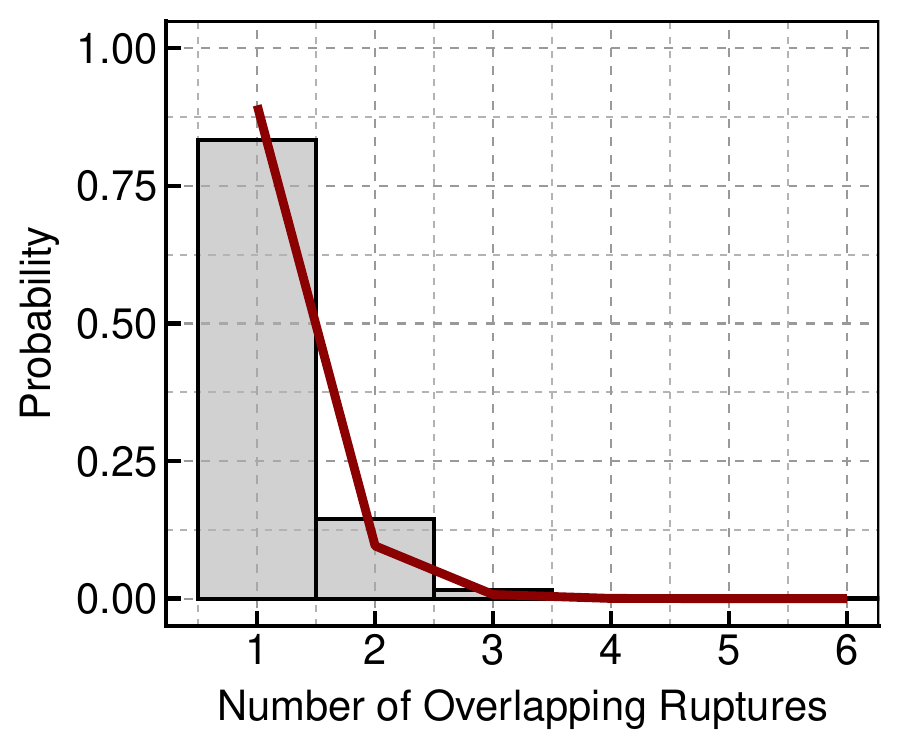}
        \label{fig:Fitted_Hurdle_NR-M2-C2}
    \end{subfigure}

    \caption{Comparison of the observed and modeled distributions of the number of overlapping principal ruptures for different faulting styles and magnitude ($M$) - curvature ($\log(\kappa)$) bins. Histograms represent the empirical distributions, and red curves denote the probability mass functions predicted by the fitted hurdle distribution. (a) strike-slip with $6.5 < M$ and $-10 < \log(\kappa)$, (b) strike-slip with $M < 6.5$ and $-10 < \log(\kappa)$, (c) strike-slip with $6.5 < M$ and $\log(\kappa) < -10$, and (d) normal and reverse with $6.5 < M$ and $-10 < \log(\kappa)$.}
    \label{fig:Fitted_Hurdle}
\end{figure}

The fitted distribution for different fault styles and different magnitude–curvature bins is shown in Figure \ref{fig:Fitted_Hurdle}. In each panel, the histogram represents the observed distribution of the number of overlapping principal ruptures, and the red curve represents the probability mass function predicted by the fitted hurdle model. The model predictions were evaluated using the mean earthquake magnitude and mean fault curvature of the corresponding bin. The consistency between the observed and predicted distributions indicates that the hurdle model adequately captures both the high frequency of single-rupture observations and the decreasing probability of larger numbers of overlapping rupture strands for a range fo magnitudes, curvature, and style of faulting. 

\begin{figure}[htbp]
    \centering

    \begin{subfigure}[b]{0.48\linewidth}
        \centering
        \caption{}
        \includegraphics[width=\linewidth]{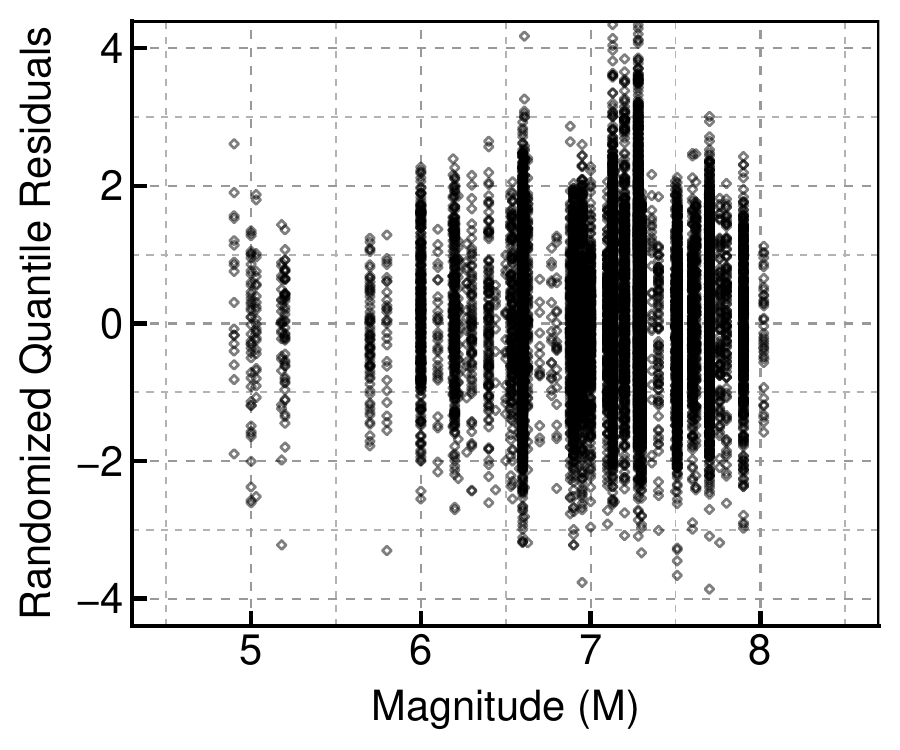}
        \label{fig:nprup_RQR_mag}
    \end{subfigure}
    \hfill
    \begin{subfigure}[b]{0.48\linewidth}
        \centering
        \caption{}
        \includegraphics[width=\linewidth]{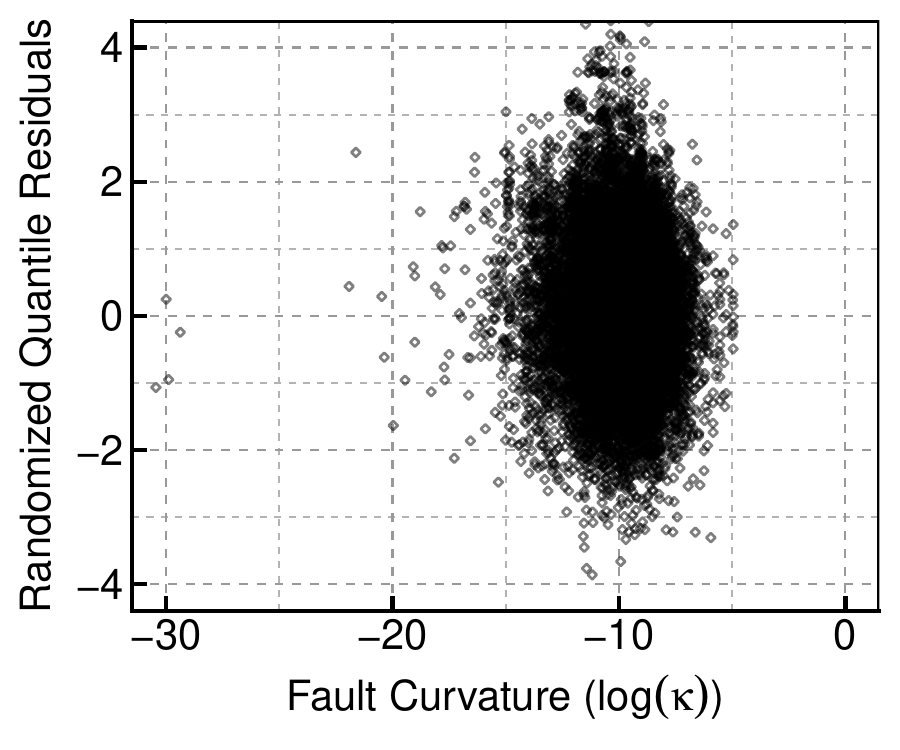}
        \label{fig:nprup_RQR_curv}
    \end{subfigure}

    \caption{Assessment of residual trends in the fitted hurdle model for the number of overlapping principal ruptures using randomized quantile residuals plotted as functions of (a) earthquake magnitude ($M$) and (b) fault curvature ($\log(\kappa)$).}
    \label{fig:nprup_RQR}
\end{figure}

The preferred model was further evaluated using randomized quantile residuals. These residuals were examined to verify that the effects of magnitude (Figure \ref{fig:nprup_RQR_mag}) and curvature (Figure \ref{fig:nprup_RQR_curv}) had been adequately captured. As shown, no remaining systematic dependence on either variable is evident. Corresponding plots for all fault styles are provided in the electronic supplement (Figure S3). Residuals were also examined against normalized location along the fault and normal distance to the fault to confirm that these excluded variables did not introduce any remaining systematic trends; the corresponding plots are provided in the electronic supplement (Figure S4).

\subsection{Amplitude ratio of displacement}

The displacement ratio describes how the aggregate displacement is partitioned among overlapping rupture strands. This allocation is physically bounded between zero and one and must satisfy the constraint that the contributions sum to unity, making the Beta distribution a suitable choice.

Initial formulations considered independent models for each combination of fault style and number of overlapping ruptures ($N_{p,~\mathrm{tot}}$), reflecting the dependence of displacement partitioning on the number of rupture strands. However, these formulations were found to be unstable due to limited data availability, especially for larger values of $N_{p,~\mathrm{tot}}$. To address this, a unified model was proposed for each fault style, which incorporated $N_{p,~\mathrm{tot}}$ directly as a parameter, allowing its influence on partitioning to be preserved without requiring separate models. Under this formulation, the mean of the Beta distribution is constrained to $1/N_{p,~\mathrm{tot}}$, so that the expectation of the displacement ratio across all overlapping ruptures is equal to unity:
\begin{equation}
    \E\left[\sum_{i=1}^{N_{p,~\mathrm{tot}}} r_{p,~i} \right] = \sum_{i=1}^{N_{p,~\mathrm{tot}}} \left( \E[r_i] \right) = N_{p,~\mathrm{tot}} \times \E[r_{p,~i}] = 1.0
\end{equation}
where $r_{p,i}$ is the displacement ratio of the $i^th$ parallel strand.

The probability density function of a Beta distribution is defined as:

\begin{equation} \label{eq:Beta_distribution}
f(r;\alpha,\beta) =
\begin{cases}
\dfrac{r^{\alpha-1}(1-r)^{\beta-1}}{B(\alpha,\beta)}, & 0 < r < 1, \\[10pt]
0, & \text{otherwise},
\end{cases}
\end{equation}

\noindent where $\alpha$ and $\beta$ are the parameters that control the shape of the distribution, while the denominator of the density function is defined as: 
\begin{equation}
    B(\alpha, \beta) = \frac{\Gamma(\alpha) \Gamma(\beta)}{\Gamma(\alpha + \beta)}
\end{equation}
where $\Gamma(\cdot)$ is the Gamma function.

The mean is equal to: 
\begin{equation}
    \mu = \frac{\alpha}{\alpha + \beta} = \frac{1}{N_{p,~\mathrm{tot}}}
\end{equation}
which is used to enforce the constraint. 
This constraint reduces the model to a single shape parameter $\alpha$, with $\beta = \alpha (N_{p,~\mathrm{tot}} - 1)$. The shape parameter $\alpha$ exhibited a linear increase with earthquake magnitude for events with $M > 5$. Consequently, $\alpha$ is represented as:

\begin{align}
    \alpha = \max\left(d_0 + d_1(M-5),\,0\right)
\end{align}

To maintain model stability and avoid overfitting, the intercept term ($d_0$) is assumed constant while the slope ($d_1$) varies across fault styles to capture differences in partitioning behavior. The event-level weighting factor defined in Equation~\ref{eq:Weight_factor} is also incorporated to balance the contribution of individual events. The model parameters were estimated using maximum likelihood estimation. The corresponding model coefficients are summarized in Table~\ref{tab:beta_distribution_coefficients}.

\begin{table}[htbp]
\centering
\caption{Estimated coefficients for the model governing the amplitude ratio of the principal displacement.}
\label{tab:beta_distribution_coefficients}

\begin{tabular}{c|cc}
\hline
SOF & $d_0$ & $d_1$ \\
\hline

All & \multirow{5}{*}{1.00}
    & 0.035 \\

SS  & & 0.006 \\

NR  & & 0.081 \\

N   & & 0.067 \\

R   & & 0.137 \\

\hline
\end{tabular}

\vspace{2mm}
\small SOF: style of faulting; SS: strike-slip; NR: normal-reverse; N: normal; R: reverse.

\end{table}

The uncertainty associated with parameter estimation was evaluated using 95\% confidence intervals obtained from the profile likelihood approach described in Equations~(\ref{eq:negative_fixed_NLL})-(\ref{eq:95percent_interval}). The resulting lower and upper confidence bounds are summarized in Table~\ref{tab:beta_profilelikelihood}.

\begin{table}[htbp]
\centering
\caption{Lower and upper bounds of the 95\% profile-likelihood confidence intervals for Beta distribution parameters describing the amplitude ratio of principal displacement for different faulting styles.}
\label{tab:beta_profilelikelihood}

\begin{tabular}{c|cc}
\hline
SOF & $d_0$ & $d_1$ \\
\hline

All & \multirow{5}{*}{$[1.00, 1.45]$}
    & $[0.032, 0.356]$ \\

SS  & & $[0.004, 0.462]$ \\

NR  & & $[0.065, 0.712]$ \\

N   & & $[0.065, 0.776]$ \\

R   & & $[0.130, 2.070]$ \\

\hline
\end{tabular}

\vspace{2mm}
\small SOF: style of faulting; SS: strike-slip; NR: normal-reverse; N: normal; R: reverse.

\end{table}

\begin{figure}[htbp]
    \centering

    \begin{subfigure}[b]{0.48\linewidth}
        \centering
        \caption{}
        \includegraphics[width=\linewidth]{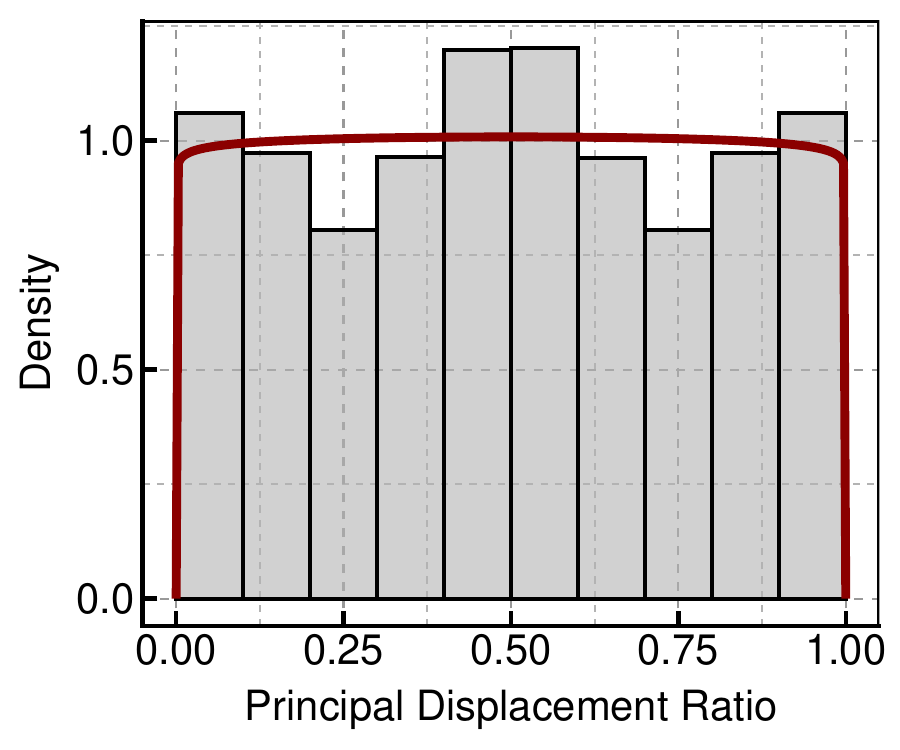}
        \label{fig:ratio_SS-N2}
    \end{subfigure}
    \hfill
    \begin{subfigure}[b]{0.48\linewidth}
        \centering
        \caption{}
        \includegraphics[width=\linewidth]{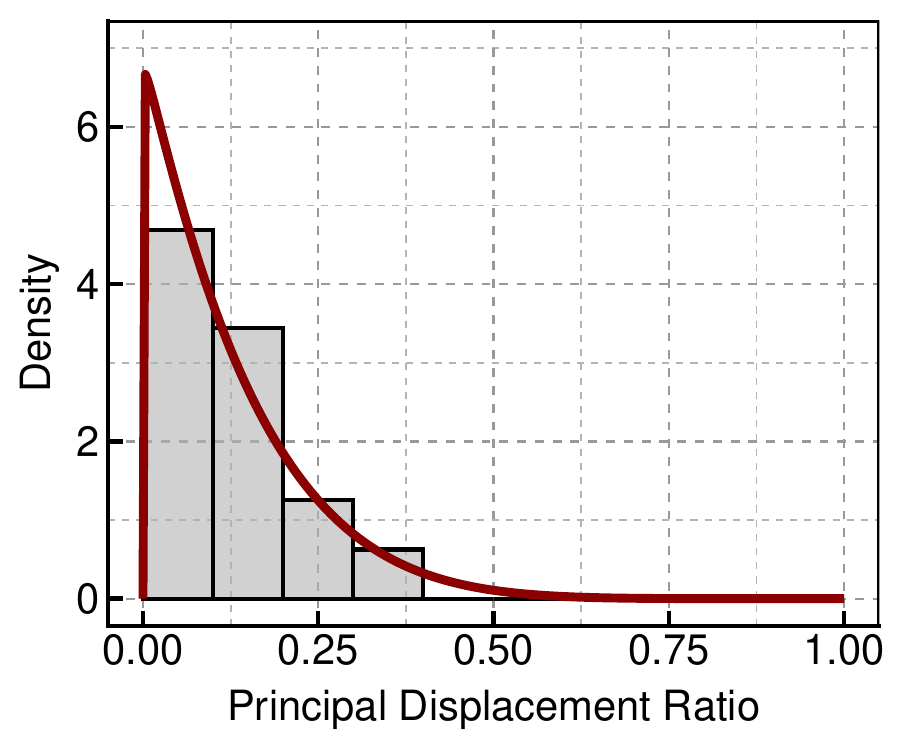}
        \label{fig:ratio_SS-N8}
    \end{subfigure}
    \begin{subfigure}[c]{0.48\linewidth}
        \centering
        \caption{}
        \includegraphics[width=\linewidth]{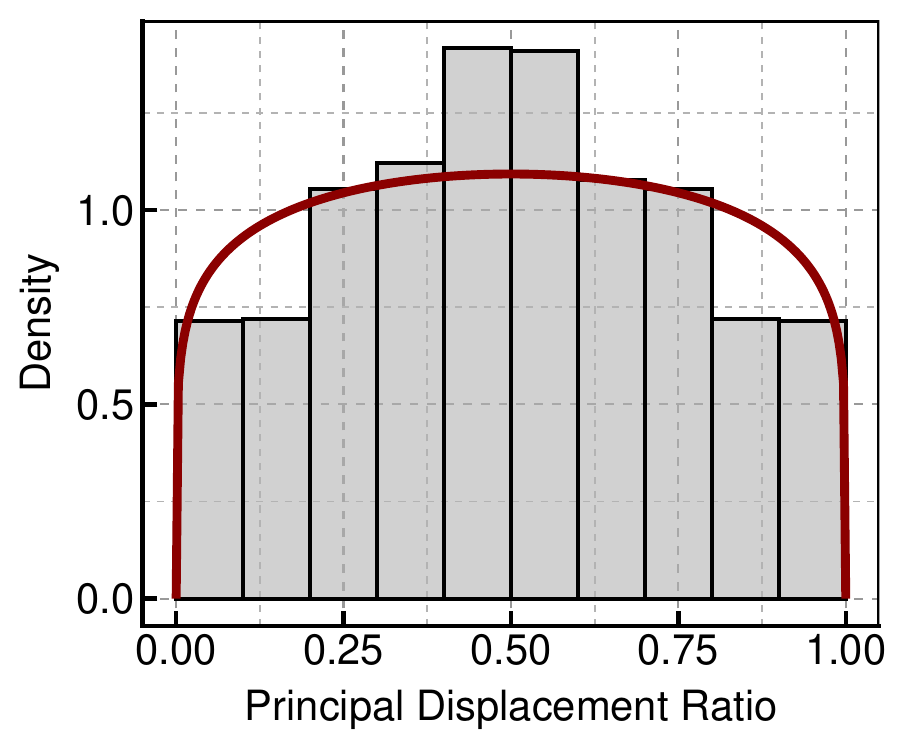}
        \label{fig:ratio_NR-N2}
    \end{subfigure}
    \begin{subfigure}[c]{0.48\linewidth}
        \centering
        \caption{}
        \includegraphics[width=\linewidth]{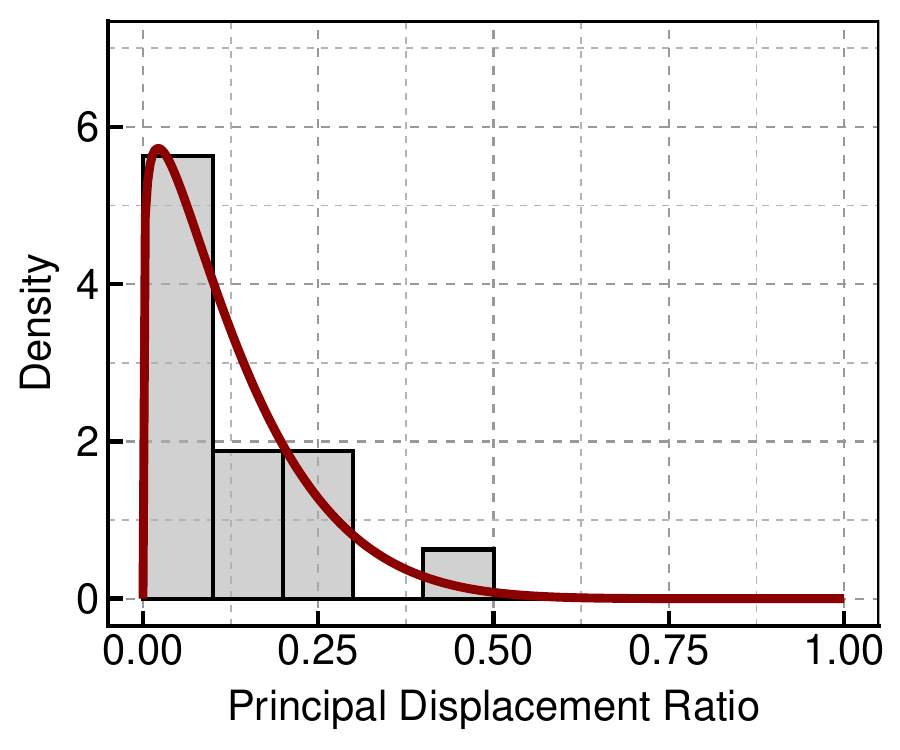}
        \label{fig:ratio_NR-N8}
    \end{subfigure}

    \caption{Comparison of the observed and modeled distributions of the principal displacement amplitude ratio for different faulting styles and numbers of overlapping ruptures ($N_{p,~\mathrm{tot}}$). Histograms represent the empirical distributions, and red curves denote the fitted Beta distribution. (a) strike-slip with $N_{p,~\mathrm{tot}} = 2$, (b) strike-slip with $N_{p,~\mathrm{tot}} = 8$, (c) normal and reverse with $N_{p,~\mathrm{tot}} = 2$, and (d) normal and reverse with $N_{p,~\mathrm{tot}} = 8$.}
    \label{fig:Fitted_Beta_distribution}
\end{figure}

Figure~\ref{fig:Fitted_Beta_distribution} compares the observed distributions of the principal displacement amplitude ratio with those predicted by the fitted Beta distribution for different numbers of overlapping rupture. The histograms represent the empirical distributions, while the solid lines denote the fitted Beta distributions evaluated at the mean earthquake magnitude corresponding to each rupture-count category. The overall agreement between the observed and fitted distributions indicates that the proposed model adequately captures the displacement partitioning behavior for different numbers of overlapping ruptures and fault styles.


\begin{figure}[htbp]
    \centering

    \begin{subfigure}[b]{0.48\linewidth}
        \centering
        \caption{}
        \includegraphics[width=\linewidth]{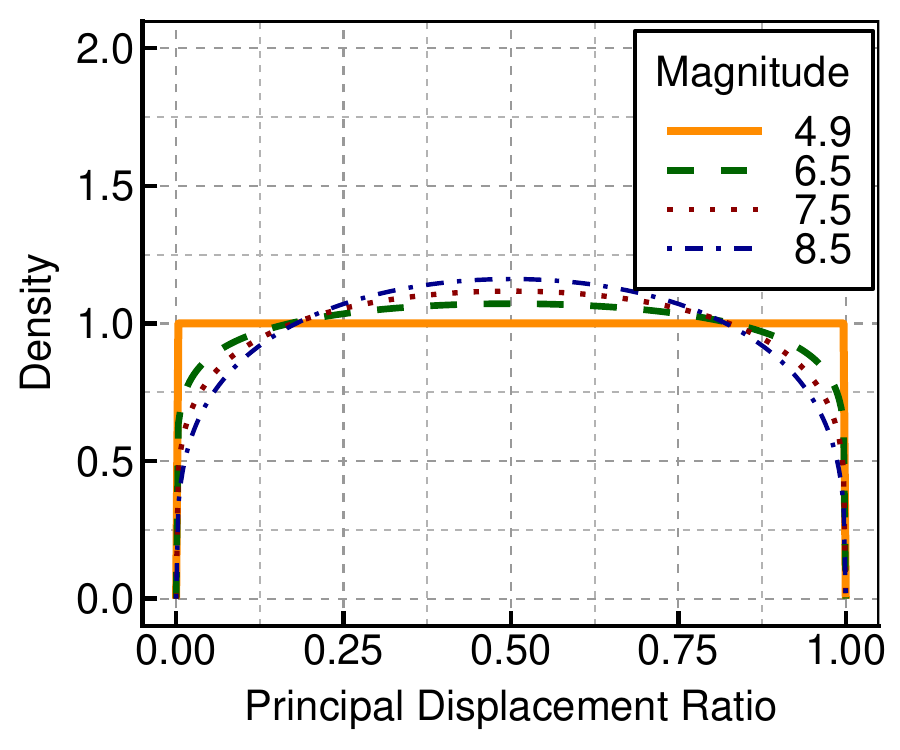}
        \label{fig:ratio_NR_n2}
    \end{subfigure}
    \hfill
    \begin{subfigure}[b]{0.48\linewidth}
        \centering
        \caption{}
        \includegraphics[width=\linewidth]{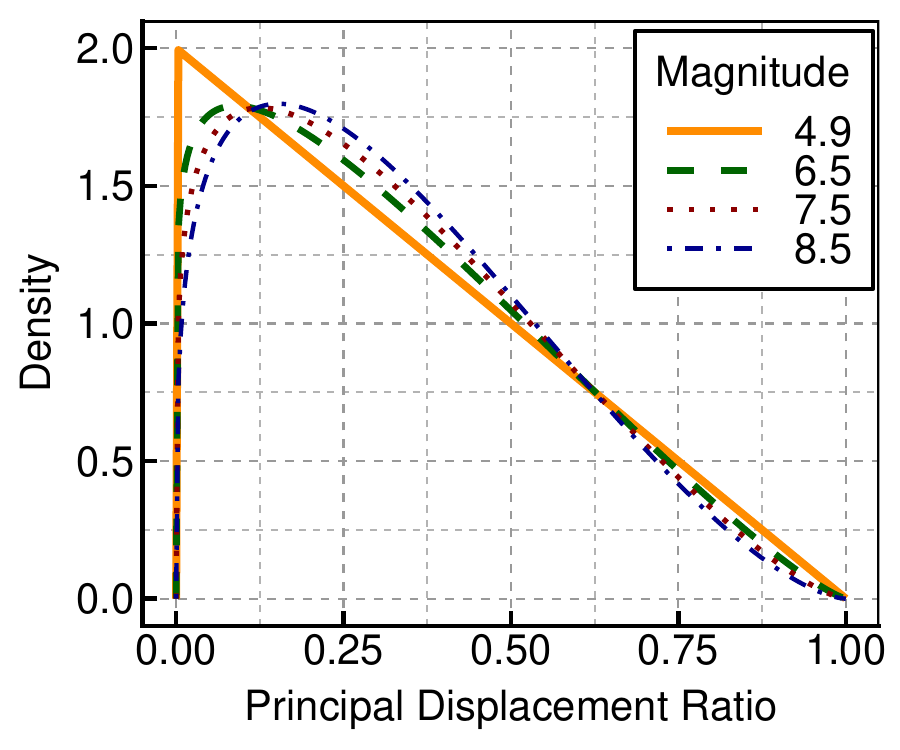}
        \label{fig:ratio_NR_n3}
    \end{subfigure}

    \caption{Influence of the number of overlapping ruptures ($N_{p,~\mathrm{tot}}$) and earthquake magnitude ($M$) on the modeled distribution of the principal displacement ratio. (a) normal and reverse with $N_{p,~\mathrm{tot}} = 2$ and (b) normal and reverse with $N_{p,~\mathrm{tot}} = 3$.}
    \label{fig:ratio_mag}
\end{figure}

The number of rupture strands directly controls the mean displacement ratio, with larger values of $N_{p,~\mathrm{tot}}$ shifting the distribution toward smaller ratios. This behavior reflects the partitioning of the aggregate displacement among a greater number of principal ruptures. Magnitude influences the distribution through its effect on $\alpha$ and, indirectly, on $\beta$, although its overall impact is comparatively modest. As earthquake magnitude increases, the variance of displacement ratios reduces and tend to converge toward the mean value, indicating a more equal partitioning of displacement among individual principal ruptures. This behavior is illustrated in Figure~\ref{fig:ratio_mag}, where the number of rupture strands primarily controls the overall shape of the distribution, while magnitude governs its variability.

The extension of the displacement-ratio model from a single rupture strand to multiple overlapping rupture strands crossed is achieved through the use of a Dirichlet distribution. The Dirichlet distribution is a multivariate probability distribution defined on a simplex, where each component is bounded between zero and one and the components collectively satisfy the constraint that their sum is equal to one ($\sum_{i=1}^{N_{p,~\mathrm{tot}}} r_{p,~i} = 1.0$).
These properties make the Dirichlet distribution particularly well suited for modeling the partitioning of aggregate principal displacement among multiple rupture strands.

Since in the proposed formulation, the displacement ratio associated with each rupture strand is assumed to follow an marginal Beta distribution,
\[
r_{p,~i} \sim \mathrm{Beta}\left(\alpha,
\beta\right),
\]
with the same parameters. 
For a configuration containing $N_{p,~\mathrm{tot}}$ rupture strands, the vector of displacement ratios is modeled as a symmetric Dirichlet distribution:
\begin{equation}
[r_1,r_2, \ldots, r_{N_{p,~\mathrm{cross}}} ,\ldots,r_{N_{p,~\mathrm{tot}}}]
\sim
\mathrm{Dir}
\left(
\alpha,\ldots,\alpha
\right),
\end{equation}
with the cumulative displacement ratio associated with the subset of  $N_{p,~\mathrm{tot}}$ rupture strands intersecting the structure,
\begin{equation}
r_{p,~\mathrm{cross}}= 
\sum_{i=1}^{N_{p,~\mathrm{cross}}}
r_{p,~i},
\end{equation}
also following a Beta distribution. Specifically,
\begin{equation}
r_{p,~\mathrm{cross}}
\sim
\mathrm{Beta}
\left(
N_{p,~\mathrm{cross}}~\alpha,
\left(
N_{p,~\mathrm{tot}}-N_{p,~\mathrm{cross}}
\right)~\alpha
\right).
\end{equation}
which provides an exact closed-form representation of the fraction of aggregate principal displacement associated with the rupture strands intersecting a structure

\subsection{Normal distance to the fault}

The final component of the proposed framework is the distribution of the distance of a principal rupture relative to a nominal fault trace, measured normal to the strike of the fault. This distribution is subsequently used to compute the probability that a rupture strand intersects the footprint of a structure.

Because the nominal rupture trace neither uniquely defined nor its location is generally known prior to an earthquake, three alternative formulations are proposed, allowing users to select the most appropriate representation based on the available information and application.

In the first formulation, the nominal rupture trace is defined using the Segment Coordinate System (SCS), representing the fault as a smooth line that captures the large-scale geometry of the rupture segment. In the second formulation, the nominal trace is represented by a randomly selected principal rupture strand among those overlapping a given along-strike location. In the third formulation, the rupture strand associated with the largest displacement at a given along-strike location is selected to define the nominal fault trace. These three alternatives represent different levels of confidence in the pre-earthquake characterization of the fault trace, ranging from a broad large-scale representation of fault geometry to a localized representation based on the dominant rupture strand.

Further research is needed to quantify the accuracy with which future surface rupture locations can be inferred from pre-earthquake geological and geomorphological observations.

Preliminary analysis of the dataset indicates that the spatial distribution of principal rupture strands is controlled primarily by earthquake magnitude and displacement amplitude. Larger-magnitude earthquakes tend to produce deformation across broader zones, whereas large displacement amplitudes are preferentially concentrated near the nominal fault trace. To capture the latter effect, a normalized displacement metric is introduced, defined as the ratio of the principal displacement to the mean aggregate principal displacement predicted for the same scenario and location by \cite{lavrentiadis2025fault}:

\begin{equation}
\hat{D}_p =
\frac{D_p}
{\overline{D}_{p,~\mathrm{agg}}}
\label{eq:normalized_displacement}
\end{equation}

\begin{figure}[htbp]
    \centering

    \begin{subfigure}[b]{0.48\linewidth}
        \centering
        \caption{}
        \includegraphics[width=\linewidth]{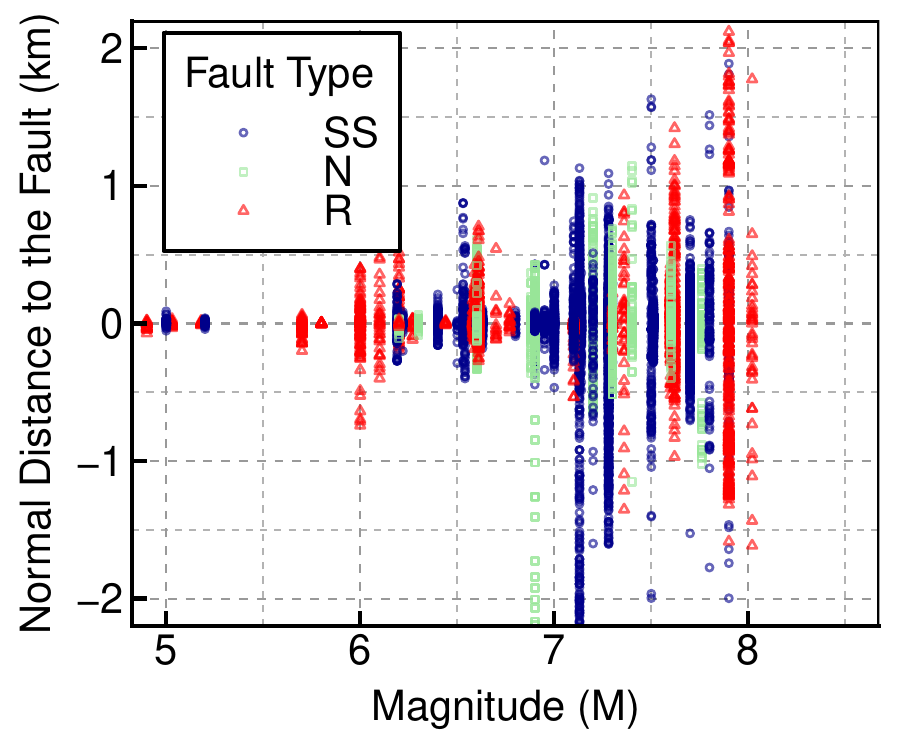}
    \end{subfigure}
    \hfill
    \begin{subfigure}[b]{0.48\linewidth}
        \centering
        \caption{}
        \includegraphics[width=\linewidth]{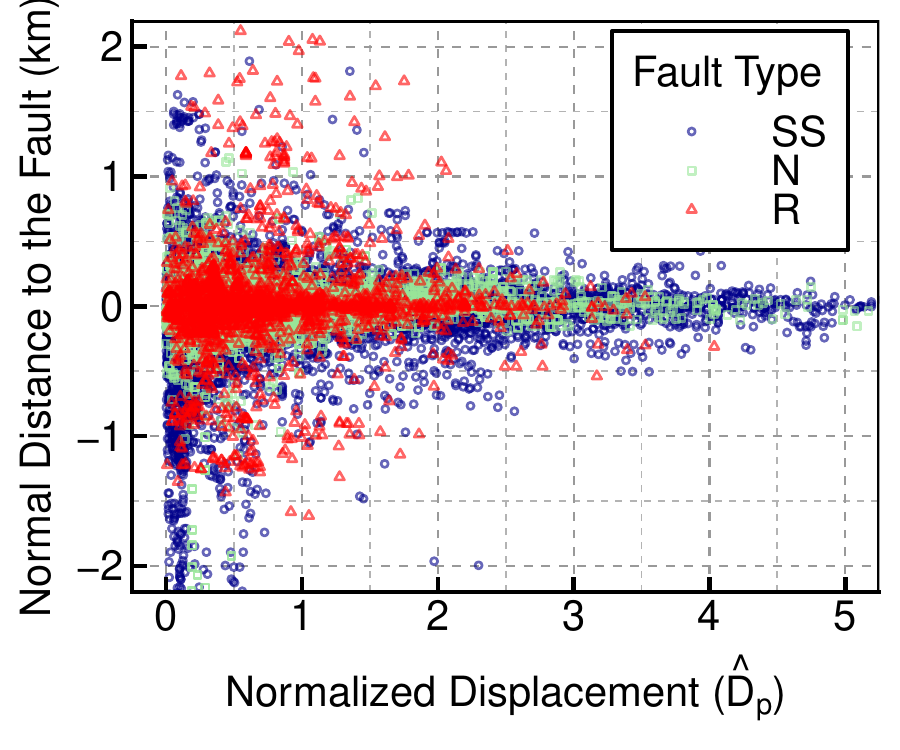}
    \end{subfigure}

    \vspace{0.3cm}

    \caption{Influence of earthquake magnitude and normalized principal displacement on the normal distance to SCS fault trace for strike-slip (SS), normal (N), and reverse (R) faulting styles. (a) illustrates the distribution of normal distance to fault trace as a function of earthquake magnitude ($M$) and (b) corresponding distribution as a function of normalized principal displacement $(\hat{D}_p)$.}
    
    \label{fig:dist_mag_norm_grid}
\end{figure}

Figure~\ref{fig:dist_mag_norm_grid} illustrates the dependence of normal distance to the SCS trace on earthquake magnitude and normalized displacement. The corresponding distributions for the alternative fault-trace definitions are provided in the electronic supplement (Figure S5). Together, earthquake magnitude and normalized displacement define the primary statistical structure of the distance distributions.

In addition to the previous terms, curvature was also examined as a potential controlling parameter. However, no consistent relationship was observed between curvature and the perpendicular distance of displacement from the fault across the dataset. The corresponding trends for each trace definition are provided in the electronic supplement (Figure S6).

\begin{figure}[htbp]
    \centering

    \begin{subfigure}[t]{0.32\linewidth}
        \centering
        \caption{}
        \includegraphics[width=\linewidth]{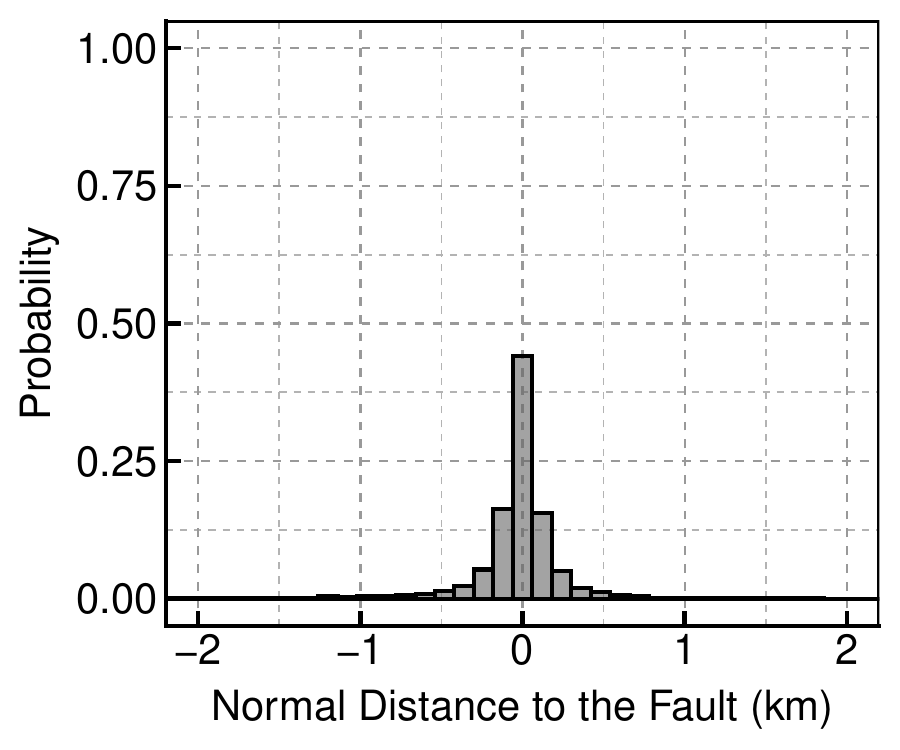}
        \label{fig:dist_hist_scs}
    \end{subfigure}
    \hfill
    \begin{subfigure}[t]{0.32\linewidth}
        \centering
        \caption{}
        \includegraphics[width=\linewidth]{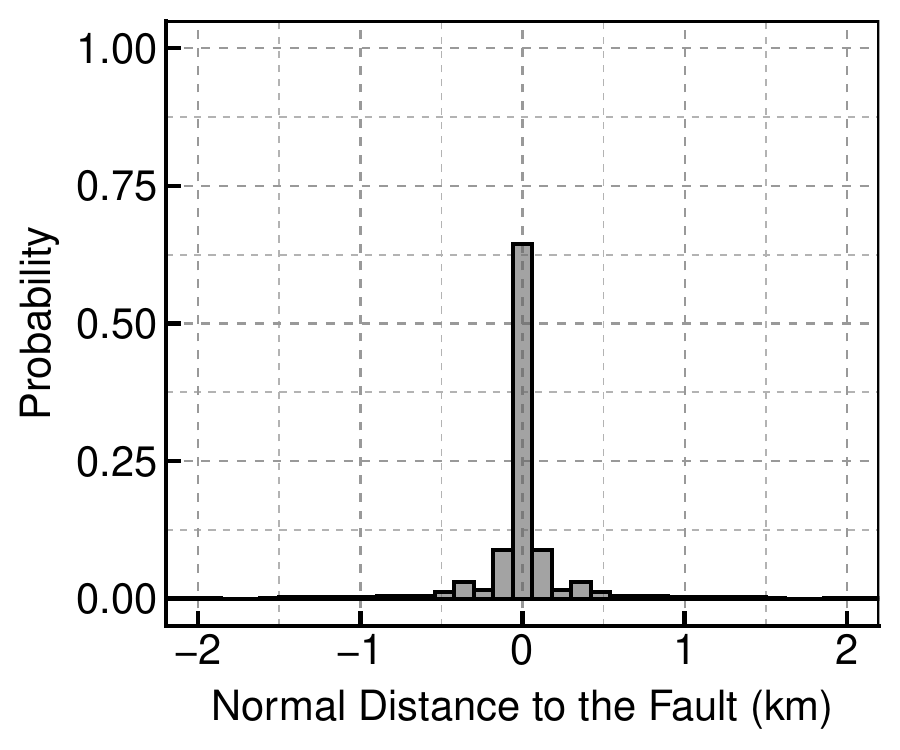}
        \label{fig:dist_hist_random}
    \end{subfigure}
    \hfill
    \begin{subfigure}[t]{0.32\linewidth}
        \centering
        \caption{}
        \includegraphics[width=\linewidth]{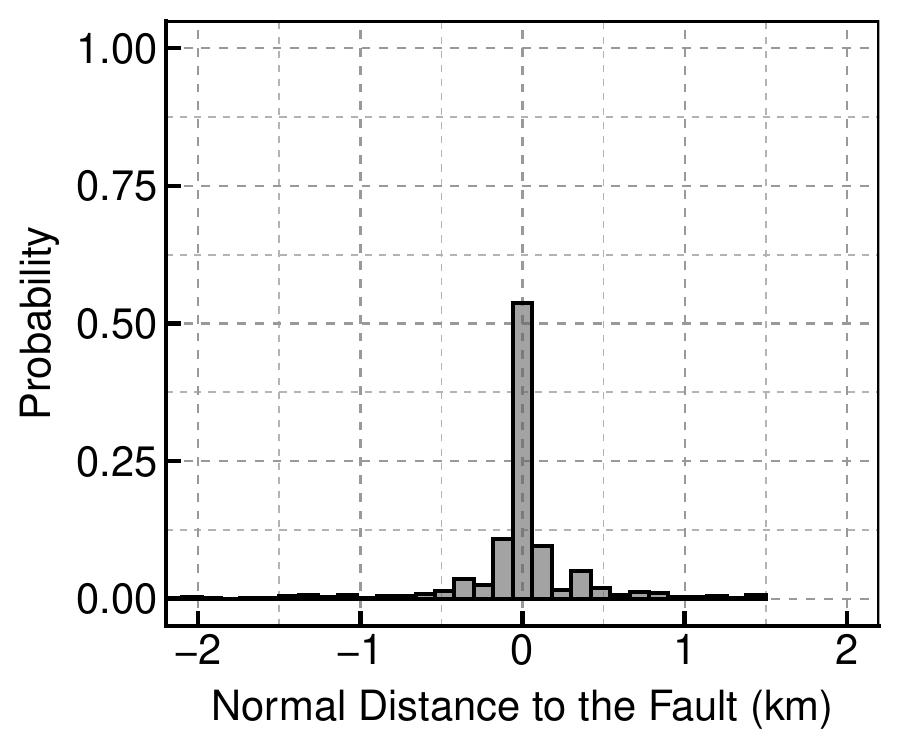}
        \label{fig:dist_hist_max}
    \end{subfigure}

    \caption{Normal distance to the fault histogram for different trace definitions: (a) SCS trace, (b) random trace, and (c) maximum principal displacement trace.}
    \label{fig:dist_hist}
\end{figure}

The histogram of perpendicular distance from the fault for the three trace definitions are shown in Figure \ref{fig:dist_hist}. In all cases, the distributions exhibit a pronounced concentration near zero, indicating that most displacements occur close to the fault trace. At the same time, the presence of non-negligible tails shows that displacement is not confined to the immediate vicinity of the fault and can extend to larger distances. While the overall behavior is consistent across the three definitions, differences in the spread and tail behavior reflect the influence of the chosen reference trace on the measured distances. Based on these observed characteristics, separate statistical formulations were adopted for the three trace definitions.

For the SCS trace definition, the empirical distribution of normal distance relative to the fault trace (Figure \ref{fig:dist_hist_scs}) is characterized by a broad central peak and heavy tails extending to distances of approximately 2 km. This behavior is represented by a Student's t distribution as:

\begin{equation} \label{eq:Student's_t_distribution_scs}
f(t_{SCS}) = \frac{1}{\sigma}\, t_{\nu}\!\left(\frac{t_{SCS}}{\sigma}, \, \nu\right)
\end{equation}

The scale parameter $\sigma$ is modeled as a function of magnitude and normalized displacement to capture their influence on the spread of the distance distribution:

\begin{equation} \label{eq:Student's_t_distribution_sigma_scs}
\sigma = k_0 + k_2 M + k_4 \exp(-\lambda\, \hat{D}_p) (m)
\end{equation}

Compared to the SCS trace definition, the distributions associated with the randomly selected trace and maximum principal displacement trace definitions (Figures \ref{fig:dist_hist_random} and \ref{fig:dist_hist_max}) exhibit a sharper concentration near zero and heavier tails. To capture both features simultaneously, the normal distance is modeled using a mixture of two Student's t distributions,

\begin{equation} \label{eq:Student's_t_distribution_random/max}
f(t_{rand/max}) = a *\frac{1}{\sigma_1}\, t_{\nu}\!\left(\frac{t_{rand/max}}{\sigma_1}, \, \nu_1\right)+ (1-a) * \frac{1}{\sigma_2}\, t_{\nu}\!\left(\frac{t_{rand/max}}{\sigma_2}, \, \nu_2\right)
\end{equation}

Because earthquake magnitude and normalized displacement exhibit trends similar to those observed for the SCS trace definition, the scale parameters $\sigma_1$ and $\sigma_2$ are modeled using the same functional dependence on these variables as:

\begin{equation} \label{eq:Student's_t_distribution_sigma_random/max}
\begin{aligned}
\sigma_1 &= k_0 + k_2 M + k_4 \exp\!\left(-\lambda\, \hat{D}_p\right) \,(m) \\
\sigma_2 &= k_1 + k_2 M + k_4 \exp\!\left(-\lambda\, \hat{D}_p\right) \,(m)
\end{aligned}
\end{equation}

To ensure stable parameter estimation given the limited dataset, the coefficients controlling the magnitude and displacement dependence ($k_2$, $k_4$, and $\lambda$) are shared across all faulting styles. Likewise, the degrees-of-freedom parameter of the primary Student's t component ($\nu_1$) is assumed to be common across faulting styles to avoid overfitting. The intercept parameters ($k_0$ and, for the mixture models, $k_1$) are allowed to vary by faulting style to capture differences in the overall scale of the distance distributions. For the mixture models, the mixing weight ($a$) and the degrees-of-freedom parameter of the second component ($\nu_2$) are estimated separately. The resulting model coefficients are summarized in Table \ref{tab:t_distribution_coefficients}.

\begin{table}[htbp]
\centering
\caption{Estimated coefficients for the model governing the normal distance to different fault trace definitions.}
\label{tab:t_distribution_coefficients}

\begin{tabular}{c|cccccccc}
\hline
SOF & $a$ & $k_0$ & $k_1$ & $k_2$ & $k_4$ & $\lambda$ & $\nu_1$ & $\nu_2$ \\

\hline
\multicolumn{9}{c}{\textbf{SCS trace}} \\
\hline
All  
& \multirow{5}{*}{-}
& -118.04
& \multirow{5}{*}{-}
& \multirow{5}{*}{24.83}
& \multirow{5}{*}{26.81}
& \multirow{5}{*}{4.61}
& \multirow{5}{*}{1.13}
& \multirow{5}{*}{-} \\

SS & & -118.93 & & & & & & \\

NR & & -117.19 & & & & & & \\

N  & & -123.74 & & & & & & \\

R  & & -116.65 & & & & & & \\

\hline
\multicolumn{9}{c}{\textbf{Random trace}} \\
\hline
All & 0.26 & 285.64 & -87.37 & \multirow{5}{*}{1.62}
& \multirow{5}{*}{114.74}
& \multirow{5}{*}{0.001}
& \multirow{5}{*}{2.97}
& 30.95 \\

SS & 0.32 & 215.66 & -88.04 & & & & & 40.61 \\

NR & 0.25 & 283.86 & -89.36 & & & & & 12.46 \\

N  & 0.16 & 273.36 & -90.34 & & & & & 13.62 \\

R  & 0.73 & 227.76 & -66.64 & & & & & 27.88 \\

\hline
\multicolumn{9}{c}{\textbf{Maximum principal displacement trace}} \\
\hline
All & 0.30 & 375.07 & -117.88 & \multirow{5}{*}{5.85}
& \multirow{5}{*}{107.09}
& \multirow{5}{*}{0.06}
& \multirow{5}{*}{4.11}
& 1.37 \\

SS & 0.21 & 422.83 & -113.13 & & & & & 1.57 \\

NR & 0.38 & 311.47 & -122.18 & & & & & 1.27 \\

N  & 0.27 & 344.60 & -124.69 & & & & & 1.37 \\

R  & 0.77 & 280.47 & -85.07 & & & & & 35.55 \\

\hline
\end{tabular}

\vspace{2mm}
\small $a$ is the mixing weight of the first Student's t component. SOF: style of faulting; SS: strike-slip;\\NR: normal-reverse; N: normal; R: reverse.

\end{table}

The uncertainty associated with parameter estimation was evaluated using 95\% confidence intervals obtained from the profile likelihood approach described in Equations~(\ref{eq:negative_fixed_NLL})-(\ref{eq:95percent_interval}), as standard error—based confidence intervals may be unreliable for constrained parameters. The resulting lower and upper confidence bounds are summarized in Table~\ref{tab:t_profilelikelihood}. 

\begin{table}[htbp]
\centering
\caption{Lower and upper bounds of the 95\% profile-likelihood confidence intervals for the normal distance model parameters corresponding to different fault trace definitions and faulting styles.}
\label{tab:t_profilelikelihood}
\setlength{\tabcolsep}{2pt}

\begin{tabular}{c|cccccccc}
\hline
SOF & $a$ & $k_0$ & $k_1$ & $k_2$ & $k_4$ & $\lambda$ & $\nu_1$ & $\nu_2$ \\

\hline
\multicolumn{9}{c}{\textbf{SCS trace}} \\
\hline
All
& \multirow{5}{*}{-}
& {$[-158.57, -82.27]$}
& \multirow{5}{*}{-}
& \multirow{5}{*}{\shortstack{$[18.06,$\\$32.36]$}}
& \multirow{5}{*}{\shortstack{$[0,$\\$100]$}}
& \multirow{5}{*}{\shortstack{$[0,$\\$10]$}}
& \multirow{5}{*}{\shortstack{$[1.00,$\\$1.48]$}}
& \multirow{5}{*}{-} \\

SS
&
& {$[-154.98, -82.89]$}
& & & & & & \\

NR
&
& \shortstack{$[-157.44, -81.68]$}
& & & & & &  \\

N
&
& \shortstack{$[-181.24, -66.24]$}
& & & & & & \\

R
&
& \shortstack{$[-156.71, -81.30]$}
& & & & & &  \\

\hline
\multicolumn{9}{c}{\textbf{Random trace}} \\
\hline
All
& {$[0.25, 0.27]$}
& {$[256.79\phantom{-}, 406.82]$}
& {$[-131.50, -82.07]$}
& \multirow{5}{*}{\shortstack{$[0.57,$\\$7.77]$}}
& \multirow{5}{*}{\shortstack{$[93.88,$\\$180.80]$}}
& \multirow{5}{*}{\shortstack{$[0.001,$\\$0.24]$}}
& \multirow{5}{*}{\shortstack{$[2.82,$\\$8.41]$}}
& {$[2.00, 39.11]$} \\

SS
& {$[0.01, 0.34]$}
& {$[-163.38, 568.55]$}
& {$[-146.74, -79.15]$}
& & & & &
{$[1.01, 50.00]$} \\

NR
& \shortstack{$[0.23, 0.25]$}
& \shortstack{$[-146.23, 312.53]$}
& \shortstack{$[-138.10, -83.95]$}
& & & & &
\shortstack{$[6.52, 26.35]$} \\

N
& \shortstack{$[0.15, 0.87]$}
& \shortstack{$[-140.82, 378.29]$}
& \shortstack{$[-139.61, -84.86]$}
& & & & &
\shortstack{$[6.64, 29.18]$} \\

R
& \shortstack{$[0.67, 0.77]$}
& \shortstack{$[-126.53, 296.78]$}
& \shortstack{$[-119.14, -62.60]$}
& & & & &
\shortstack{$[8.43, 50.00]$} \\

\hline
\multicolumn{9}{c}{\textbf{Maximum principal displacement trace}} \\
\hline
All
& {$[0.26, 0.35]$}
& {$[322.03, 412.96]$}
& {$[-125.03, -82.16]$}
& \multirow{5}{*}{\shortstack{$[2.48,$\\$7.44]$}}
& \multirow{5}{*}{\shortstack{$[97.36,$\\$168.75]$}}
& \multirow{5}{*}{\shortstack{$[0.001,$\\$0.90]$}}
& \multirow{5}{*}{\shortstack{$[3.07,$\\$4.78]$}}
& {$[1.13,\phantom{/} 1.76]$} \\

SS
& {$[0.16, 0.25]$}
& {$[397.20, 841.39]$}
& {$[-120.00, -92.56]$}
& & & & &
{$[1.16,\phantom{/} 2.39]$} \\

NR
& {$[0.33, 0.41]$}
& {$[354.84, 342.93]$}
& {$[-129.58, -95.02]$}
& & & & &
{$[1.01,\phantom{/} 1.88]$} \\

N
& {$[0.22, 0.31]$}
& {$[281.94, 407.25]$}
& {$[-132.25, -86.91]$}
& & & & &
{$[1.01,\phantom{/} 2.04]$} \\

R
& {$[0.70, 0.90]$}
& {$[206.81, 320.13]$}
& {$[-97.10\phantom{..}, -62.72]$}
& & & & &
{$[1.50, 50.00]$} \\
\hline

\end{tabular}

\vspace{2mm}
\small $a$ is the mixing weight of the first Student's t component. SOF: style of faulting; SS: strike-slip;\\NR: normal-reverse; N: normal; R: reverse.

\end{table}

The upper profile-likelihood bounds for $k_4$ and $\lambda$ in SCS trace model were found to be weakly identifiable due to the strong tradeoff between the two parameters in the term $k_4\exp(-\lambda \hat{D}_p)$. During the profile-likelihood analysis, a broad range of $(k_4,\lambda)$ combinations satisfied the likelihood-ratio criterion, resulting in upper confidence bounds that could not be reliably constrained by the data alone. Consequently, upper search ranges for $k_4$ and $\lambda$ were defined based on the observed data range and the fitted parameter values. The reported upper bounds therefore correspond to the limits of the investigated parameter space rather than statistically identifiable confidence bounds.

\begin{figure}[htbp]
    \centering

    \begin{subfigure}[d]{0.49\linewidth}
        \centering
        \caption{}
        \includegraphics[width=\linewidth]{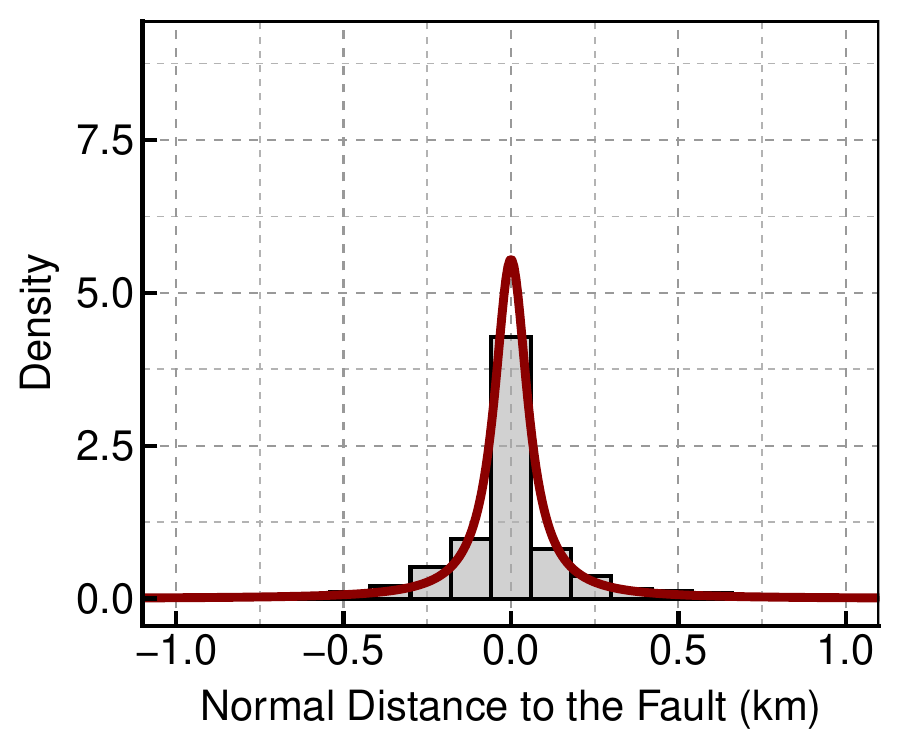}
        \label{fig:Fitted_t_SS-6.5<M<7.25-D:0-1}
    \end{subfigure}
    \hfill
    \begin{subfigure}[d]{0.49\linewidth}
        \centering
        \caption{}
        \includegraphics[width=\linewidth]{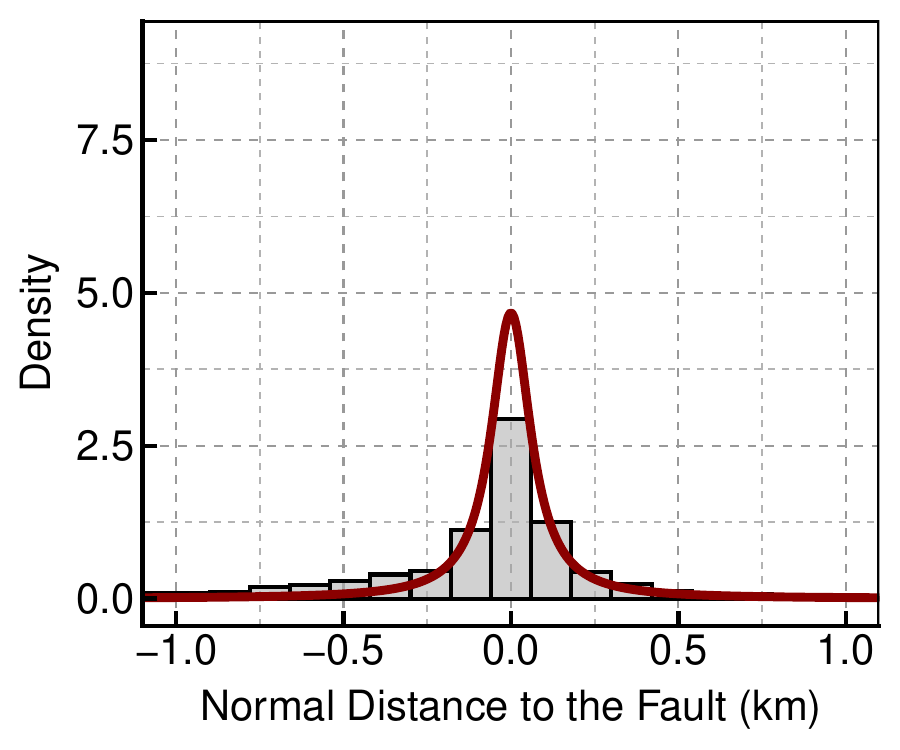}
        \label{fig:Fitted_t_SS-7.25<M-D:0-1}
    \end{subfigure}
    \begin{subfigure}[d]{0.49\linewidth}
        \centering
        \caption{}
        \includegraphics[width=\linewidth]{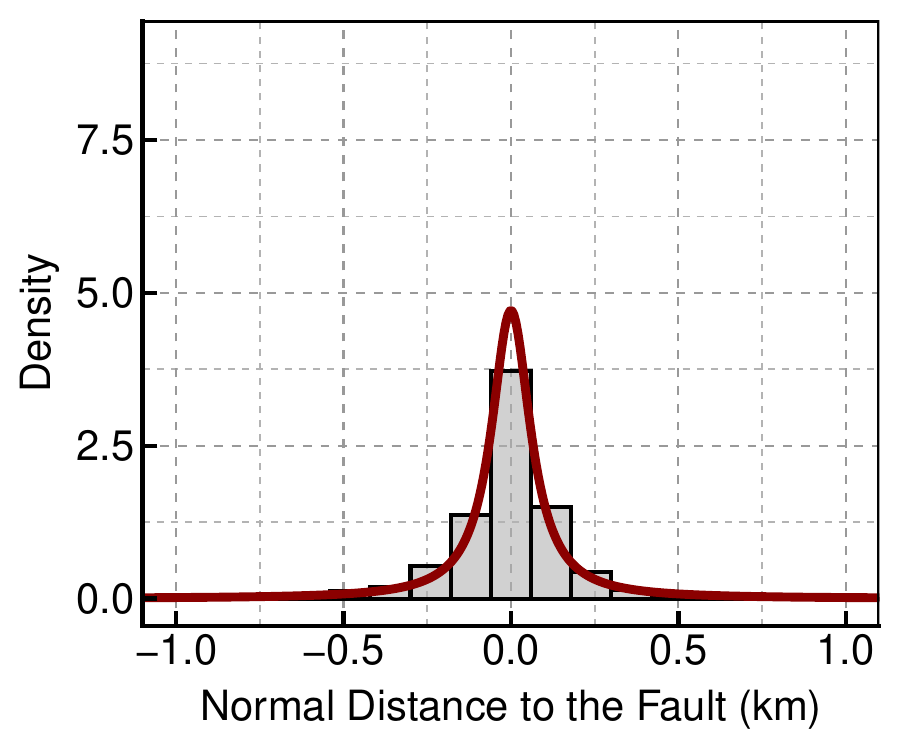}
        \label{fig:Fitted_t_SS-7.25<M-D:1-10}
    \end{subfigure}
    \begin{subfigure}[d]{0.49\linewidth}
        \centering
        \caption{}
        \includegraphics[width=\linewidth]{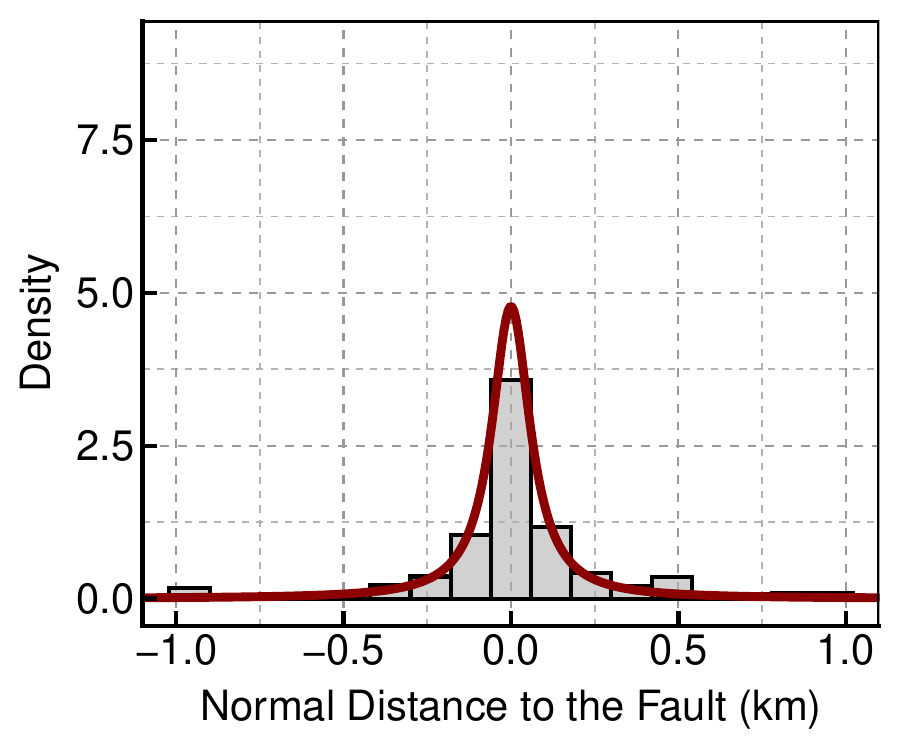}
        \label{fig:Fitted_t_NR-7.25<M-D:1-10}
    \end{subfigure}

    \caption{Comparison of the observed and modeled distributions of the normal distance to the SCS fault trace for different faulting styles and magnitude ($M$) - displacement ($\hat{D}_p$) bins. Histograms represent the empirical distributions, and red curves denote the probability density functions predicted by the fitted Student's t distribution. (a) strike-slip faults with $6.5 < M < 7.25$ and $\hat{D_p} < 1$, (b) strike-slip faults with $7.25 < M$ and $\hat{D_p} < 1$, (c) strike-slip faults with $7.25 < M$ and $1 < \hat{D_p}$, and (d) normal and reverse faults with $7.25 <M$ and $1 < \hat{D_p}$.}
    \label{fig:Fitted_t_distribution_scs}
\end{figure}

Figure \ref{fig:Fitted_t_distribution_scs} compares the fitted model with the observed distributions of normal distance to the SCS fault trace for different faulting styles and magnitude-displacement bins. The fitted model provides a reasonable representation of the observed. Distances between overlapping ruptures were also compared with the observed rupture separations, and the fitted Student's t distribution was found to reproduce these distances effectively. Equivalent comparisons between the fitted models and empirical histograms for the randomly selected trace and maximum principal displacement trace definitions are presented in the electronic supplement (Figures S7 and S8).

\begin{figure}[htbp]
    \centering

    \begin{subfigure}[b]{0.48\linewidth}
        \centering
        \caption{}
        \includegraphics[width=\linewidth]{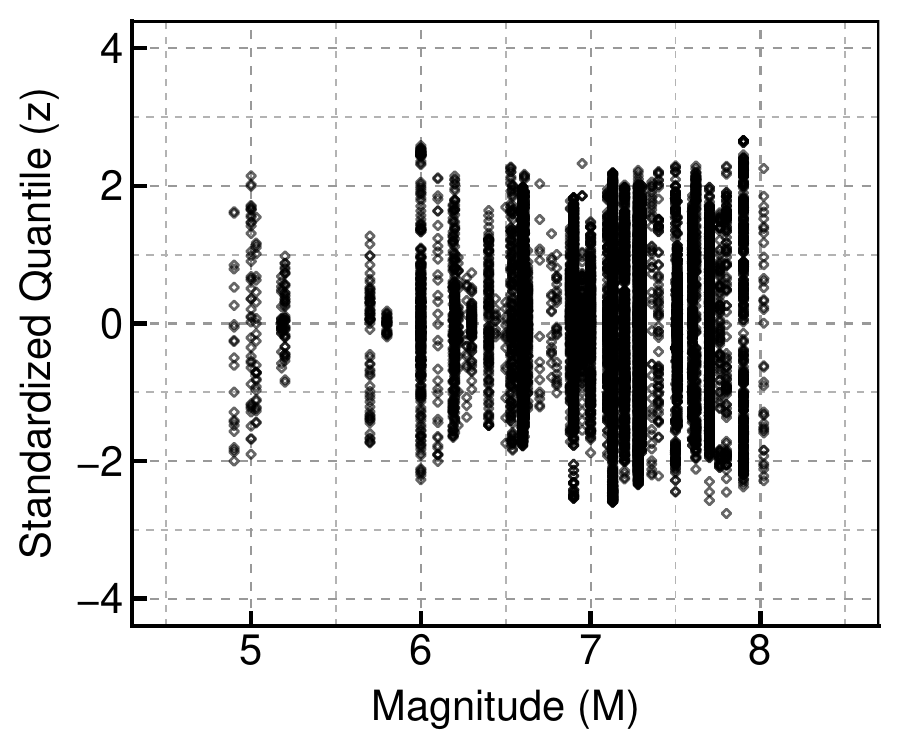}
        \label{fig:dist_z_mag}
    \end{subfigure}
    \hfill
    \begin{subfigure}[b]{0.48\linewidth}
        \centering
        \caption{}
        \includegraphics[width=\linewidth]{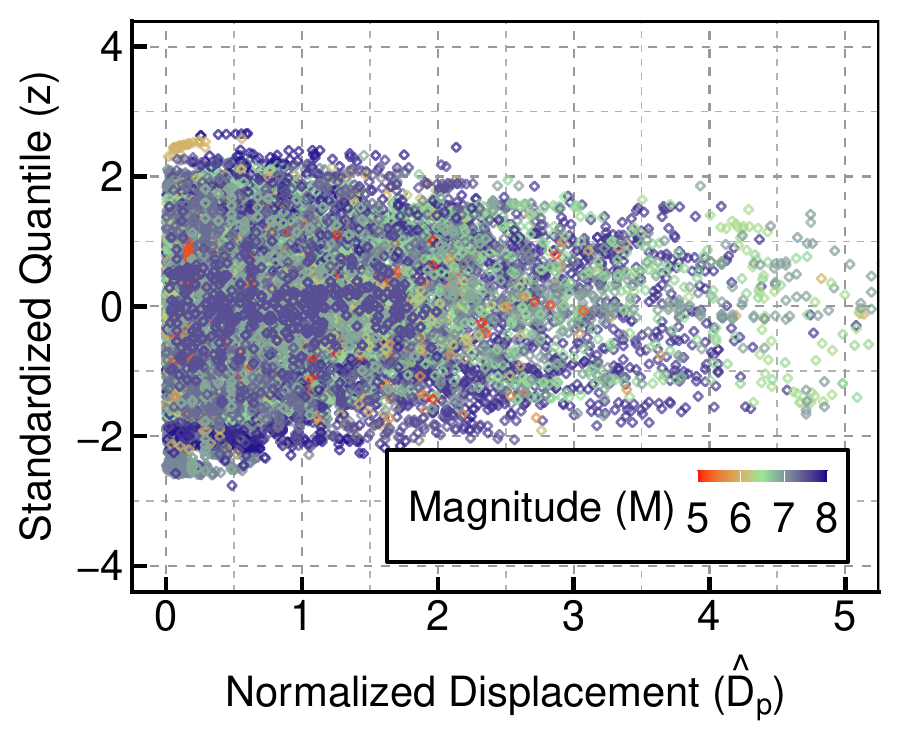}
        \label{fig:dist_z_disp}
    \end{subfigure}

    \caption{Assessment of residual trends in the fitted Student's t model for normal distance to the SCS fault trace using standardized quantiles plotted as functions of (a) earthquake magnitude ($M$) and  (b) normalized displacement ($\hat D_p$).}
    \label{fig:dist_z}
\end{figure}

The standardized quantile residuals confirm that both magnitude and normalized displacement effects are adequately captured by the model for normal distance to the SCS trace. As shown in Figures \ref{fig:dist_z_mag} and \ref{fig:dist_z_disp}, the residuals exhibit no systematic trends with respect to either variable, indicating that no additional dependence remains unmodeled. Corresponding plots for all faulting styles are provided in Figure S9 of the electronic supplement. Equivalent residual analyses for the randomly selected trace and maximum principal displacement trace models are provided in Figures S10 and S11, respectively.
The residuals for all three normal distance models were also examined as a function of fault curvature to further verify the absence of any systematic relationship between curvature and normal distance. The corresponding plots are provided in the electronic supplement (Figure S12).

The distributions defined above describe the probability that a rupture is located at a given perpendicular distance from the fault. For a structure with span $[t_1, t_2]$, the probability that an individual rupture lies within this interval is obtained from the corresponding distributions and is denoted by $P_t$. For a total of $N_{p,{\text{tot}}}$ overlapping ruptures, each rupture is treated as an independent Bernoulli trial with success probability $P_t$, where ``success'' corresponds to the rupture intersecting the structural span. In the case of a single crossing rupture, exactly one rupture out of $N_{p,{\text{tot}}}$ is assumed to fall within the span, while the remaining ruptures lie outside. More generally, the probability that exactly $N_{p,{\text{cross}}}$ ruptures intersect the structure is given by the binomial formulation described in Equation (\ref{eq:P_t}) and (\ref{eq:multirupture_distance_distribution}).

\begin{equation} \label{eq:P_t}
P_t = P\!\left(t \in [t_{1}, t_{2}]\right)
     = F(t_2)-F(t_1)
\end{equation}

where $F(t)$ denotes the cumulative distribution function (CDF)
of the normal distance distribution corresponding to the selected
fault-trace definition.

\begin{equation} \label{eq:multirupture_distance_distribution}
P\!\left(t \in [t_{1}, t_{2}] \mid M, SOF, \hat{D_{p}}\right)
= \binom{N_{p,{\text{tot}}}}{N_{p,{\text{cross}}}}\, P_{t}^{\,N_{p,{\text{cross}}}}\, (1 - P_{t})^{\,N_{p,{\text{tot}}} - N_{p,{\text{cross}}}}
\end{equation}

\section{Hazard evaluation}

To facilitate comparison with existing fault-displacement hazard methodologies, the benchmark cases developed by the International Atomic Energy Agency (IAEA) \citep{valentini2021probabilistic,valentini2025iaea} were used to evaluate the proposed framework. These benchmark scenarios have been widely adopted for inter-model comparisons and therefore provide a consistent basis for assessing differences between the proposed and existing approaches. Within this study, the single-rupture framework considers the hazard associated with an individual rupture strand, whereas the multi-rupture framework accounts for the cumulative contribution of one or more rupture strands intersecting the structure. 

As discussed in \hyperref[sec:Methodology]{Section: Methodology}, separate fault-type grouping were developed to account for the limited number of normal and reverse faulting events. Rather than selecting a single regression model for hazard evaluation, the proposed framework propagates the resulting epistemic uncertainty through a logic-tree weighting approach. For strike-slip faulting, the hazard is evaluated using weighted contributions from the strike-slip (SS) and All fault type (All) models, whereas for normal and reverse faulting the hazard is computed from weighted combinations of the fault type specific (N or R), combination of normal and reverse (NR), and All fault type (All) models. The adopted logic-tree weights are summarized in Table~\ref{tab:logic_tree_sof}.

\begin{table}[htbp]
\centering
\caption{Logic tree weights for faulting style epistemic uncertainty used in hazard evaluation.}
\label{tab:logic_tree_sof}

\begin{tabular}{c|c|c}
\hline
{Faulting style} & {Regression model} & {Weight} \\
\hline

\multirow{2}{*}{Strike-slip}
& SS  & 0.75 \\
& All & 0.25 \\
\hline

\multirow{3}{*}{Normal}
& N   & 0.50 \\
& NR  & 0.25 \\
& All & 0.25 \\
\hline

\multirow{3}{*}{Reverse}
& R   & 0.50 \\
& NR  & 0.25 \\
& All & 0.25 \\
\hline
\end{tabular}

\end{table}

The resulting hazard curves are presented down to annual exceedance probabilities of $10^{-8}$ to illustrate differences between the considered hazard formulations and rupture-crossing assumptions over a broad range of hazard levels. While such low exceedance probabilities are associated with substantial uncertainty due to the limited observational constraints available at very long return periods, a detailed assessment of this uncertainty is beyond the scope of the present study. The low hazard levels obtained are primarily a consequence of the selected benchmark scenarios rather than the proposed methodology itself.

Table~\ref{tab:IAEA_Benchmarks} summarizes the benchmark cases considered for principal displacement hazard analysis, spanning all styles of faulting. Because fault-curvature information was not available for the benchmark faults, a representative value of $\kappa = 0.002$ was adopted for all scenarios.

\begin{table}[htbp] 

\centering
\caption{IAEA Benchmark Fault Sources and Parameters Used for Hazard Evaluation \citep{valentini2021probabilistic, valentini2025iaea}.}
\resizebox{\textwidth}{!}{
\begin{tabular}{lccccccccc}
\hline
\textbf{Case} & 
\textbf{Fault type} & 
\textbf{Fault source} &
\textbf{Weight} &
\textbf{Max. rup. length (km)} &
\textbf{Magnitude} &
\textbf{Mean Rate ($\times 10^{-5}$/yr)} &
\textbf{u/SRL} &
\textbf{Curvature} &
\textbf{Distance (m)} \\
\hline

\multirow{5}{*}{Kumamoto Case 2} 
& \multirow{5}{*}{SS}
& Floating (1+2+3)               & 0.5 & 20 & 6.5 & 64.2 & variable & \multirow{5}{*}{0.002} & \multirow{5}{*}{[0--500]} \\
& & Uto (2)                       & 0.5 & 22 & 6.5 & 18.9 & 0.23     & & \\
& & Futagawa + Uto (1+2)          & 1   & 46 & 6.9 & 1.28 & 0.37     & & \\
& & Uto + Uto-Hanto-North (2+3)   & 1   & 54 & 7.0 & 3.53 & 0.10     & & \\
& & Futagawa + Uto + Uto-Hanto-North (1+2+3) & 1 & 78 & 7.2 & 1.28 & 0.37 & & \\
\hline
Kumamoto Case 3 & SS & SFZ & 1 & 5.4 & 5.8 & 23.3 & 0.39 & 0.002 & [0--500] \\
\hline

\multirow{7}{*}{Norcia Case 1}
& \multirow{7}{*}{N}
& \multirow{7}{*}{MVFS}
& \multirow{7}{*}{1}
& \multirow{7}{*}{34}
& 6.4 & 4.29 & \multirow{7}{*}{0.05} & \multirow{7}{*}{0.002} & \multirow{7}{*}{[0--500]} \\
& & & & & 6.5 & 5.66 & & & \\
& & & & & 6.6 & 6.69 & & & \\
& & & & & 6.7 & 7.07 & & & \\
& & & & & 6.8 & 6.69 & & & \\
& & & & & 6.9 & 5.66 & & & \\
& & & & & 7.0 & 4.29 & & & \\
\hline

Le~Teil Case 2 & R & LRF & 1 & 5.8 & 5.52 & 4.6 & 0.46 & 0.002 & [5--500] \\

\hline
\end{tabular}
}
\vspace{2mm}
\small SOF: style of faulting; SS: strike-slip; N: normal; R: reverse.
\label{tab:IAEA_Benchmarks}
\end{table}

Hazard curves for each IAEA benchmark scenario are provided in the electronic supplement (Figures S13--S20), including the aggregate principal displacement, single-rupture, and multi-rupture formulations for all three fault-trace (SCS trace, randomly selected trace, and maximum-principal-displacement trace) definitions.

\begin{figure}[htbp]
    \centering

    \begin{subfigure}[d]{0.49\linewidth}
        \centering
        \caption{}
        \includegraphics[width=\linewidth]{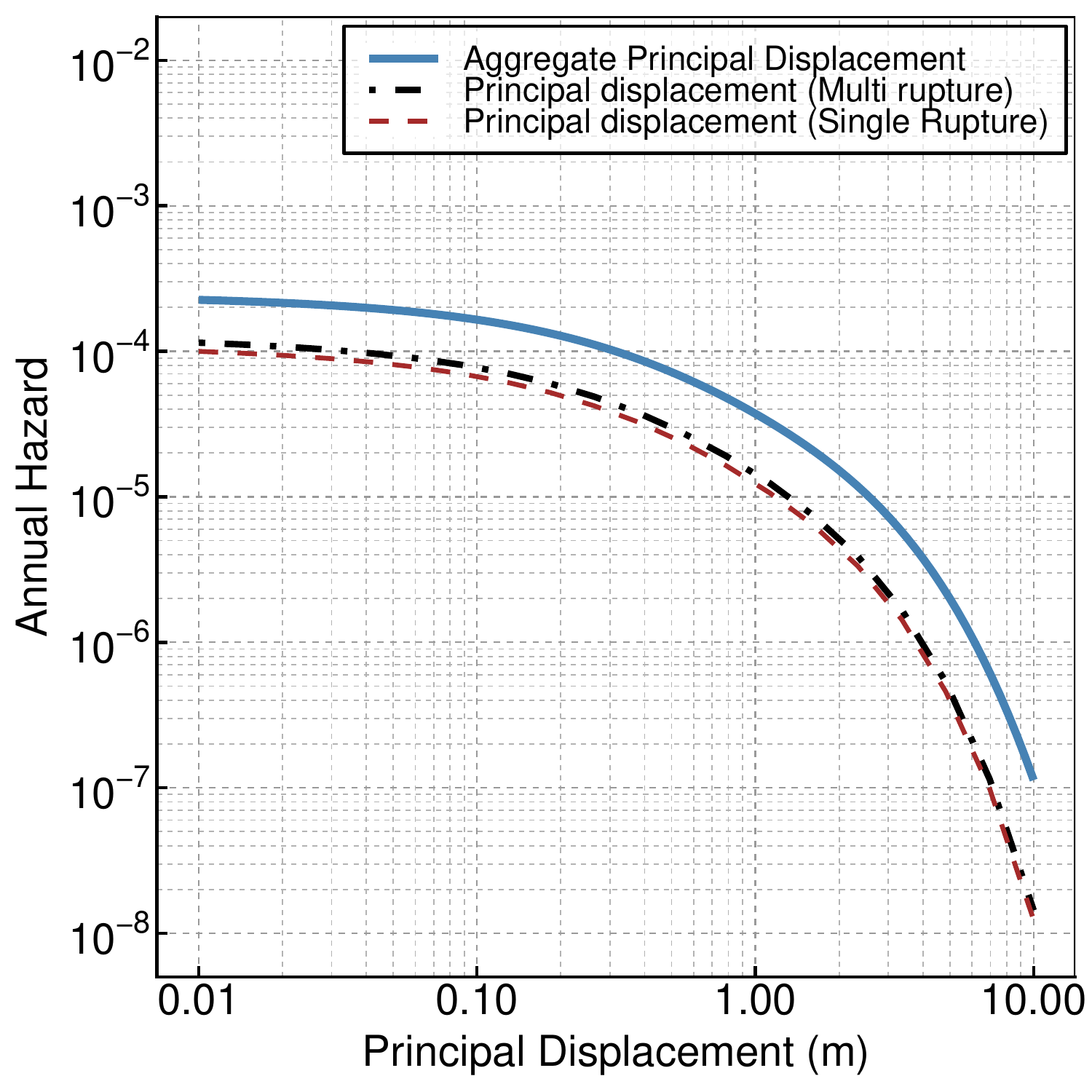}
        \label{fig:haz_agg_prncp_SCS_Kumamoto_2}
    \end{subfigure}
    \hfill
    \begin{subfigure}[d]{0.49\linewidth}
        \centering
        \caption{}
        \includegraphics[width=\linewidth]{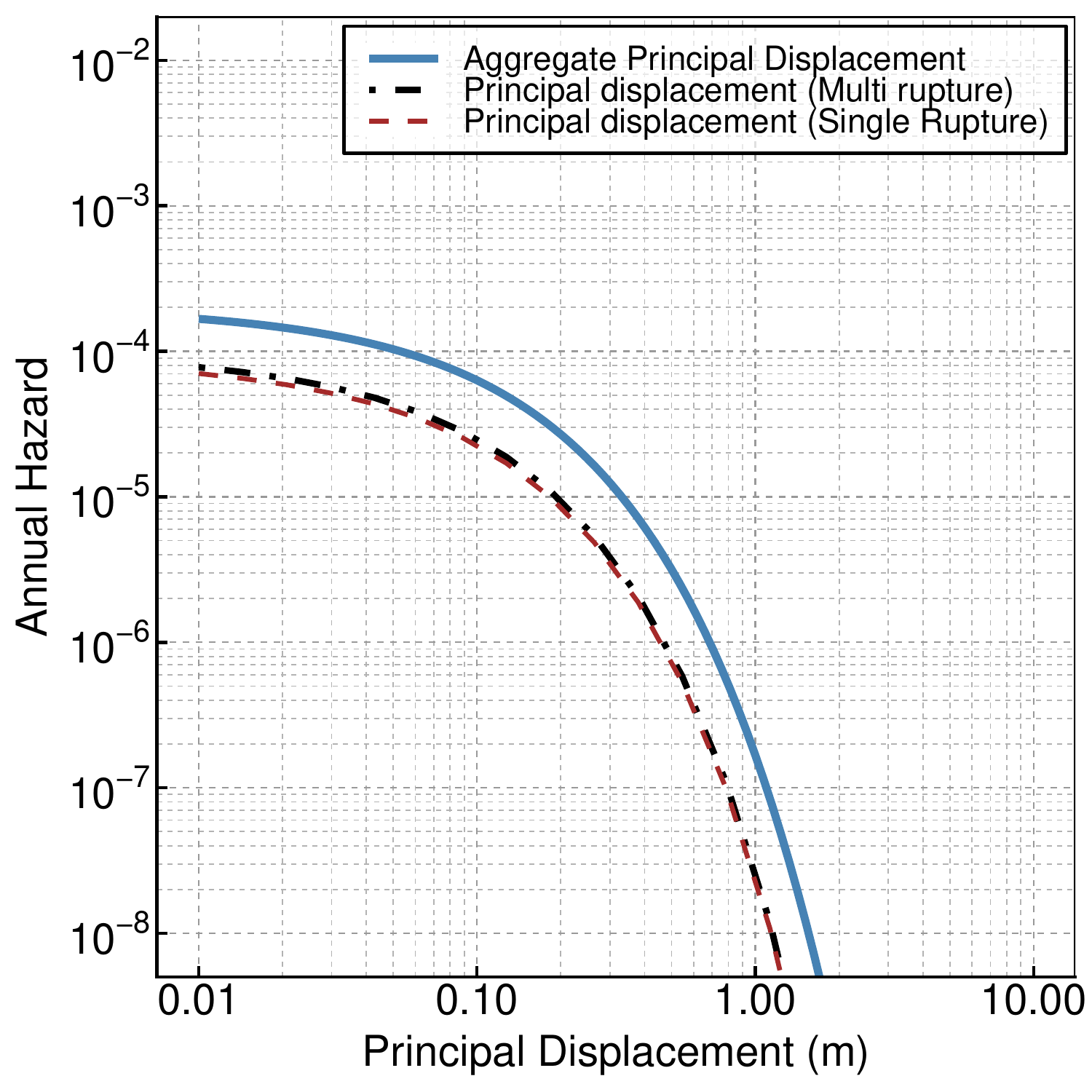}
        \label{fig:haz_agg_prncp_SCS_Kumamoto_3}
    \end{subfigure}
    \begin{subfigure}[d]{0.49\linewidth}
        \centering
        \caption{}
        \includegraphics[width=\linewidth]{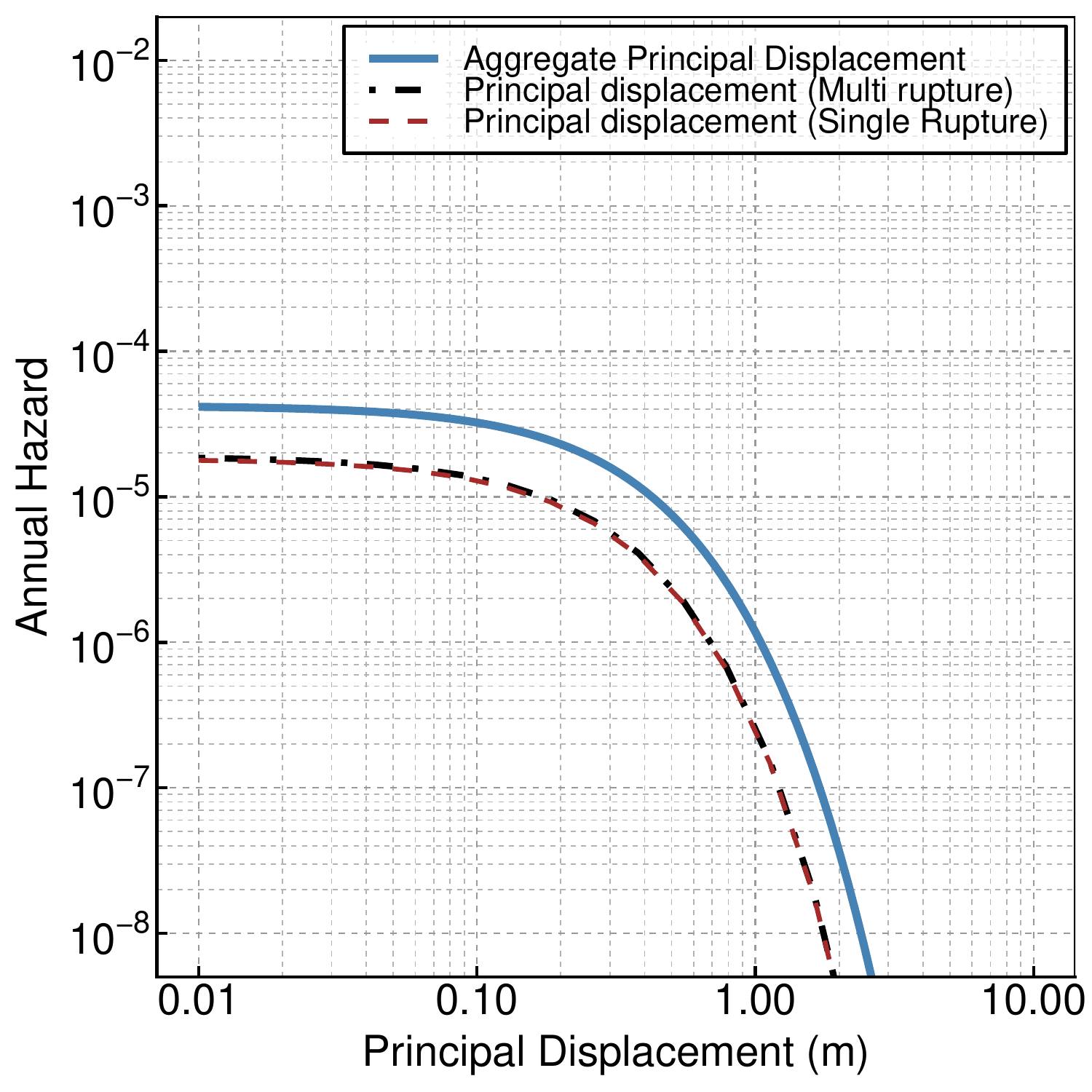}
        \label{fig:haz_agg_prncp_SCS_LeTeil_2}
    \end{subfigure}
    \begin{subfigure}[d]{0.49\linewidth}
        \centering
        \caption{}
        \includegraphics[width=\linewidth]{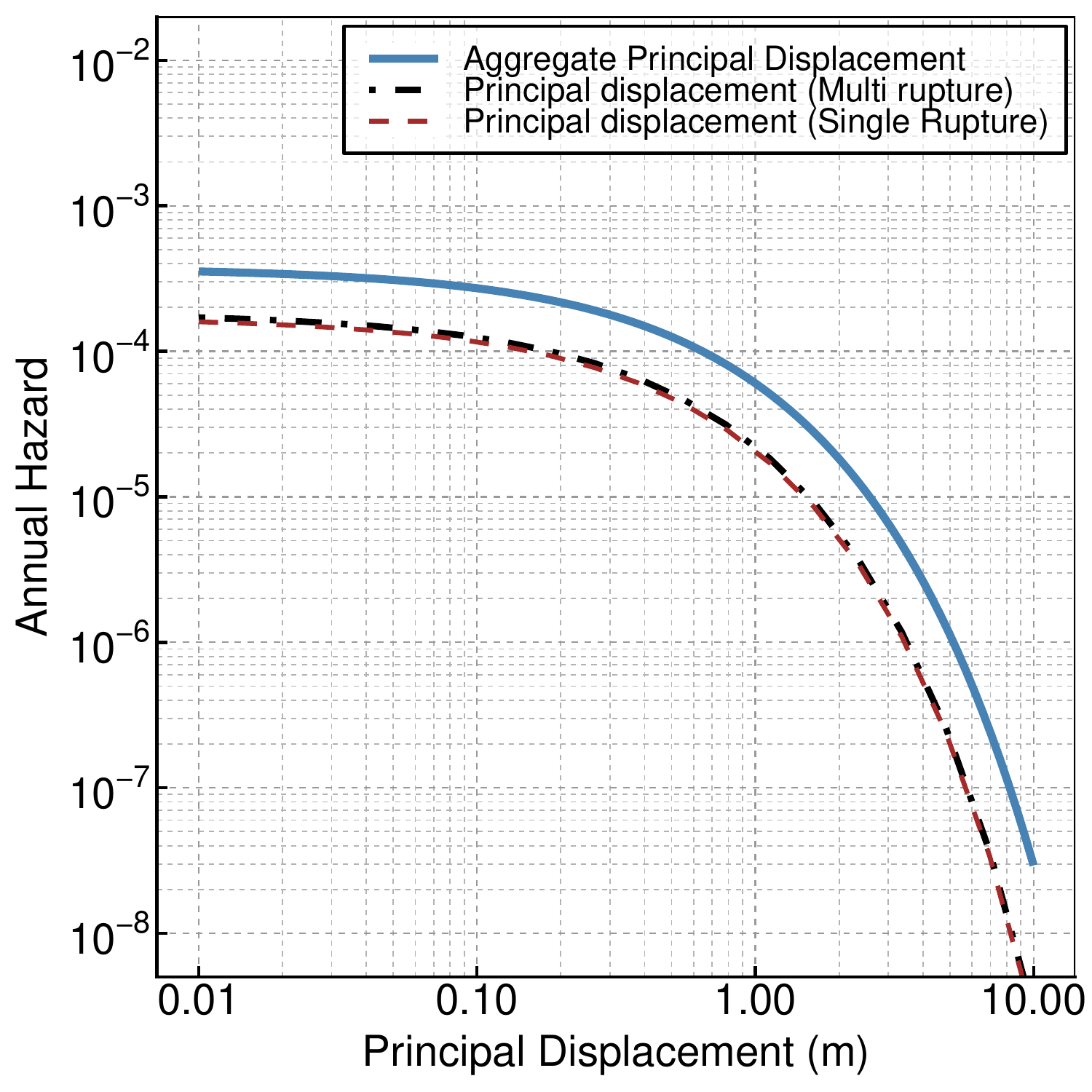}
        \label{fig:haz_agg_prncp_SCS_norcia_1}
    \end{subfigure}

    \caption{Comparison of hazard curves for aggregate principal displacement and principal displacement considering both single-rupture (a single rupture strand intersecting the structure) and multi-rupture (the cumulative contribution of one or more rupture strands intersecting the structure) hazard formulations under the SCS fault trace definition for the IAEA benchmark cases \citep{valentini2021probabilistic, valentini2025iaea}: (a) Kumamoto Case 2, (b) Kumamoto Case 3, (c) Le Teil Case 2, and (d) Norcia Case 1.}
    \label{fig:Hazardcurves_agg_prncp_SCS}
\end{figure}

Figure~\ref{fig:Hazardcurves_agg_prncp_SCS} compares principal-displacement hazard curves obtained using three alternative hazard formulations: (i) the aggregate principal displacement framework, and the proposed (ii) single-rupture and (iii) multi-rupture probabilistic framework. Results are shown for all benchmark scenarios using the SCS fault-trace definition to compute distance normal to the fault. Equivalent comparisons using the randomly selected trace and maximum-principal-displacement trace definitions are provided in the electronic supplement (Figures S20 and S21), and exhibit similar trends. 

A consistent ordering is observed across all benchmark cases. The aggregate principal displacement framework produces the highest hazard estimates because it implicitly assumes that all overlapping rupture strands intersect the structure. As a result, it provides an upper-bound estimate of hazard. The single-rupture framework yields the lowest hazard estimates because it assumes that only a single rupture strand intersects the structure, while multi-rupture framework falls between the other two cases. The reduction in hazard relative to the aggregate principal displacement framework reflects the finite extent of the structure and the possibility that some rupture strands contributing to the aggregate displacement occur outside the structure footprint and therefore do not contribute to the hazard.

These results highlight the conservative nature of hazard estimates based on aggregate principal displacement and demonstrate the importance of explicitly accounting for rupture-intersection probabilities in fault-displacement hazard analysis.

\begin{figure}[htbp]
    \centering

    \begin{subfigure}[d]{0.49\linewidth}
        \centering
        \caption{}
        \includegraphics[width=\linewidth]{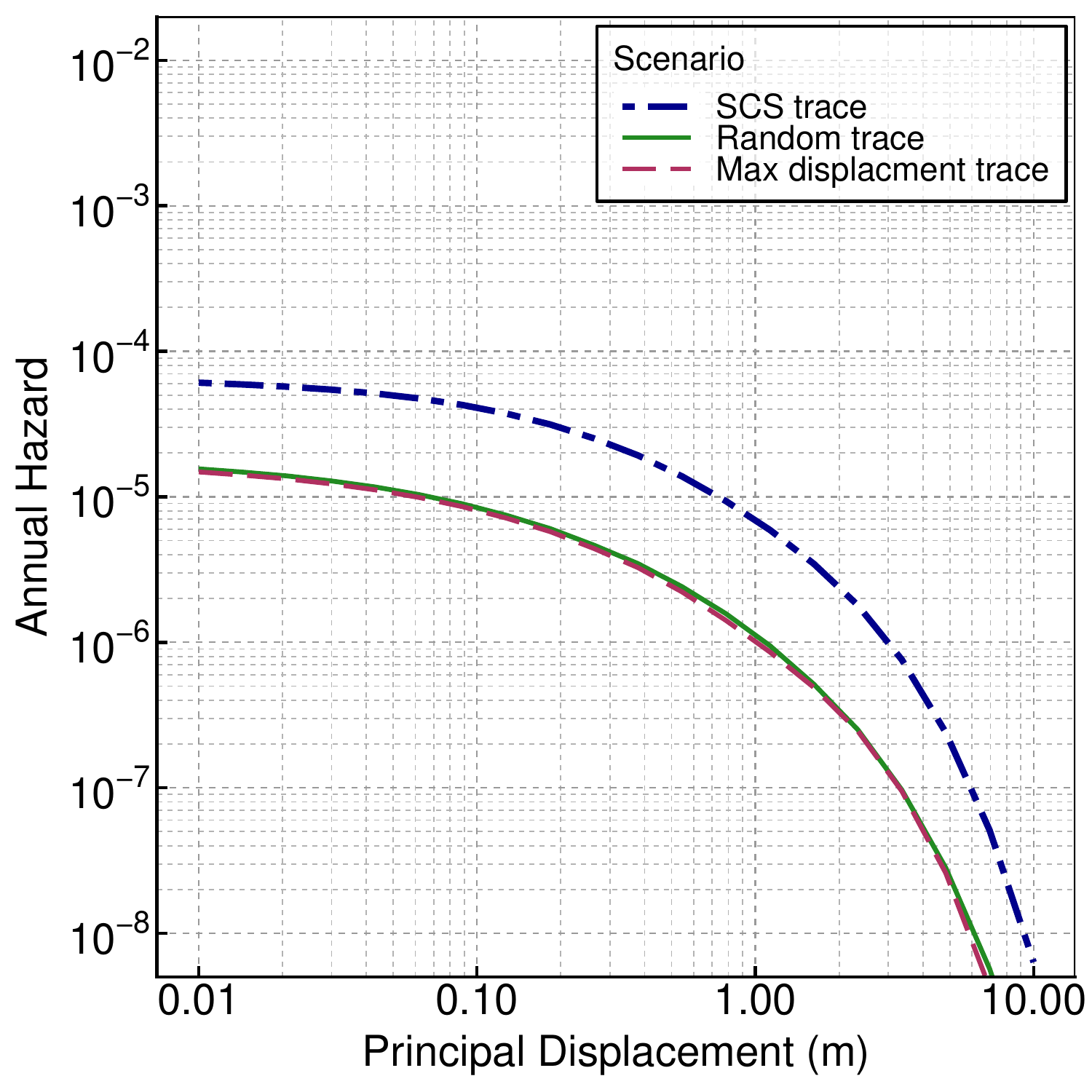}
        \label{fig:haz_trace_0050_Kumamoto_2}
    \end{subfigure}
    \hfill
    \begin{subfigure}[d]{0.49\linewidth}
        \centering
        \caption{}
        \includegraphics[width=\linewidth]{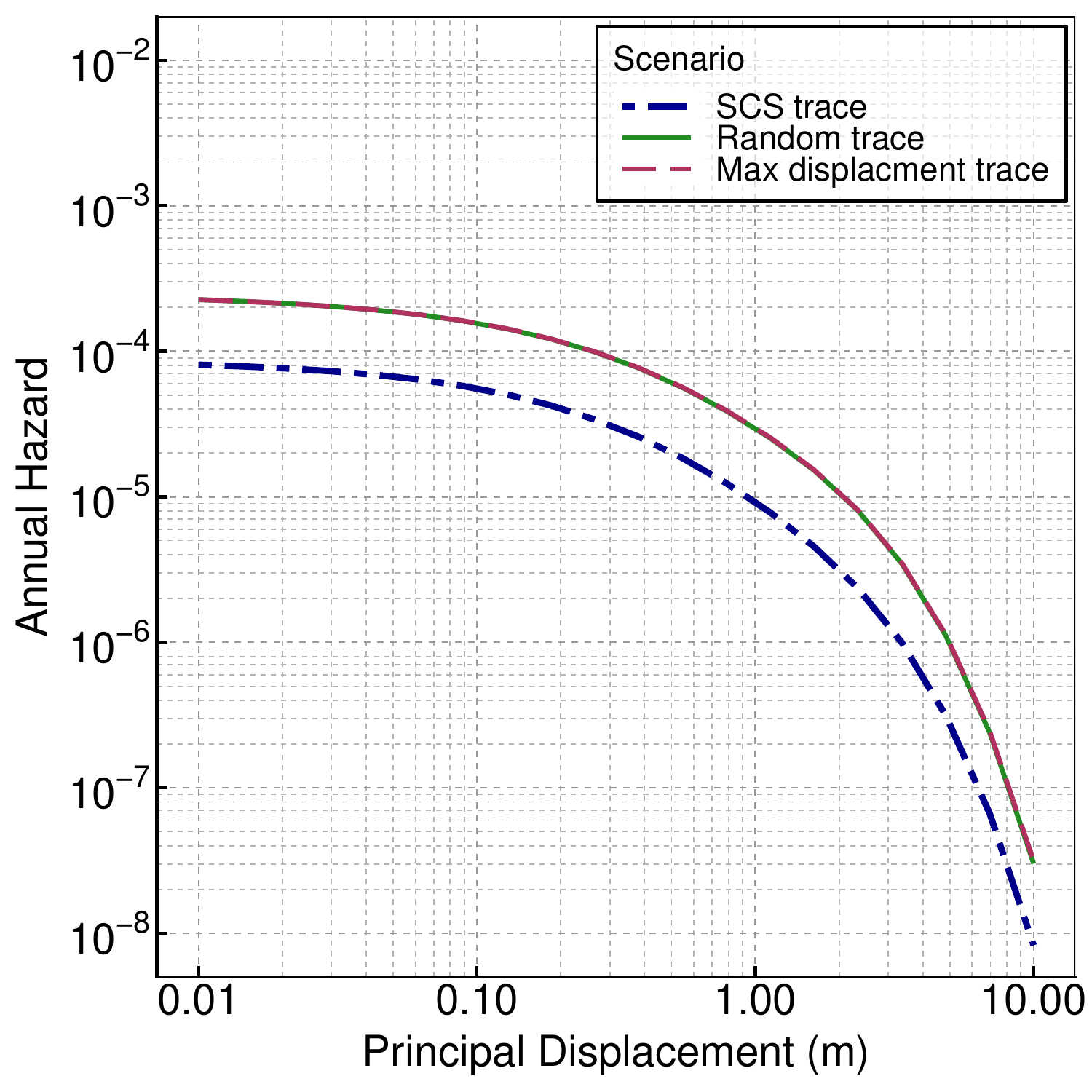}
        \label{fig:haz_trace_2525_Kumamoto_2}
    \end{subfigure}

    \caption{Comparison of hazard curves for the multi-rupture principal displacement framework (cumulative contribution of one or more rupture strands intersecting the structure) considering different fault trace representations for the IAEA benchmark (Kumamoto Case 2) \citep{valentini2021probabilistic,valentini2025iaea} for structures located within distance ranges of: (a) $[5~to~50]~m$ from the fault and (b) $[-25~to~25]~m$ from the fault.}
    \label{fig:Hazardcurves_traces_kumaoto_2}
\end{figure}

Figure \ref{fig:Hazardcurves_traces_kumaoto_2} presents a comparison of hazard curves for the Kumamoto case 2 scenario across different fault trace definitions for a stuctural footprint spancing $5~to~50~m$ of the nominal trace (Figure \ref{fig:Hazardcurves_traces_kumaoto_2}a) and $-25~to~25~m$ of the nomial trace (Fgure \ref{fig:Hazardcurves_traces_kumaoto_2}b). The results highlight a clear dependence of the hazard on the assumed spatial distribution of rupture locations. When the structure is located away from the fault trace, the random trace and maximum-displacement trace models yield lower hazard levels compared to the SCS-based model, due to their narrower distance distributions. 
However, when the structure intersects the fault trace, these two models produce higher hazard estimates. This is because a larger proportion of their probability mass is concentrated at shorter distances from the fault, increasing the likelihood of significant displacement at the structure location. As a result, the hazard becomes more sensitive to the assumed trace definition in near-fault conditions.
Equivalent comparisons for the remaining scenarios, which exhibit consistent trends, are provided in the electronic supplement (Figure S22 and S23).

\begin{figure}[htbp]
    \centering

    \begin{subfigure}[d]{0.49\linewidth}
        \centering
        \caption{}
        \includegraphics[width=\linewidth]{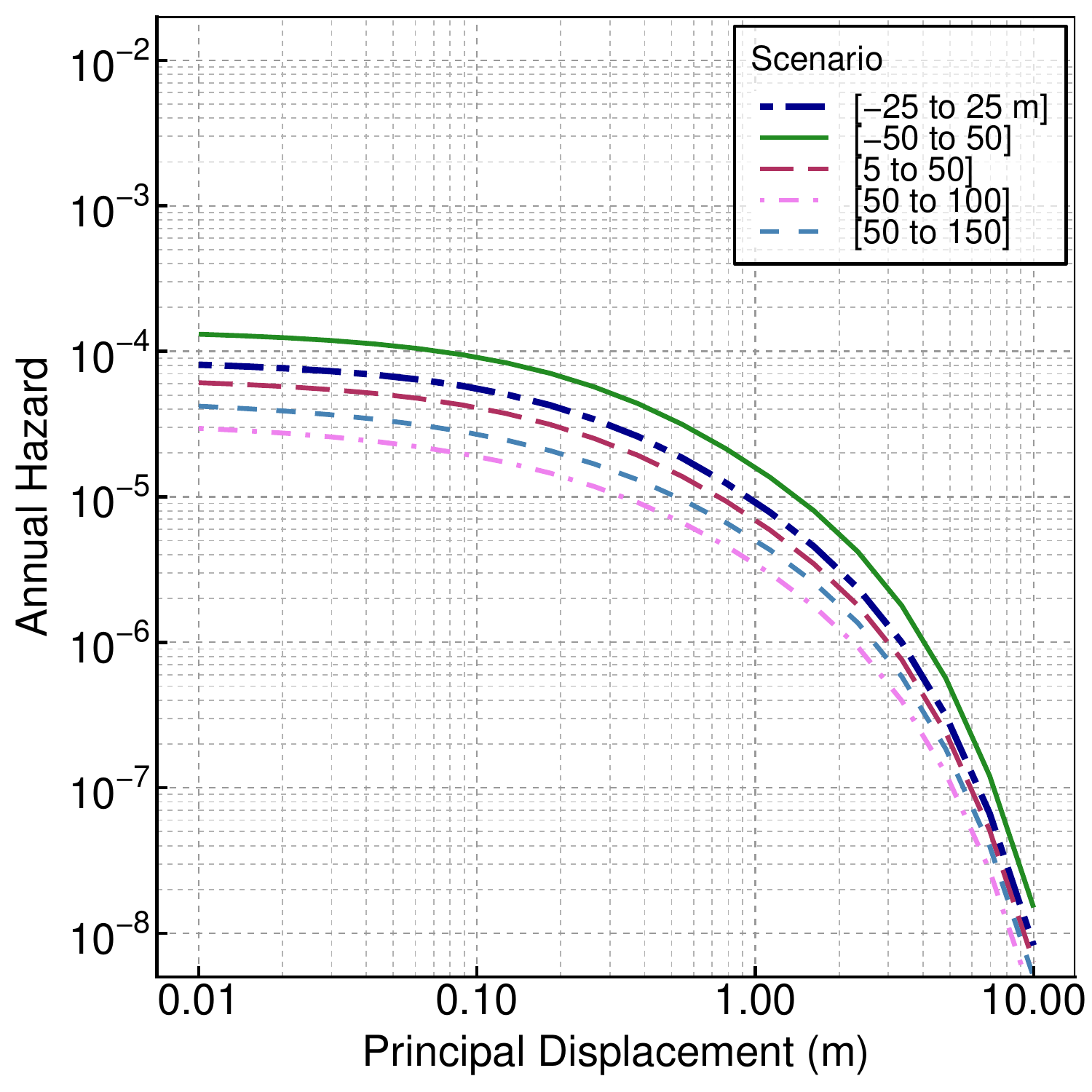}
        \label{fig:haz_dist_SCS_Kumamoto_2}
    \end{subfigure}
    \hfill
    \begin{subfigure}[d]{0.49\linewidth}
        \centering
        \caption{}
        \includegraphics[width=\linewidth]{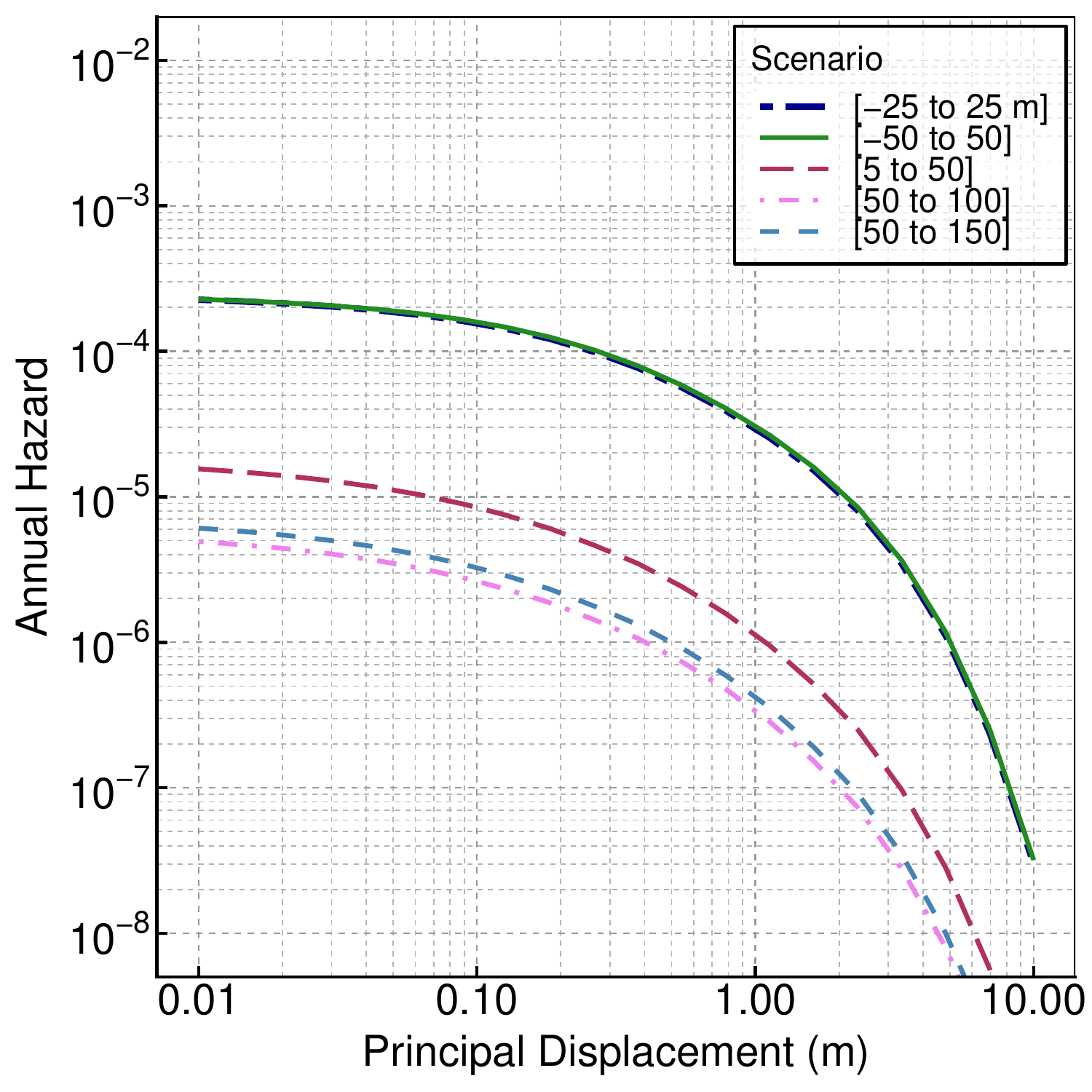}
        \label{fig:haz_dist_random_Kumamoto_2}
    \end{subfigure}
    \begin{subfigure}[d]{0.49\linewidth}
        \centering
        \caption{}
        \includegraphics[width=\linewidth]{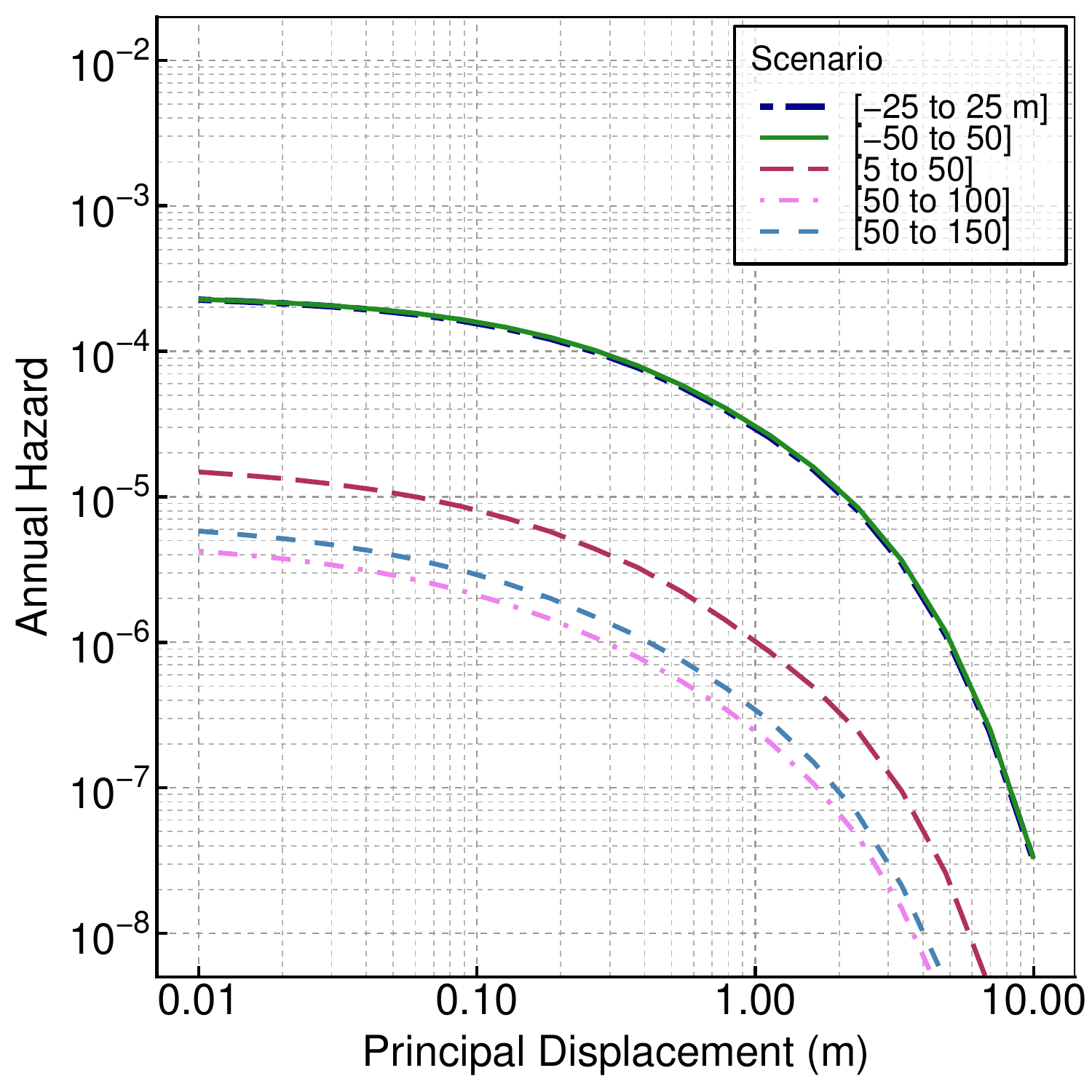}
        \label{fig:haz_dist_max_Kumamoto_2}
    \end{subfigure}

    \caption{Comparison of hazard curves for multi-rupture principal displacement framework (cumulative contribution of one or more rupture strands intersecting the structure) considering different structure-to-fault distance ranges for the IAEA benchmark (Kumamoto Case 2) \citep{valentini2021probabilistic, valentini2025iaea} for: (a) SCS fault trace, (b) random fault trace, and (c) max displacement fault trace.}
    \label{fig:Hazardcurves_dist_kumaoto_2}
\end{figure}

Figure~\ref{fig:Hazardcurves_dist_kumaoto_2} illustrates the influence of structural footprint on fault-displacement hazard for the Kumamoto Case 2 scenario under the different fault-trace definitions. The considered cases include structures centered on the nominal fault trace with widths of 50 m ($t=-25$ to $25 m$) and 100 m ($t=-50$ to $50 m$), a structure located on one side of the fault ($t=5$ to $50 m$), and two structures located farther from the fault with different widths ($t=50$ to $100 m$ and $t=50$ to $150 m$).

For the SCS trace definition (Figure~\ref{fig:haz_dist_SCS_Kumamoto_2}), the highest hazard levels are observed for structures centered on and located close to the fault trace. This behavior reflects the approximately symmetric distribution of rupture distances about the nominal trace, with the highest probability density concentrated near zero distance. For structures located at similar distances from the fault, hazard increases with structural width because a larger footprint increases the probability of intersecting one or more rupture strands. However, proximity to the fault remains the dominant controlling factor, as a shorter structure located near the fault may exhibit higher hazard than a longer structure located farther away.

Similar trends are observed for the randomly selected trace and maximum-principal-displacement trace definitions (Figures~\ref{fig:haz_dist_random_Kumamoto_2} and \ref{fig:haz_dist_max_Kumamoto_2}). However, these trace definitions produce a stronger concentration of rupture occurrence near zero distance, resulting in a more rapid decay of hazard with increasing distance from the fault. Consequently, the influence of structural width becomes less significant beyond short distances from the fault, and hazard is primarily controlled by proximity to the trace and the likelihood of rupture intersection. This reflects the fact that, in these definitions, one of the overlapping ruptures effectively defines the trace, leading to a higher likelihood of rupture occurrence at very short distances.

Equivalent comparisons for the remaining benchmark scenarios are provided in the electronic supplement (Figures S24-S26).

\section{Discussion and Conclusions}

A new probabilistic framework was introduced for modeling principal surface fault displacement within probabilistic fault displacement hazard analysis (PFDHA), moving beyond conventional approaches based on aggregate principal displacement. Evaluation against the IAEA benchmark \cite{valentini2021probabilistic, valentini2025iaea} cases demonstrated that aggregate formulations can lead to higher hazard estimates because they implicitly assume that a structure is simultaneously affected by the cumulative displacement of all overlapping ruptures. In contrast, the proposed framework models displacement at the individual rupture level, recognizing that a structure may intersect a single rupture without necessarily intersecting all overlapping ruptures at the same time. Even though the proposed methodology accounts for the possibility of crossing multiple overlapping principal ruptures, the benchmark evaluations showed that the dominant contribution to the hazard generally arises from single-rupture intersections, while simultaneous crossings of multiple ruptures occur with substantially lower probability. 
These results demonstrate the importance of incorporating the probability that a structure is affected by each individual rupture into PFDHA, rather than assuming that all overlapping ruptures contribute simultaneously, as assumed in aggregate principal displacement formulations that effectively represent a conservative upper-bound estimate of the hazard.

To define principal displacement at the rupture level, the framework was developed around three main components: the number of overlapping ruptures, the amplitude ratio of principal displacement, and the perpendicular distance from the fault trace. Using the expanded FDHI database \cite{sarmiento2025database}, the proposed formulation incorporates physically meaningful predictors, including earthquake magnitude, rupture curvature, and normalized displacement. In addition, the framework explicitly incorporates the spatial uncertainty of the fault trace through three alternative trace definitions, allowing the hazard to account for variability in rupture location relative to the structure. The IAEA benchmark evaluations demonstrated that these trace definitions can produce notably different hazard estimates depending on the structure location and fault geometry, highlighting the importance of representing spatial uncertainty in project-specific fault displacement assessments.

Overall, the proposed framework provides a physically informed and probabilistically robust basis for advancing PFDHA, while offering a scalable formulation that can be applied from single-rupture to multi-rupture scenarios. By modeling displacement at the level of individual ruptures, the methodology provides a more practical basis for incorporating rupture geometry, structural footprint, and crossing location into project-specific hazard evaluations for infrastructure systems located near or across active faults. A key direction for future work is the expansion of the proposed rupture-level displacement model to distributed ruptures and integration with probabilistic seismic hazard analysis (PSHA), enabling a unified hazard framework that accounts simultaneously for both ground shaking and permanent ground deformation. Such an approach would support more realistic seismic design and risk assessment for critical infrastructure exposed to surface fault rupture.

\section{Notation}

\begin{longtable}{l|p{10cm}}
\caption{Table of notation.}
\label{tab:notation}\\

\hline
{Symbol} & {Description} \\
\hline
\endfirsthead

\multicolumn{2}{c}{\tablename\ \thetable\ (continued)}\\
\hline
{Symbol} & {Description} \\
\hline
\endhead

\hline
\multicolumn{2}{r}{Continued on next page}\\
\endfoot

\hline
\endlastfoot

\multicolumn{2}{c}{\textbf{Coordinates}}\\
\hline
$u$ & Distance along the fault measured from one end of the rupture \\
$t$ & Normal distance to the nominal fault trace \\
\hline
\multicolumn{2}{c}{\textbf{Earthquake variables}}\\
\hline
$M$ & Earthquake magnitude \\
$SOF$ & Style of faulting \\
$\kappa$ & Fault curvature \\
$SRL$ & Surface rupture length\\
\hline
\multicolumn{2}{c}{\textbf{Hazard parameters}}\\
\hline
$N_{eq}$ & Number of earthquake scenarios\\
$\nu$ & Annual earthquake occurrence rate\\
$SR$ & Surface rupture\\
$GAP$ & Gap between fault segments\\
$u/SRL$ & Normalized along-fault location\\
$[t_1, t_2]$ & Structure span normal to the fault\\
\hline
\multicolumn{2}{c}{\textbf{Displacements}}\\
\hline
$D_{p,~\mathrm{agg}}$ & Aggregate principal displacement\\
${\bar D_{p,~\mathrm{agg}}}$ & Mean aggregate principal displacement\\
$D_p$ & Principal displacement \\
$D_{p,~\mathrm{cross}}$ & Sum of Principal displacements intersecting the structure\\
$\hat{D}_p$ & Normalized principal displacement \\
\hline
\multicolumn{2}{c}{\textbf{Rupture variables}}\\
\hline
$N_{p,~\mathrm{tot}}$ & Total number of overlapping principal ruptures \\
$N_{p,~{\mathrm{cross}}}$ & Number of overlapping principal ruptures intersecting the structure \\
$r_p$ & Ratio of principal displacement to aggregate principal displacement \\
$r_{p,~\mathrm{cross}}$ & Ratio of intersecting principal displacement to aggregate principal displacement\\
\hline
\multicolumn{2}{c}{\textbf{Regression parameters}}\\
\hline
\multicolumn{2}{@{}l}{{1. Number of overlapping ruptures}}\\
\hline
$P_1$ & Probability of a single overlapping principal rupture.\\
$f_{+}(N_{p,~\mathrm{tot}})$ & Zero-truncated Poisson probability mass function.\\
$\lambda$ & Poisson rate parameter for the number of overlapping principal ruptures.\\
$\theta$ & Bernoulli distribution parameter controlling the probability of a single overlapping rupture.\\
$a_0, a_1, a_2$ & Regression coefficients for the Poisson rate parameter $\lambda$.\\
$b_0, b_1, b_2$ & Regression coefficients for the Bernoulli probability parameter $\theta$.\\
\hline
\multicolumn{2}{@{}l}{{2. Amplitude ratio of displacement}}\\
\hline
$\alpha$ & First shape parameter of the Beta distribution.\\
$\beta$ & Second shape parameter of the Beta distribution.\\
$B(\alpha,\beta)$ & Beta function.\\
$\mu$ & Mean of the Beta distribution.\\
$d_0$ & Intercept of the shape-parameter model.\\
$d_1$ & Magnitude coefficient of the shape-parameter model.\\
\hline
\multicolumn{2}{@{}l}{{3. Normal distance to the fault}}\\
\hline
$\sigma$      & Scale parameter of the Student's $t$ distribution.\\
$\sigma_1$    & Scale parameter of the first Student's $t$ component.\\
$\sigma_2$    & Scale parameter of the second Student's $t$ component.\\
$\nu$         & Degrees of freedom of the Student's $t$ distribution.\\
$\nu_1$       & Degrees of freedom of the first Student's $t$ component.\\
$\nu_2$       & Degrees of freedom of the second Student's $t$ component.\\
$a$           & Mixing weight of the first Student's $t$ component.\\
$k_0$ & Intercept of the Student's $t$ scale-parameter model.\\
$k_1$ & Intercept of the second Student's $t$ component scale-parameter model.\\
$k_2$ & Magnitude coefficient of the Student's $t$ scale-parameter model.\\
$k_4$ & Normalized displacement coefficient of the Student's $t$ scale-parameter model.\\
$\lambda$ & Exponential decay coefficient of the Student's $t$ scale-parameter model.\\
\hline
\multicolumn{2}{c}{\textbf{Dataset variables}}\\
\hline
$N_e$ & Number of observations in an event\\
$w_{e,~i}$ & Weight assigned to observation $i$ in event $e$\\
\hline
\multicolumn{2}{c}{\textbf{Likelihood measures}}\\
\hline
$NLL_{min}$ & Minimum negative log-likelihood\\
$NLL_{p} (\theta_i)$ & Profile negative log-likelihood evaluated for parameter $\theta_i$\\
\hline

\end{longtable}

\section*{Acknowledgements}

Financial support provided by the Department of Civil, Structural and Environmental Engineering and the Institute of Bridge Engineering at the University at Buffalo is gratefully acknowledged.

The authors would like to thank Alexandra Sarmiento, Stephen Thompson, Christopher Madugo, and Albert Kottke for valuable discussions and suggestions during the early stages of this work.

\section*{Declaration of Competing Interests}

The authors declared no potential conflicts of interest with respect to the research, authorship, and/or publication of this article.

\section*{ORCID iD}

\href{https://orcid.org/0009-0000-4221-5200}{\includegraphics[scale=0.06]{orcid.pdf}\hspace{1mm}Saba Marmarchinia} \\
\href{https://orcid.org/0000-0001-6546-1340} {\includegraphics[scale=0.06]{orcid.pdf}\hspace{1mm}Grigorios Lavrentiadis}

\section*{Data and Resources}

The project code is available at:
\url{https://github.com/UB-AEGIS-R/ml26-principal-fault-rupture-model}. The processed data are publicly accessible through DesignSafe at:
\url{https://www.designsafe-ci.org/data/browser/public/designsafe.storage.published/PRJ-6366}

\bibliography{references_PrincipalDisplacement}

\ifnum\flagreview>0
    \listoffigures
\fi

\end{document}


\maketitle


This electronic supplement provides additional figures and evaluations related to the probabilistic modeling framework and hazard calculations presented in the main manuscript. 
For the number of overlapping ruptures component, the supplement includes evaluations of the influence of different parameters, including magnitude and curvature, as well as additional examined parameters that were found to have no influence on number  of ruptures. Model evaluations for different faulting styles are also presented. 
For the distance from the fault component, the supplement provides evaluations of the effects of earthquake magnitude and normalized displacement for all three fault-trace definitions (SCS, random trace, and maximum-displacement trace), together with additional evaluations showing the comparatively limited influence of fault curvature. The supplement further includes examples of fitted distributions for the alternative models and evaluations of the final selected models against all of the parameters. 
In addition, hazard-curve calculations are provided for all International Atomic Energy Agency benchmark (IAEA) scenarios \citep{valentini2021probabilistic,valentini2025iaea}, including comparisons across different fault-trace definitions and structural footprint configurations.

\begin{figure}[htbp!]
    \centering
    \includegraphics[width=0.5\linewidth]{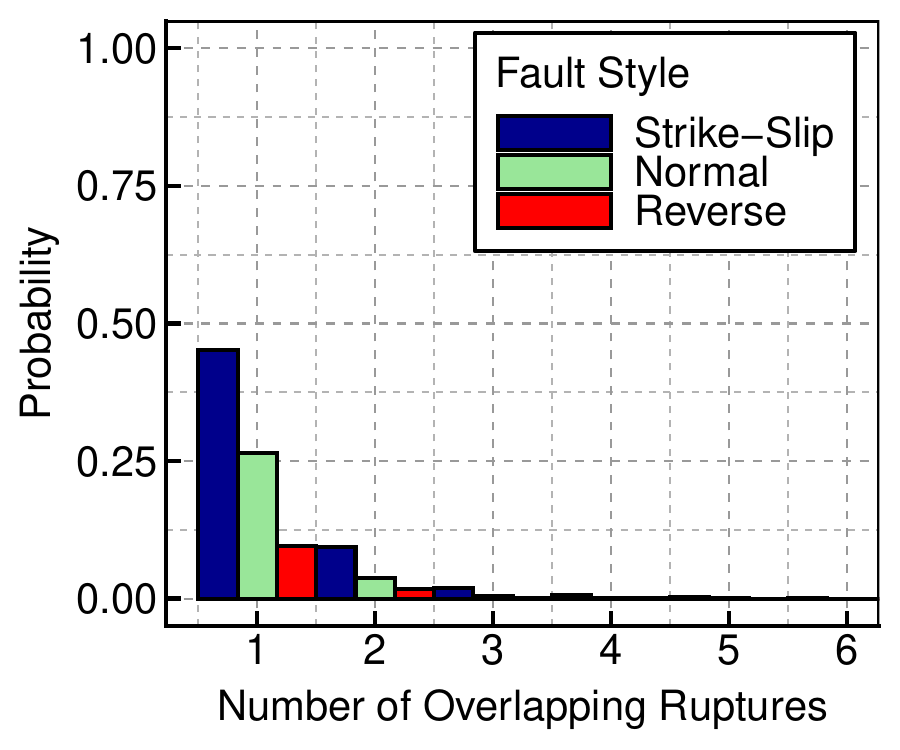}
    \caption{Empirical distributions of the number of overlapping principal ruptures for strike-slip, normal, and reverse faulting styles. The distributions are dominated by single-rupture observations, while the frequency of multiple overlapping rupture strands decreases rapidly with increasing rupture count.}
    \label{fig:Number_of_overlapping_principal_ruptures_by_fault_style}
\end{figure}

Figure \ref{fig:Number_of_overlapping_principal_ruptures_by_fault_style} displays the histogram of number of overlapping ruptures for each styles of fault.

\begin{figure}[htbp]
    \centering

    \begin{subfigure}[t]{0.48\linewidth}
        \centering
         \caption{}
        \includegraphics[width=\linewidth]{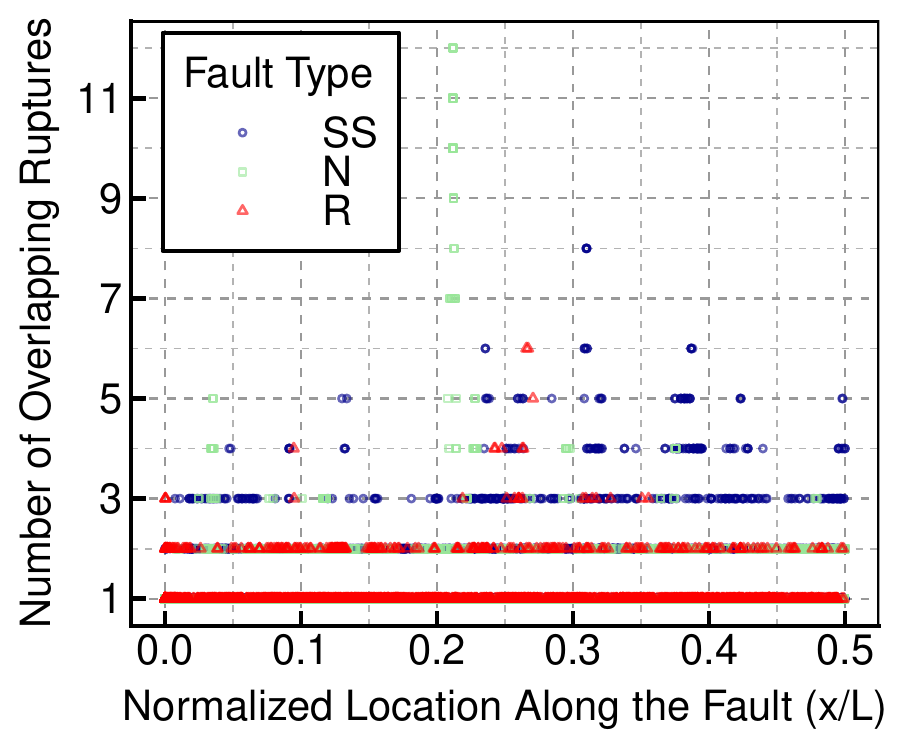}
        \label{fig:nprup_distnce_along_the_fault}
    \end{subfigure}
    \hfill
    \begin{subfigure}[t]{0.48\linewidth}
        \centering
        \caption{}
        \includegraphics[width=\linewidth]{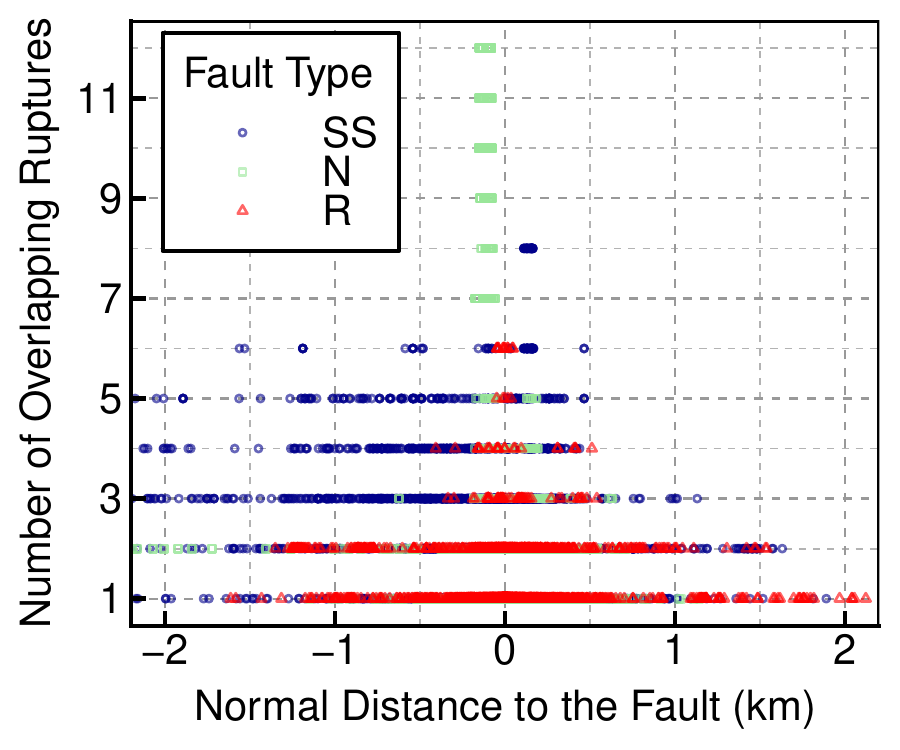}
        
        \label{fig:nprup_distance_from_the_fault}
    \end{subfigure}

    \caption{Influence of normalized along-strike distance and strike-normal distance on the number of overlapping principal ruptures for strike-slip (SS), normal (N), and reverse (R) faulting styles. (a) variation with normalized along-strike distance ($x/L$) and (b) variation with normal distance to the fault.}
    \label{fig:nprup_combined_along_and_away_distance}
\end{figure}

Figure \ref{fig:nprup_combined_along_and_away_distance} explores the influence of different parameters on the number of overlapping ruptures.

Figure \ref{fig:z_Nr} presents the randomized quantile residuals ($z$) for the number of overlapping ruptures model, including their histogram and their trends with respect to earthquake magnitude ($M$) and fault curvature ($\log(\kappa)$).

Figure \ref{fig:z_Nr_ut} presents the randomized quantile residuals ($z$) for the number of overlapping ruptures model, including their trends with respect to normalized location along the fault ($x/L$) and normal distance from the fault.

\begin{figure}[htbp!]
    \centering

    \begin{subfigure}[b]{0.48\linewidth}
        \centering
        \caption{}
        \includegraphics[width=\linewidth]{figures/Distance/Distance_SCS/Magnitude_vs_distance_different_fault_type.pdf}
    \end{subfigure}
    \hfill
    \begin{subfigure}[b]{0.48\linewidth}
        \centering
        \caption{}
        \includegraphics[width=\linewidth]{figures/Distance/Distance_SCS/Normalized_displacement_vs_distance_different_fault_type.pdf}
    \end{subfigure}

    \vspace{0.3cm}

    \begin{subfigure}[b]{0.48\linewidth}
        \centering
        \caption{}
        \includegraphics[width=\linewidth]{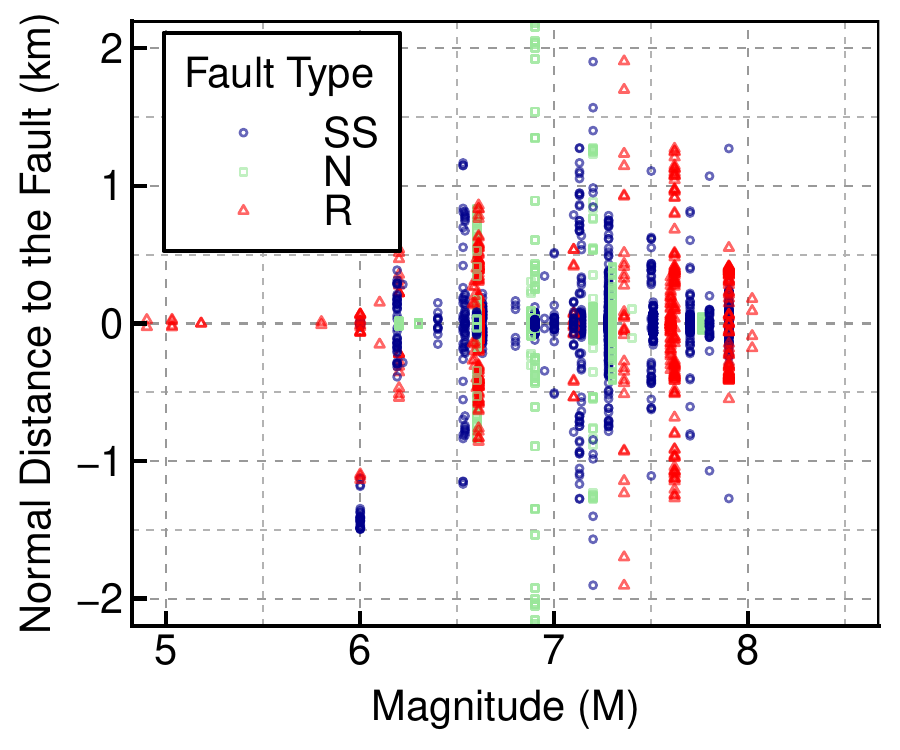}
    \end{subfigure}
    \hfill
    \begin{subfigure}[b]{0.48\linewidth}
        \centering
        \caption{}
        \includegraphics[width=\linewidth]{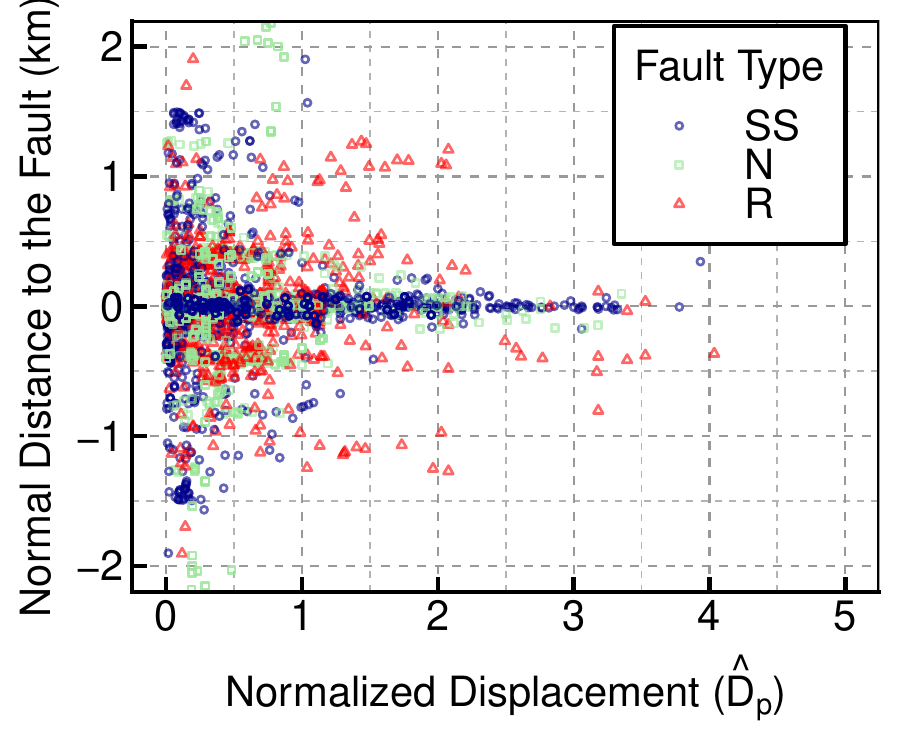}
    \end{subfigure}

    \vspace{0.3cm}

    \begin{subfigure}[b]{0.48\linewidth}
        \centering
        \caption{}
        \includegraphics[width=\linewidth]{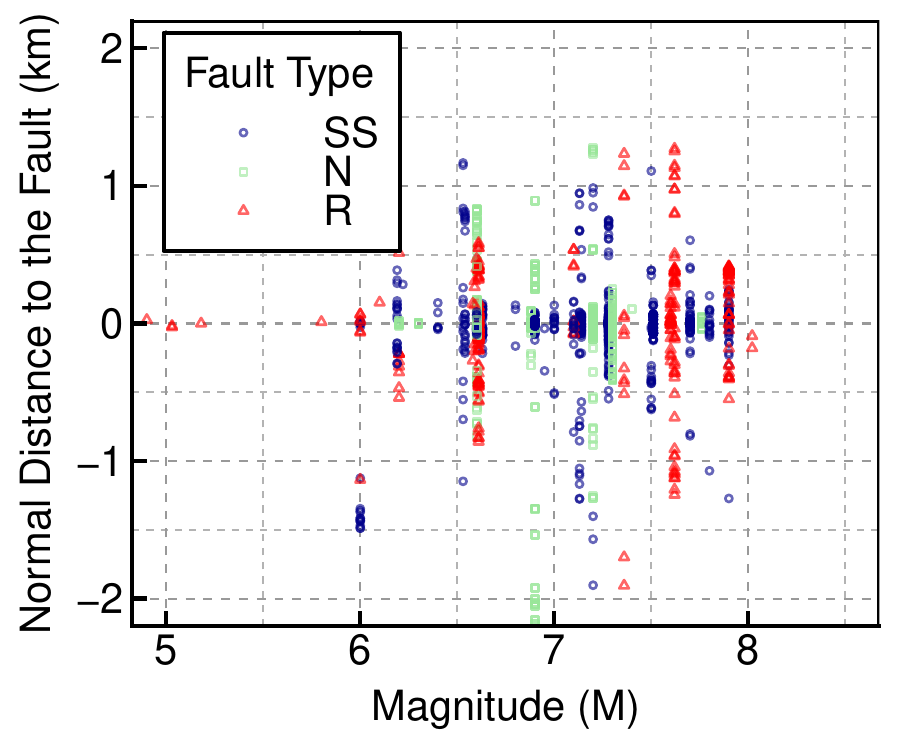}
    \end{subfigure}
    \hfill
    \begin{subfigure}[b]{0.48\linewidth}
        \centering
        \caption{}
        \includegraphics[width=\linewidth]{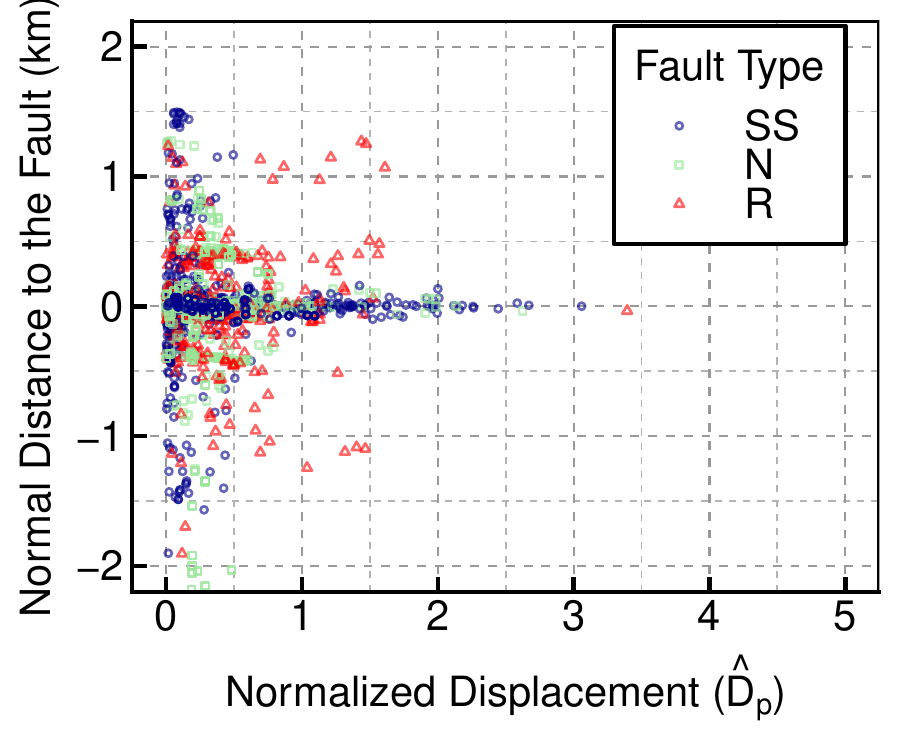}
    \end{subfigure}

    \caption{Influence of earthquake magnitude and normalized principal displacement on the normal distance to different definition of fault trace for strike-slip (SS), normal (N), and reverse (R) faulting styles. Rows correspond to trace definitions (SCS, randomly selected trace, and maximum principal displacement trace), while columns represent the influencing parameters: earthquake magnitude ($M$) (left) and normalized displacement $(\hat{D}_p)$ (right).}
    
    \label{fig:dist_mag_norm_grid}
\end{figure}

Figure \ref{fig:dist_mag_norm_grid} illustrates the influence of earthquake magnitude and normalized displacement on the perpendicular distance of principal displacement from the fault for different fault-trace definitions.

\begin{figure}[htbp]
    \centering

    \begin{subfigure}[b]{0.48\linewidth}
        \centering
        \caption{}
        \includegraphics[width=\linewidth]{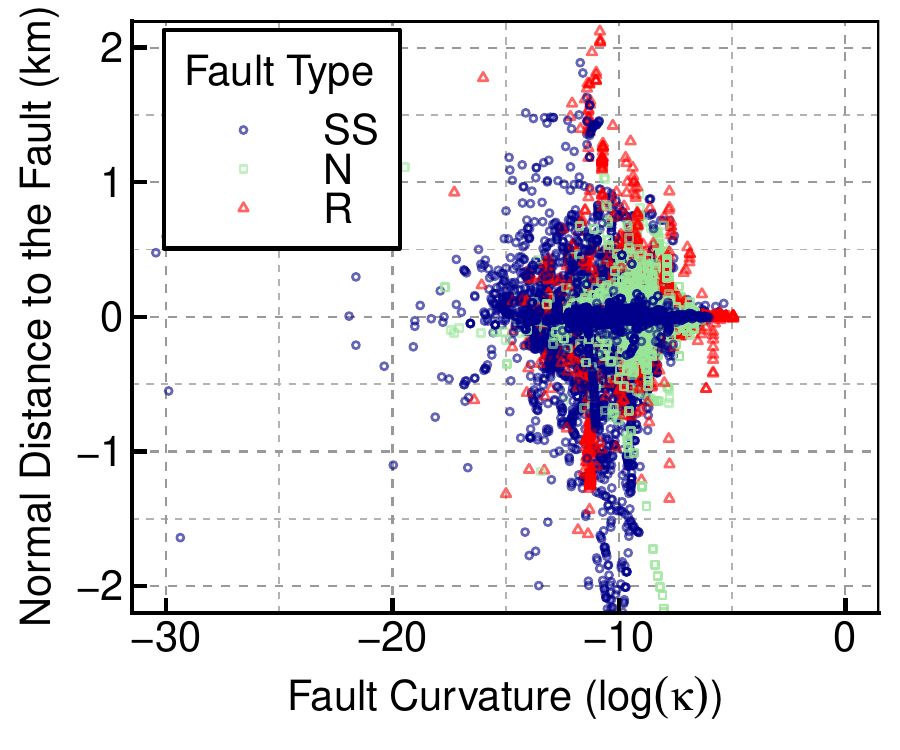}
    \end{subfigure}
    \hfill
    \begin{subfigure}[b]{0.48\linewidth}
        \centering
        \caption{}
        \includegraphics[width=\linewidth]{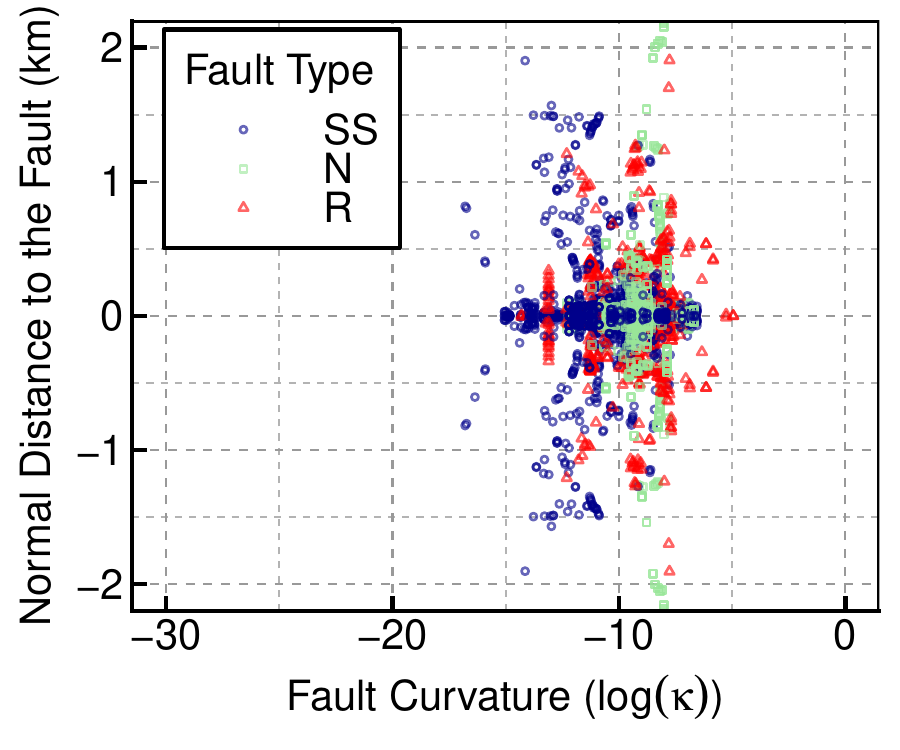}
    \end{subfigure}

    \vspace{0.3cm}

    \begin{subfigure}[b]{0.48\linewidth}
        \centering
        \caption{}
        \includegraphics[width=\linewidth]{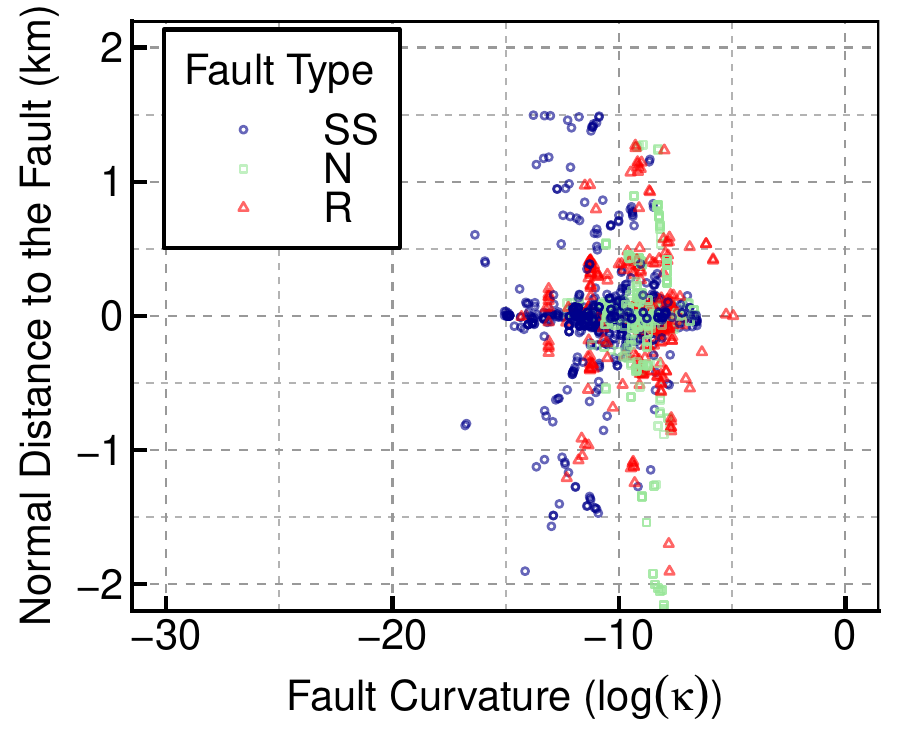}
    \end{subfigure}

    \caption{Influence of curvature $(\log(\kappa))$ on the normal distance to different definition of fault trace for strike-slip (SS), normal (N), and reverse (R) faulting styles. (a) SCS trace, (b) randomly selected trace, and (c) maximum principal displacement trace.}
    
    \label{fig:dist_curvature_grid}
\end{figure}

Figure \ref{fig:dist_curvature_grid} presents the relationship between fault curvature and the perpendicular distance of principal displacement from the fault for different fault-trace definitions.

\begin{figure}[htbp]
    \centering

    \begin{subfigure}[d]{0.49\linewidth}
        \centering
        \caption{}
        \includegraphics[width=\linewidth]{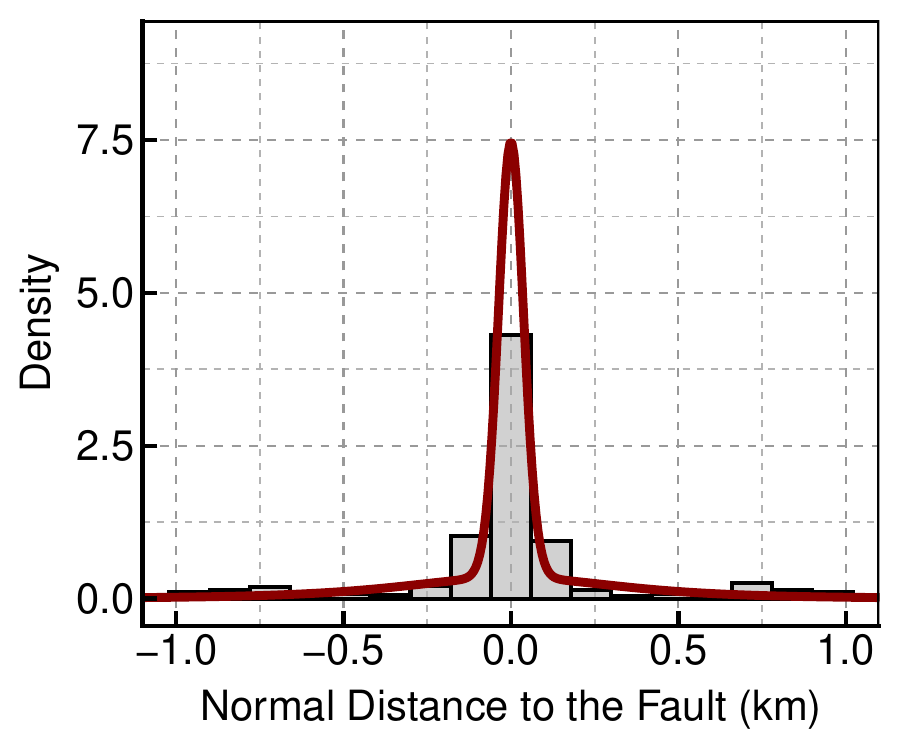}
        \label{fig:Fitted_t_SS-6.5<M<7.25-D:0-1}
    \end{subfigure}
    \hfill
    \begin{subfigure}[d]{0.49\linewidth}
        \centering
        \caption{}
        \includegraphics[width=\linewidth]{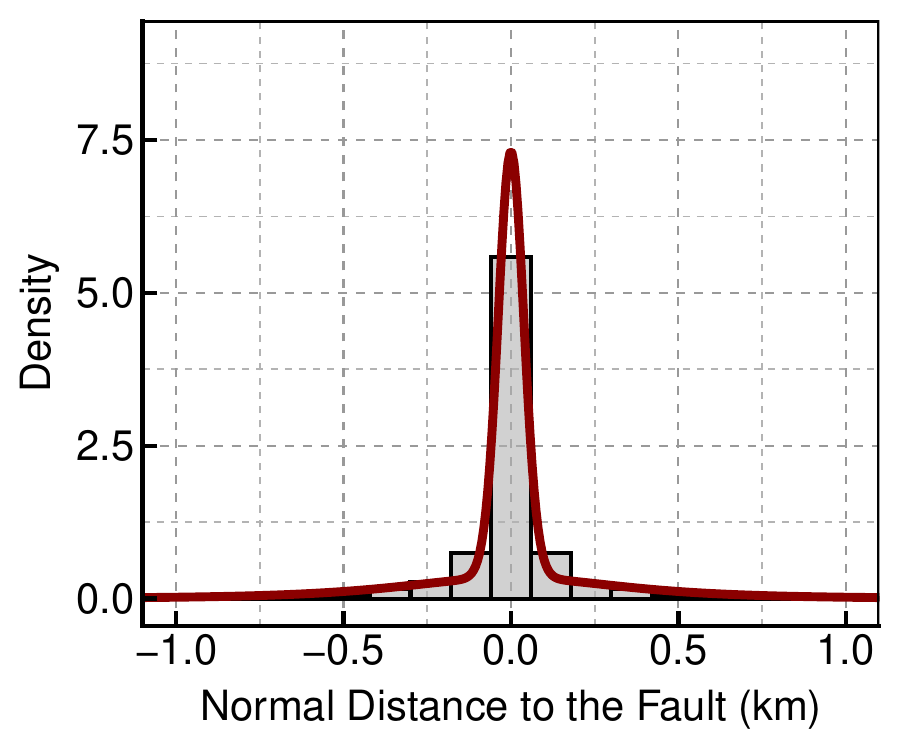}
        \label{fig:Fitted_t_SS-7.25<M-D:0-1}
    \end{subfigure}
    \begin{subfigure}[d]{0.49\linewidth}
        \centering
        \caption{}
        \includegraphics[width=\linewidth]{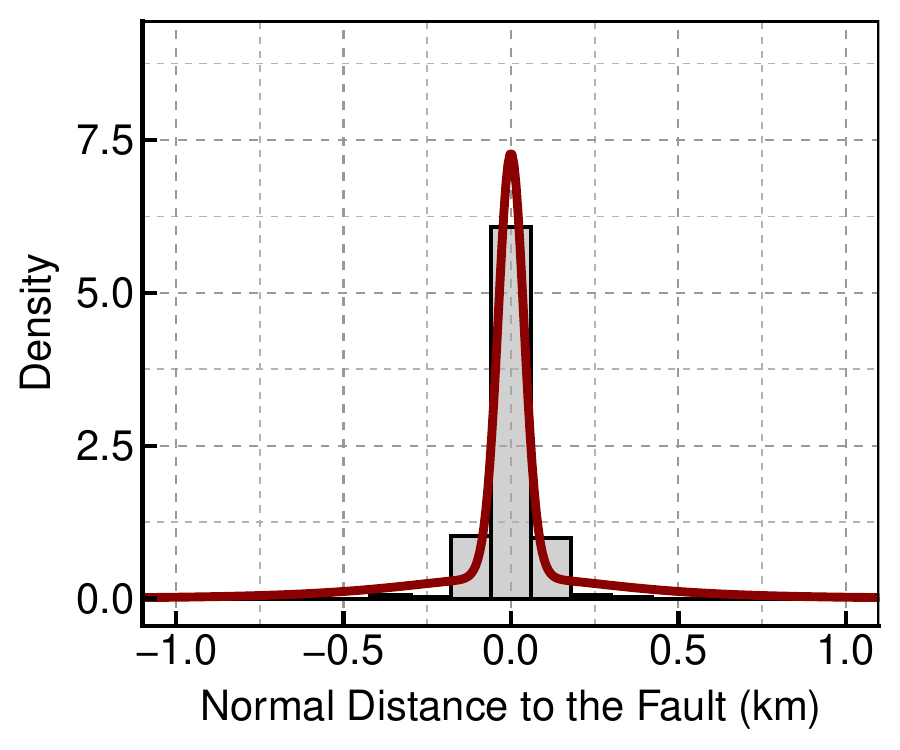}
        \label{fig:Fitted_t_SS-7.25<M-D:1-10}
    \end{subfigure}
    \begin{subfigure}[d]{0.49\linewidth}
        \centering
        \caption{}
        \includegraphics[width=\linewidth]{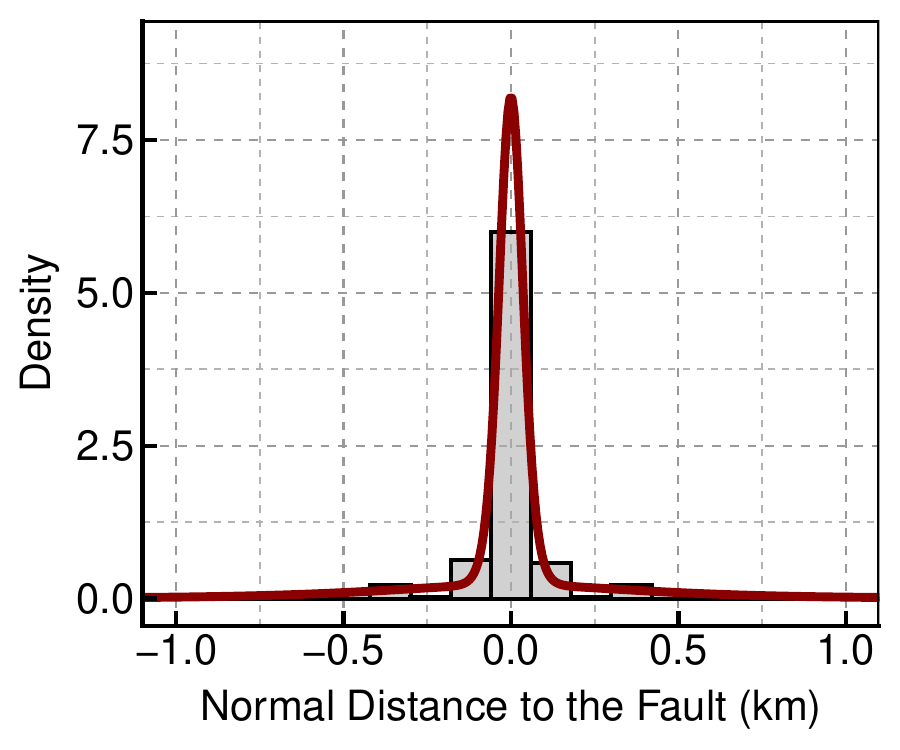}
        \label{fig:Fitted_t_NR-6.5<M<7.25-D:0-1}
    \end{subfigure}

    \caption{Comparison of the observed and modeled distributions of the normal distance to the randomly selected fault trace for different faulting styles and magnitude ($M$) - displacement $\hat{(D_p)}$ bins. Histograms represent the empirical distributions, and red curves denote the probability density functions predicted by the fitted mix of two Student's t distributions. (a) strike-slip faults with $6.5 < M < 7.25$ and $\hat{(D_p)} < 1$, (b) strike-slip faults with $7.25 < M$ and $\hat{(D_p)} < 1$, (c) strike-slip faults with $7.25 < M$ and $1 < \hat{(D_p)}$, and (d) normal and reverse faults with $6.5 < M < 7.25$ and $\hat{(D_p)} < 1$.}
    \label{fig:Fitted_t_distribution_random}
\end{figure}

Figure \ref{fig:Fitted_t_distribution_random} illustrates examples of fitted Student’s t-distributions for rupture location under the random trace definition for different magnitude, normalized displacement, and fault-style bins.

\begin{figure}[htbp]
    \centering

    \begin{subfigure}[d]{0.49\linewidth}
        \centering
        \caption{}
        \includegraphics[width=\linewidth]{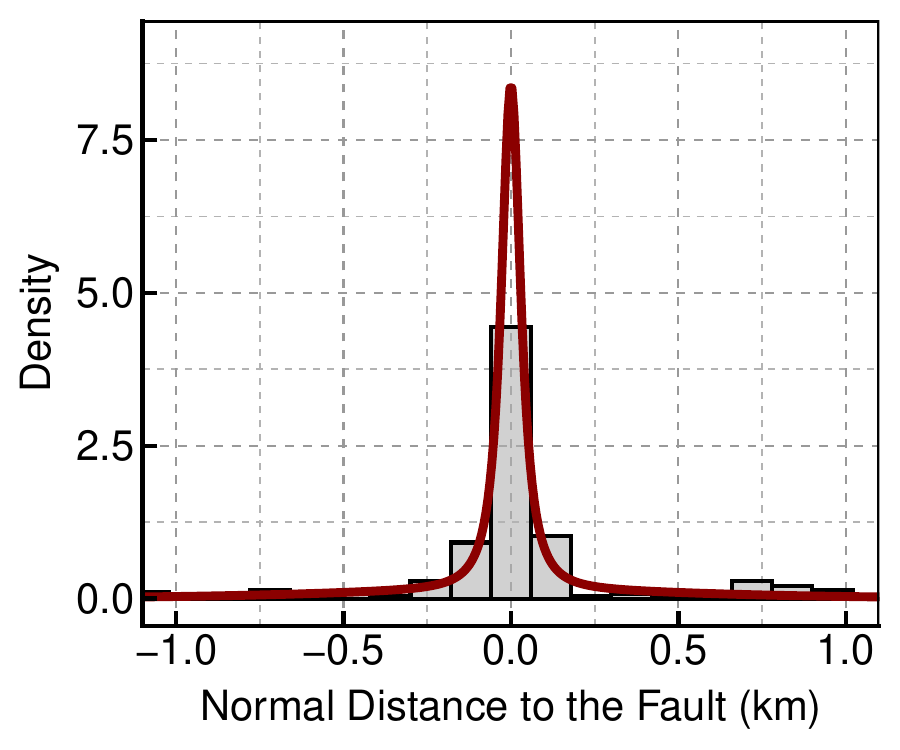}
        \label{fig:Fitted_t_SS-6.5<M<7.25-D:0-1}
    \end{subfigure}
    \hfill
    \begin{subfigure}[d]{0.49\linewidth}
        \centering
        \caption{}
        \includegraphics[width=\linewidth]{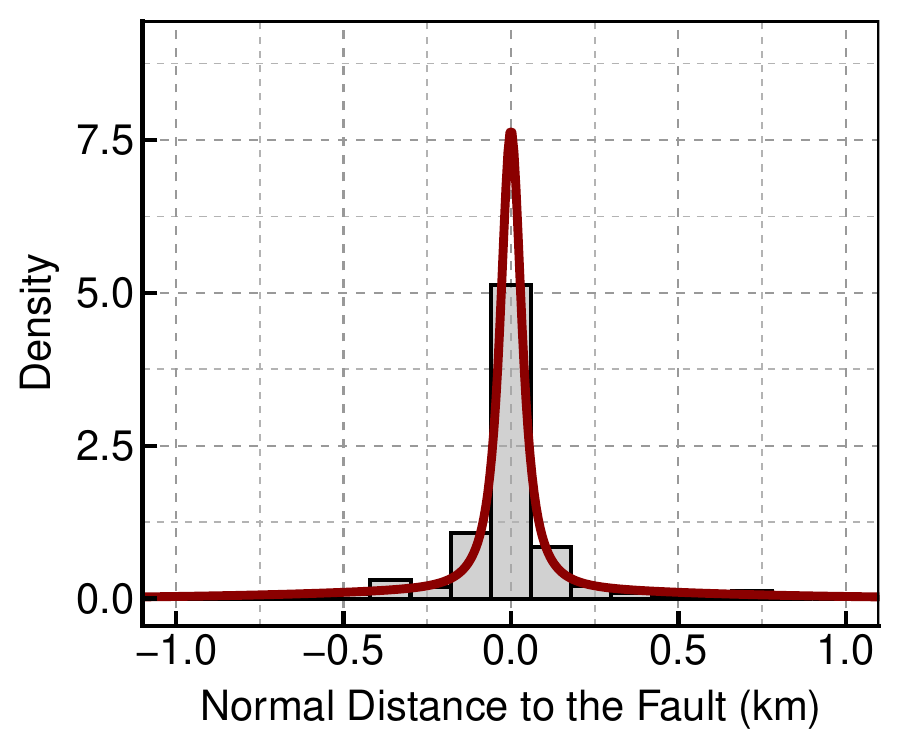}
        \label{fig:Fitted_t_SS-7.25<M-D:0-1}
    \end{subfigure}
    \begin{subfigure}[d]{0.49\linewidth}
        \centering
        \caption{}
        \includegraphics[width=\linewidth]{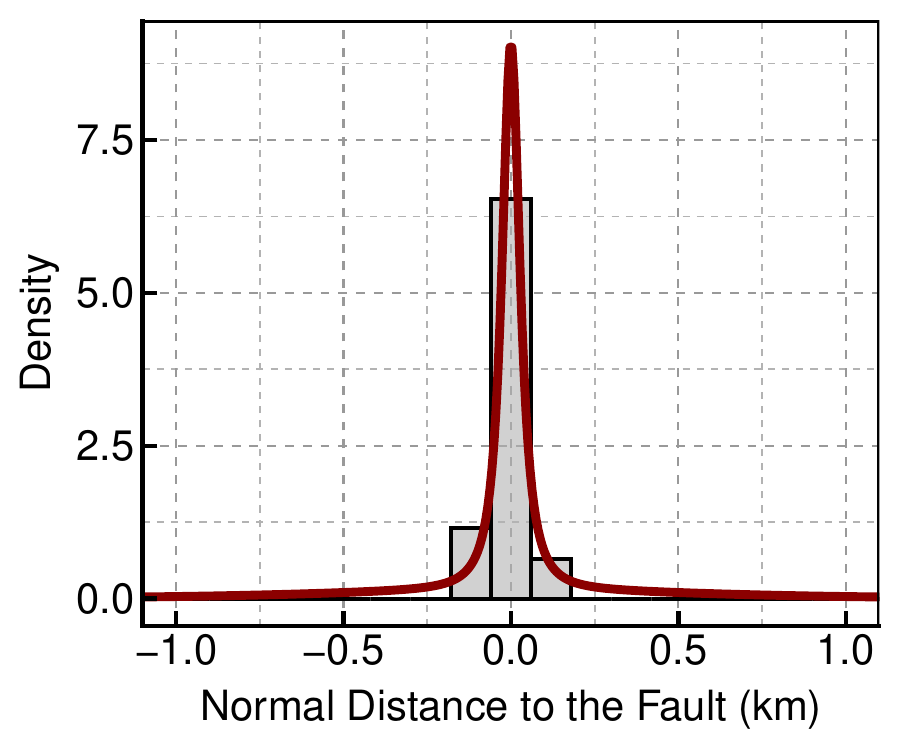}
        \label{fig:Fitted_t_SS-7.25<M-D:1-10}
    \end{subfigure}
    \begin{subfigure}[d]{0.49\linewidth}
        \centering
        \caption{}
        \includegraphics[width=\linewidth]{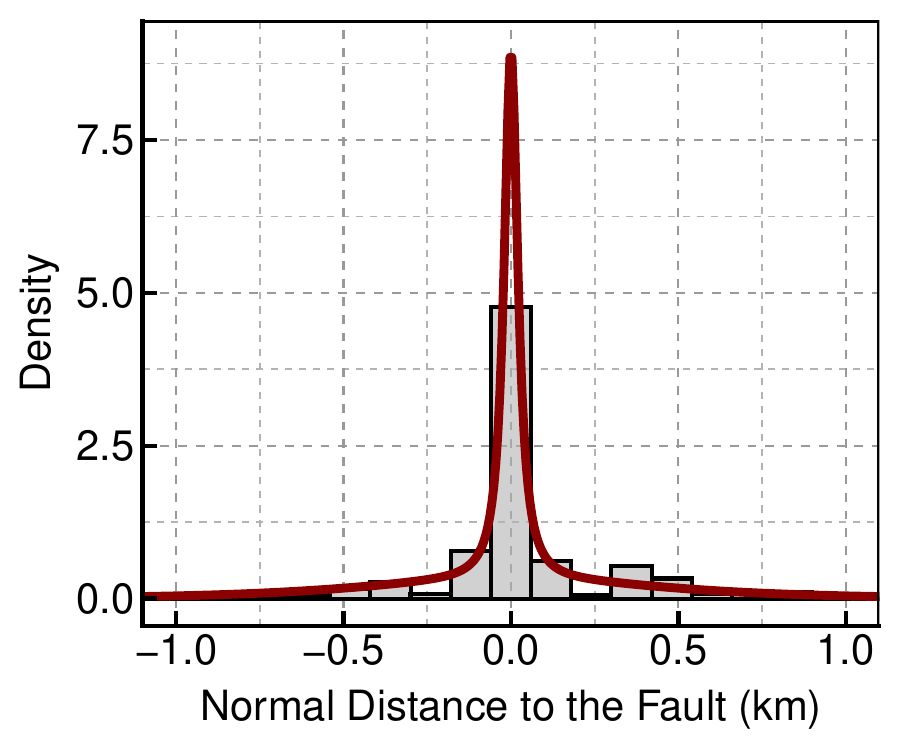}
        \label{fig:Fitted_t_NR-6.5<M<7.25-D:0-1}
    \end{subfigure}

    \caption{Comparison of the observed and modeled distributions of the normal distance to the maximum principal displacement fault trace for different faulting styles and magnitude ($M$) - displacement $\hat{(D_p)}$ bins. Histograms represent the empirical distributions, and red curves denote the probability density functions predicted by the fitted mix of two Student's t distributions. (a) strike-slip fault with $6.5 < M < 7.25$ and $\hat{(D_p)} < 1$, (b) strike-slip fault with $7.25 < M$ and $\hat{(D_p)} < 1$, (c) strike-slip fault with $7.25 < M$ and $1 < \hat{(D_p)}$, and (d) normal and reverse fault with $6.5 < M < 7.25$ and $\hat{(D_p)} < 1$.}
    \label{fig:Fitted_t_distribution_max}
\end{figure}

Figure \ref{fig:Fitted_t_distribution_max} illustrates examples of fitted Student’s t-distributions for rupture location under the max displacement trace definition for different magnitude, normalized displacement, and fault-style bins.

Figure \ref{fig:dist_z_scs} presents the evaluation of standardized quantiles ($z$) for the SCS trace distance model including their histogram and their trends with respect to earthquake magnitude ($M$) and normalized displacement ($\hat{D_p}$).

Figure \ref{fig:dist_z_random} presents the evaluation of standardized quantiles ($z$) for the randomly selected trace distance model including their histogram and their trends with respect to earthquake magnitude ($M$) and normalized displacement ($\hat{D_p}$).

Figure \ref{fig:dist_z_max} presents the evaluation of standardized quantiles ($z$) for the maximum principal displacement trace distance model including their histogram and their trends with respect to earthquake magnitude ($M$) and normalized displacement ($\hat{D_p}$).

Figure \ref{fig:dist_z_curv} presents the evaluation of standardized quantile residuals ($z$) for the three normal distance models as a function of fault curvature.

\begin{figure}[htbp]
    \centering

    \begin{subfigure}[d]{0.49\linewidth}
        \centering
        \caption{}
        \includegraphics[width=\linewidth]{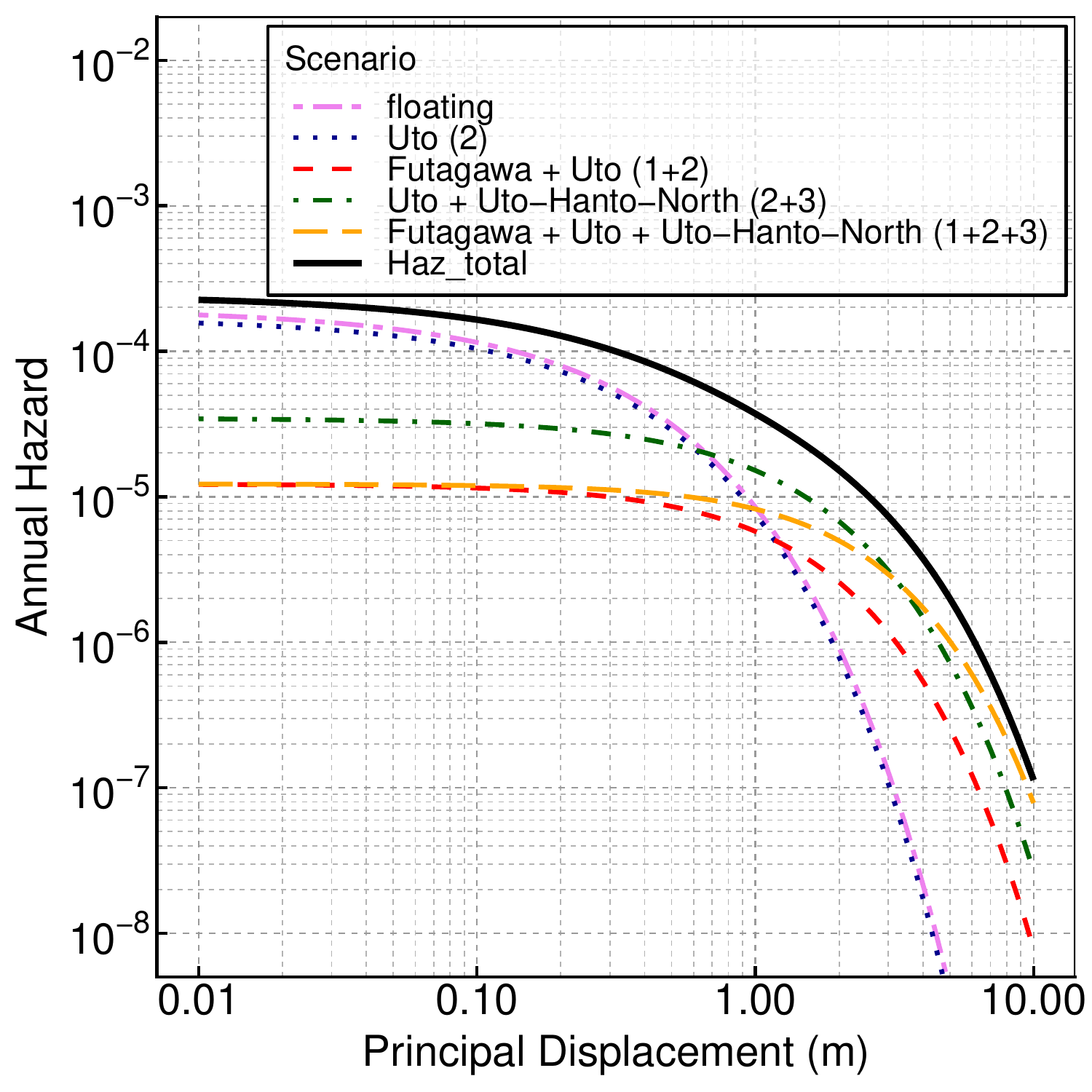}
        \label{fig:haz_agg_kumamoto_2}
    \end{subfigure}
    \hfill
    \begin{subfigure}[d]{0.49\linewidth}
        \centering
        \caption{}
        \includegraphics[width=\linewidth]{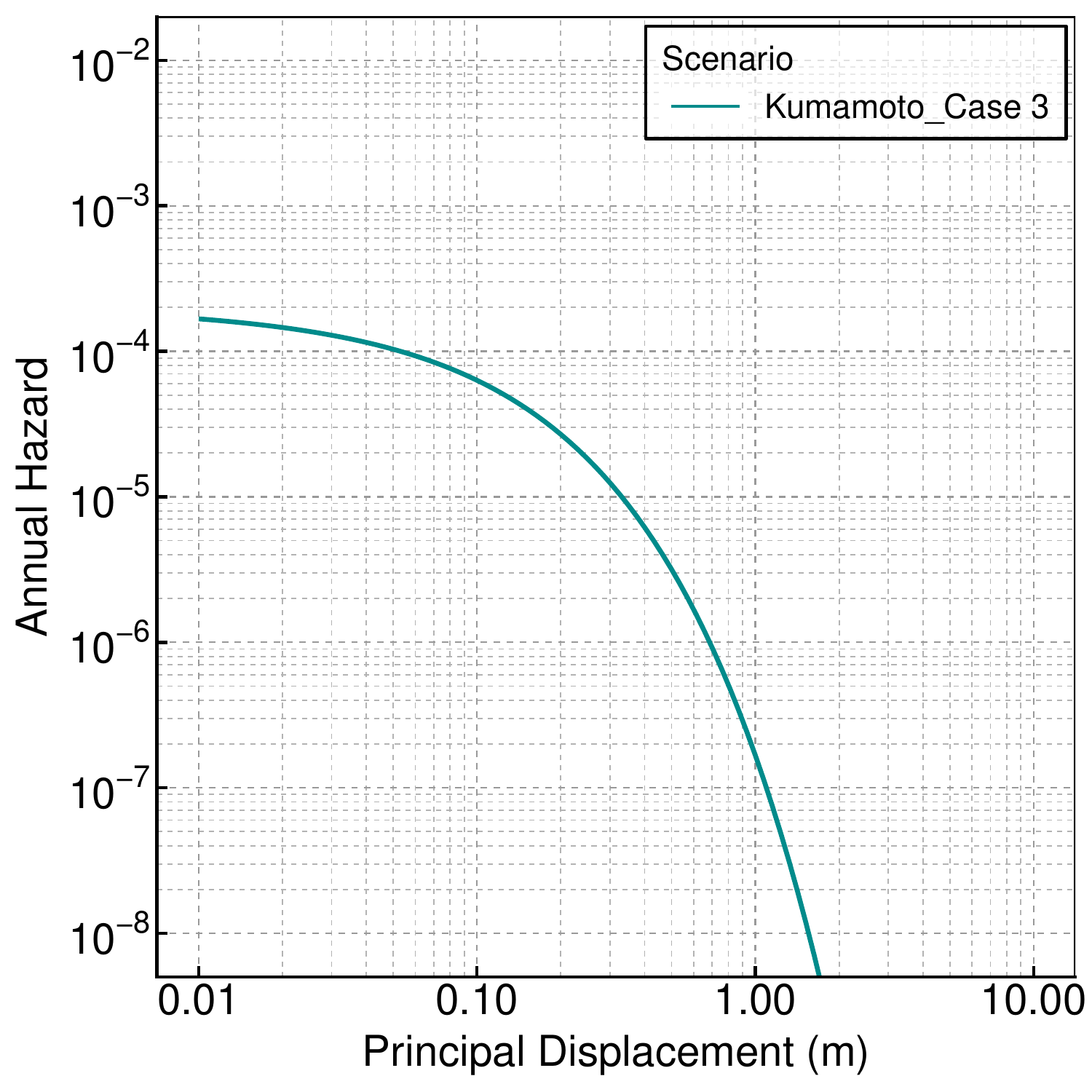}
        \label{fig:haz_agg_kumamoto_2}
    \end{subfigure}
    \begin{subfigure}[d]{0.49\linewidth}
        \centering
        \caption{}
        \includegraphics[width=\linewidth]{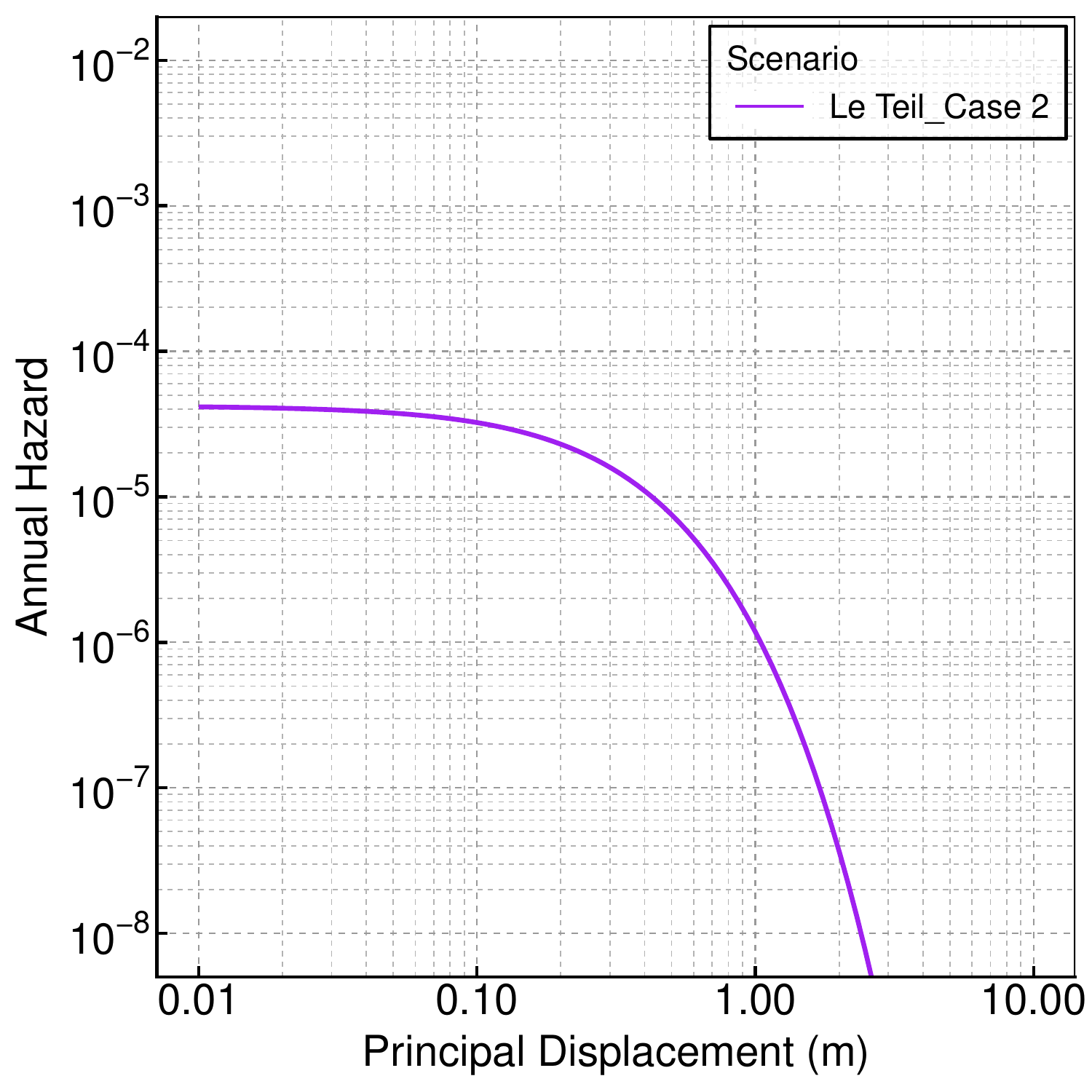}
        \label{fig:haz_agg_LeTeil_2}
    \end{subfigure}
    \begin{subfigure}[d]{0.49\linewidth}
        \centering
        \caption{}
        \includegraphics[width=\linewidth]{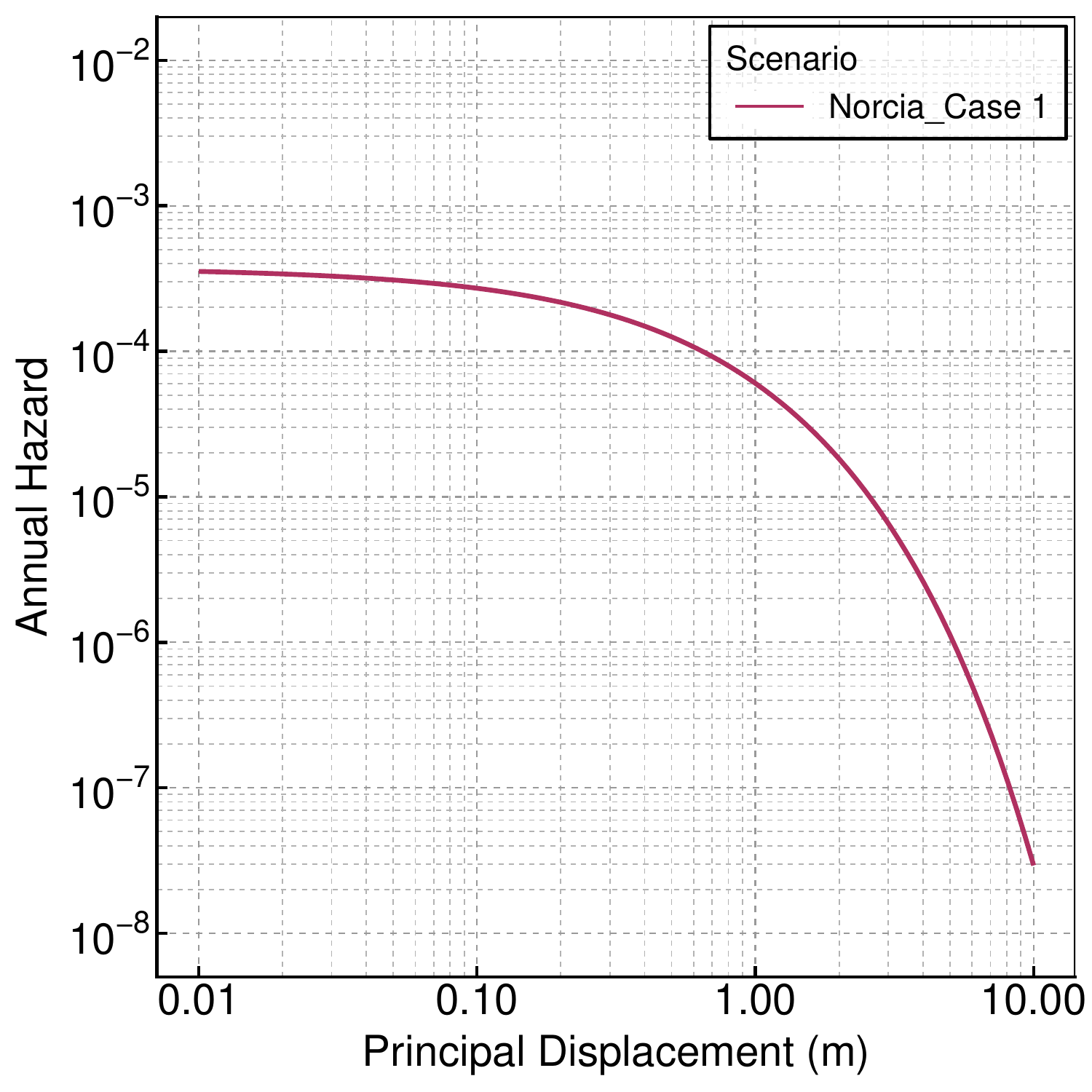}
        \label{fig:haz_agg_Norcia_1}
    \end{subfigure}

    \caption{Evaluation of the IAEA benchmark hazard curves for aggregate principal displacement based on \cite{valentini2021probabilistic, valentini2025iaea}: (a) Kumamoto case 2, (b) Kumamoto case 3, (c) Le Teil case 2, and (d) Norcia case 1.}
    \label{fig:Hazardcurves_Aggregate}
\end{figure}

Figure \ref{fig:Hazardcurves_Aggregate} presents the hazard curve evaluation of the IAEA benchmark (\cite{valentini2021probabilistic} and \cite{valentini2025iaea}) for aggregate principal displacement.

\begin{figure}[htbp]
    \centering

    \begin{subfigure}[d]{0.49\linewidth}
        \centering
        \caption{}
        \includegraphics[width=\linewidth]{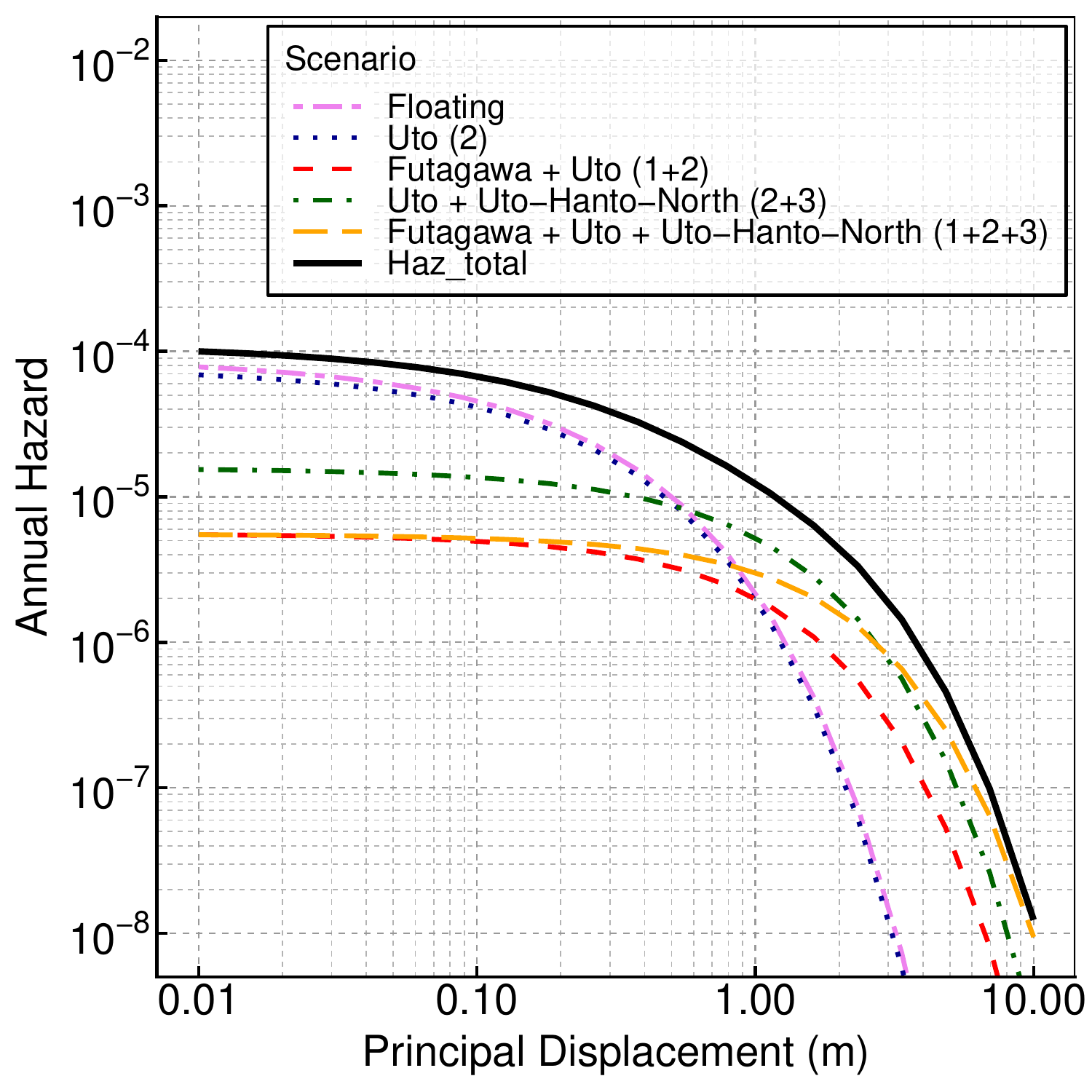}
        \label{fig:haz_prncp_single_SCS_kumamoto_2}
    \end{subfigure}
    \hfill
    \begin{subfigure}[d]{0.49\linewidth}
        \centering
        \caption{}
        \includegraphics[width=\linewidth]{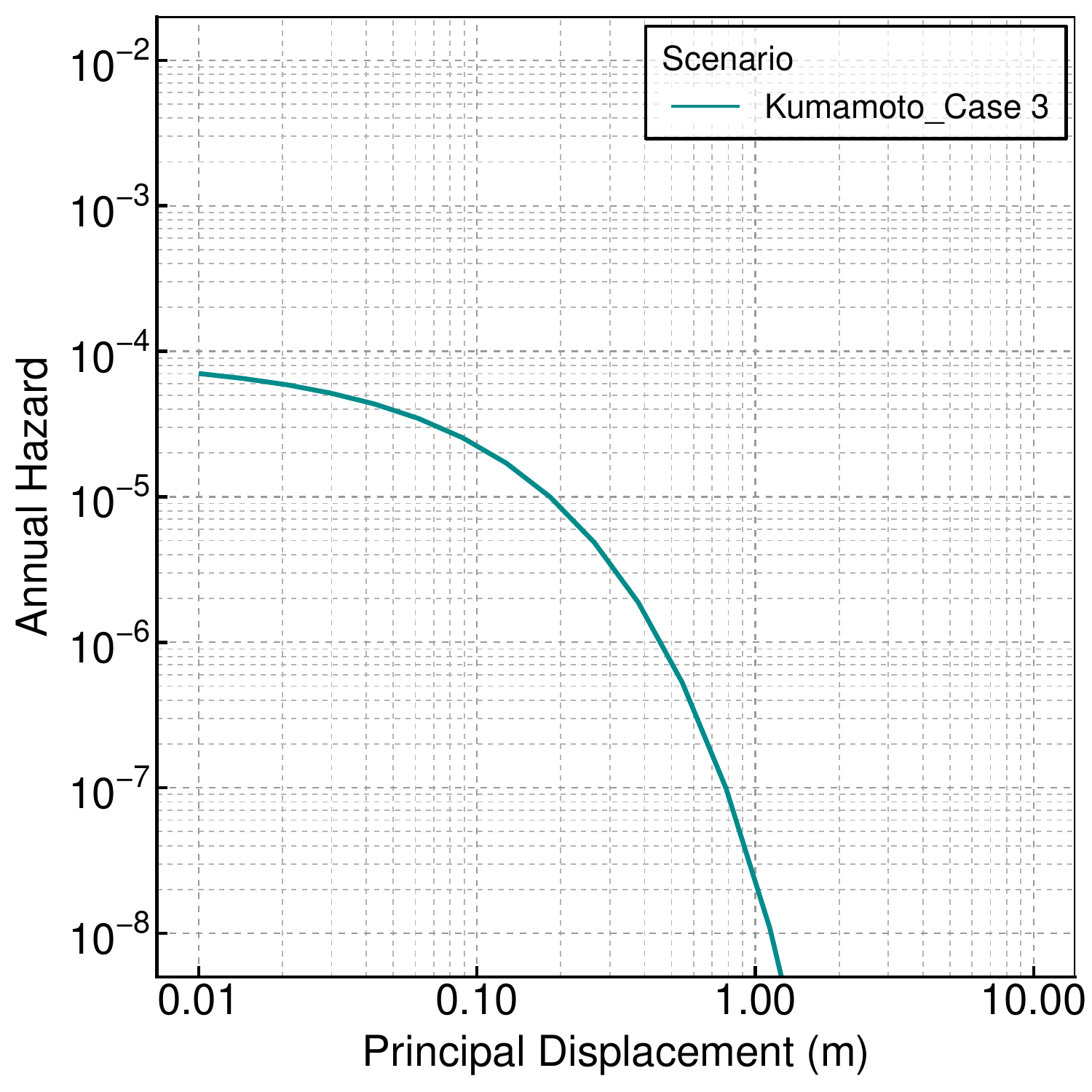}
        \label{fig:haz_prncp_single_SCS_kumamoto_3}
    \end{subfigure}
    \begin{subfigure}[d]{0.49\linewidth}
        \centering
        \caption{}
        \includegraphics[width=\linewidth]{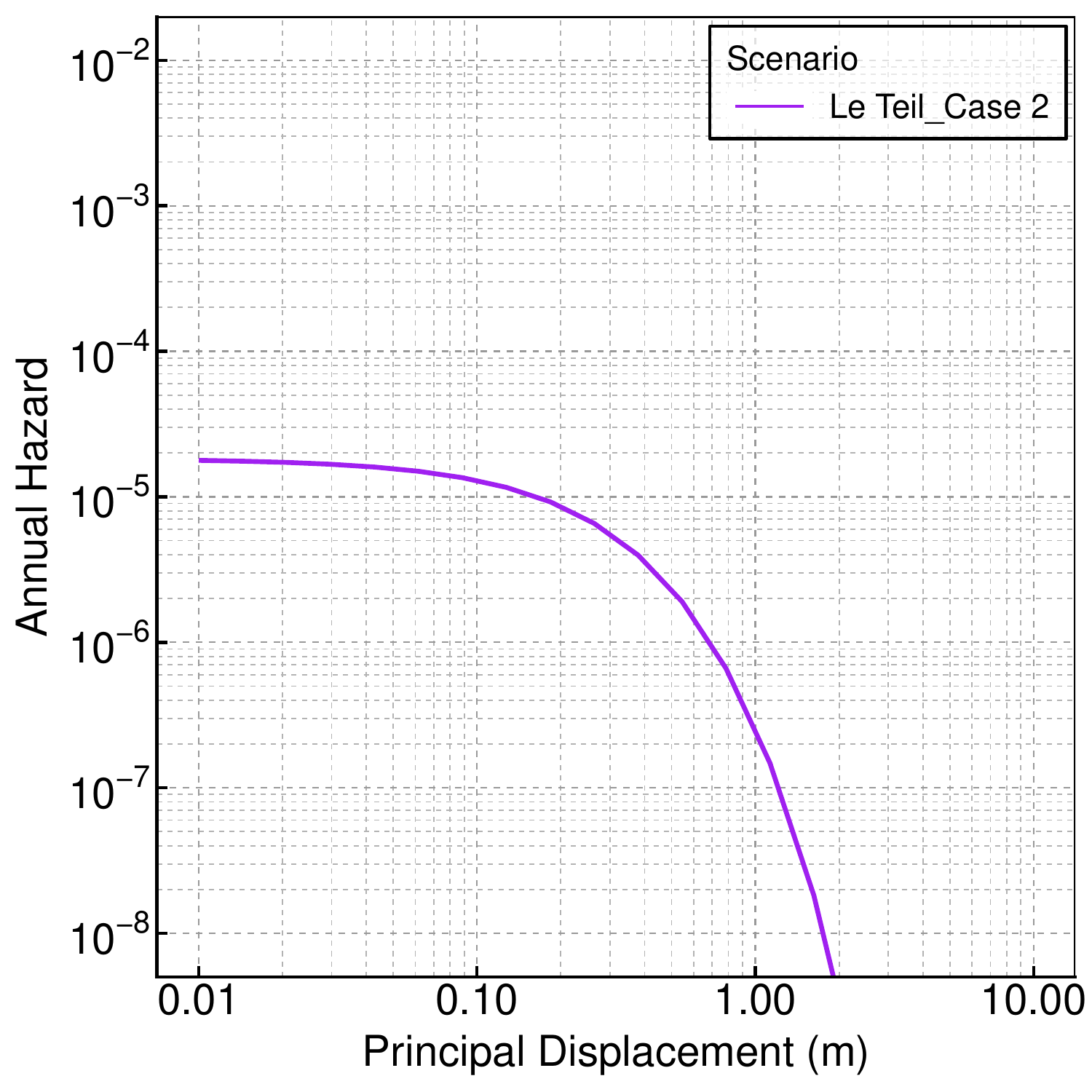}
        \label{fig:haz_prncp_single_SCS_LeTeil_2}
    \end{subfigure}
    \begin{subfigure}[d]{0.49\linewidth}
        \centering
        \caption{}
        \includegraphics[width=\linewidth]{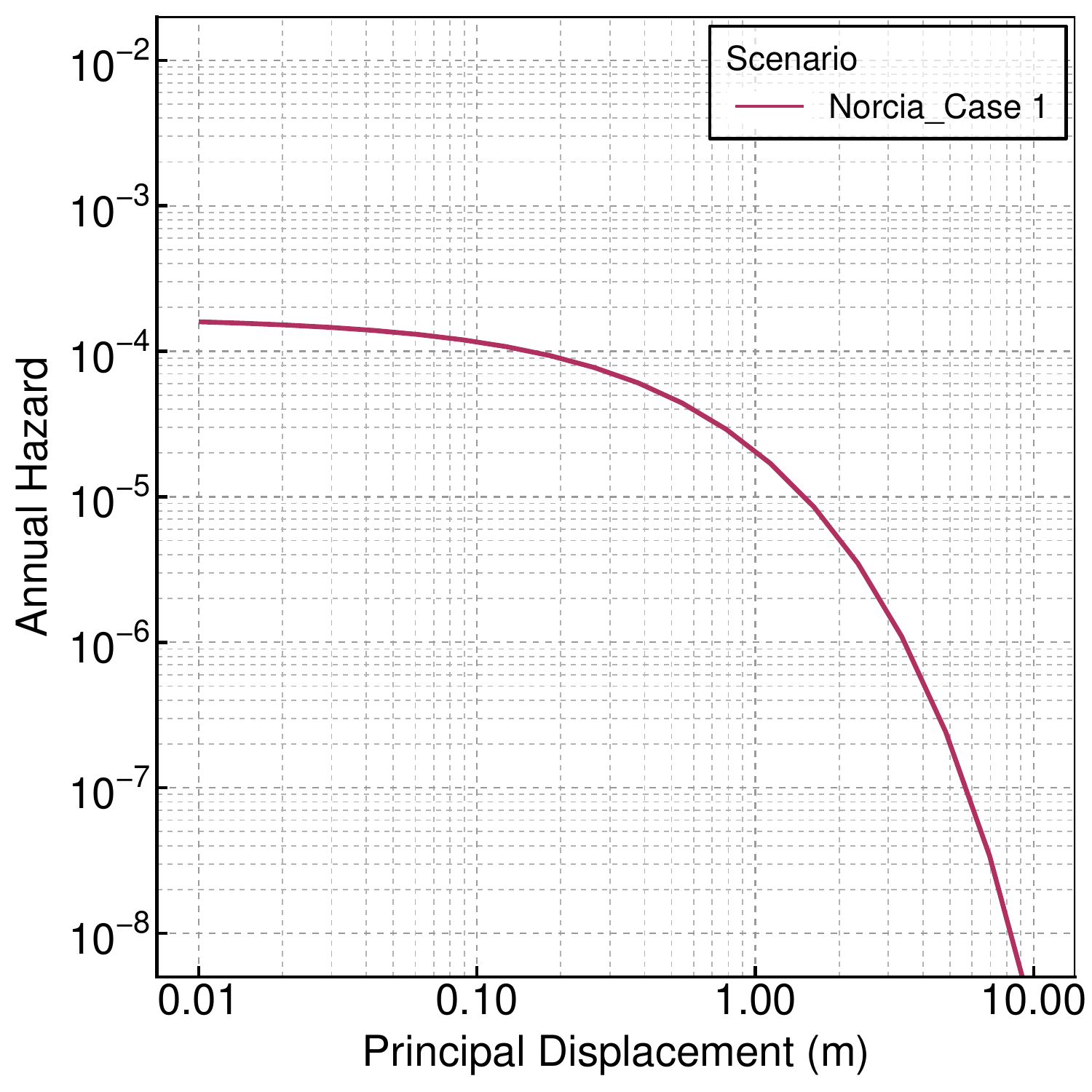}
        \label{fig:haz_prncp_single_SCS_norcia_1}
    \end{subfigure}

    \caption{Evaluation of the IAEA benchmark hazard curves for single-rupture  principal displacement (a single rupture strand intersecting the structure) considering SCS fault traces and structures located at distances ranging from 5 to 500 m, based on \cite{valentini2021probabilistic, valentini2025iaea}: (a) Kumamoto case 2, (b) Kumamoto case 3, (c) Le Teil case 2, and (d) Norcia case 1.}
    \label{fig:Hazardcurves_prncp_single_SCS}
\end{figure}

Figure \ref{fig:Hazardcurves_prncp_single_SCS} presents the hazard curve evaluation of the IAEA benchmark \citep{valentini2021probabilistic, valentini2025iaea} for single-rupture principal displacement (a single rupture strand intersecting the structure) considering the SCS fault trace definition.

\begin{figure}[htbp]
    \centering

    \begin{subfigure}[d]{0.49\linewidth}
        \centering
        \caption{}
        \includegraphics[width=\linewidth]{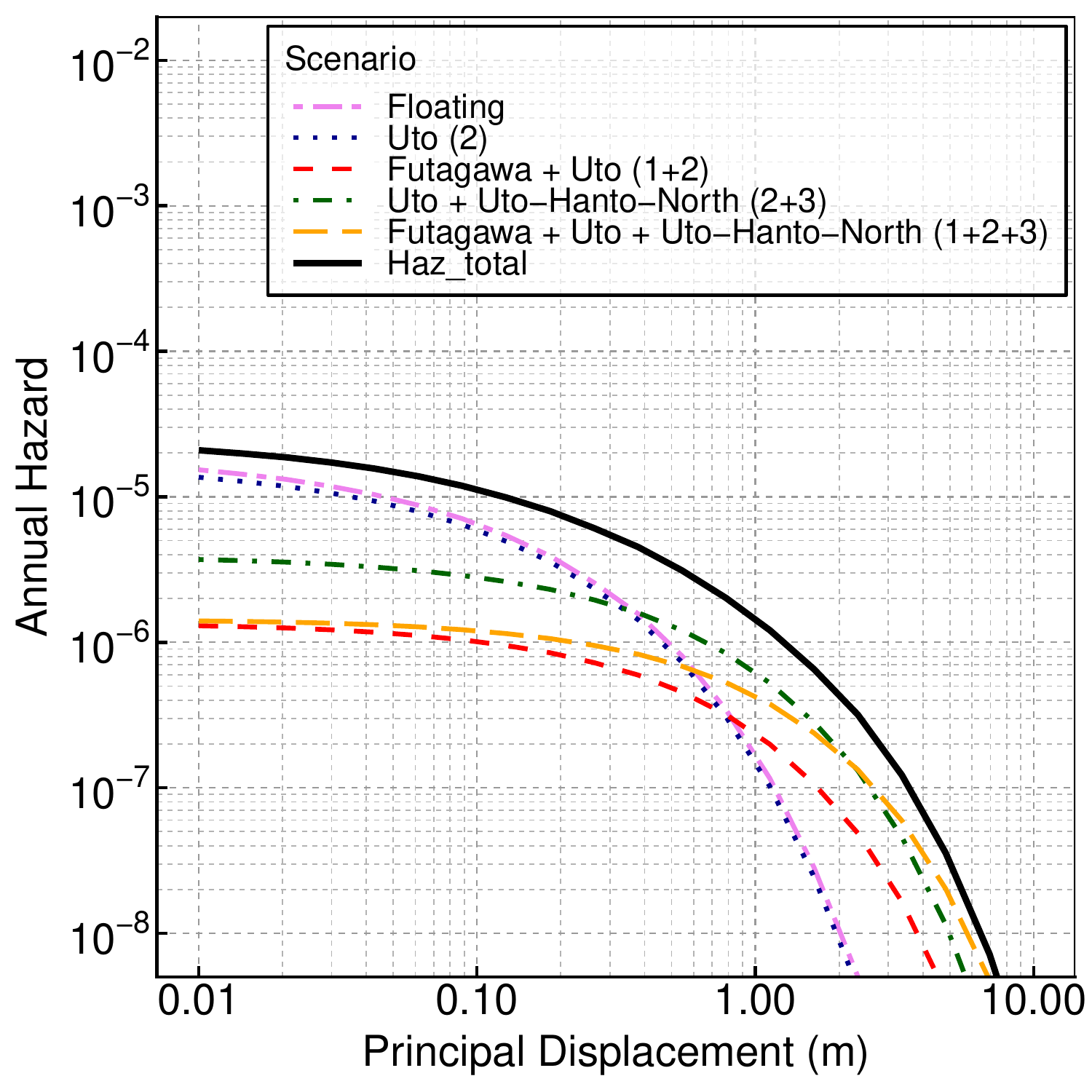}
        \label{fig:haz_prncp_single_random_kumamoto_2}
    \end{subfigure}
    \hfill
    \begin{subfigure}[d]{0.49\linewidth}
        \centering
        \caption{}
        \includegraphics[width=\linewidth]{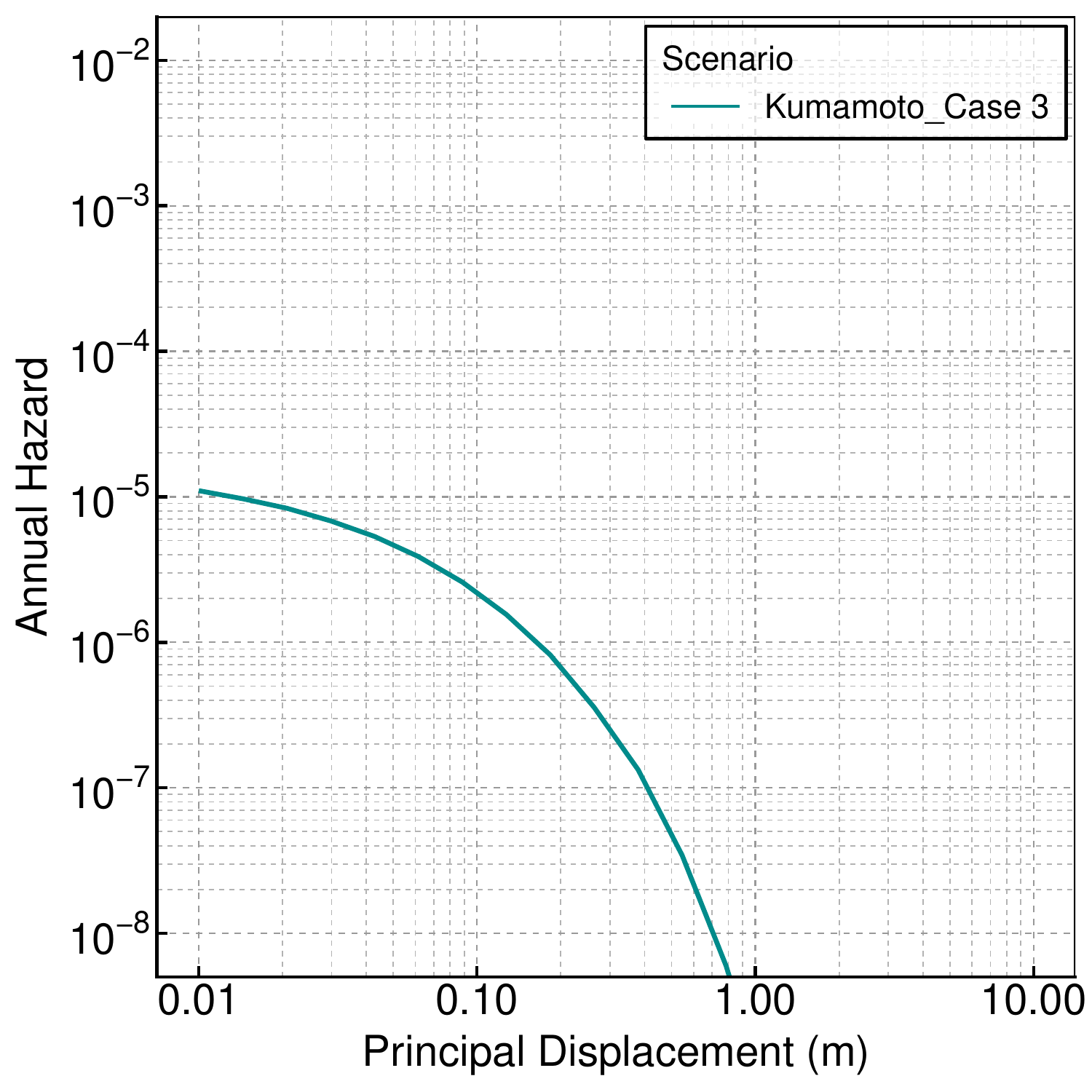}
        \label{fig:haz_prncp_single_random_kumamoto_3}
    \end{subfigure}
    \begin{subfigure}[d]{0.49\linewidth}
        \centering
        \caption{}
        \includegraphics[width=\linewidth]{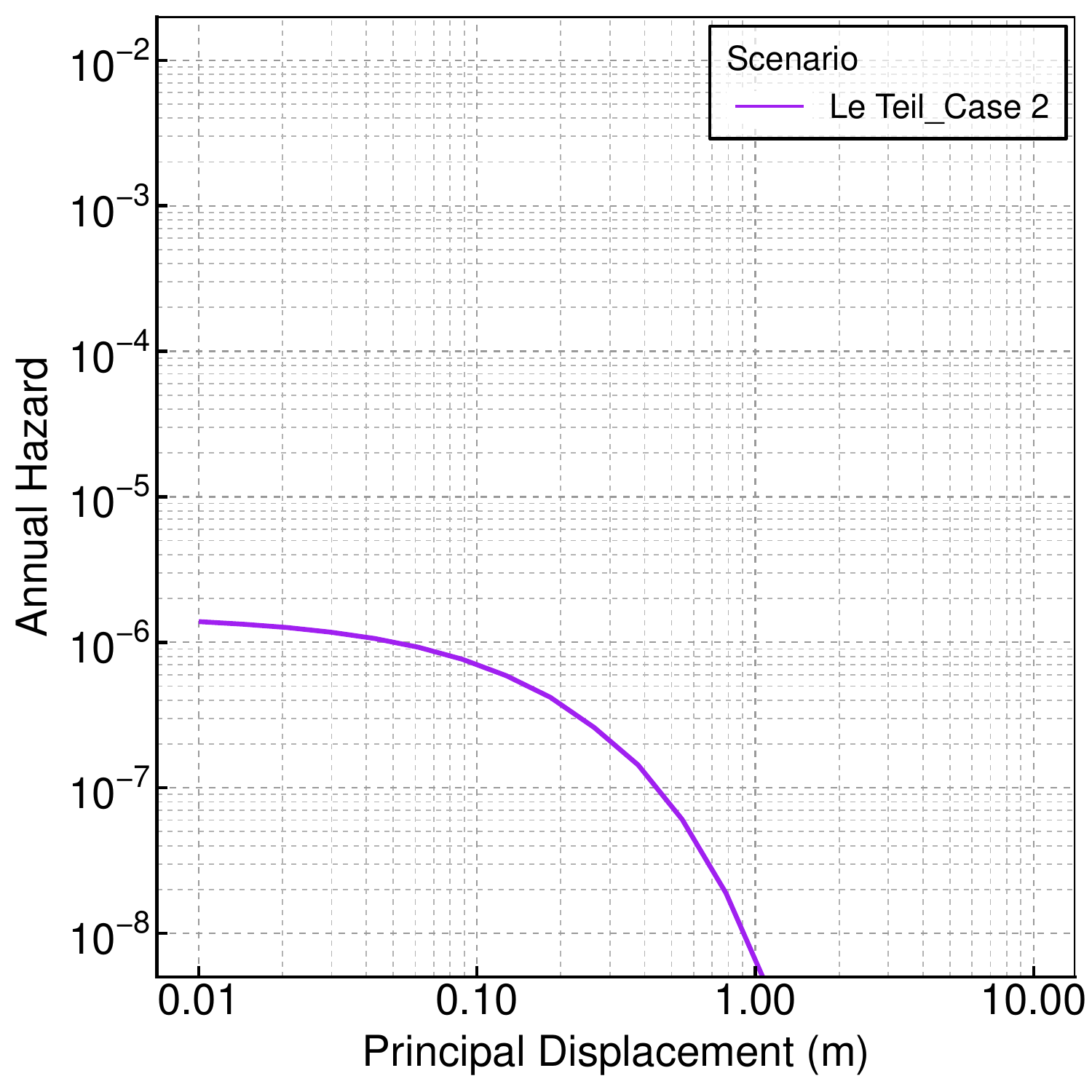}
        \label{fig:haz_prncp_single_random_LeTeil_2}
    \end{subfigure}
    \begin{subfigure}[d]{0.49\linewidth}
        \centering
        \caption{}
        \includegraphics[width=\linewidth]{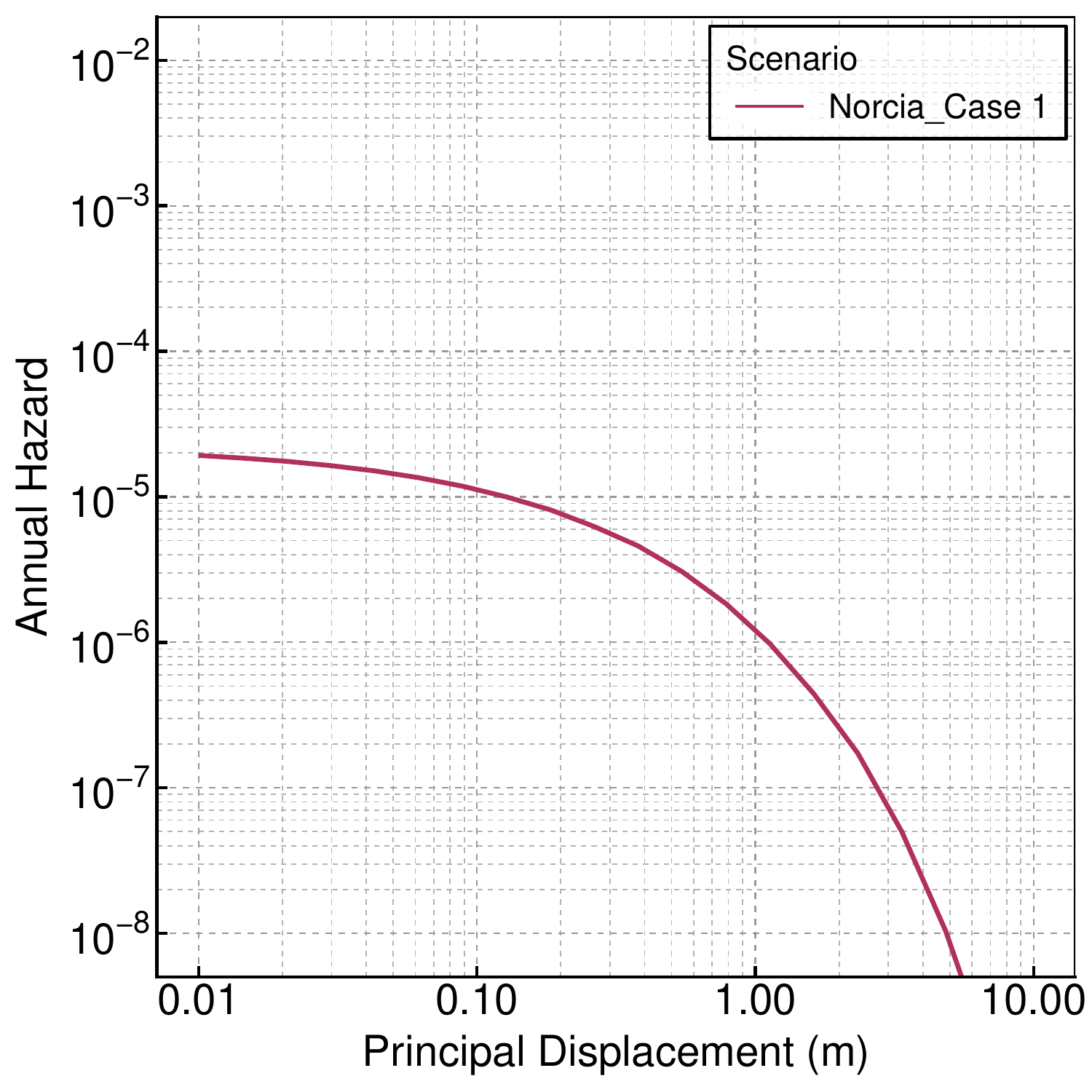}
        \label{fig:haz_prncp_single_random_norcia_1}
    \end{subfigure}

    \caption{Evaluation of the IAEA benchmark hazard curves for single-rupture  principal displacement (a single rupture strand intersecting the structure) considering random fault traces and structures located at distances ranging from 5 to 500 m, based on \cite{valentini2021probabilistic, valentini2025iaea}: (a) Kumamoto case 2, (b) Kumamoto case 3, (c) Le Teil case 2, and (d) Norcia case 1.}
    \label{fig:Hazardcurves_prncp_single_random}
\end{figure}

Figure \ref{fig:Hazardcurves_prncp_single_random} presents the hazard curve evaluation of the IAEA benchmark \citep{valentini2021probabilistic, valentini2025iaea} for single-rupture principal displacement (a single rupture strand intersecting the structure) considering the randomly selected fault trace definition.

\begin{figure}[htbp]
    \centering

    \begin{subfigure}[d]{0.49\linewidth}
        \centering
        \caption{}
        \includegraphics[width=\linewidth]{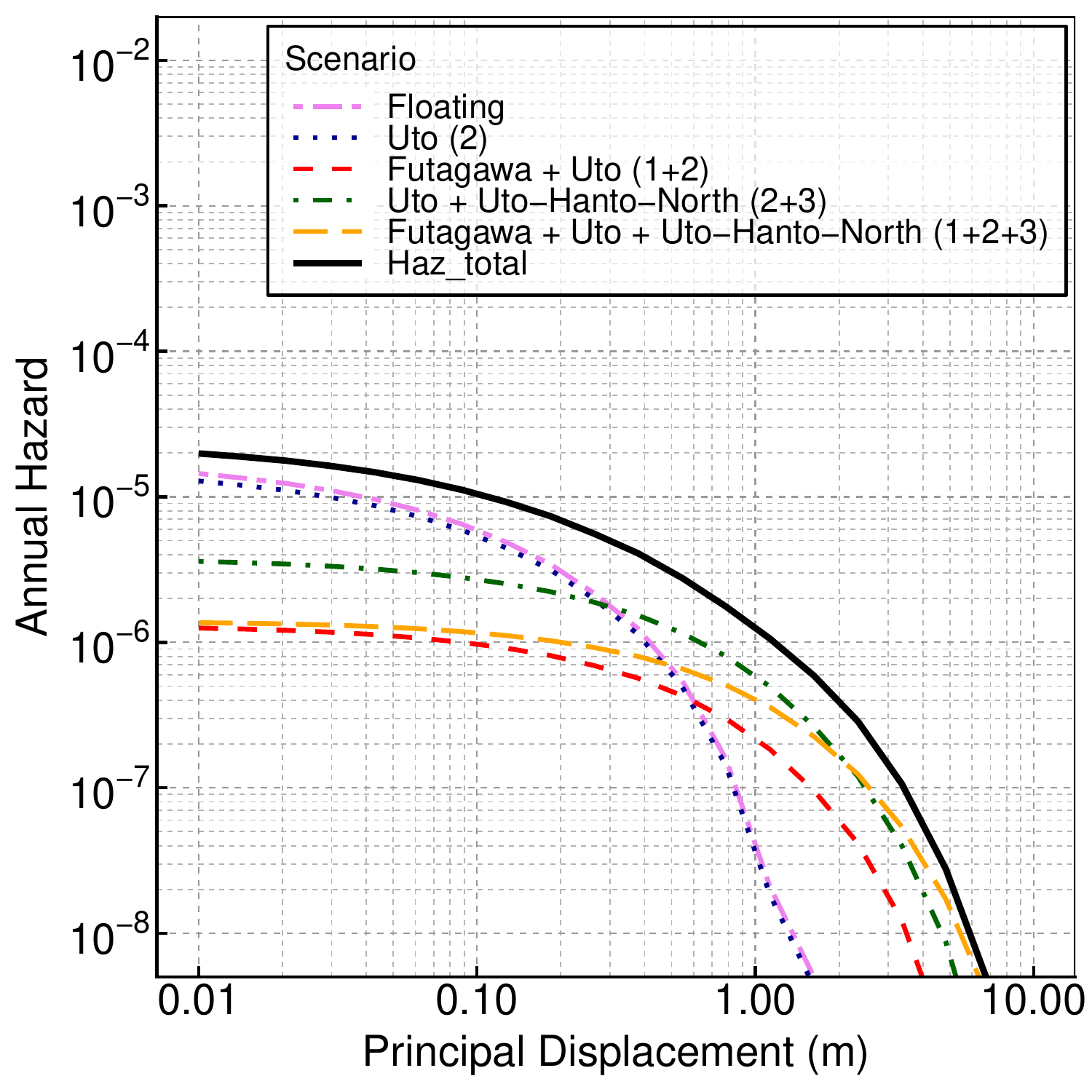}
        \label{fig:haz_prncp_single_max_kumamoto_2}
    \end{subfigure}
    \hfill
    \begin{subfigure}[d]{0.49\linewidth}
        \centering
        \caption{}
        \includegraphics[width=\linewidth]{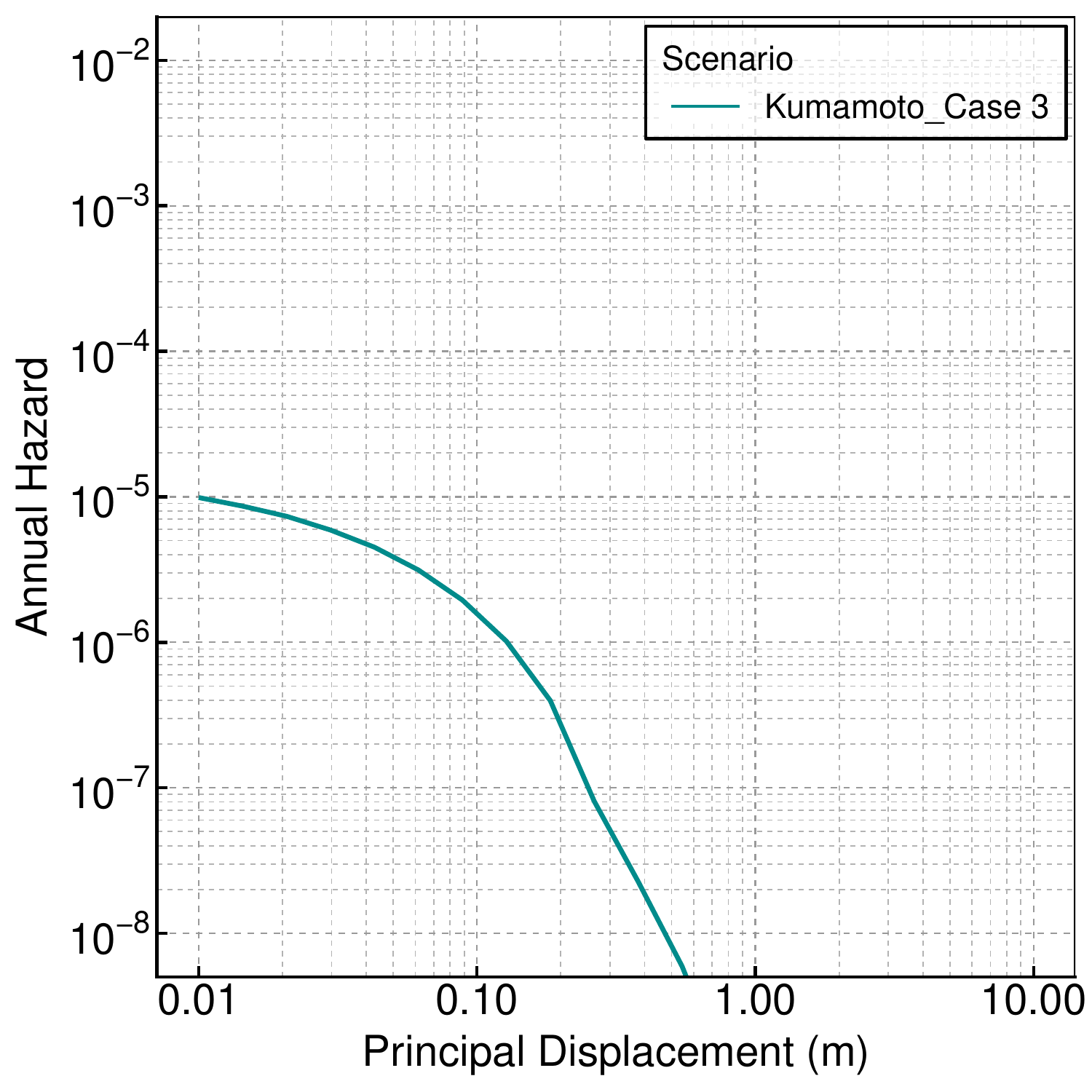}
        \label{fig:haz_prncp_single_max_kumamoto_3}
    \end{subfigure}
    \begin{subfigure}[d]{0.49\linewidth}
        \centering
        \caption{}
        \includegraphics[width=\linewidth]{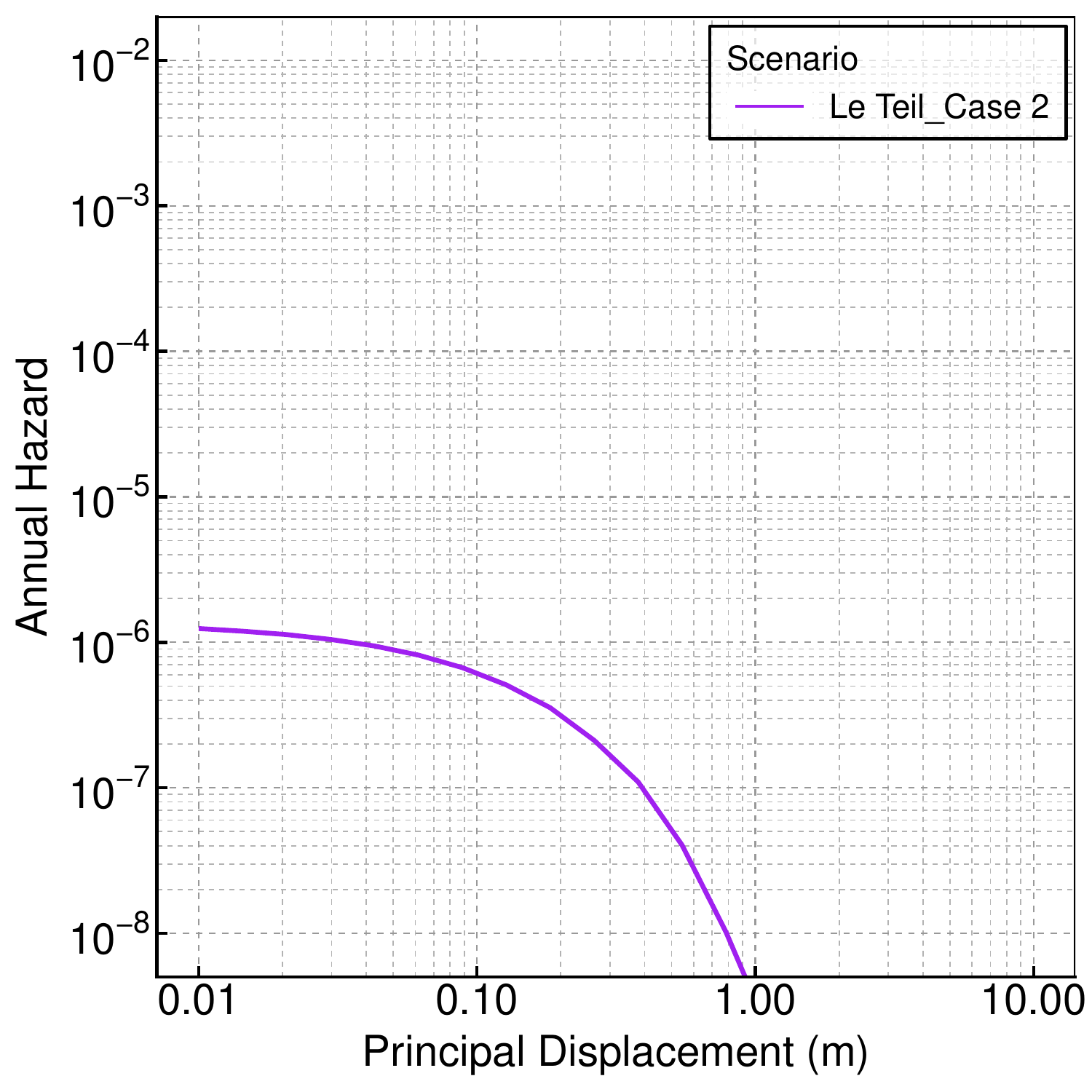}
        \label{fig:haz_prncp_single_max_LeTeil_2}
    \end{subfigure}
    \begin{subfigure}[d]{0.49\linewidth}
        \centering
        \caption{}
        \includegraphics[width=\linewidth]{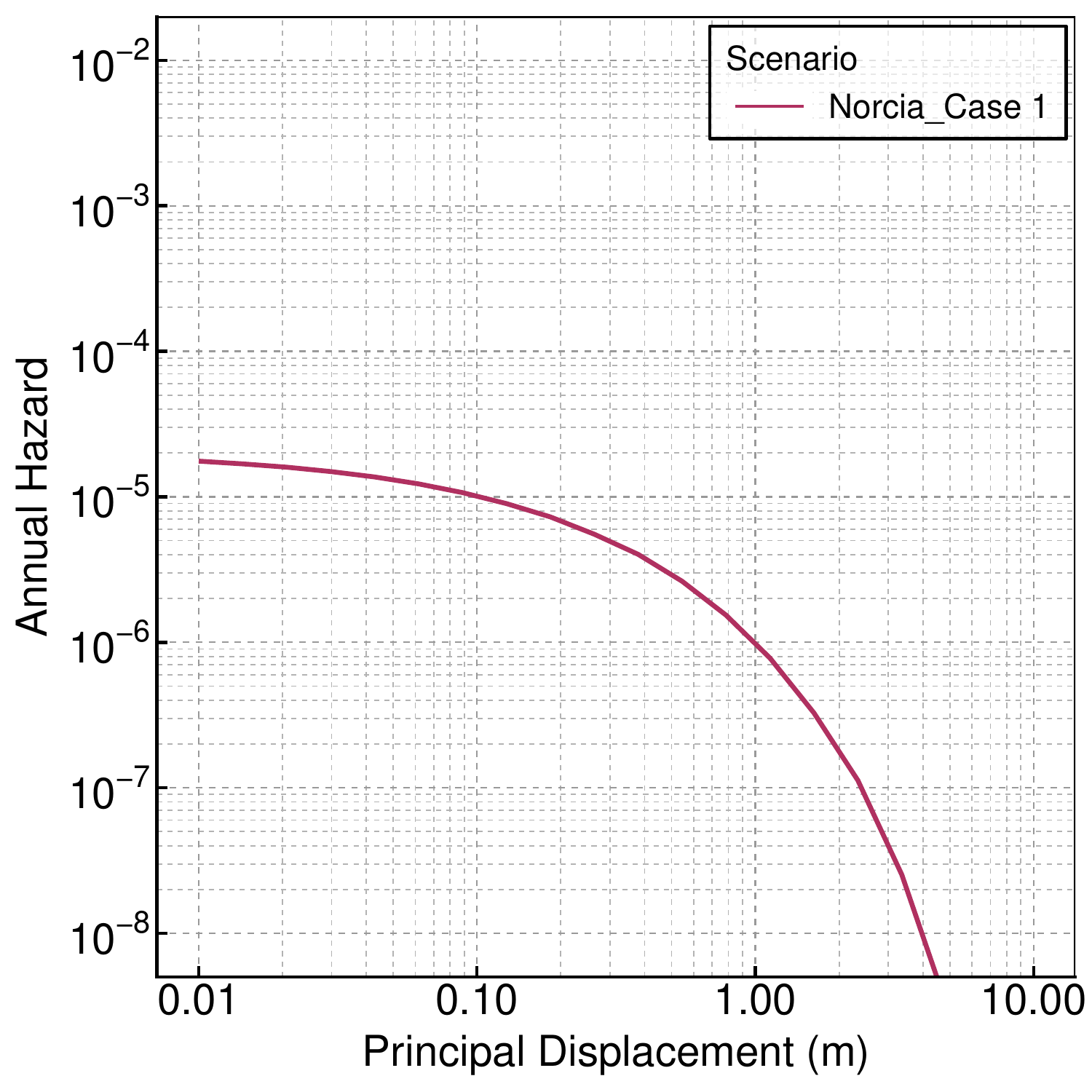}
        \label{fig:haz_prncp_single_max_norcia_1}
    \end{subfigure}

    \caption{Evaluation of the IAEA benchmark hazard curves for single-rupture  principal displacement (a single rupture strand intersecting the structure) considering max displacement fault traces and structures located at distances ranging from 5 to 500 m, based on \cite{valentini2021probabilistic, valentini2025iaea}: (a) Kumamoto case 2, (b) Kumamoto case 3, (c) Le Teil case 2, and (d) Norcia case 1.}
    \label{fig:Hazardcurves_prncp_single_max}
\end{figure}

Figure \ref{fig:Hazardcurves_prncp_single_max} presents the hazard curve evaluation of the IAEA benchmark \citep{valentini2021probabilistic, valentini2025iaea} for single-rupture principal displacement (a single rupture strand intersecting the structure) considering the maximum principal displacement fault trace definition.

\begin{figure}[htbp]
    \centering

    \begin{subfigure}[d]{0.49\linewidth}
        \centering
        \caption{}
        \includegraphics[width=\linewidth]{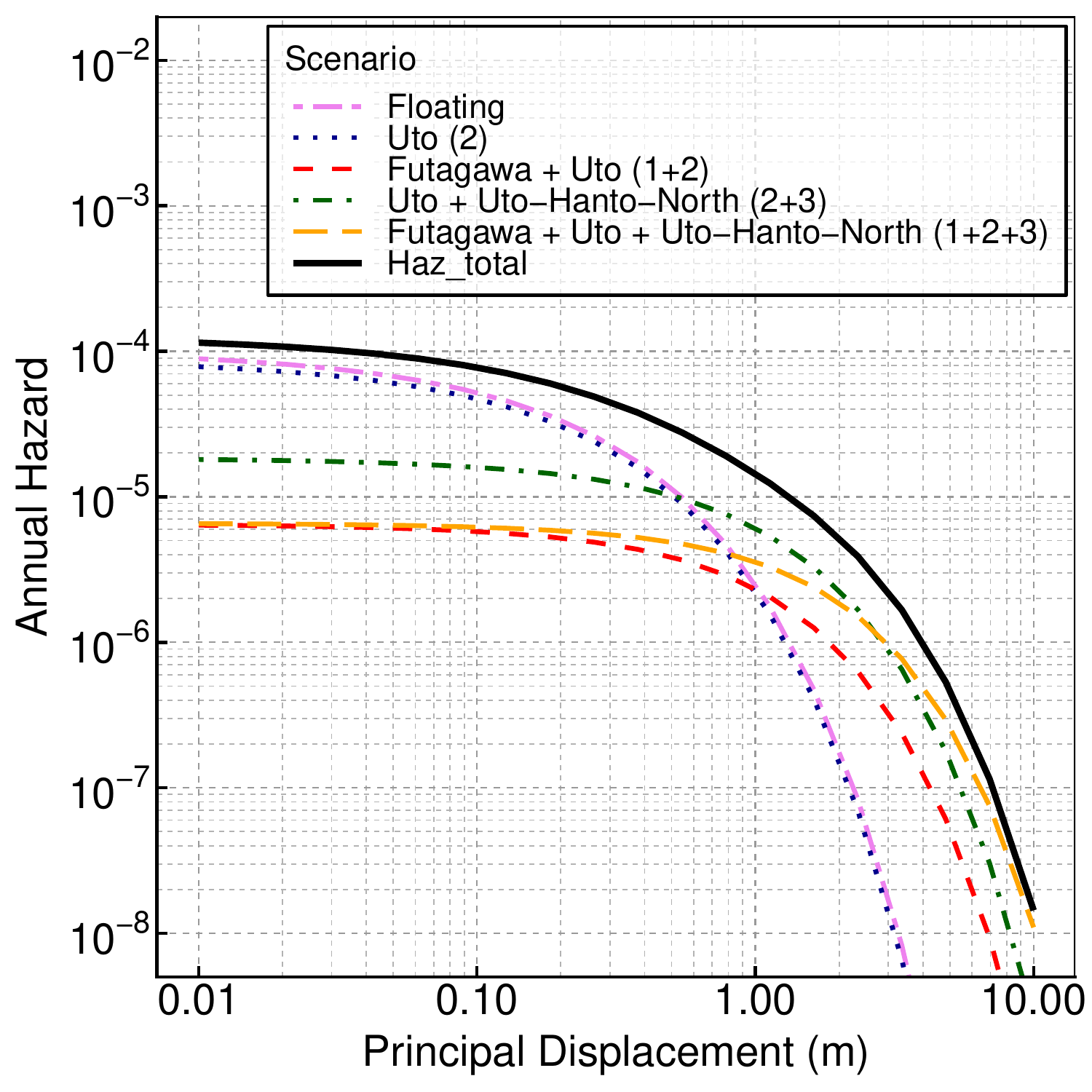}
        \label{fig:haz_prncp_multi_SCS_kumamoto_2}
    \end{subfigure}
    \hfill
    \begin{subfigure}[d]{0.49\linewidth}
        \centering
        \caption{}
        \includegraphics[width=\linewidth]{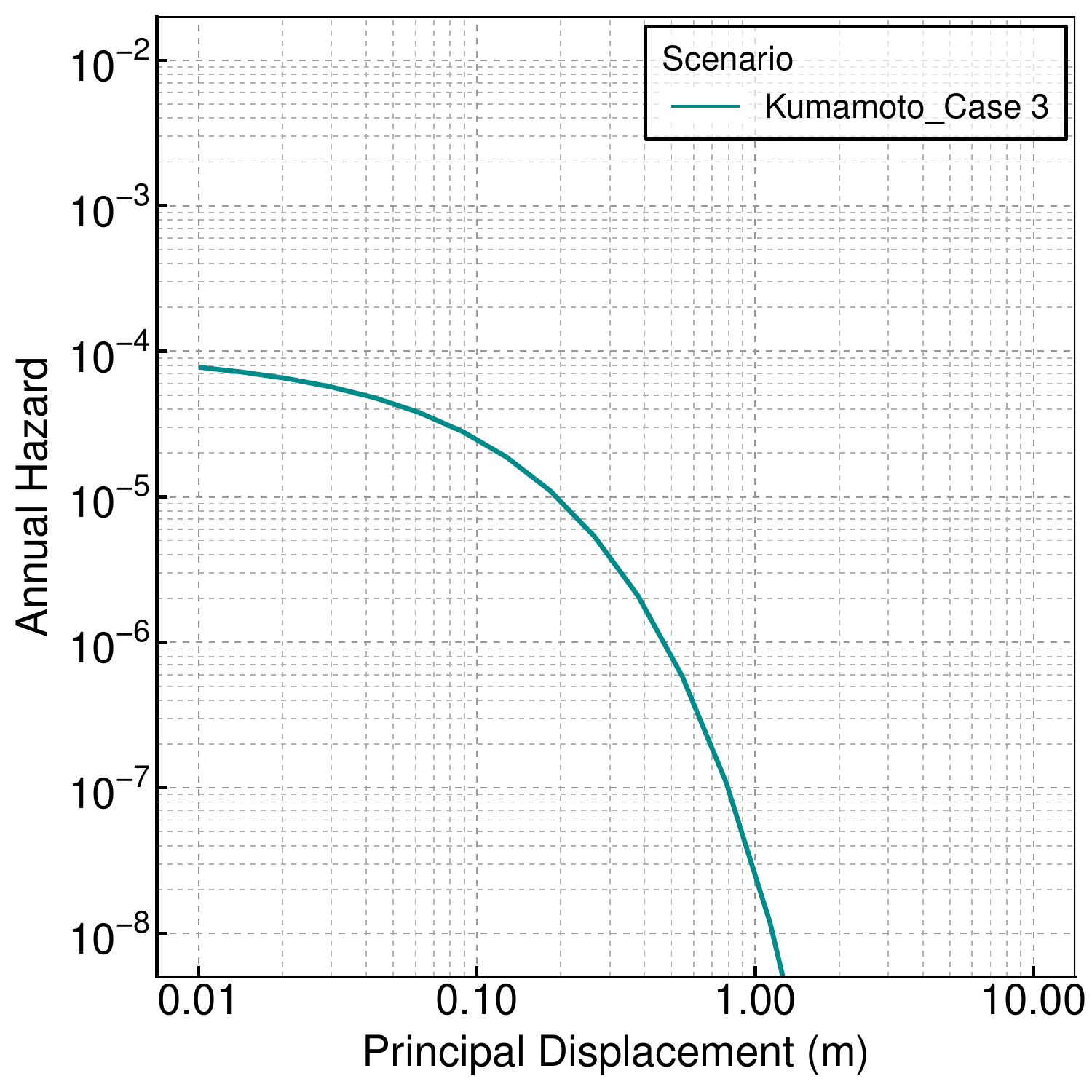}
        \label{fig:haz_prncp_multi_SCS_kumamoto_3}
    \end{subfigure}
    \begin{subfigure}[d]{0.49\linewidth}
        \centering
        \caption{}
        \includegraphics[width=\linewidth]{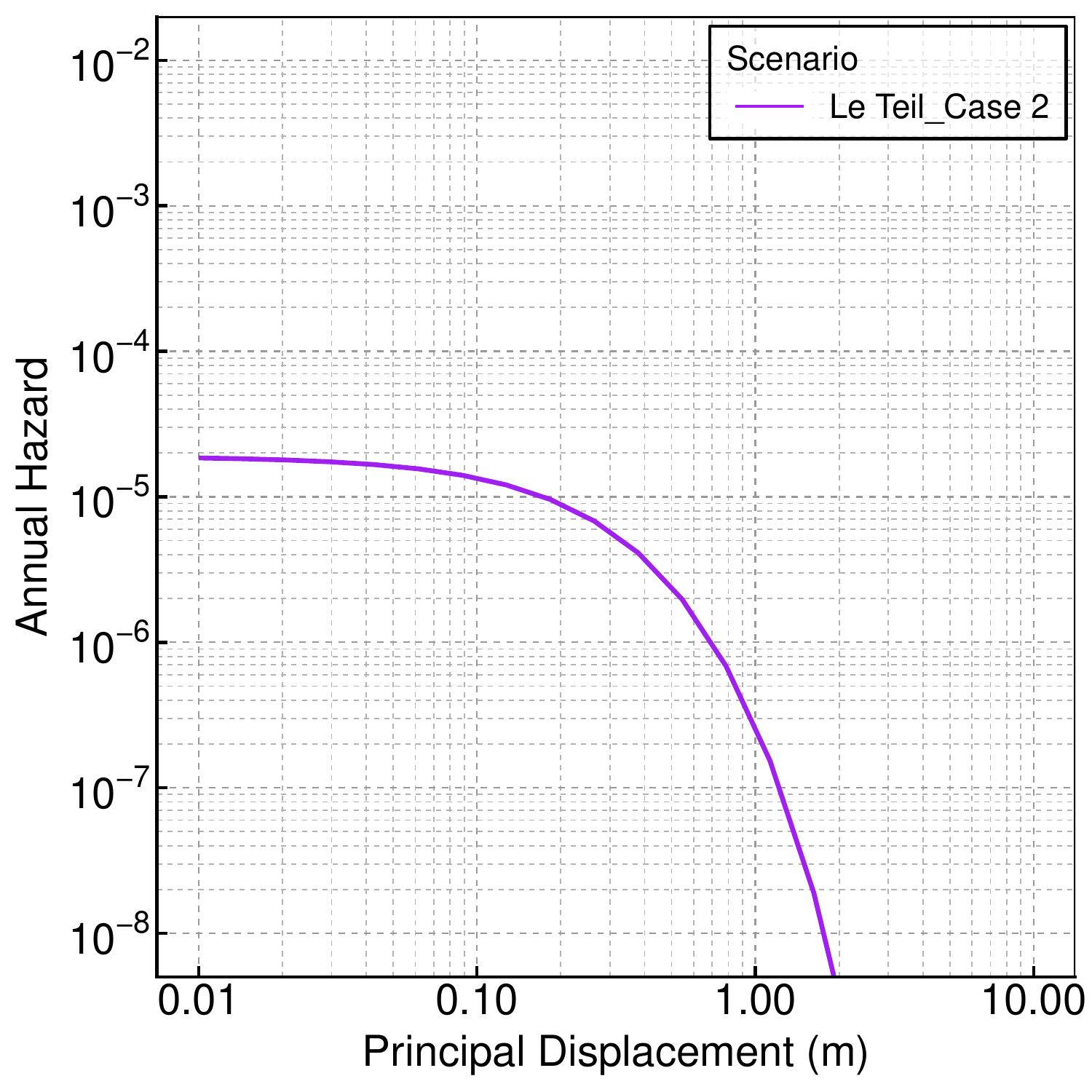}
        \label{fig:haz_prncp_multi_SCS_LeTeil_2}
    \end{subfigure}
    \begin{subfigure}[d]{0.49\linewidth}
        \centering
        \caption{}
        \includegraphics[width=\linewidth]{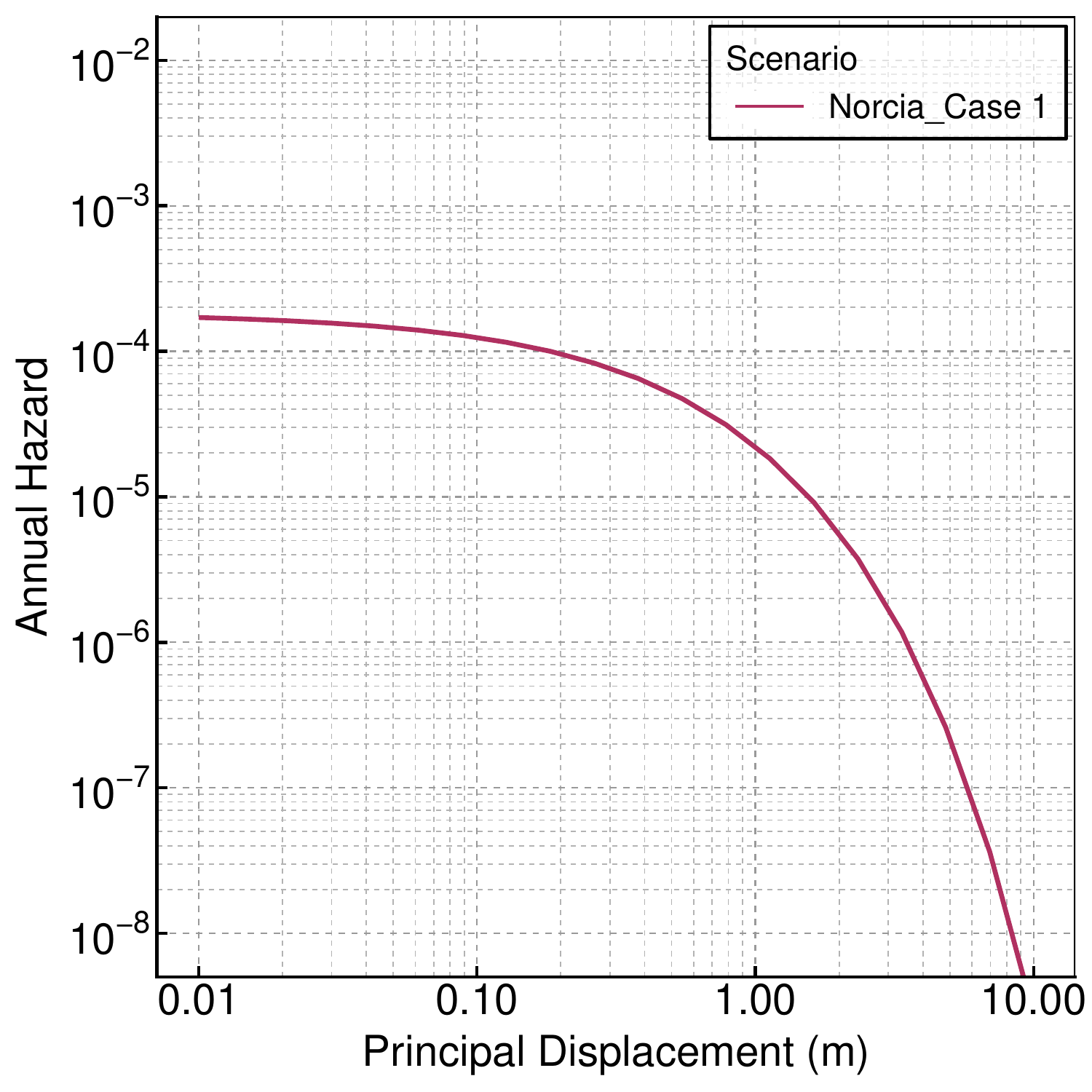}
        \label{fig:haz_prncp_multi_SCS_norcia_1}
    \end{subfigure}

    \caption{Evaluation of the IAEA benchmark hazard curves for multi-rupture principal displacement (the cumulative contribution of one or more rupture strands intersecting the structure) considering SCS fault traces and structures located at distances ranging from 5 to 500 m, based on \cite{valentini2021probabilistic, valentini2025iaea}: (a) Kumamoto case 2, (b) Kumamoto case 3, (c) Le Teil case 2, and (d) Norcia case 1.}
    \label{fig:Hazardcurves_prncp_multi_SCS}
\end{figure}

Figure \ref{fig:Hazardcurves_prncp_multi_SCS} presents the hazard curve evaluation of the IAEA benchmark \citep{valentini2021probabilistic, valentini2025iaea} for multi-rupture principal displacement (the cumulative contribution of one or more rupture strands intersecting the structure) considering the SCS fault trace definition.

\begin{figure}[htbp]
    \centering

    \begin{subfigure}[d]{0.49\linewidth}
        \centering
        \caption{}
        \includegraphics[width=\linewidth]{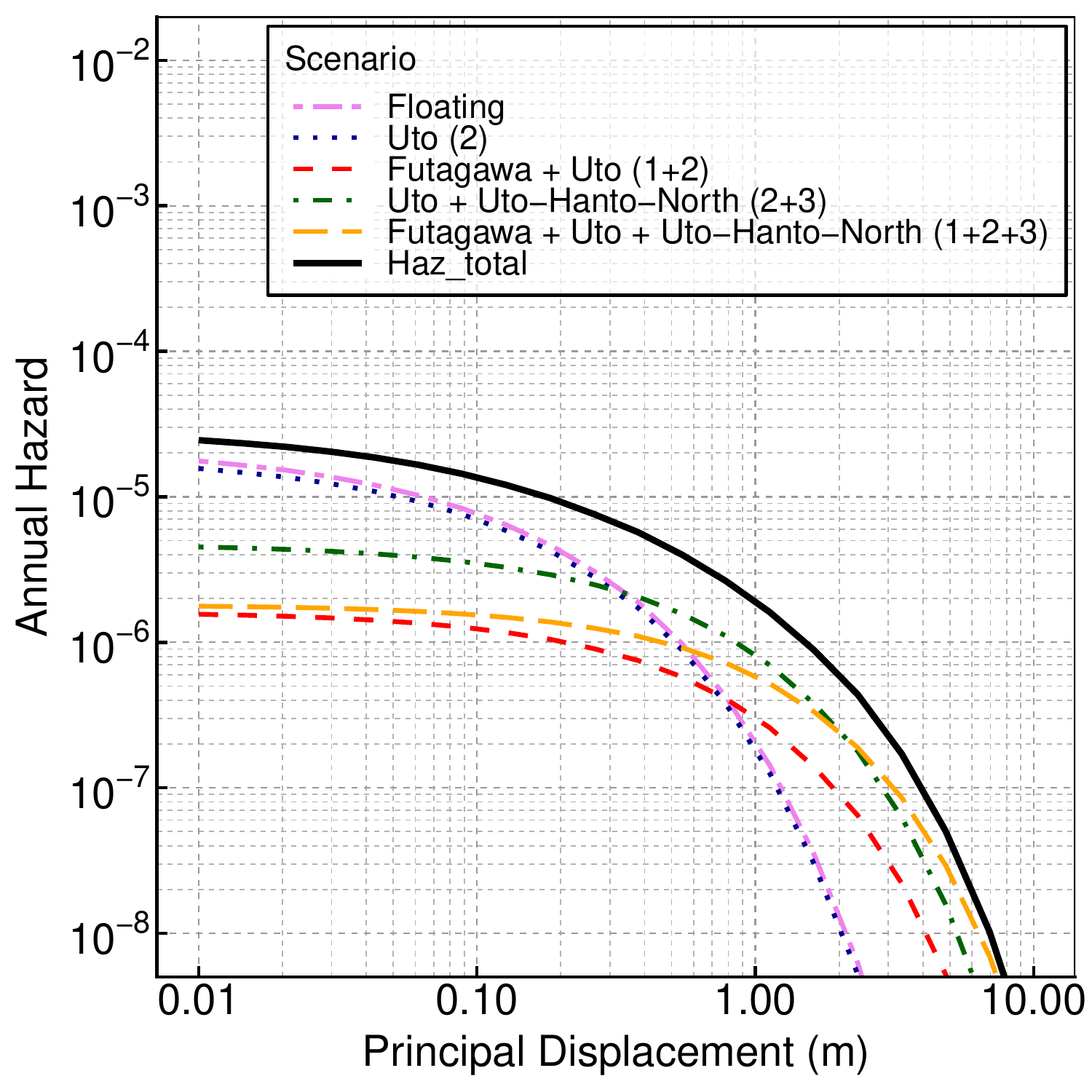}
        \label{fig:haz_prncp_multi_random_kumamoto_2}
    \end{subfigure}
    \hfill
    \begin{subfigure}[d]{0.49\linewidth}
        \centering
        \caption{}
        \includegraphics[width=\linewidth]{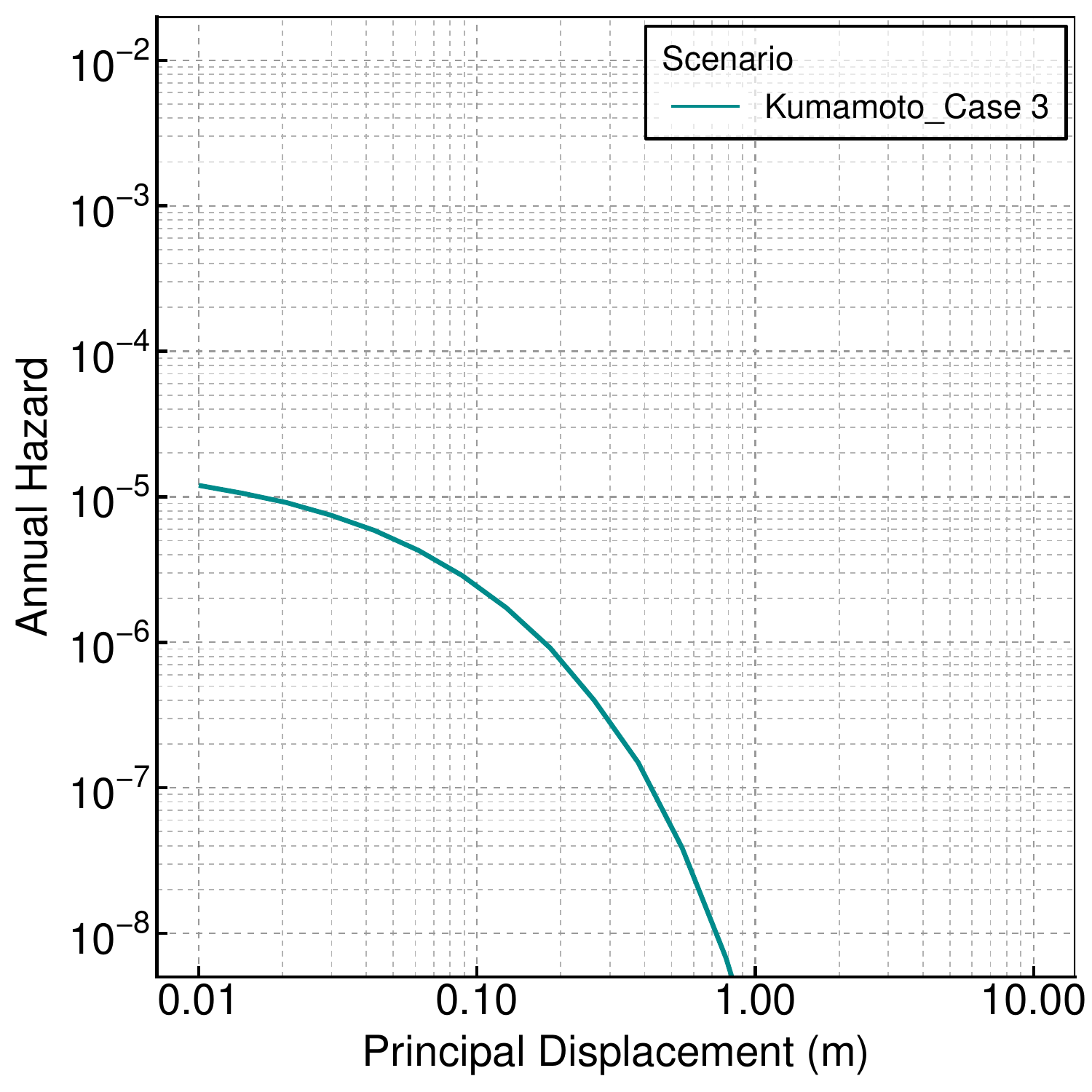}
        \label{fig:haz_prncp_multi_random_kumamoto_3}
    \end{subfigure}
    \begin{subfigure}[d]{0.49\linewidth}
        \centering
        \caption{}
        \includegraphics[width=\linewidth]{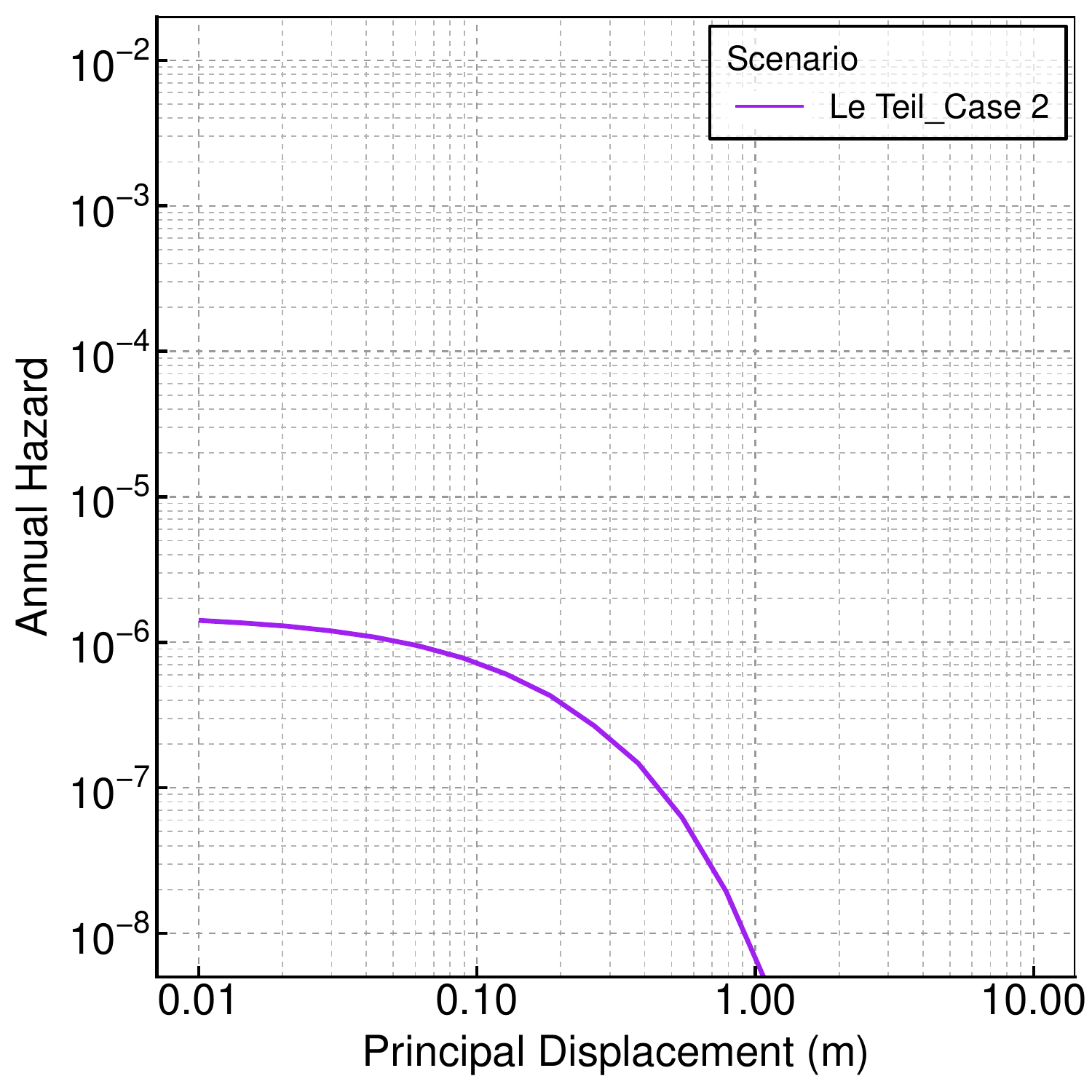}
        \label{fig:haz_prncp_multi_random_LeTeil_2}
    \end{subfigure}
    \begin{subfigure}[d]{0.49\linewidth}
        \centering
        \caption{}
        \includegraphics[width=\linewidth]{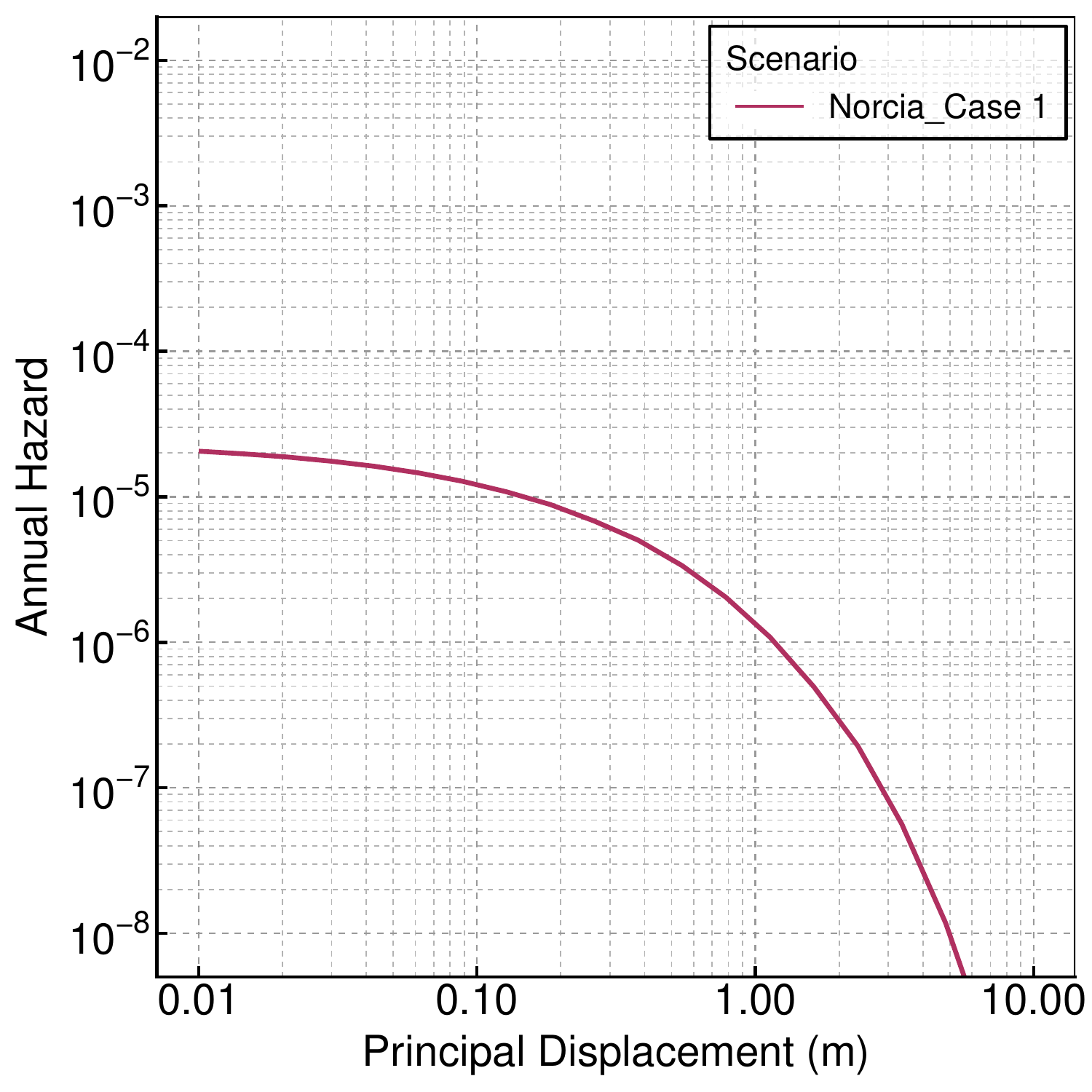}
        \label{fig:haz_prncp_multi_random_norcia_1}
    \end{subfigure}

    \caption{Evaluation of the IAEA benchmark hazard curves for multi-rupture principal displacement (the cumulative contribution of one or more rupture strands intersecting the structure) considering random fault traces and structures located at distances ranging from 5 to 500 m, based on \cite{valentini2021probabilistic, valentini2025iaea}: (a) Kumamoto case 2, (b) Kumamoto case 3, (c) Le Teil case 2, and (d) Norcia case 1.}
    \label{fig:Hazardcurves_prncp_multi_random}
\end{figure}

Figure \ref{fig:Hazardcurves_prncp_multi_random} presents the hazard curve evaluation of the IAEA benchmark \citep{valentini2021probabilistic, valentini2025iaea} for multi-rupture principal displacement (the cumulative contribution of one or more rupture strands intersecting the structure) considering the randomly selected fault trace definition.

\begin{figure}[htbp]
    \centering

    \begin{subfigure}[d]{0.49\linewidth}
        \centering
        \caption{}
        \includegraphics[width=\linewidth]{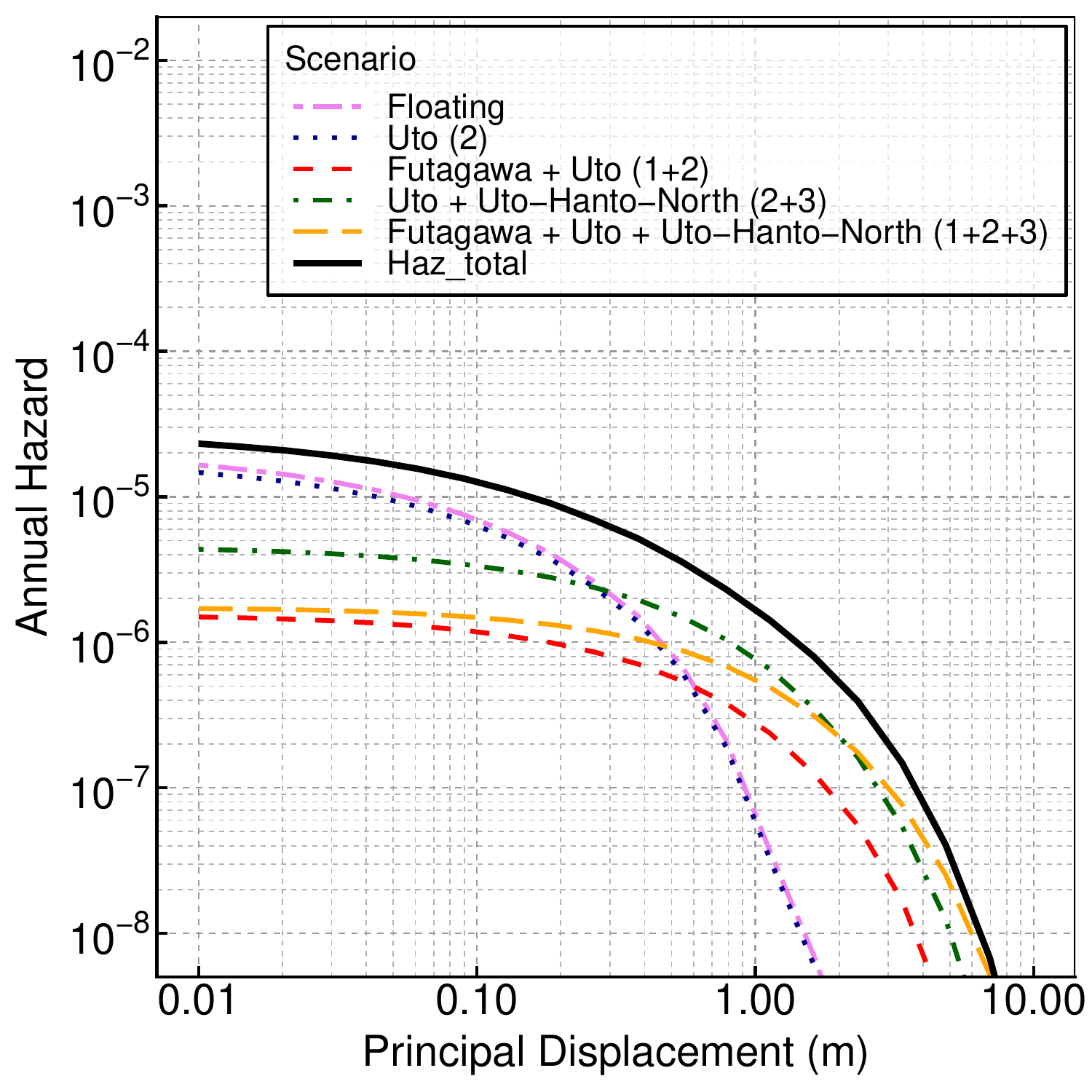}
        \label{fig:haz_prncp_multi_max_kumamoto_2}
    \end{subfigure}
    \hfill
    \begin{subfigure}[d]{0.49\linewidth}
        \centering
        \caption{}
        \includegraphics[width=\linewidth]{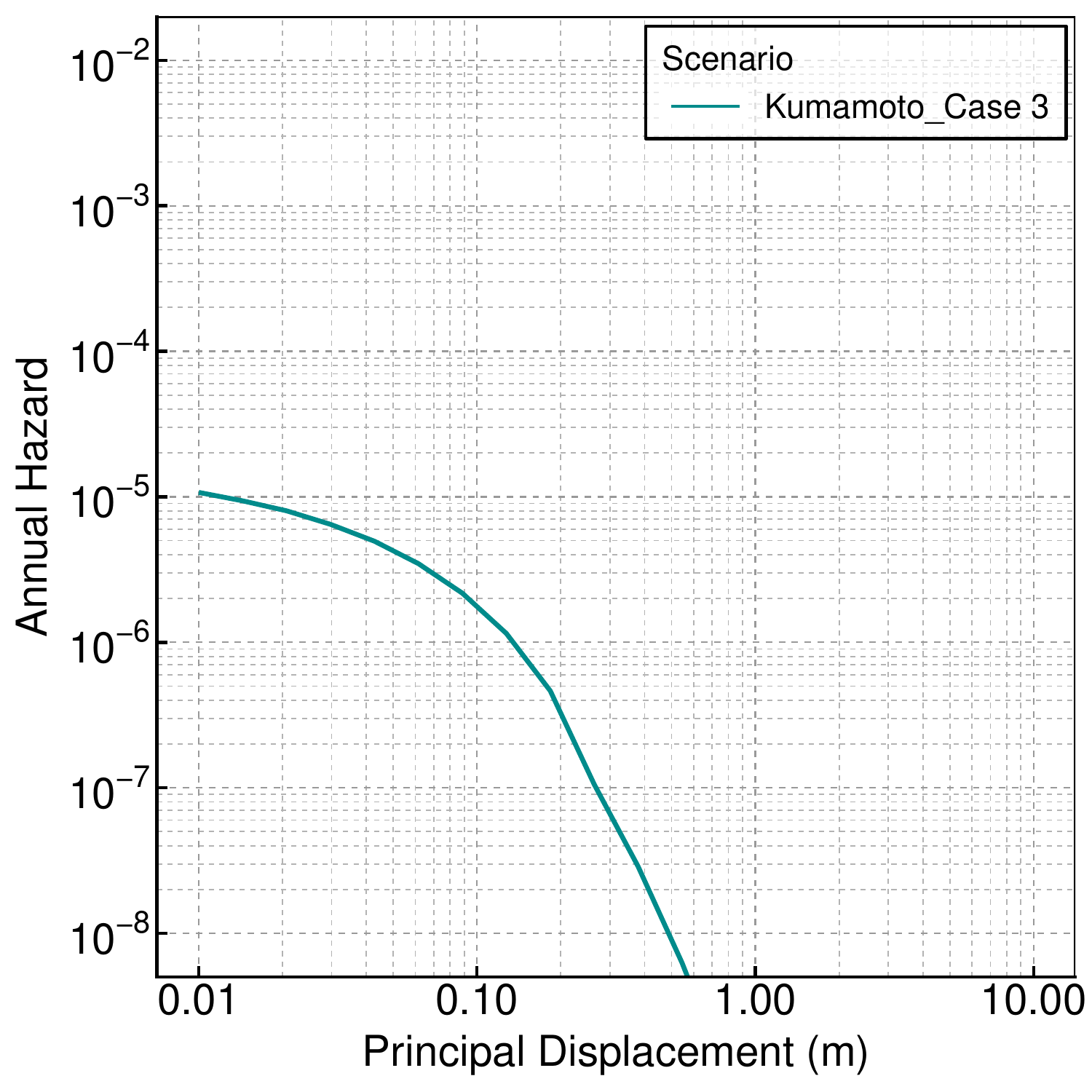}
        \label{fig:haz_prncp_multi_max_kumamoto_3}
    \end{subfigure}
    \begin{subfigure}[d]{0.49\linewidth}
        \centering
        \caption{}
        \includegraphics[width=\linewidth]{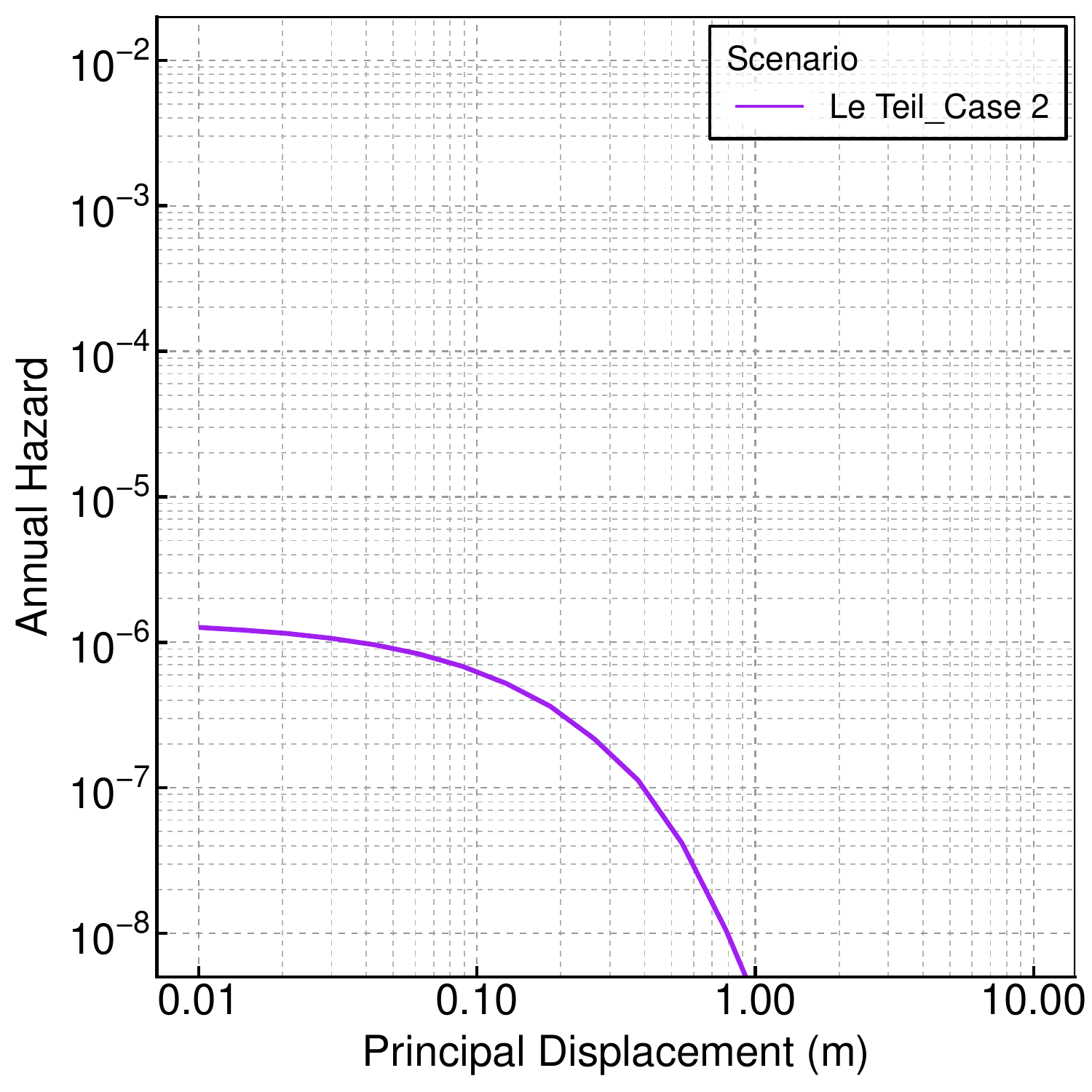}
        \label{fig:haz_prncp_multi_max_LeTeil_2}
    \end{subfigure}
    \begin{subfigure}[d]{0.49\linewidth}
        \centering
        \caption{}
        \includegraphics[width=\linewidth]{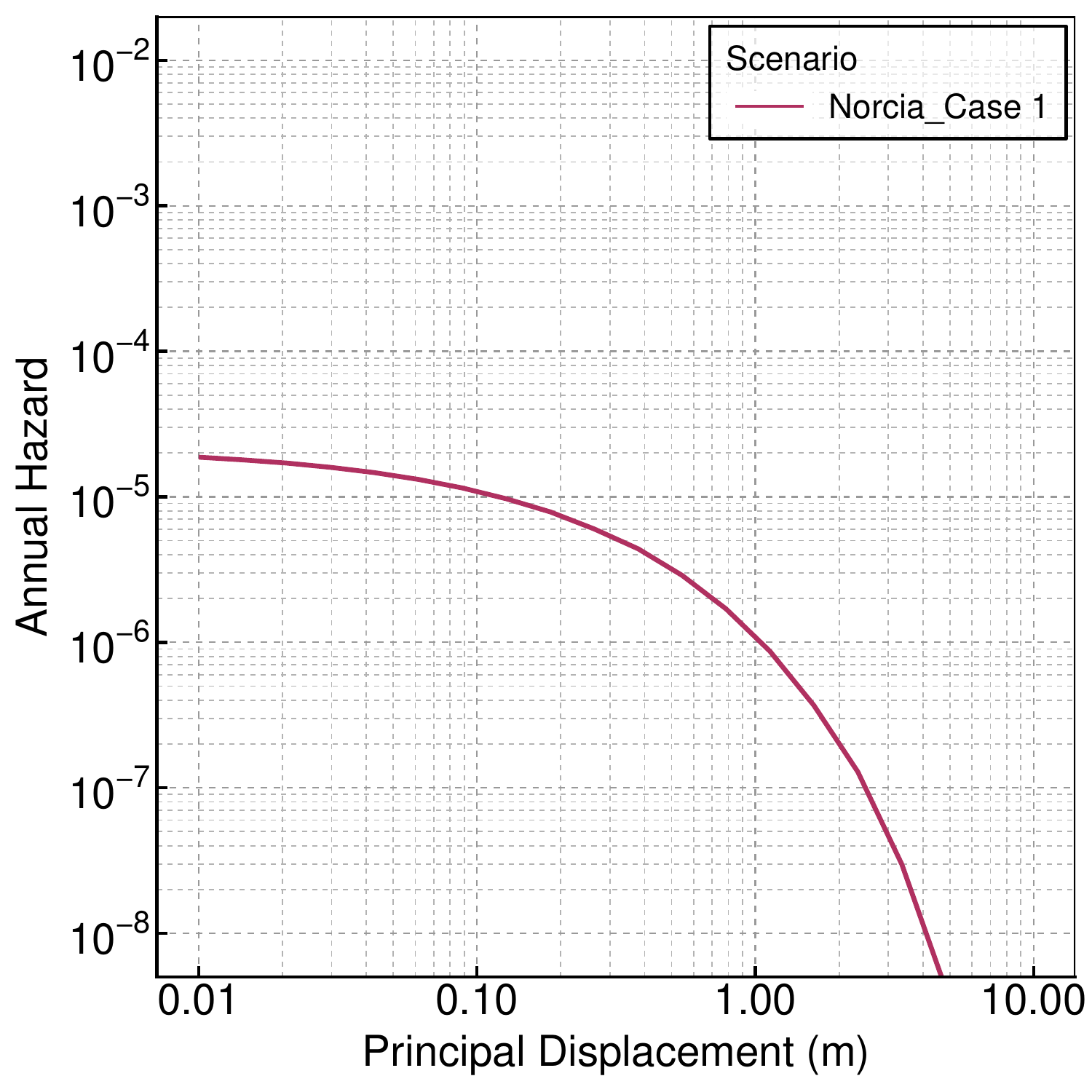}
        \label{fig:haz_prncp_multi_max_norcia_1}
    \end{subfigure}

    \caption{Evaluation of the IAEA benchmark hazard curves for multi-rupture principal displacement (the cumulative contribution of one or more rupture strands intersecting the structure) considering max displacement fault traces and structures located at distances ranging from 5 to 500 m, based on \cite{valentini2021probabilistic, valentini2025iaea}: (a) Kumamoto case 2, (b) Kumamoto case 3, (c) Le Teil case 2, and (d) Norcia case 1.}
    \label{fig:Hazardcurves_prncp_multi_max}
\end{figure}

Figure \ref{fig:Hazardcurves_prncp_multi_max} presents the hazard curve evaluation of the IAEA benchmark \citep{valentini2021probabilistic, valentini2025iaea} for multi-rupture principal displacement (the cumulative contribution of one or more rupture strands intersecting the structure) considering the maximum principal displacement fault trace definition.

\begin{figure}[htbp]
    \centering

    \begin{subfigure}[d]{0.49\linewidth}
        \centering
        \caption{}
        \includegraphics[width=\linewidth]{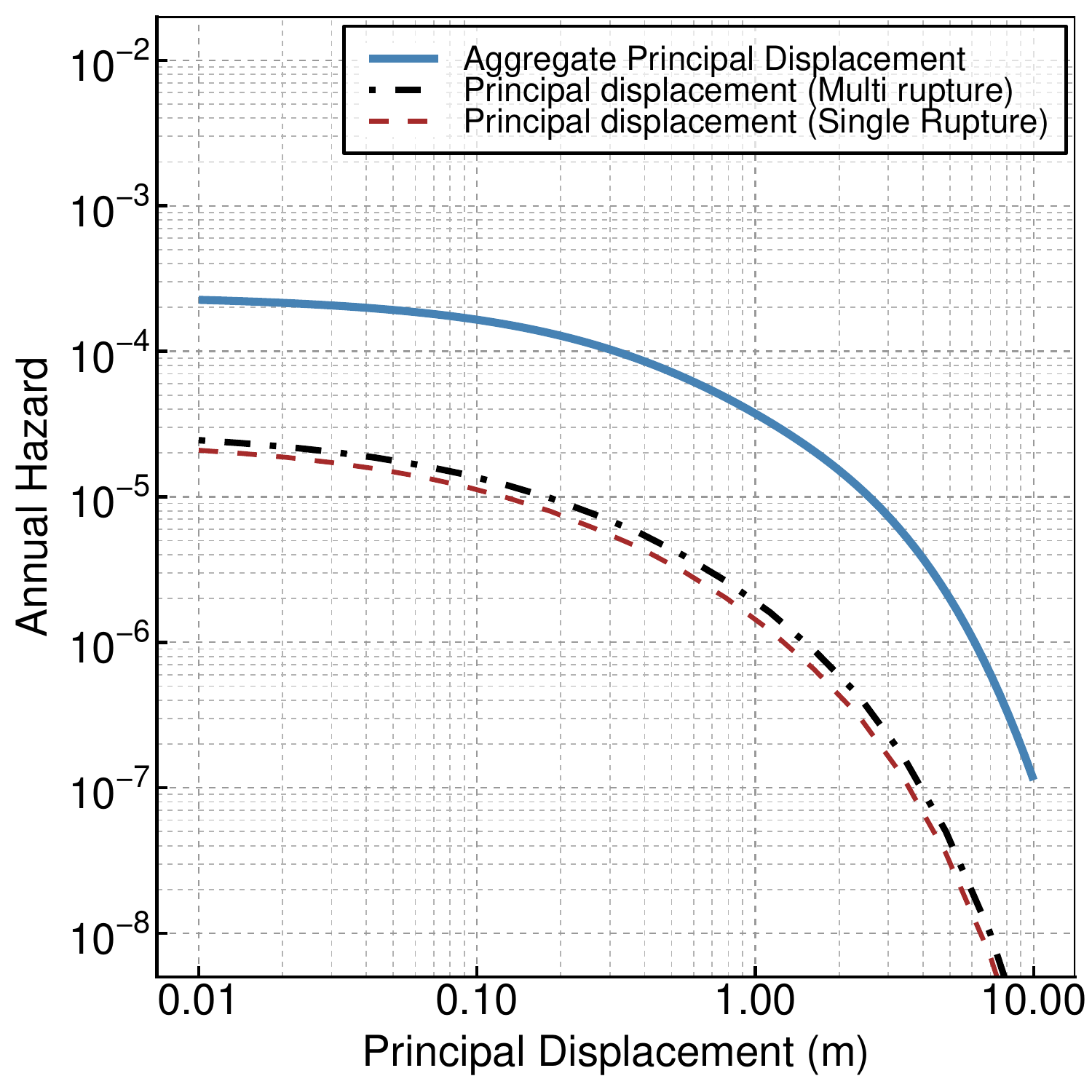}
        \label{fig:haz_agg_prncp_random_Kumamoto_2}
    \end{subfigure}
    \hfill
    \begin{subfigure}[d]{0.49\linewidth}
        \centering
        \caption{}
        \includegraphics[width=\linewidth]{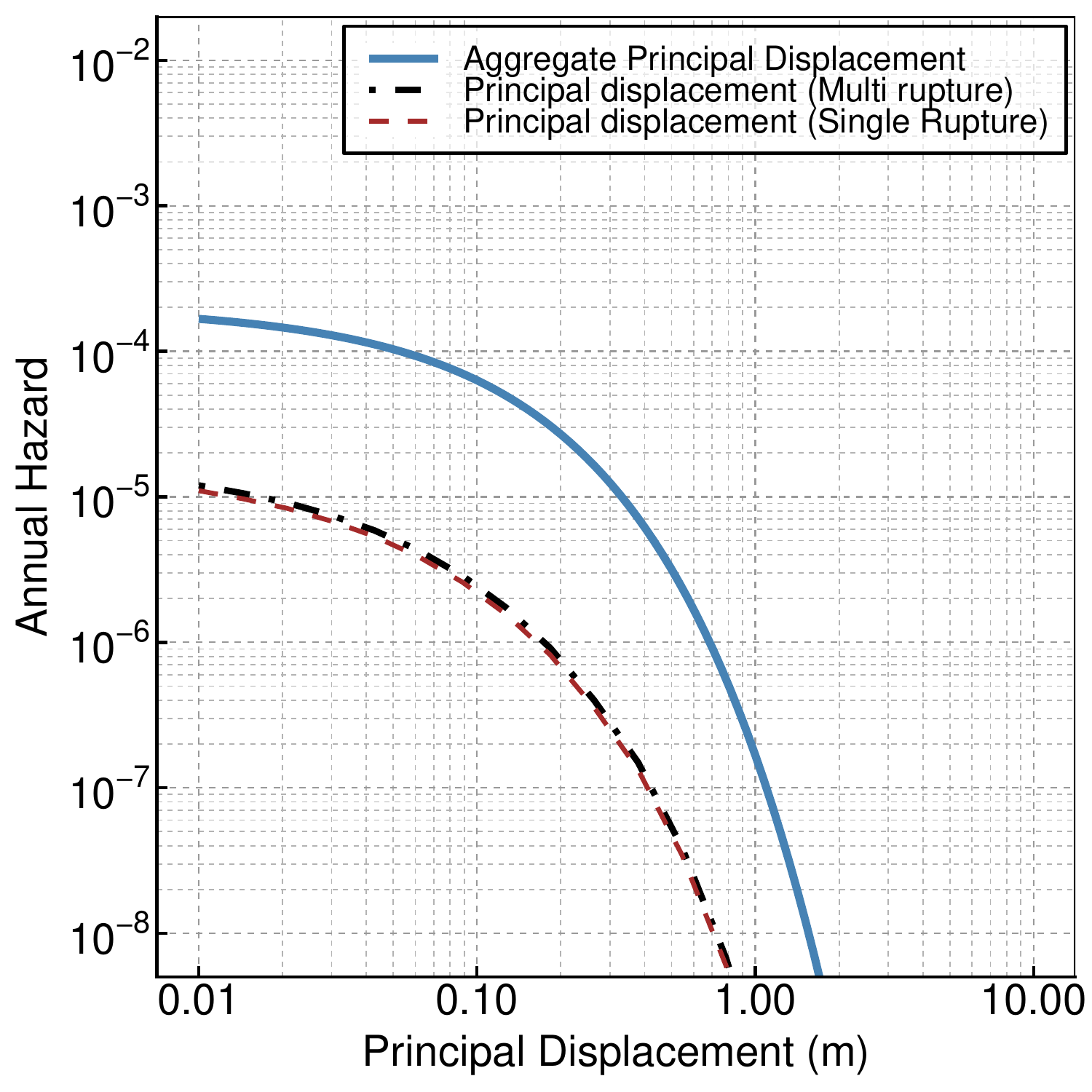}
        \label{fig:haz_agg_prncp_random_Kumamoto_3}
    \end{subfigure}
    \begin{subfigure}[d]{0.49\linewidth}
        \centering
        \caption{}
        \includegraphics[width=\linewidth]{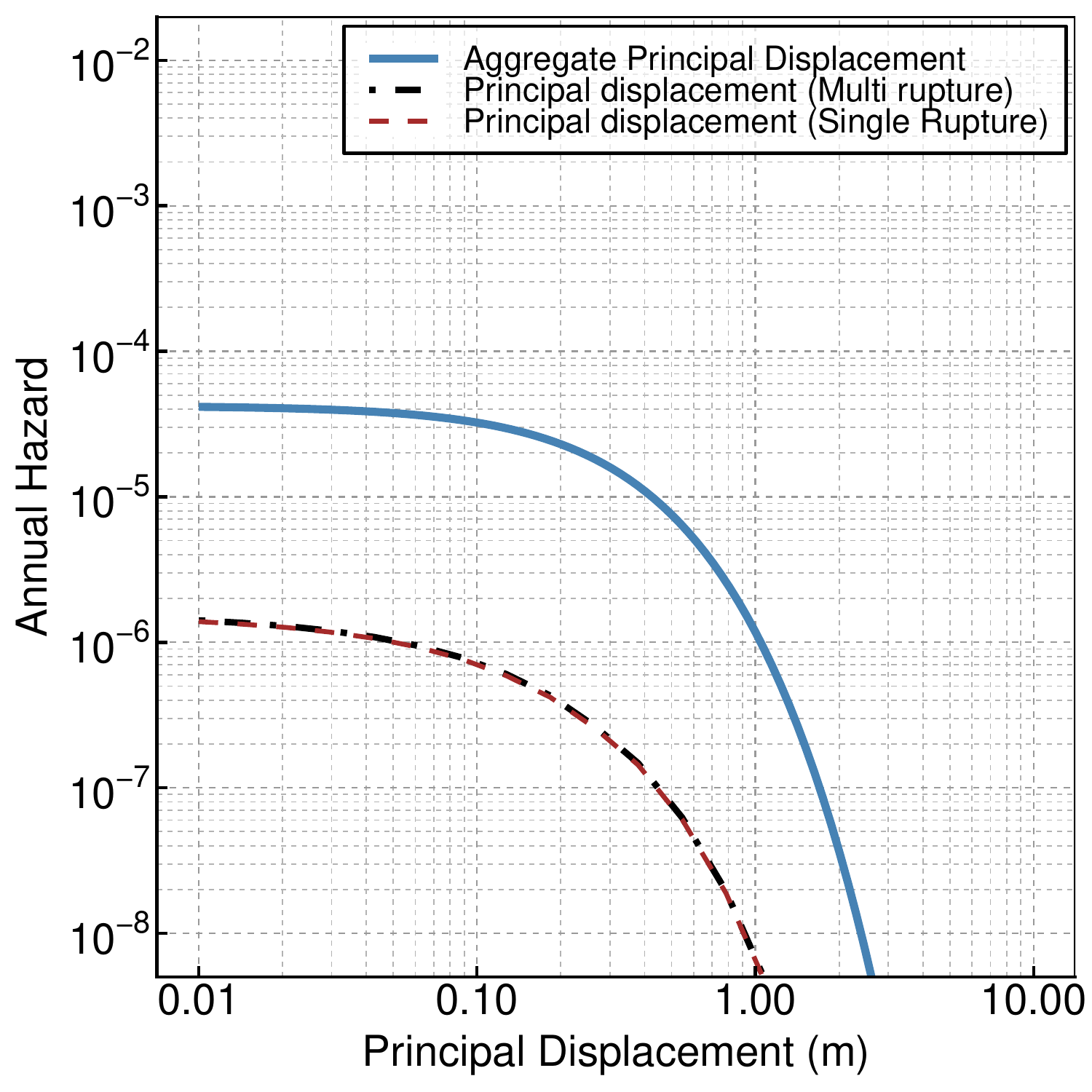}
        \label{fig:haz_agg_prncp_random_LeTeil_2}
    \end{subfigure}
    \begin{subfigure}[d]{0.49\linewidth}
        \centering
        \caption{}
        \includegraphics[width=\linewidth]{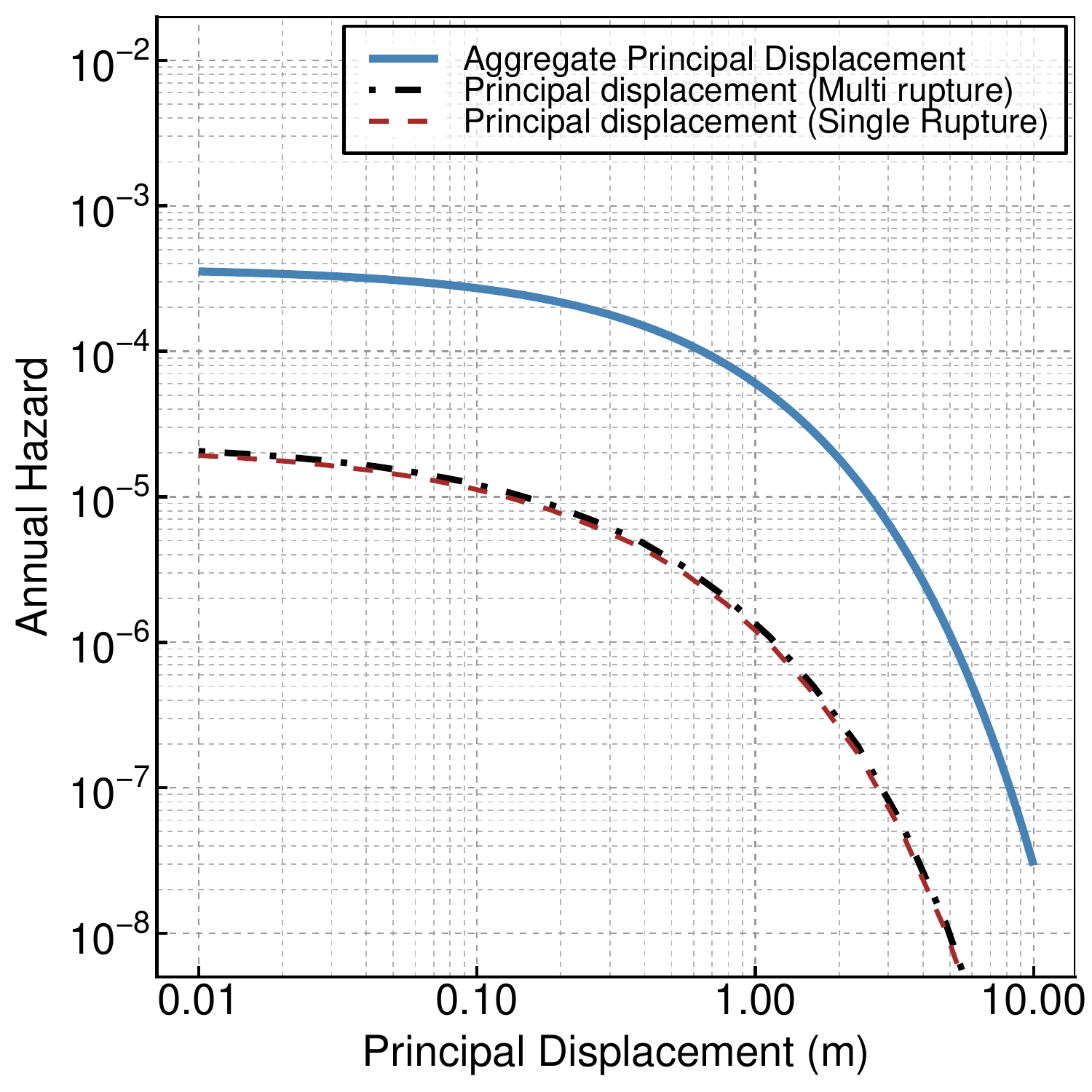}
        \label{fig:haz_agg_prncp_random_norcia_1}
    \end{subfigure}

    \caption{Comparison of hazard curves for aggregate principal displacement and principal displacement considering both single-rupture (a single rupture strand intersecting the structure) and multi-rupture (the cumulative contribution of one or more rupture strands intersecting the structure) hazard formulations under the randomly selected fault trace definition for the IAEA benchmark cases \cite{valentini2021probabilistic, valentini2025iaea}: (a) Kumamoto Case 2, (b) Kumamoto Case 3, (c) Le Teil Case 2, and (d) Norcia Case 1.}
    \label{fig:Hazardcurves_agg_prncp_random}
\end{figure}

Figure \ref{fig:Hazardcurves_agg_prncp_random} presents a comparison of hazard curve evaluations for the IAEA benchmark \citep{valentini2021probabilistic, valentini2025iaea} between aggregate principal displacement and principal displacement considering both single-rupture (a single rupture strand intersecting the structure) and multi-rupture (the cumulative contribution of one or more rupture strands intersecting the structure) with the randomly selected fault trace definition.

\begin{figure}[htbp]
    \centering

    \begin{subfigure}[d]{0.49\linewidth}
        \centering
        \caption{}
        \includegraphics[width=\linewidth]{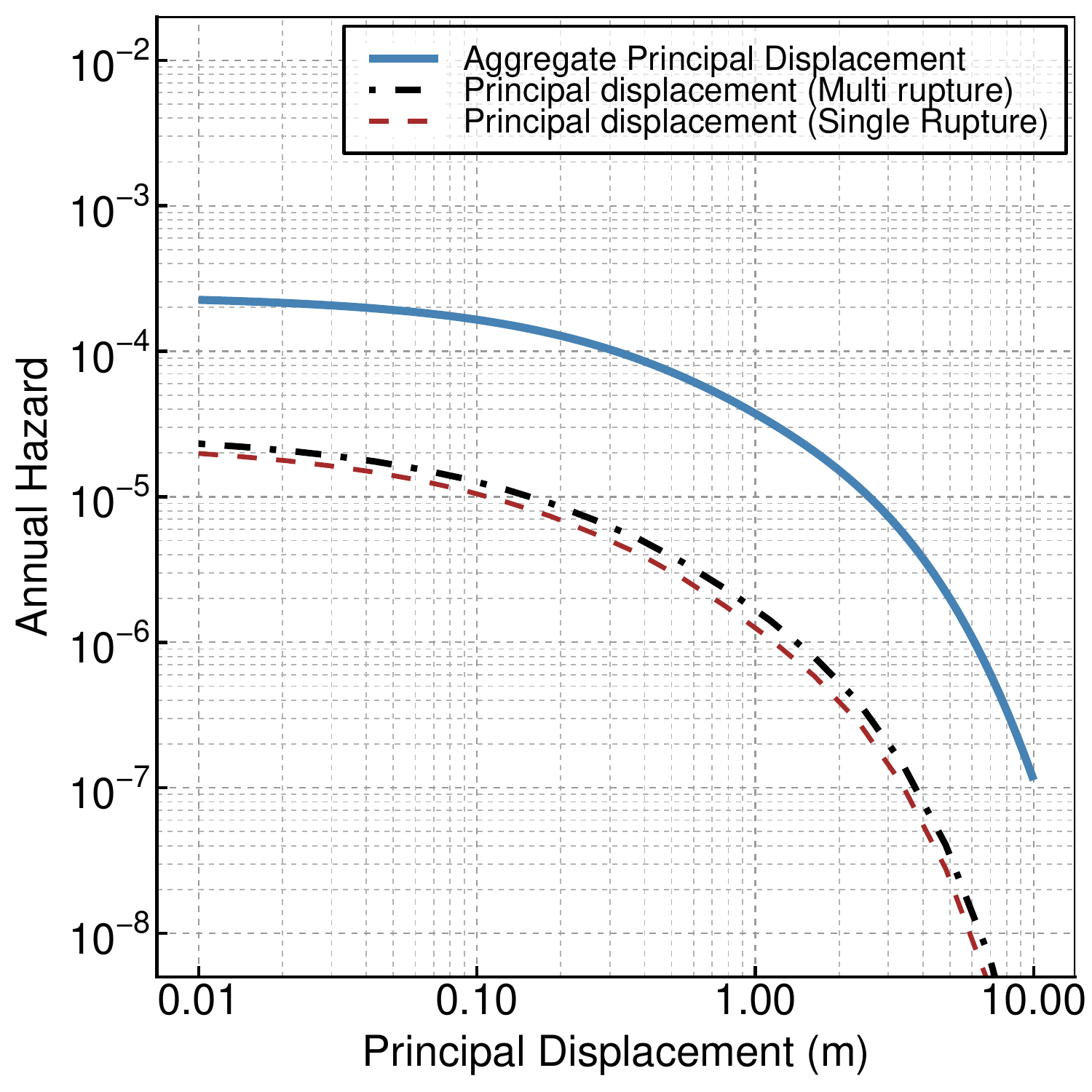}
        \label{fig:haz_agg_prncp_max_Kumamoto_2}
    \end{subfigure}
    \hfill
    \begin{subfigure}[d]{0.49\linewidth}
        \centering
        \caption{}
        \includegraphics[width=\linewidth]{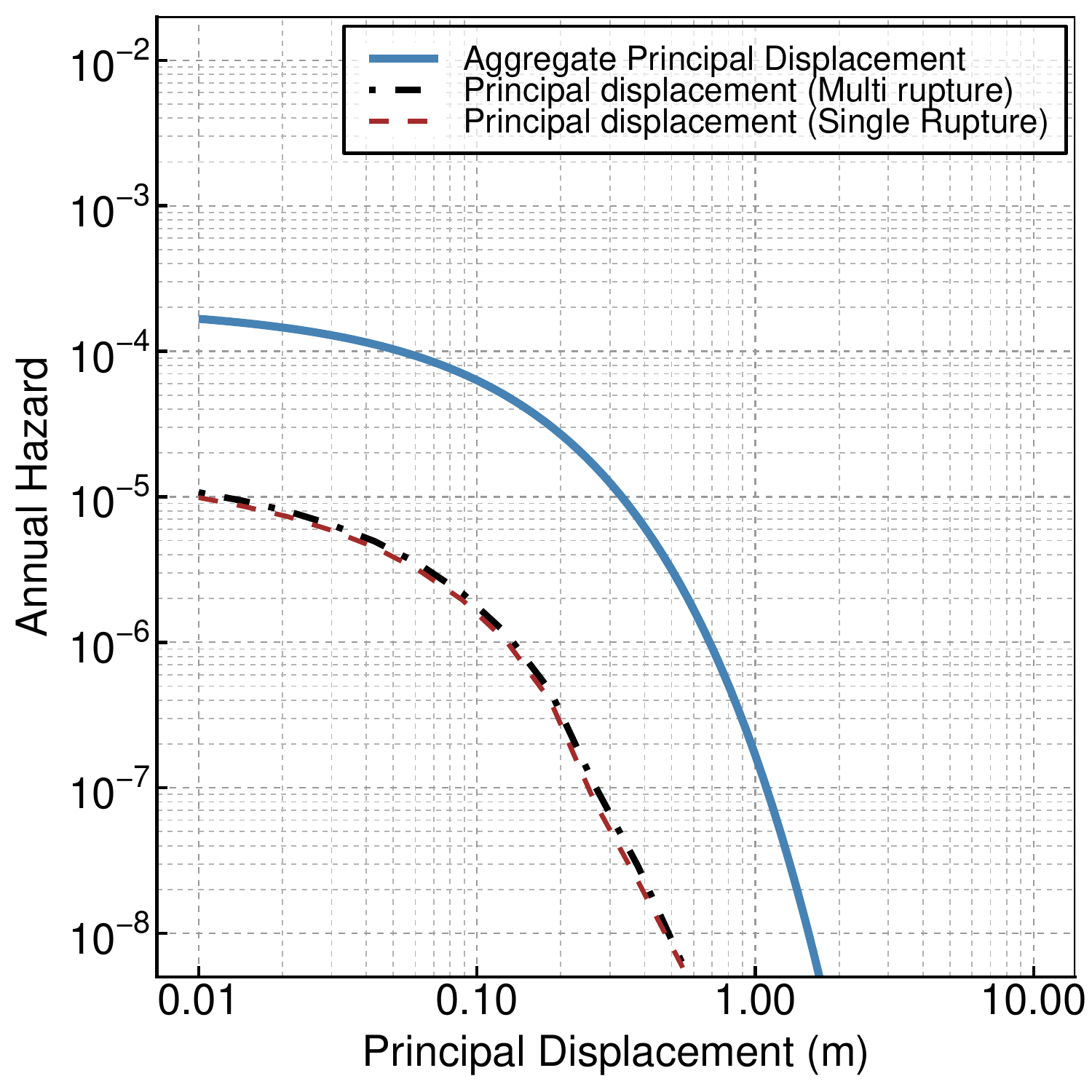}
        \label{fig:haz_agg_prncp_max_Kumamoto_3}
    \end{subfigure}
    \begin{subfigure}[d]{0.49\linewidth}
        \centering
        \caption{}
        \includegraphics[width=\linewidth]{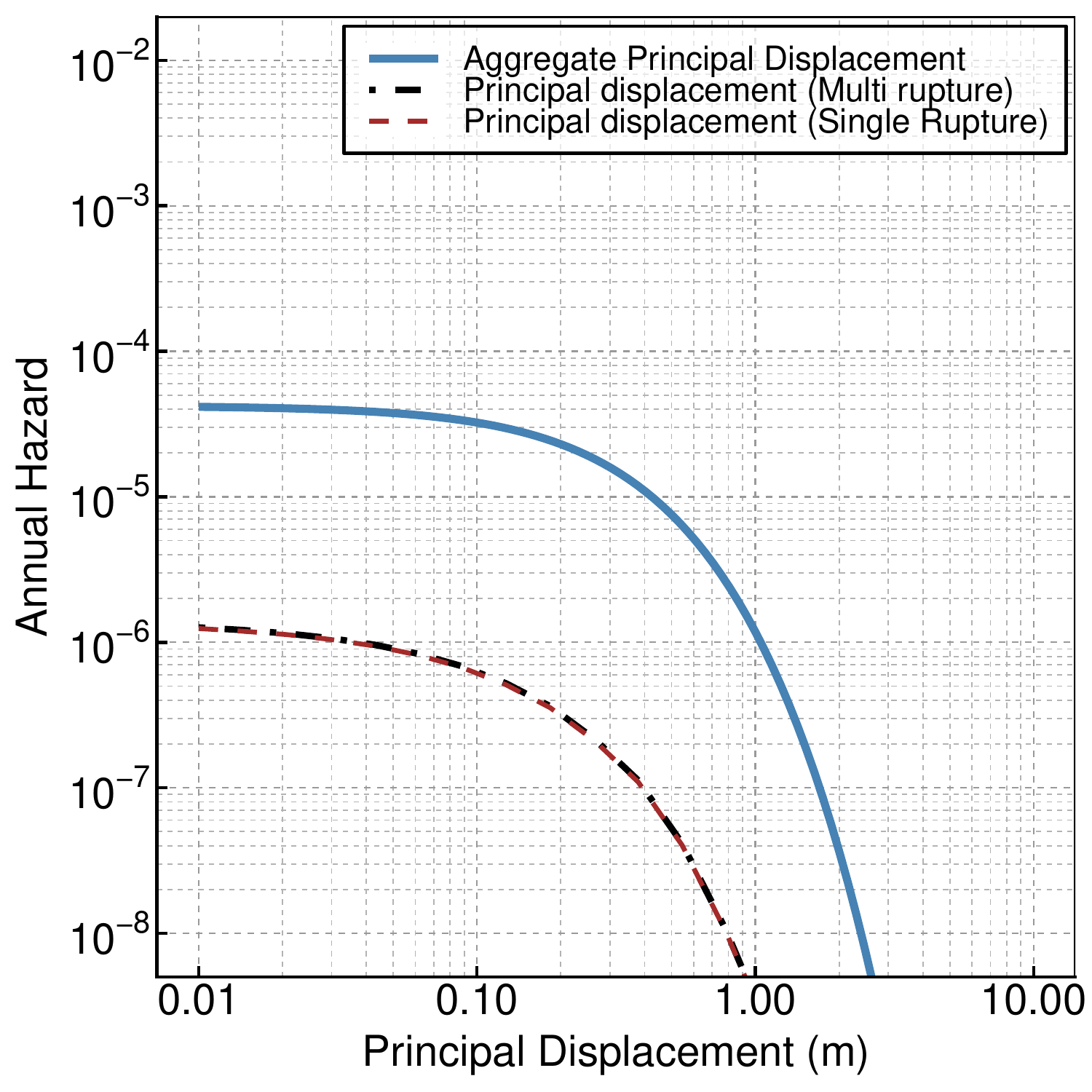}
        \label{fig:haz_agg_prncp_max_LeTeil_2}
    \end{subfigure}
    \begin{subfigure}[d]{0.49\linewidth}
        \centering
        \caption{}
        \includegraphics[width=\linewidth]{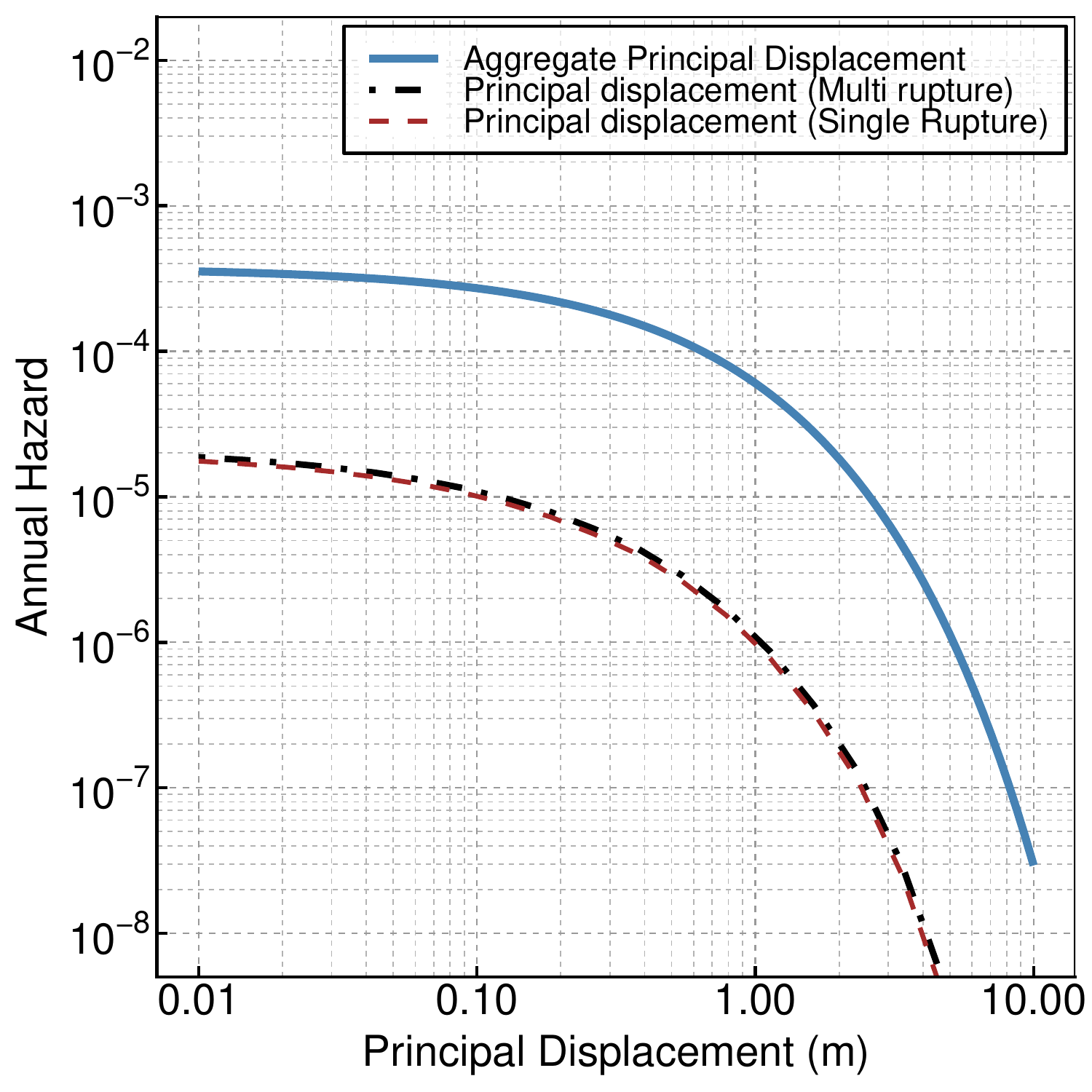}
        \label{fig:haz_agg_prncp_max_norcia_1}
    \end{subfigure}

    \caption{Comparison of hazard curves for aggregate principal displacement and principal displacement considering both single-rupture (a single rupture strand intersecting the structure) and multi-rupture (the cumulative contribution of one or more rupture strands intersecting the structure) hazard formulations under the maximum principal displacement fault trace definition for the IAEA benchmark cases \cite{valentini2021probabilistic, valentini2025iaea}: (a) Kumamoto Case 2, (b) Kumamoto Case 3, (c) Le Teil Case 2, and (d) Norcia Case 1.}
    \label{fig:Hazardcurves_agg_prncp_max}
\end{figure}

Figure \ref{fig:Hazardcurves_agg_prncp_max} presents a comparison of hazard curve evaluations for the IAEA benchmark \citep{valentini2021probabilistic, valentini2025iaea} between aggregate principal displacement and principal displacement considering both single-rupture (a single rupture strand intersecting the structure) and multi-rupture (the cumulative contribution of one or more rupture strands intersecting the structure) with the maximum principal displacement fault trace definition.

\begin{figure}[htbp]
    \centering

    \begin{subfigure}[d]{0.49\linewidth}
        \centering
        \caption{}
        \includegraphics[width=\linewidth]{figures/Hazard_curves/PFDHA_hazard_dist_Kumamoto_2_0050_multirupture_principal_displacement_plot.pdf}
        \label{fig:haz_trace_0050_Kumamoto_2}
    \end{subfigure}
    \hfill
    \begin{subfigure}[d]{0.49\linewidth}
        \centering
        \caption{}
        \includegraphics[width=\linewidth]{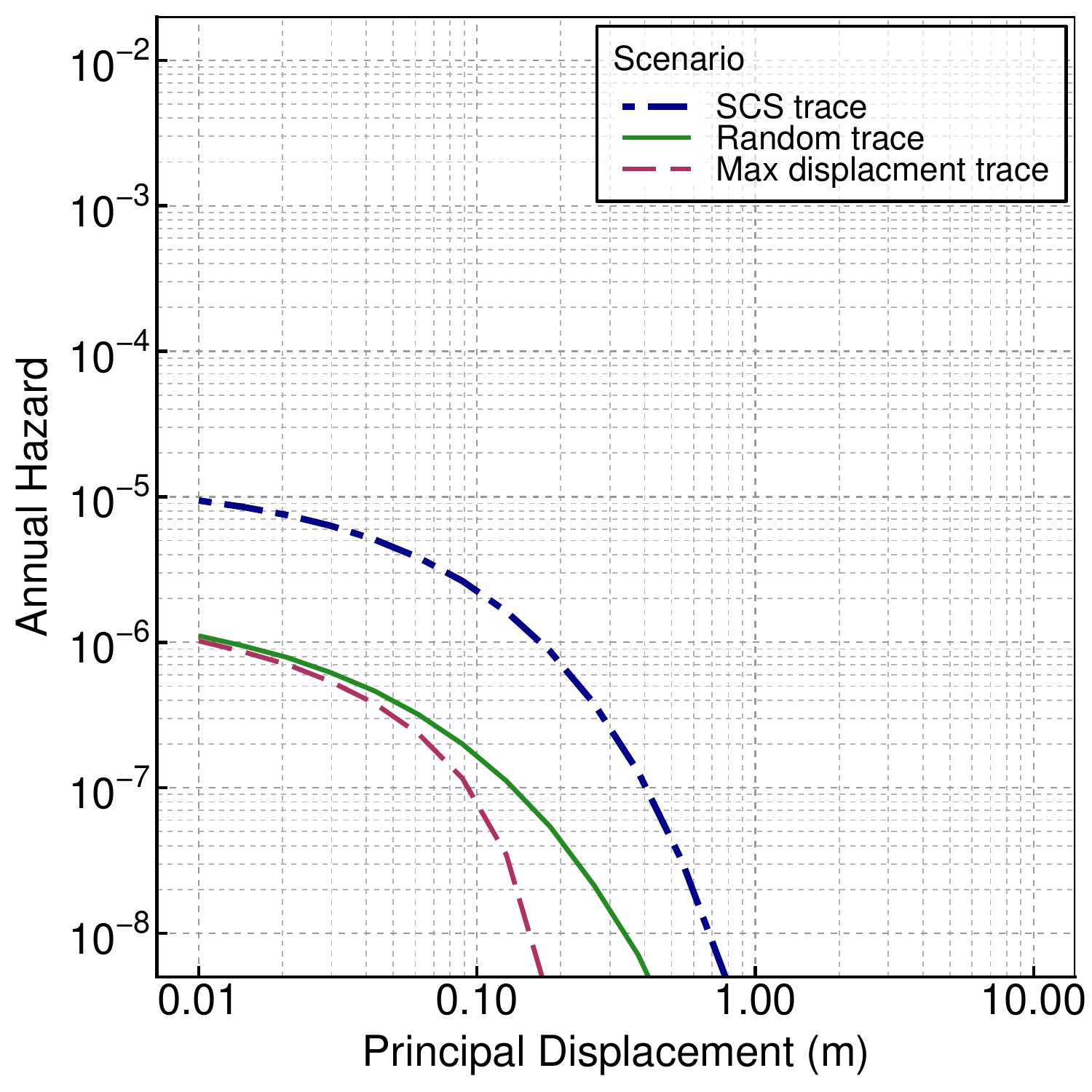}
        \label{fig:haz_trace_0050_Kumamoto_3}
    \end{subfigure}
    \begin{subfigure}[d]{0.49\linewidth}
        \centering
        \caption{}
        \includegraphics[width=\linewidth]{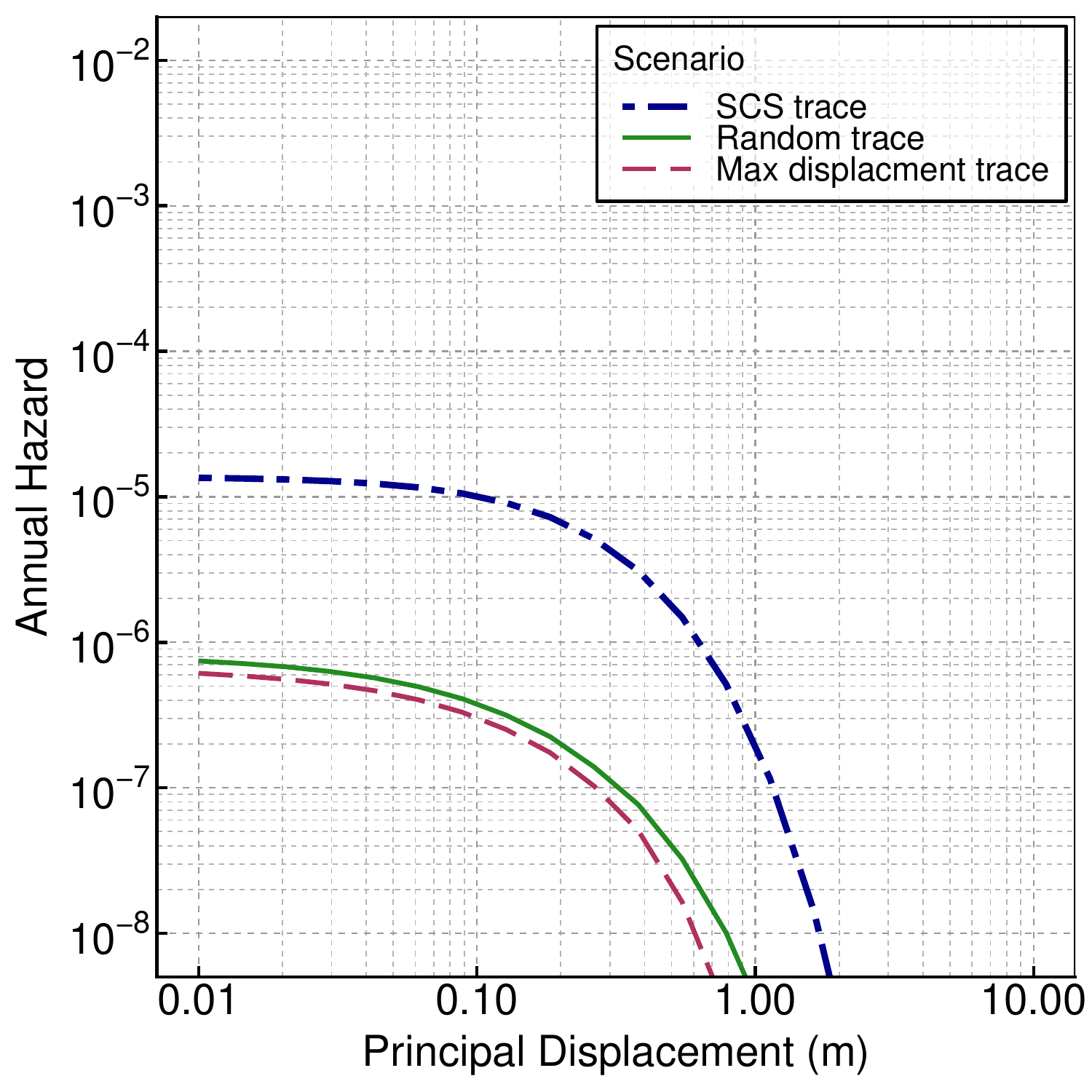}
        \label{fig:haz_trace_0050_LeTeil_2}
    \end{subfigure}
    \begin{subfigure}[d]{0.49\linewidth}
        \centering
        \caption{}
        \includegraphics[width=\linewidth]{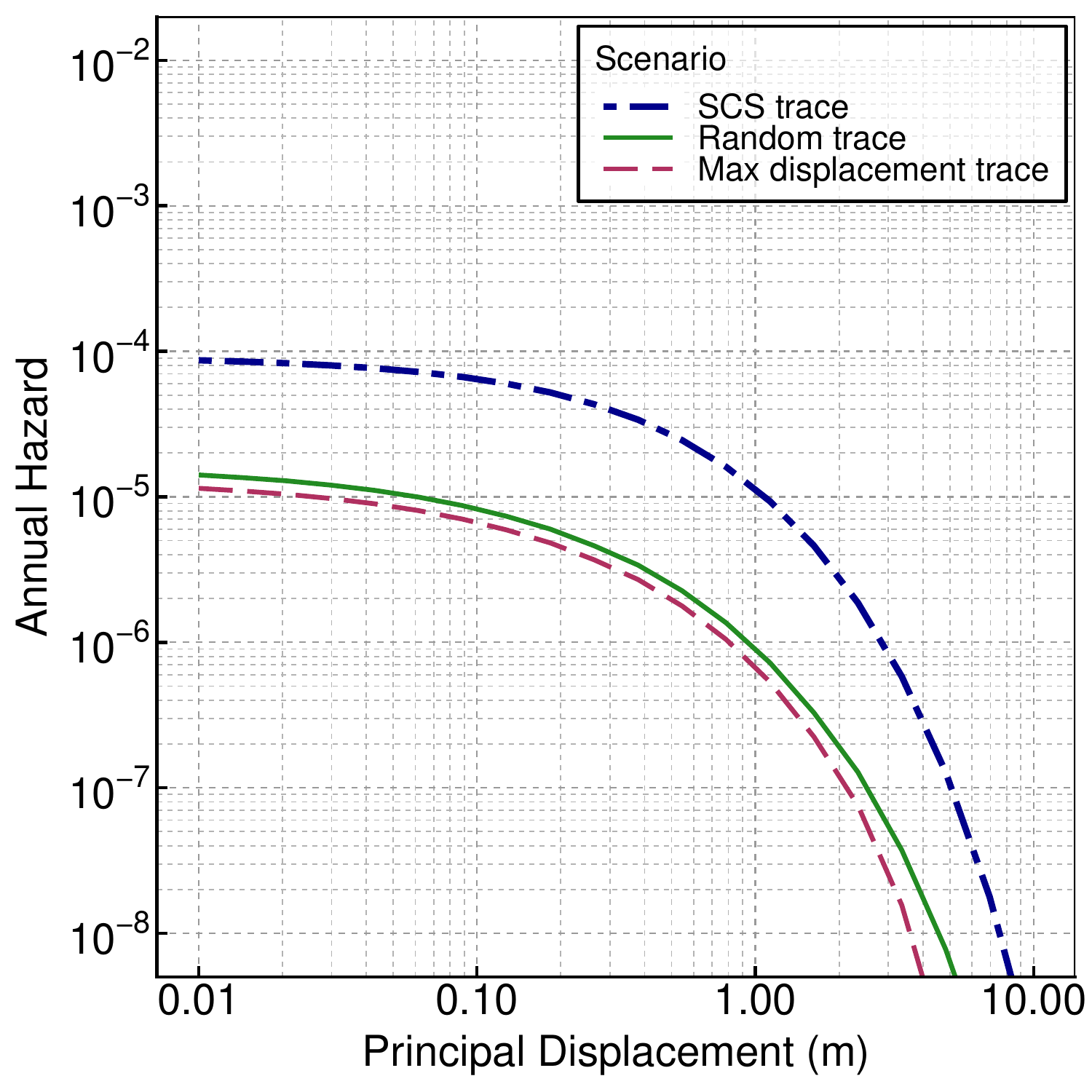}
        \label{fig:haz_trace_0050_norcia_1}
    \end{subfigure}

    \caption{Comparison of hazard curves for the multi-rupture principal displacement framework (cumulative contribution of one or more rupture strands intersecting the structure) considering different fault trace representations for structures located within a distance range of $[5~to~50]~m$ from the fault for the IAEA benchmark \cite{valentini2021probabilistic, valentini2025iaea}: (a) Kumamoto case 2, (b) Kumamoto case 3, (c) Le Teil case 2, and (d) Norcia case 1.}
    \label{fig:Hazardcurves_traces_0050}
\end{figure}

Figure \ref{fig:Hazardcurves_traces_0050} presents comparisons of hazard curve evaluations for the multi-rupture principal displacement framework (cumulative contribution of one or more rupture strands intersecting the structure) for the IAEA benchmark \citep{valentini2021probabilistic, valentini2025iaea} considering different fault trace definitions for structures located within a distance range of $[5~to~50]~m$ from the fault.

\begin{figure}[htbp]
    \centering

    \begin{subfigure}[d]{0.49\linewidth}
        \centering
        \caption{}
        \includegraphics[width=\linewidth]{figures/Hazard_curves/PFDHA_hazard_dist_Kumamoto_2_2525_multirupture_principal_displacement_plot.pdf}
        \label{fig:haz_trace_2525_Kumamoto_2}
    \end{subfigure}
    \hfill
    \begin{subfigure}[d]{0.49\linewidth}
        \centering
        \caption{}
        \includegraphics[width=\linewidth]{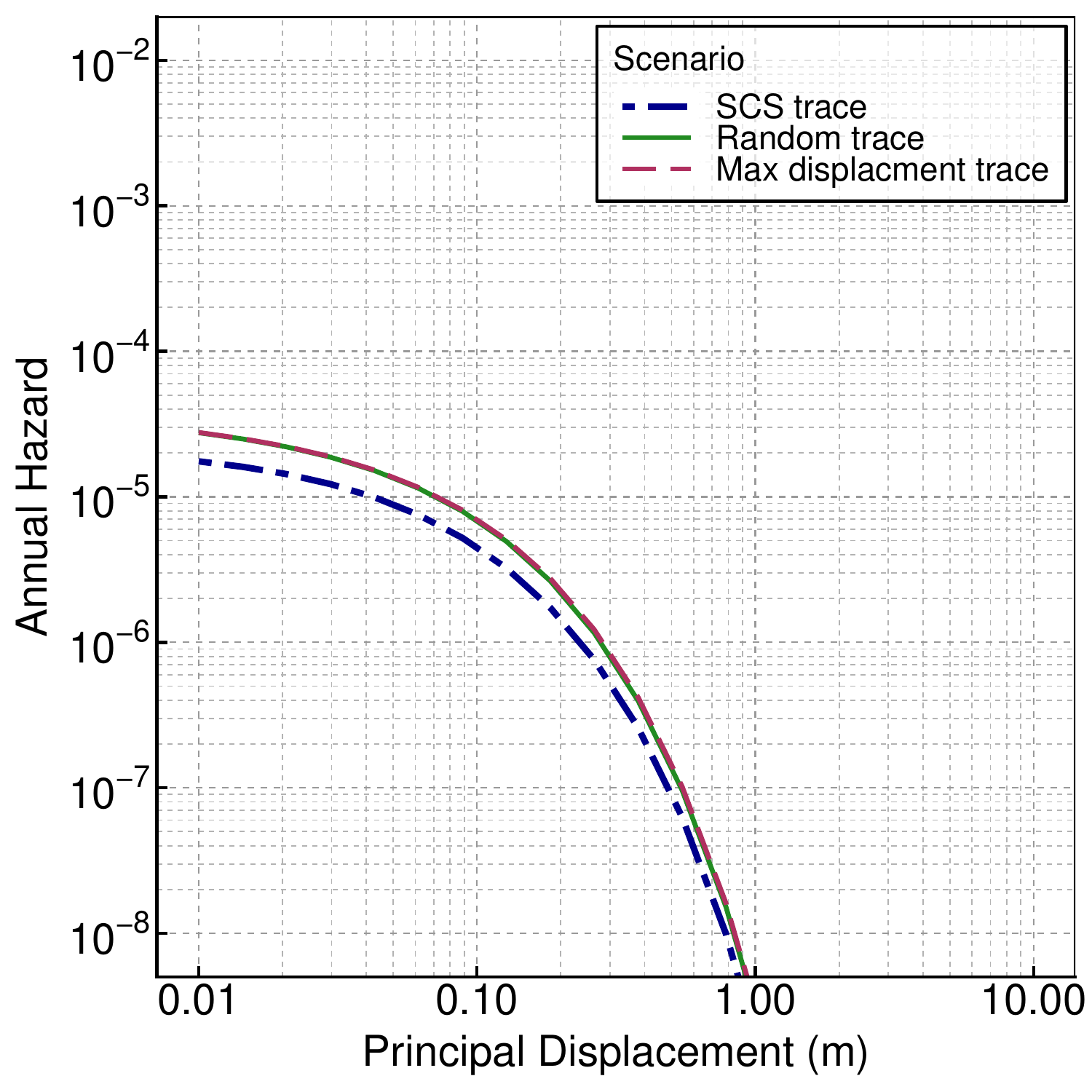}
        \label{fig:haz_trace_2525_Kumamoto_3}
    \end{subfigure}
    \begin{subfigure}[d]{0.49\linewidth}
        \centering
        \caption{}
        \includegraphics[width=\linewidth]{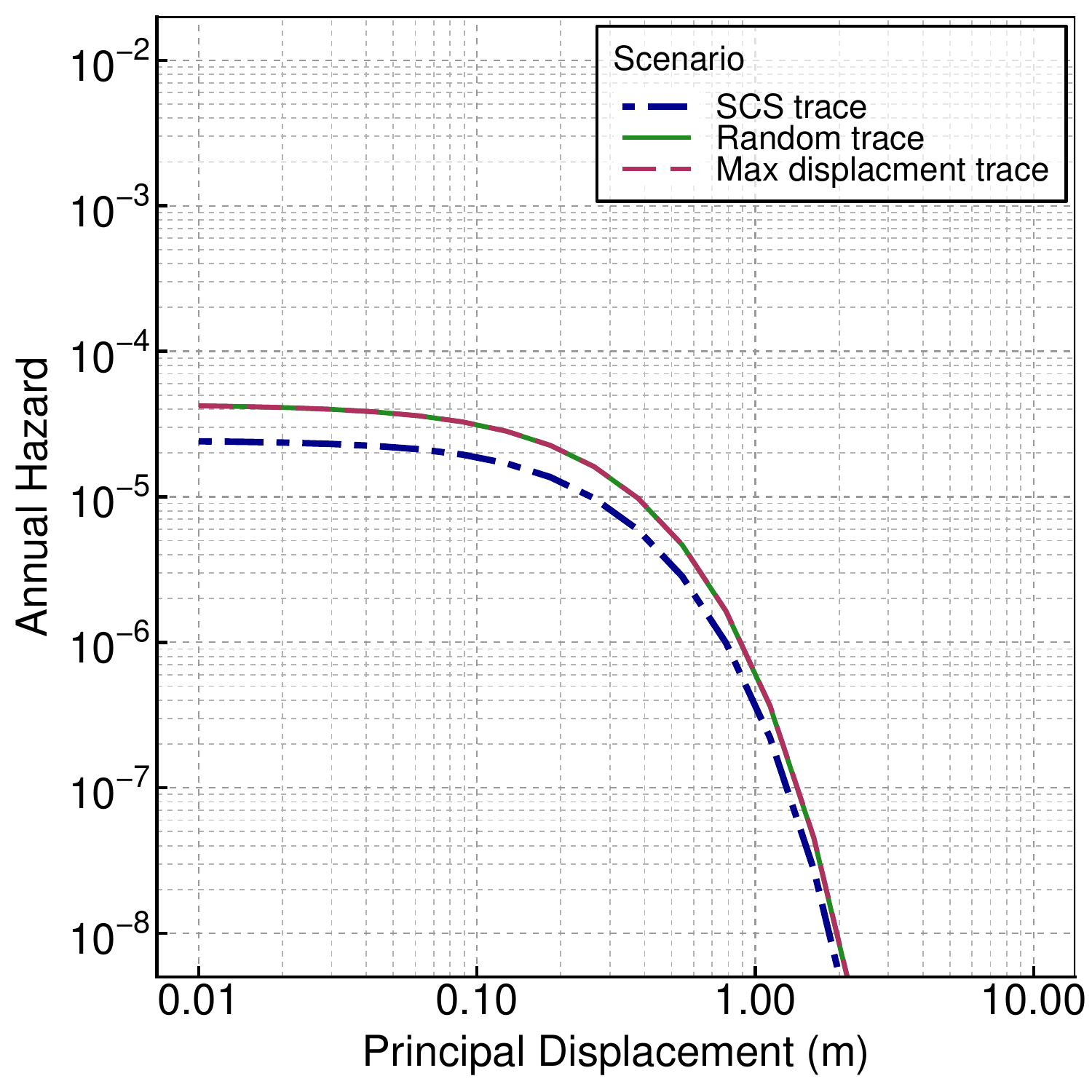}
        \label{fig:haz_trace_2525_LeTeil_2}
    \end{subfigure}
    \begin{subfigure}[d]{0.49\linewidth}
        \centering
        \caption{}
        \includegraphics[width=\linewidth]{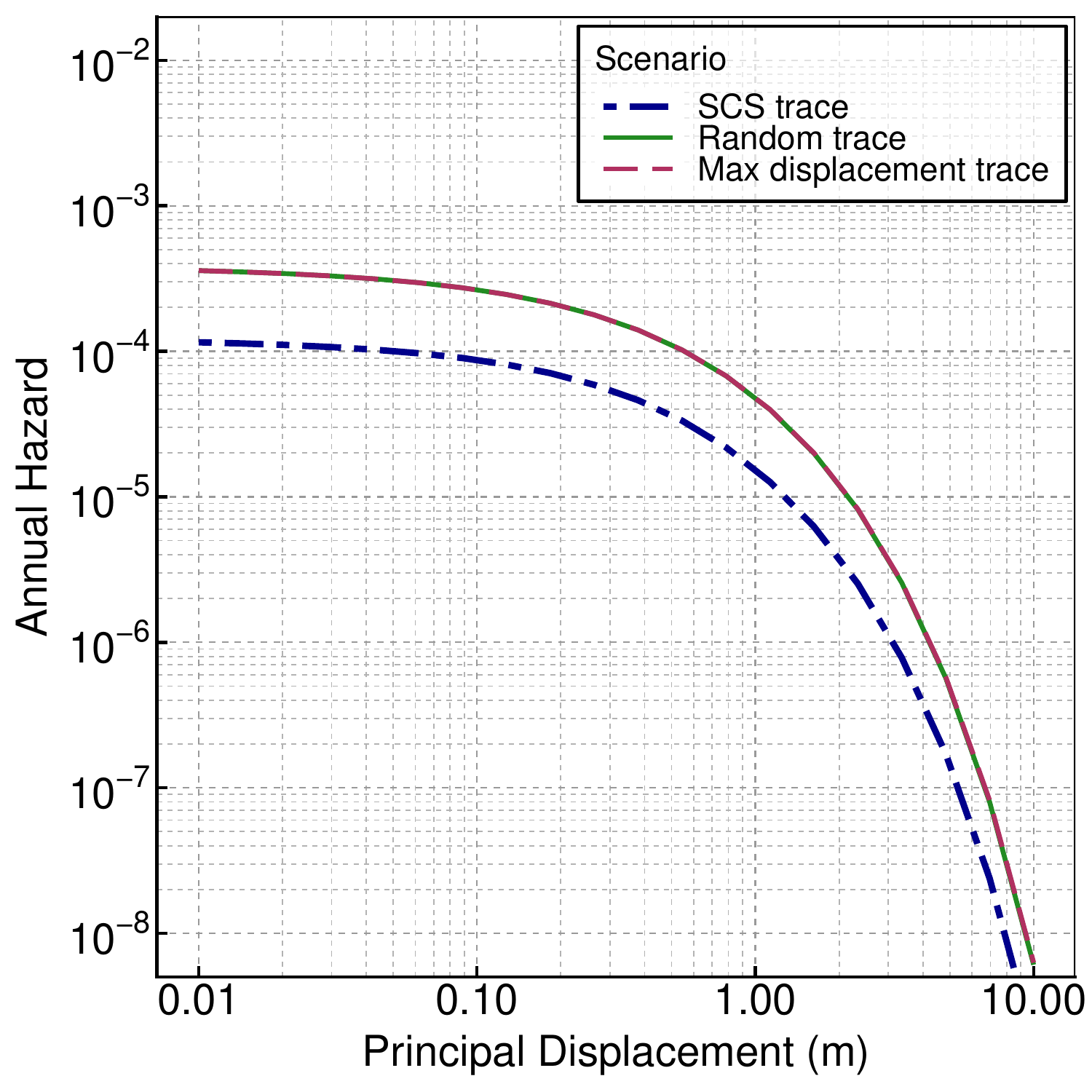}
        \label{fig:haz_trace_2525_norcia_1}
    \end{subfigure}

    \caption{Comparison of hazard curves for the multi-rupture principal displacement framework (cumulative contribution of one or more rupture strands intersecting the structure) considering different fault trace representations for structures located within a distance range of $[-25~to~25]~m$ from the fault for the IAEA benchmark \cite{valentini2021probabilistic, valentini2025iaea}: (a) Kumamoto case 2, (b) Kumamoto case 3, (c) Le Teil case 2, and (d) Norcia case 1.}
        \label{fig:Hazardcurves_traces_2525}
\end{figure}

Figure \ref{fig:Hazardcurves_traces_2525} presents comparisons of hazard curve evaluations for the multi-rupture principal displacement framework (cumulative contribution of one or more rupture strands intersecting the structure) for the IAEA benchmark \citep{valentini2021probabilistic, valentini2025iaea} considering different fault trace definitions for structures located within a distance range of $[-25~to~25]~m$ from the fault.

\begin{figure}[htbp]
    \centering

    \begin{subfigure}[d]{0.49\linewidth}
        \centering
        \caption{}
        \includegraphics[width=\linewidth]{figures/Hazard_curves/PFDHA_hazard_dist_multirupture_Kumamoto_2_SCStrace_principal_displacement_plot.pdf}
        \label{fig:haz_dist_SCS_Kumamoto_2}
    \end{subfigure}
    \hfill
    \begin{subfigure}[d]{0.49\linewidth}
        \centering
        \caption{}
        \includegraphics[width=\linewidth]{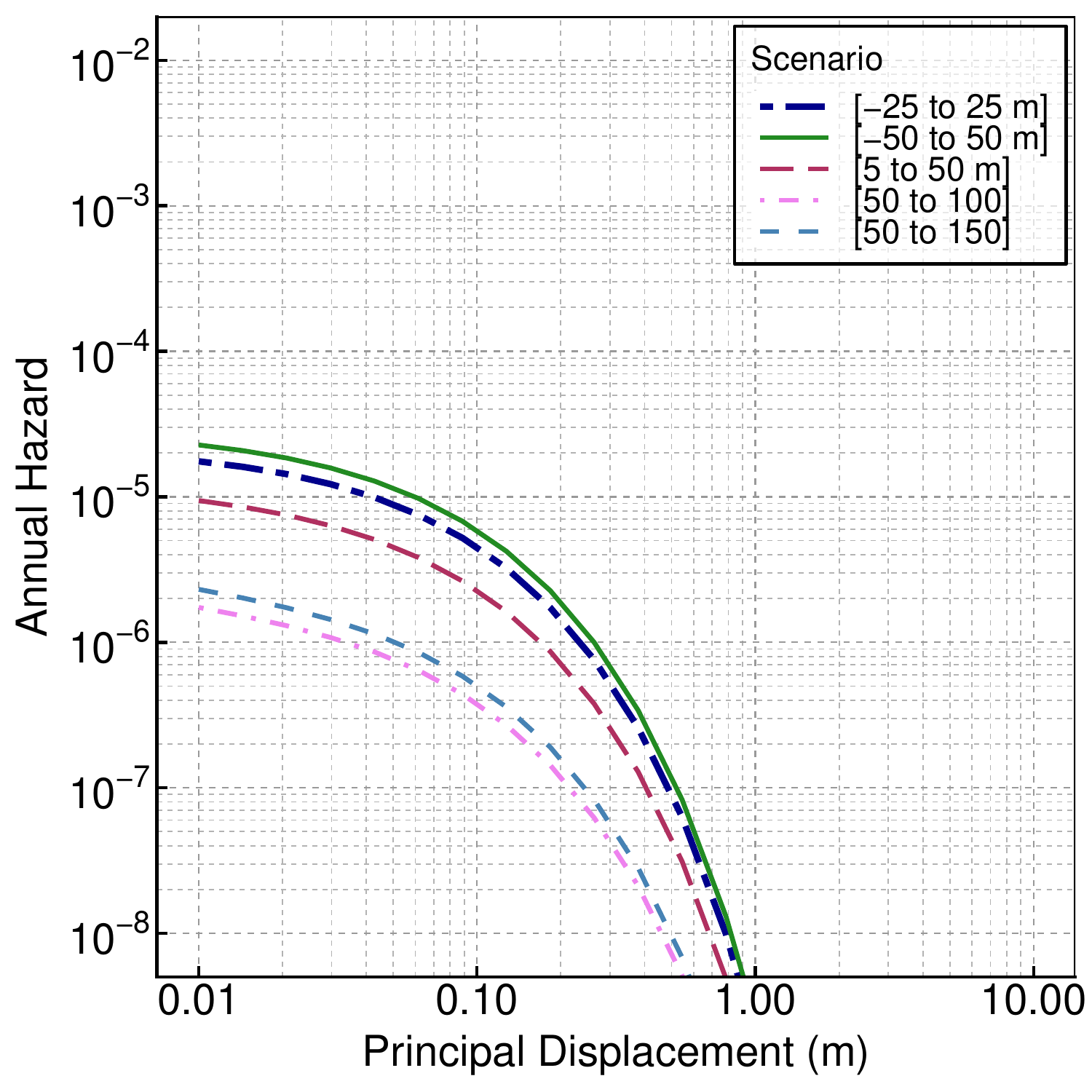}
        \label{fig:haz_dist_SCS_Kumamoto_3}
    \end{subfigure}
    \begin{subfigure}[d]{0.49\linewidth}
        \centering
        \caption{}
        \includegraphics[width=\linewidth]{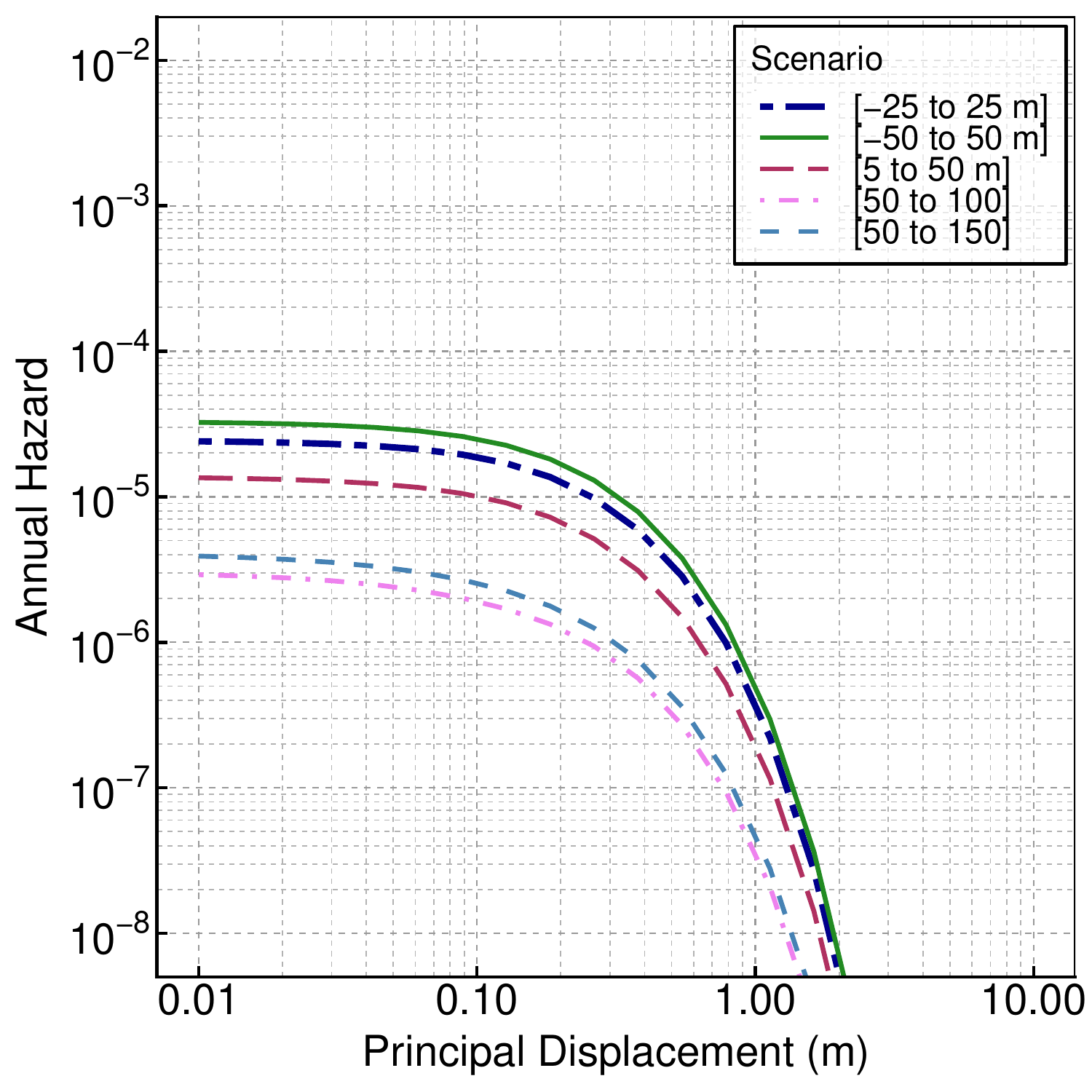}
        \label{fig:haz_dist_SCS_LeTeil_2}
    \end{subfigure}
    \begin{subfigure}[d]{0.49\linewidth}
        \centering
        \caption{}
        \includegraphics[width=\linewidth]{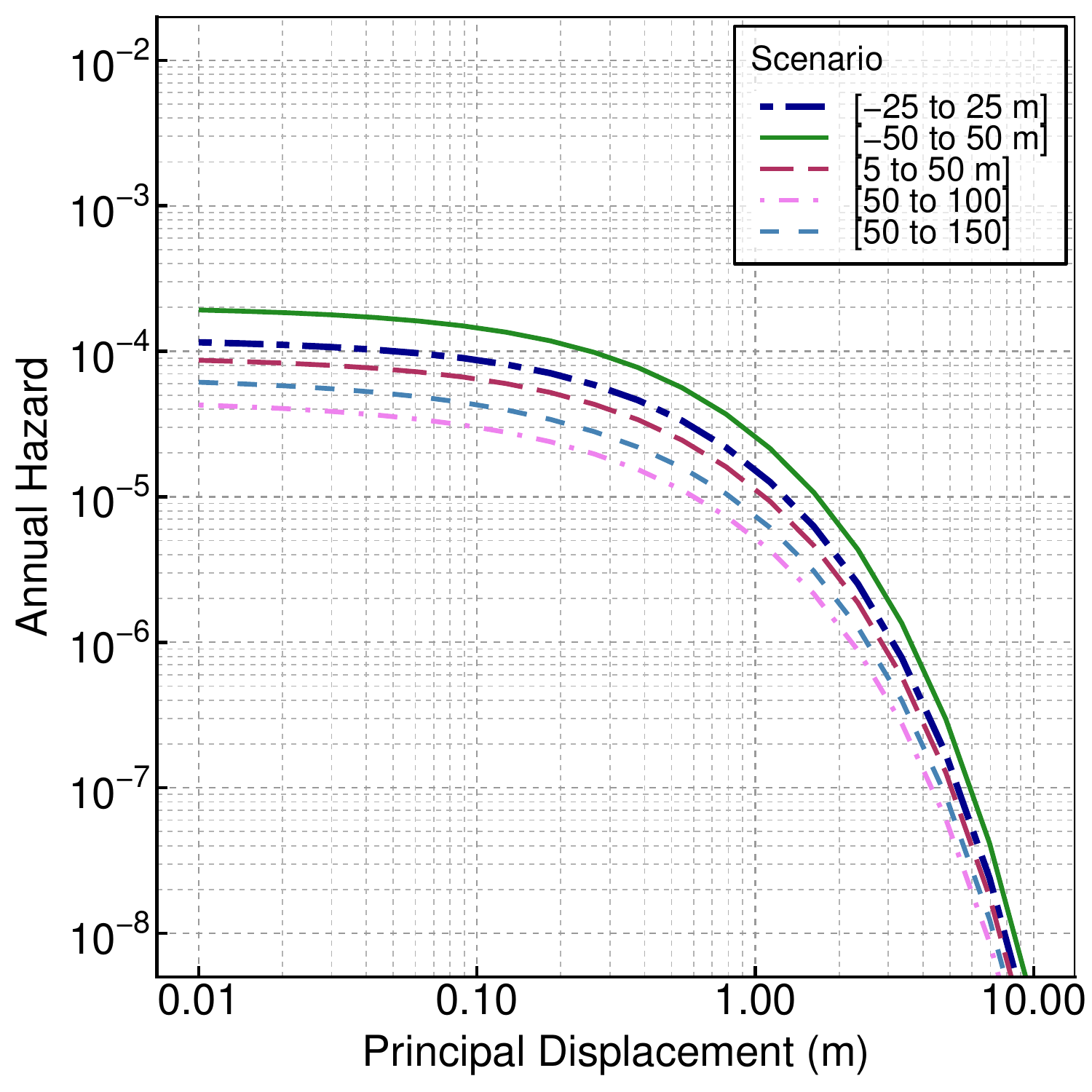}
        \label{fig:haz_dist_SCS_norcia_1}
    \end{subfigure}

    \caption{Comparison of hazard curves for multi-rupture principal displacement framework (cumulative contribution of one or more rupture strands intersecting the structure) with the SCS fault-trace framework considering different structure-to-fault distance ranges for the IAEA benchmark \cite{valentini2021probabilistic, valentini2025iaea}. The curves represent different distance intervals relative to the fault trace: (a) Kumamoto case 2, (b) Kumamoto case 3, (c) Le Teil case 2, and (d) Norcia case 1.}
    \label{fig:Hazardcurves_dist_SCS}
\end{figure}

Figure \ref{fig:Hazardcurves_dist_SCS} presents comparisons of hazard curve evaluations for the multi-rupture principal displacement framework (cumulative contribution of one or more rupture strands intersecting the structure) for the IAEA benchmark \citep{valentini2021probabilistic, valentini2025iaea} considering different structure-to-fault distance ranges under the SCS fault trace definition.

\begin{figure}[htbp]
    \centering

    \begin{subfigure}[d]{0.49\linewidth}
        \centering
        \caption{}
        \includegraphics[width=\linewidth]{figures/Hazard_curves/PFDHA_hazard_dist_multirupture_Kumamoto_2_randomtrace_principal_displacement_plot.pdf}
        \label{fig:haz_dist_random_Kumamoto_2}
    \end{subfigure}
    \hfill
    \begin{subfigure}[d]{0.49\linewidth}
        \centering
        \caption{}
        \includegraphics[width=\linewidth]{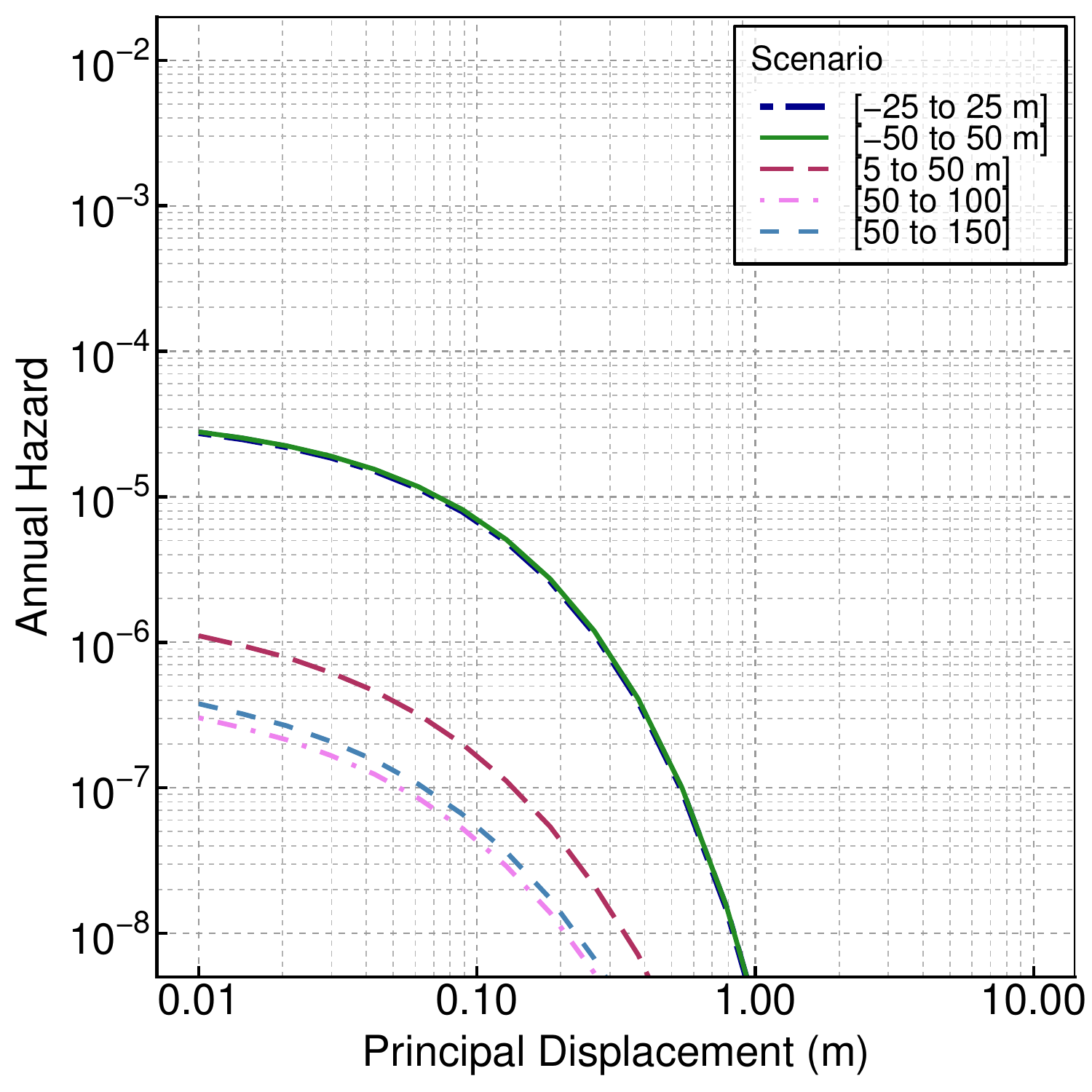}
        \label{fig:haz_dist_random_Kumamoto_3}
    \end{subfigure}
    \begin{subfigure}[d]{0.49\linewidth}
        \centering
        \caption{}
        \includegraphics[width=\linewidth]{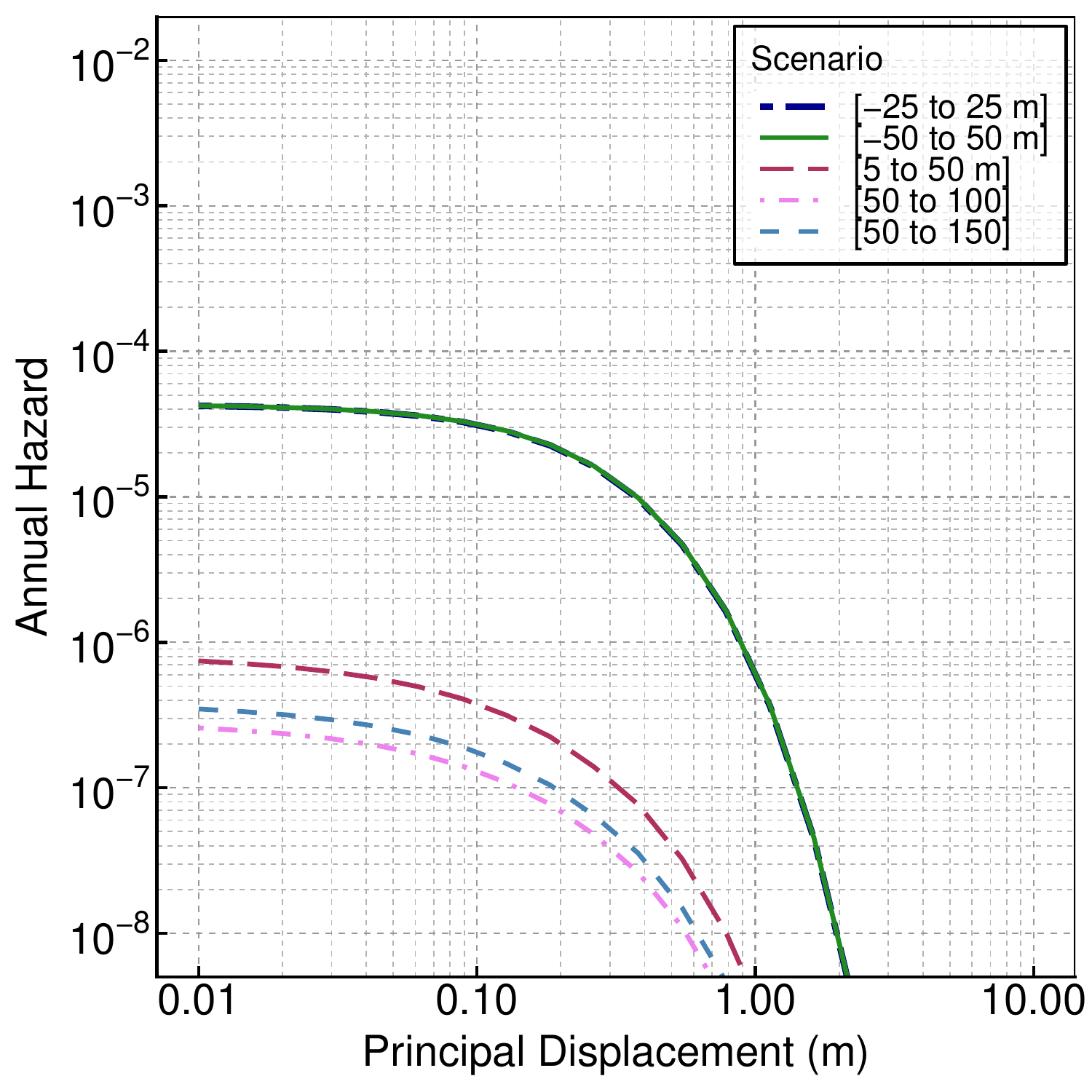}
        \label{fig:haz_dist_random_LeTeil_2}
    \end{subfigure}
    \begin{subfigure}[d]{0.49\linewidth}
        \centering
        \caption{}
        \includegraphics[width=\linewidth]{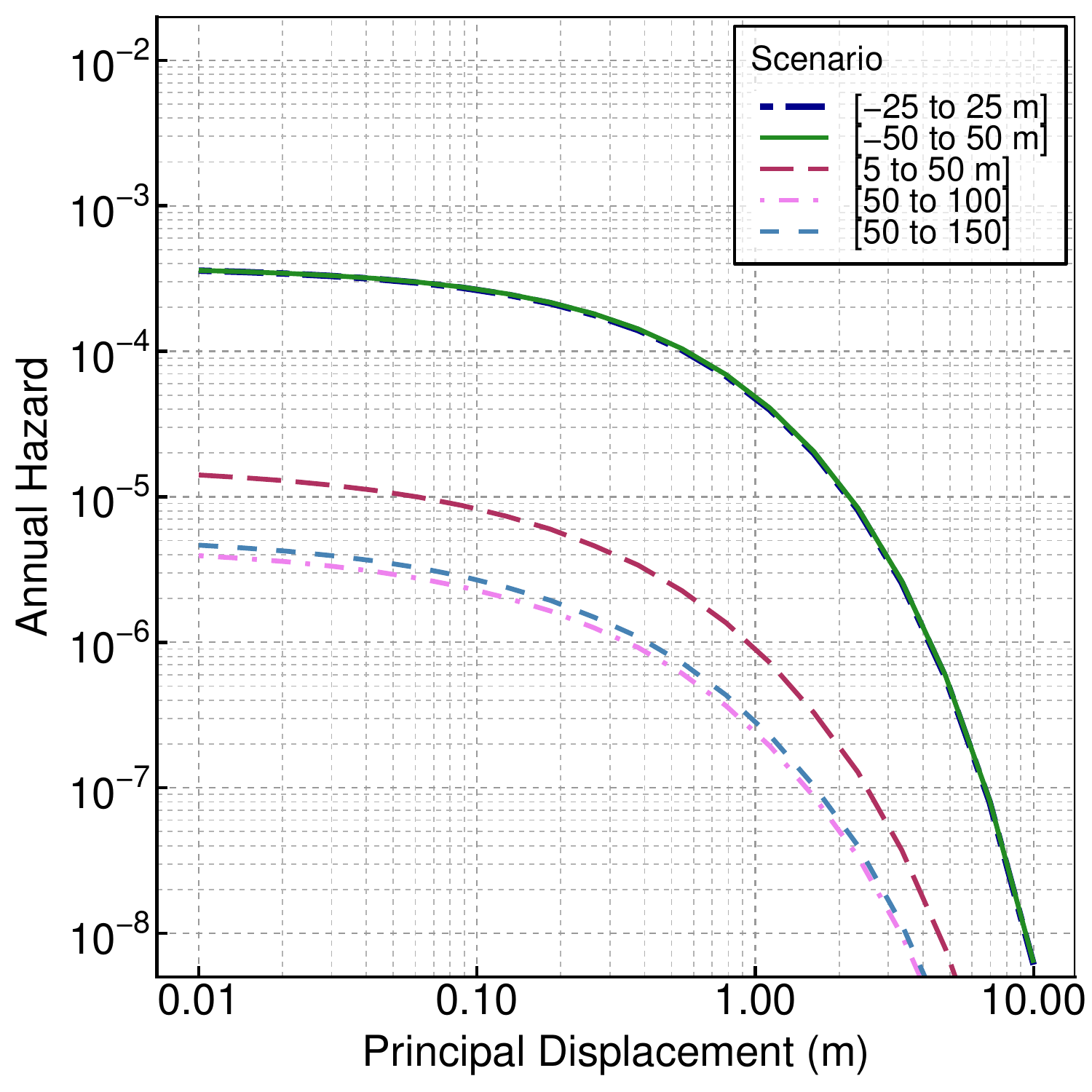}
        \label{fig:haz_dist_random_norcia_1}
    \end{subfigure}

    \caption{Comparison of hazard curves for multi-rupture principal displacement framework (cumulative contribution of one or more rupture strands intersecting the structure) with the random fault-trace framework considering different structure-to-fault distance ranges for the IAEA benchmark \cite{valentini2021probabilistic, valentini2025iaea}. The curves represent different distance intervals relative to the fault trace: (a) Kumamoto case 2, (b) Kumamoto case 3, (c) Le Teil case 2, and (d) Norcia case 1.}
    \label{fig:Hazardcurves_dist_random}
\end{figure}

Figure \ref{fig:Hazardcurves_dist_random} presents comparisons of hazard curve evaluations for the multi-rupture principal displacement framework (cumulative contribution of one or more rupture strands intersecting the structure) for the IAEA benchmark \citep{valentini2021probabilistic, valentini2025iaea} considering different structure-to-fault distance ranges under the randomly selected fault trace definition.

\begin{figure}[htbp]
    \centering

    \begin{subfigure}[d]{0.49\linewidth}
        \centering
        \caption{}
        \includegraphics[width=\linewidth]{figures/Hazard_curves/PFDHA_hazard_dist_multirupture_Kumamoto_2_maxtrace_principal_displacement_plot.pdf}
        \label{fig:haz_dist_max_Kumamoto_2}
    \end{subfigure}
    \hfill
    \begin{subfigure}[d]{0.49\linewidth}
        \centering
        \caption{}
        \includegraphics[width=\linewidth]{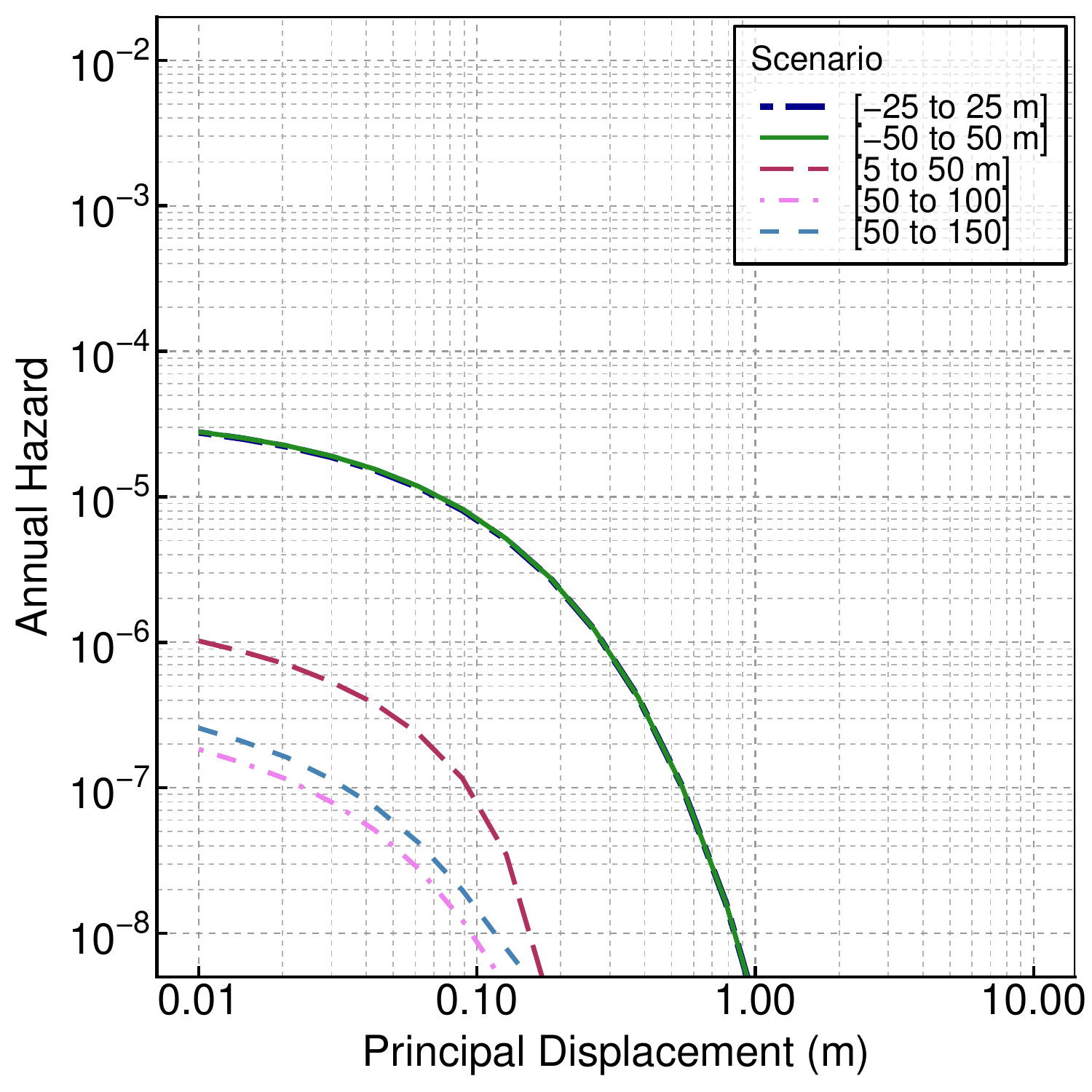}
        \label{fig:haz_dist_max_Kumamoto_3}
    \end{subfigure}
    \begin{subfigure}[d]{0.49\linewidth}
        \centering
        \caption{}
        \includegraphics[width=\linewidth]{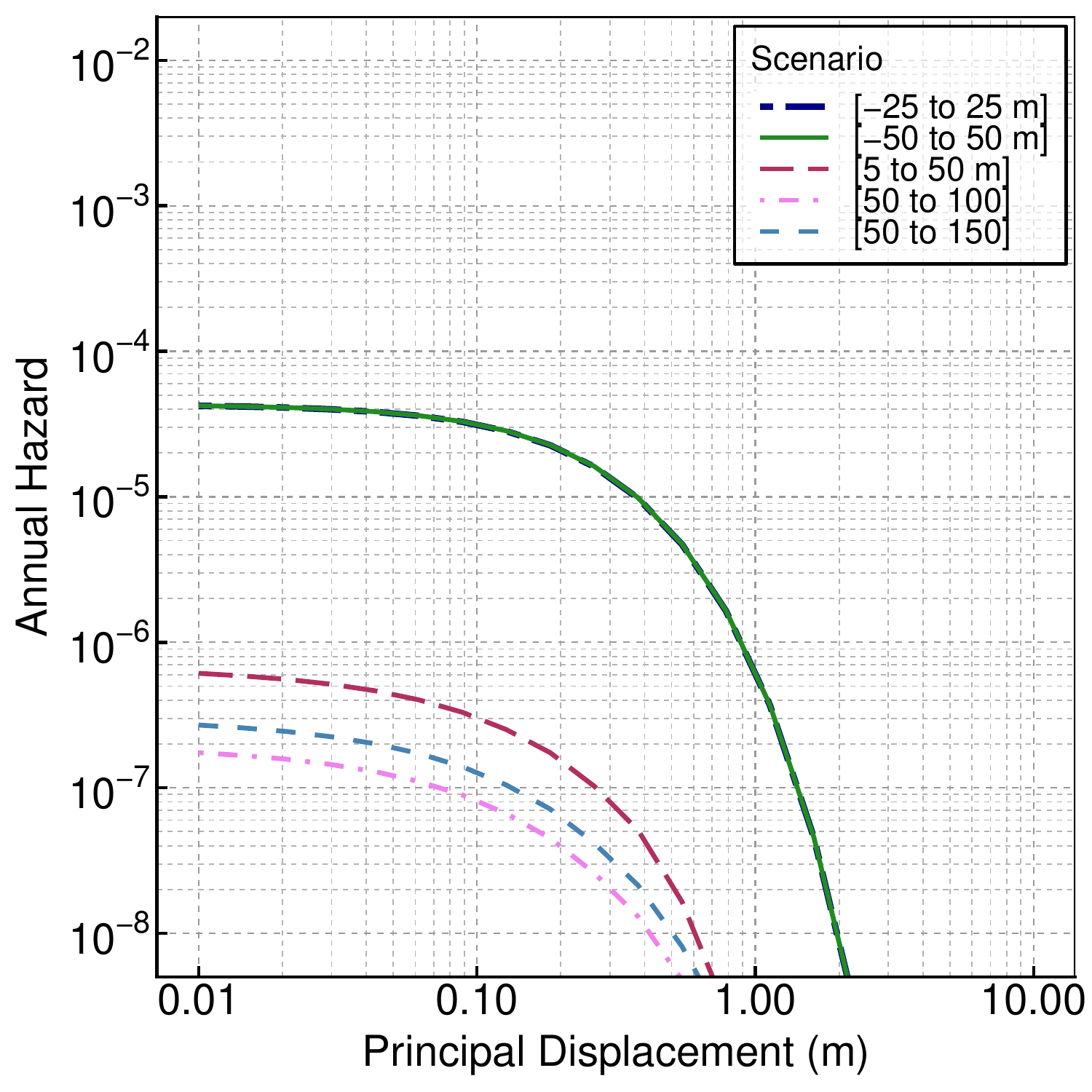}
        \label{fig:haz_dist_max_LeTeil_2}
    \end{subfigure}
    \begin{subfigure}[d]{0.49\linewidth}
        \centering
        \caption{}
        \includegraphics[width=\linewidth]{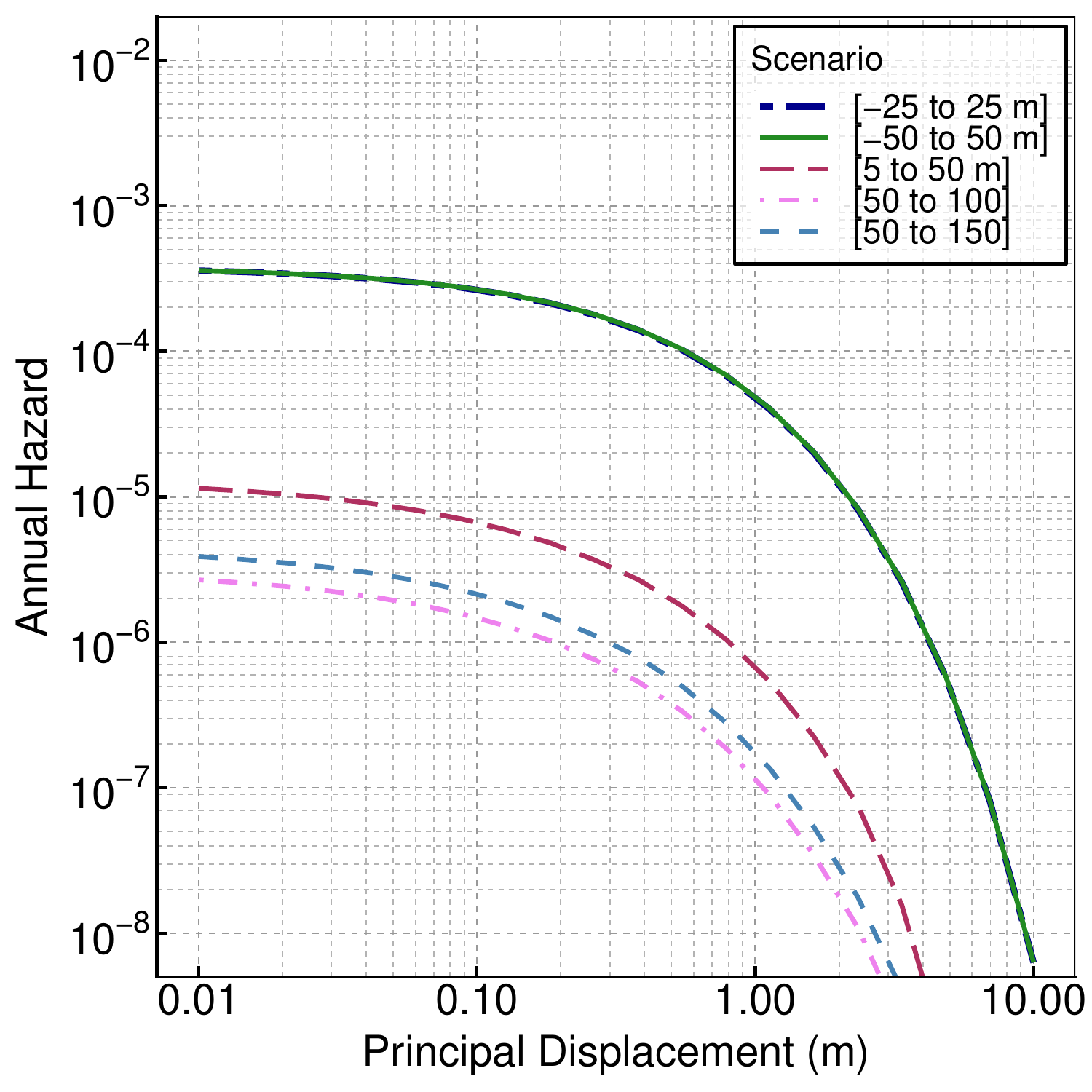}
        \label{fig:haz_dist_max_norcia_1}
    \end{subfigure}

    \caption{Comparison of hazard curves for multi-rupture principal displacement framework (cumulative contribution of one or more rupture strands intersecting the structure) with the max displacement fault-trace framework considering different structure-to-fault distance ranges for the IAEA benchmark \cite{valentini2021probabilistic, valentini2025iaea}. The curves represent different distance intervals relative to the fault trace: (a) Kumamoto case 2, (b) Kumamoto case 3, (c) Le Teil case 2, and (d) Norcia case 1.}
    \label{fig:Hazardcurves_dist_max}
\end{figure}

Figure \ref{fig:Hazardcurves_dist_max} presents comparisons of hazard curve evaluations for the multi-rupture principal displacement framework (cumulative contribution of one or more rupture strands intersecting the structure) for the IAEA benchmark \citep{valentini2021probabilistic, valentini2025iaea} considering different structure-to-fault distance ranges under the maximum principal displacement fault trace definition.

%
\newpage


\begin{figure}[p]
\centering

\begin{subfigure}[t]{0.31\linewidth}
    \centering
    \caption{}
    \includegraphics[width=\linewidth]{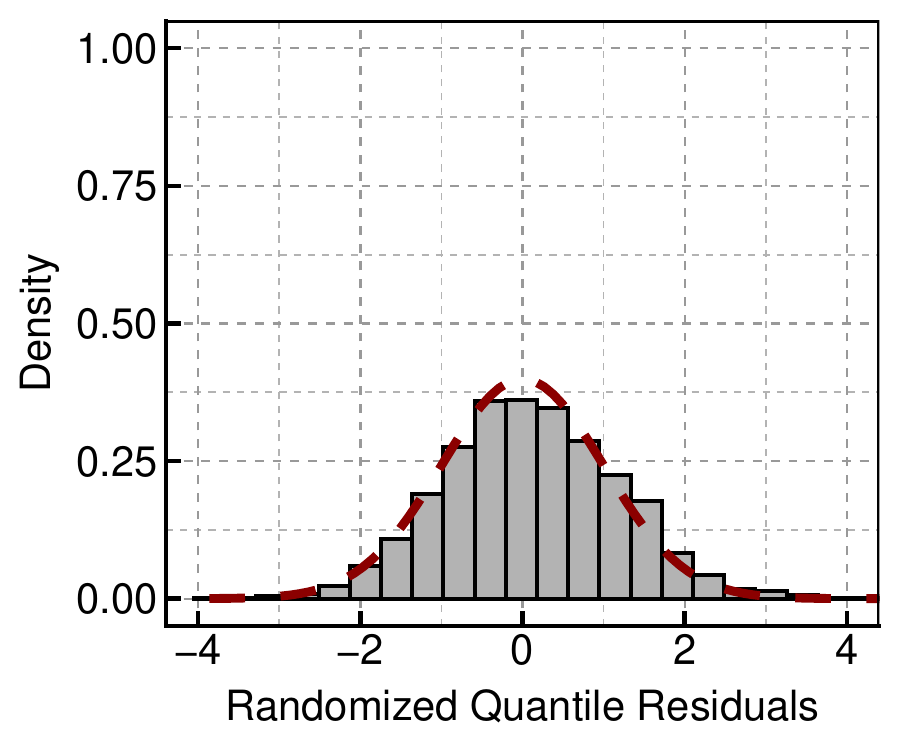}    
\end{subfigure}
\hfill
\begin{subfigure}[t]{0.31\linewidth}
    \centering
    \caption{}
    \includegraphics[width=\linewidth]{figures/Overlapping_rupture/model_nprup_All_RQR_vs_M.pdf}    
\end{subfigure}
\hfill
\begin{subfigure}[t]{0.31\linewidth}
    \centering
    \caption{}
    \includegraphics[width=\linewidth]{figures/Overlapping_rupture/model_nprup_All_RQR_vs_C.pdf}    
\end{subfigure}

\vspace{0.25cm}

\begin{subfigure}[t]{0.31\linewidth}
    \centering
    \caption{}
    \includegraphics[width=\linewidth]{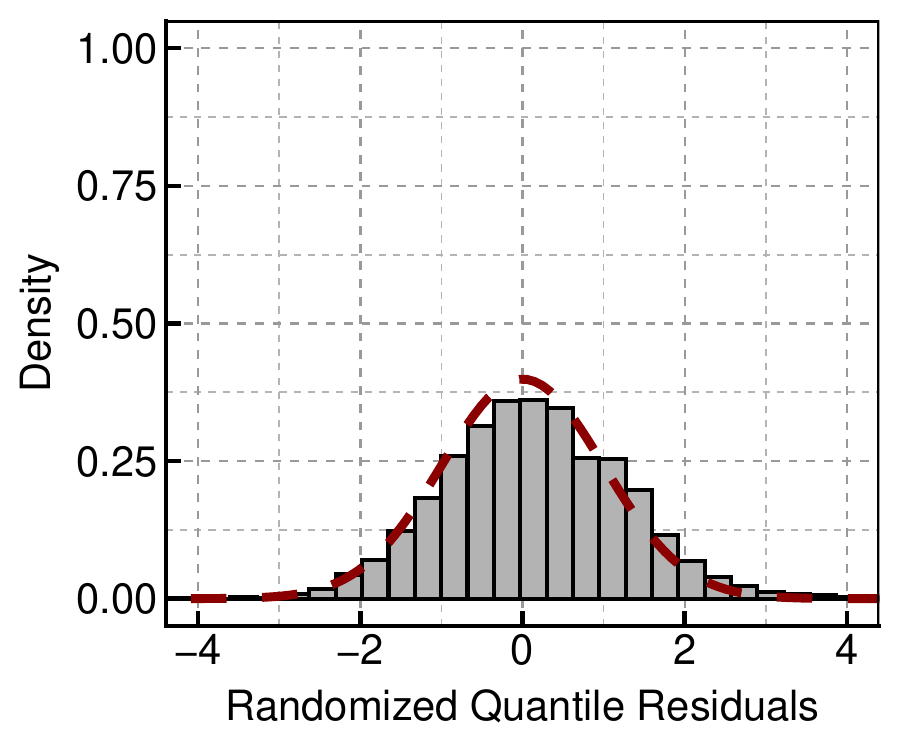}
\end{subfigure}
\hfill
\begin{subfigure}[t]{0.31\linewidth}
    \centering
     \caption{}
    \includegraphics[width=\linewidth]{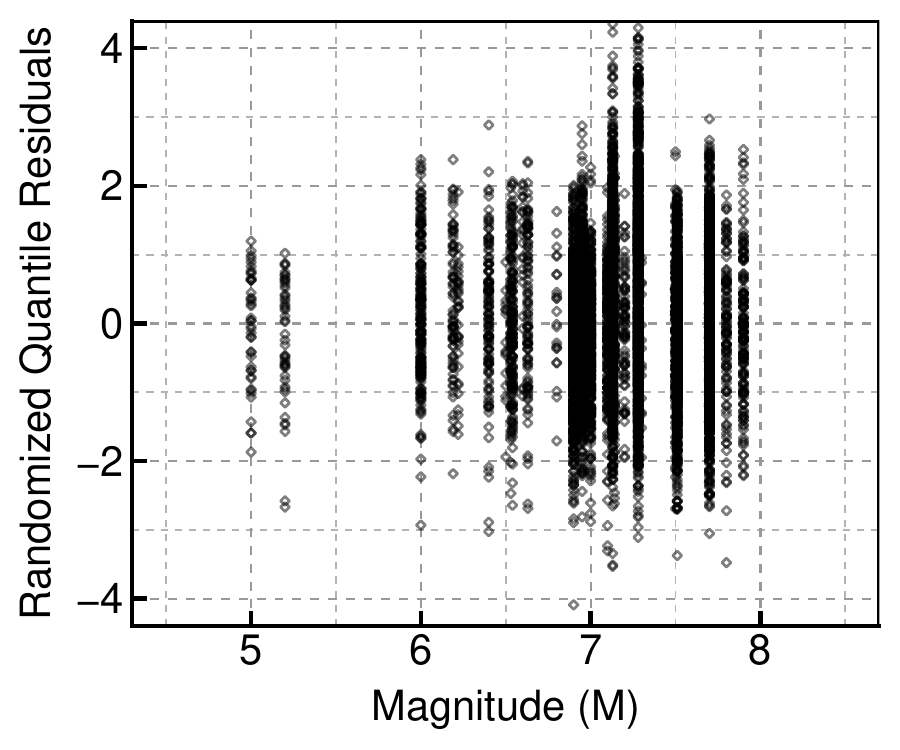}
\end{subfigure}
\hfill
\begin{subfigure}[t]{0.31\linewidth}
    \centering
    \caption{}
    \includegraphics[width=\linewidth]{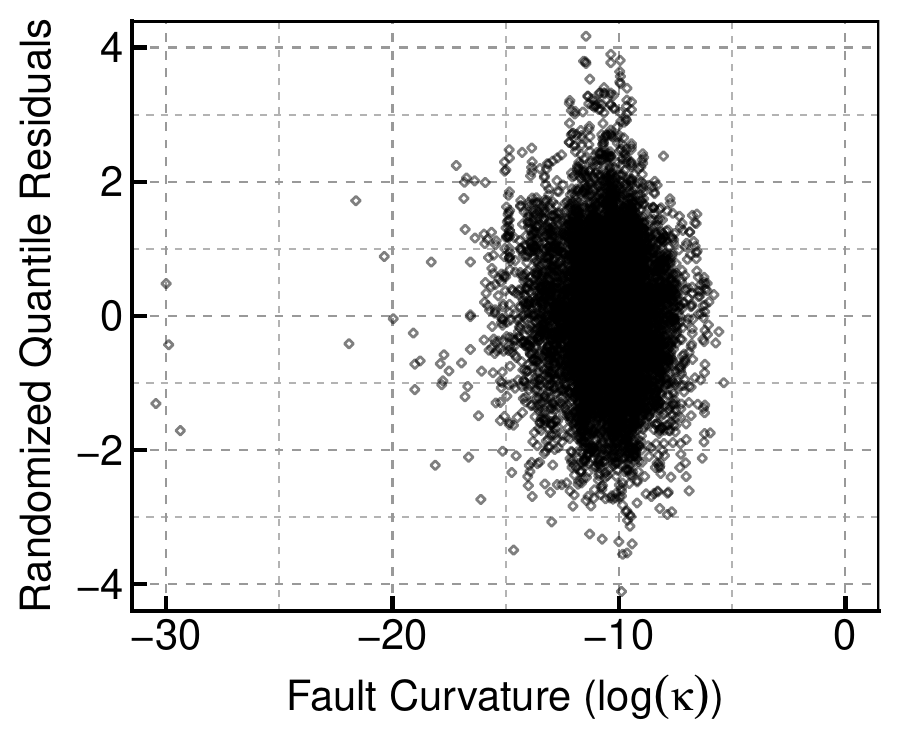}
\end{subfigure}

\vspace{0.25cm}

\begin{subfigure}[t]{0.31\linewidth}
    \centering
    \caption{}
    \includegraphics[width=\linewidth]{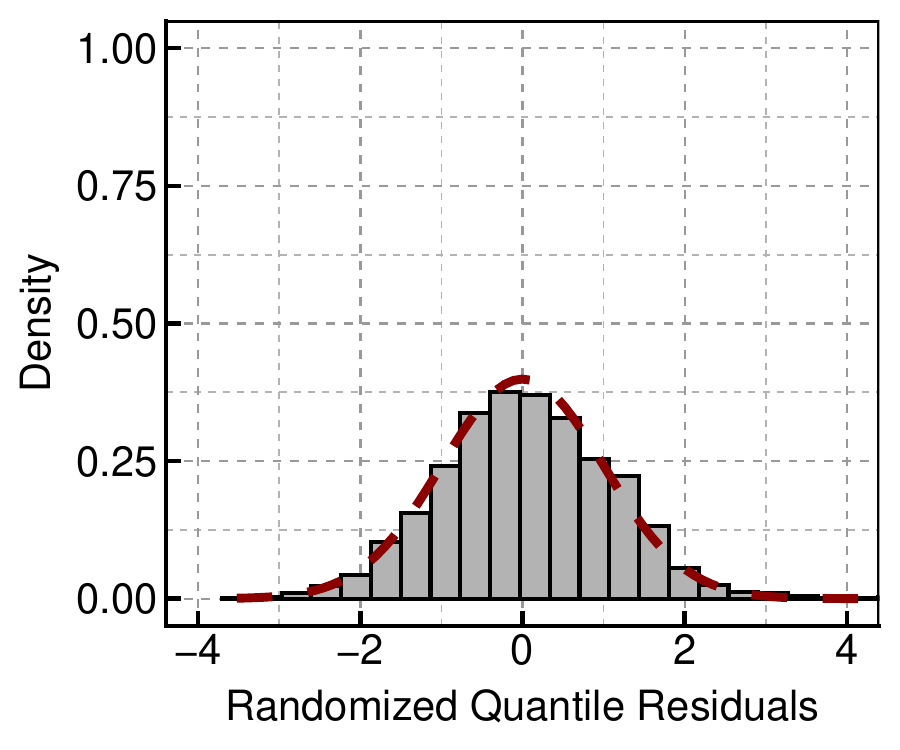}
\end{subfigure}
\hfill
\begin{subfigure}[t]{0.31\linewidth}
    \centering
    \caption{}
    \includegraphics[width=\linewidth]{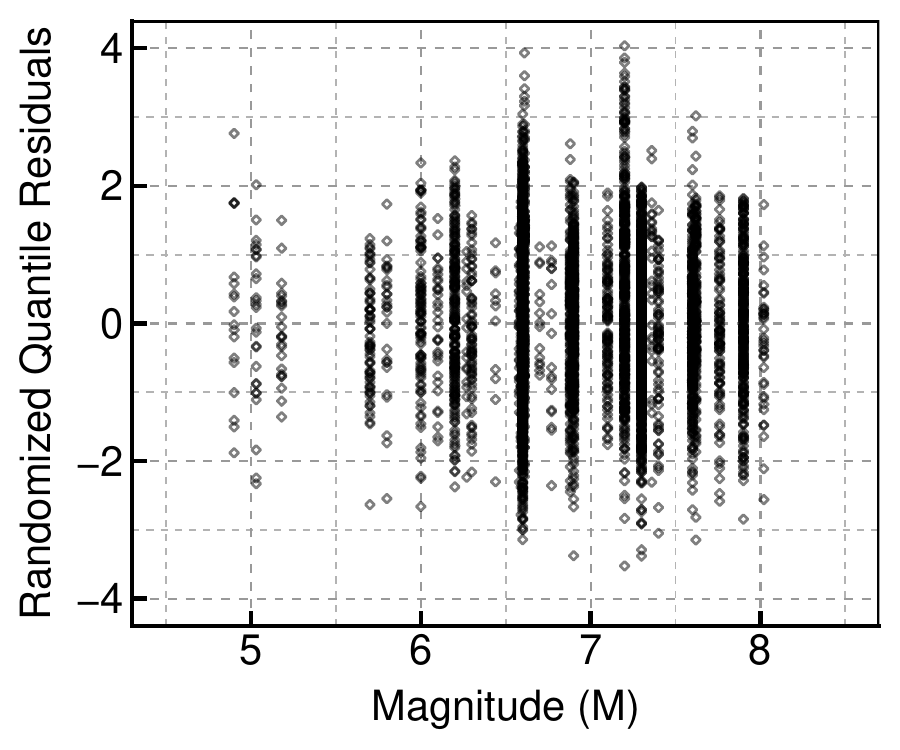}
\end{subfigure}
\hfill
\begin{subfigure}[t]{0.31\linewidth}
    \centering
    \caption{}
    \includegraphics[width=\linewidth]{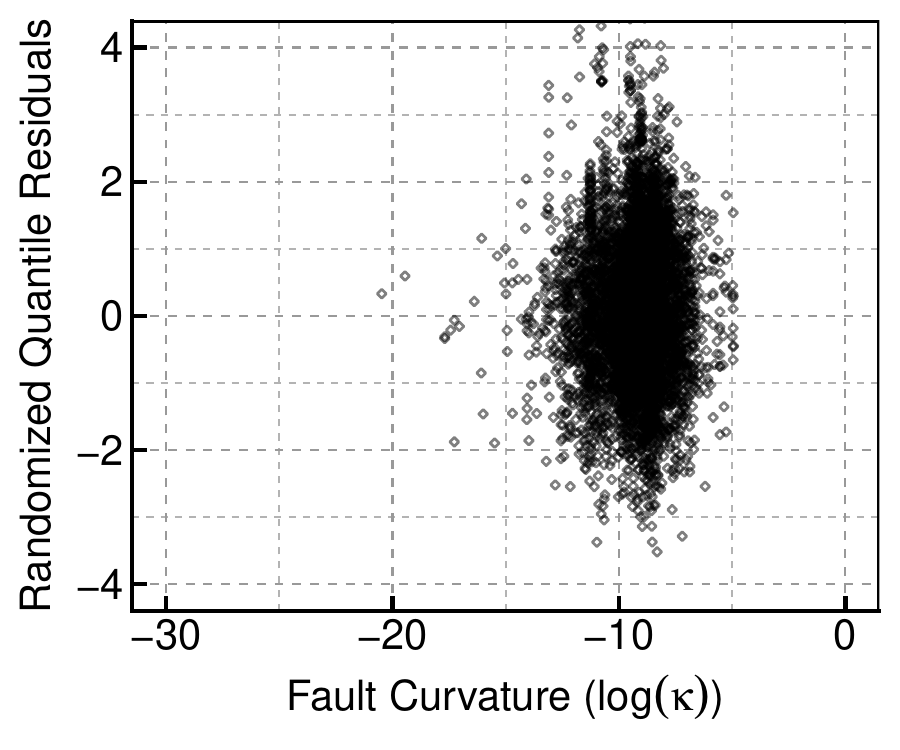}
\end{subfigure}

\caption{Assessment of residual trends in the fitted hurdle model for the number of overlapping principal ruptures using randomized quantile residuals. Rows correspond to faulting-style groups (all style of fault, strike-slip, normal and reverse, normal, and reverse), and columns show (left) histograms of randomized quantile residuals, (middle) residuals plotted against earthquake magnitude ($M$) and (right) residuals plotted against curvature ($\log(\kappa)$).}
\label{fig:z_Nr}

\end{figure}

\begin{figure}[p]\ContinuedFloat
\centering

\begin{subfigure}[t]{0.31\linewidth}
    \centering
    \caption{}
    \includegraphics[width=\linewidth]{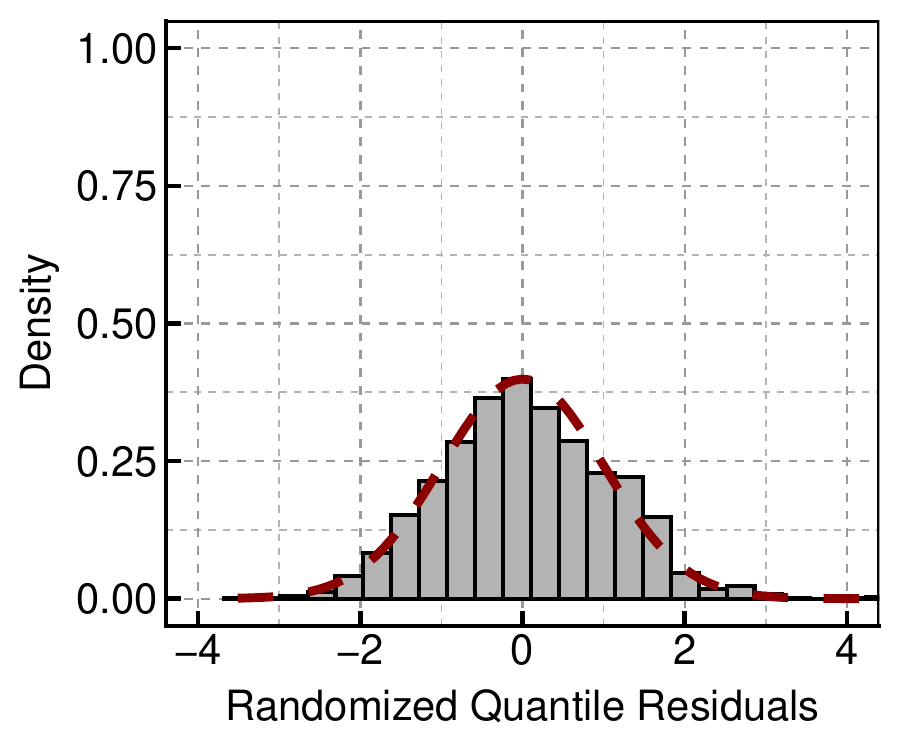}
\end{subfigure}
\hfill
\begin{subfigure}[t]{0.31\linewidth}
    \centering
    \caption{}
    \includegraphics[width=\linewidth]{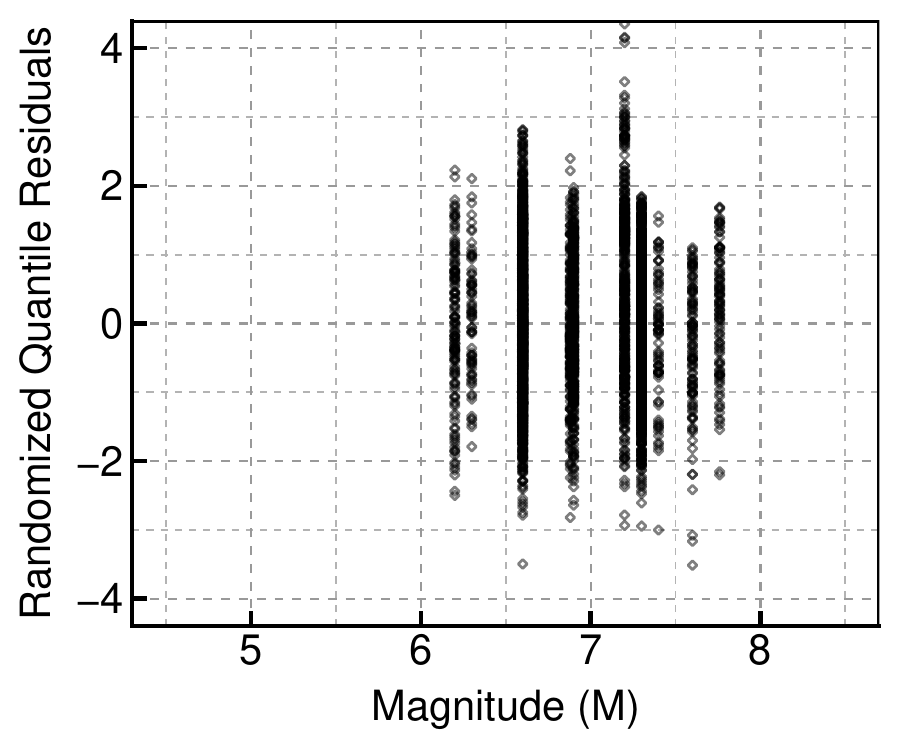}
\end{subfigure}
\hfill
\begin{subfigure}[t]{0.31\linewidth}
    \centering
    \caption{}
    \includegraphics[width=\linewidth]{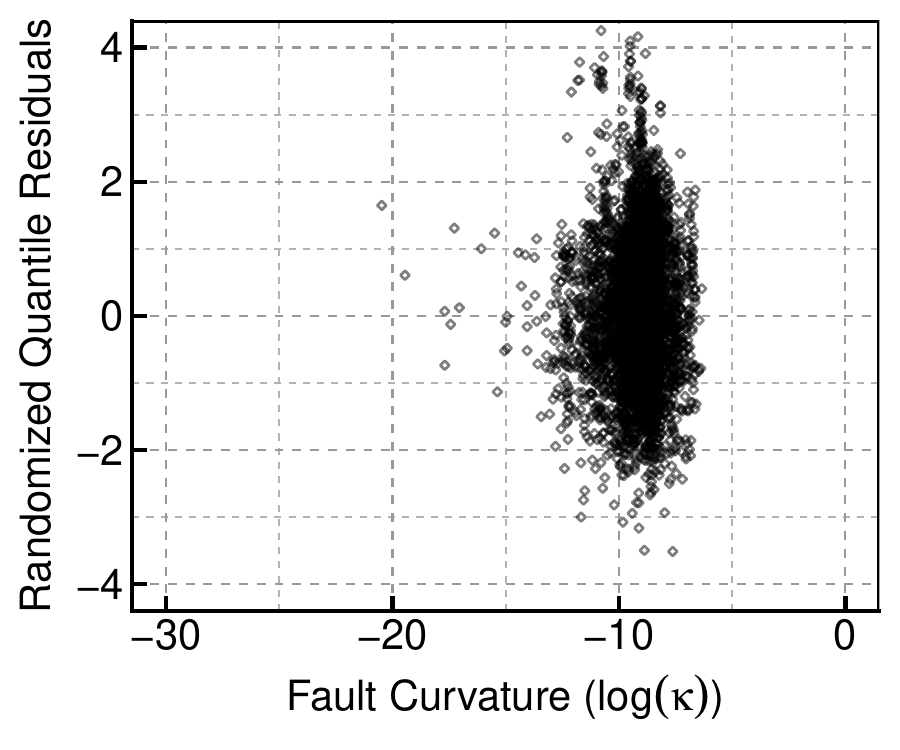}
\end{subfigure}

\vspace{0.25cm}

\begin{subfigure}[t]{0.31\linewidth}
    \centering
    \caption{}
    \includegraphics[width=\linewidth]{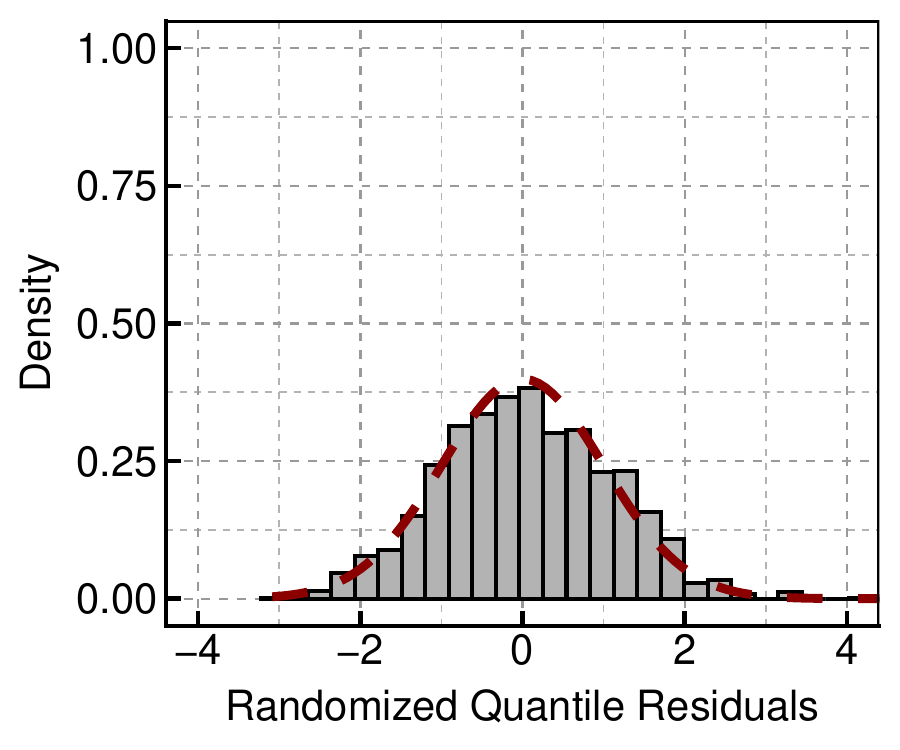}
\end{subfigure}
\hfill
\begin{subfigure}[t]{0.31\linewidth}
    \centering
    \caption{}
    \includegraphics[width=\linewidth]{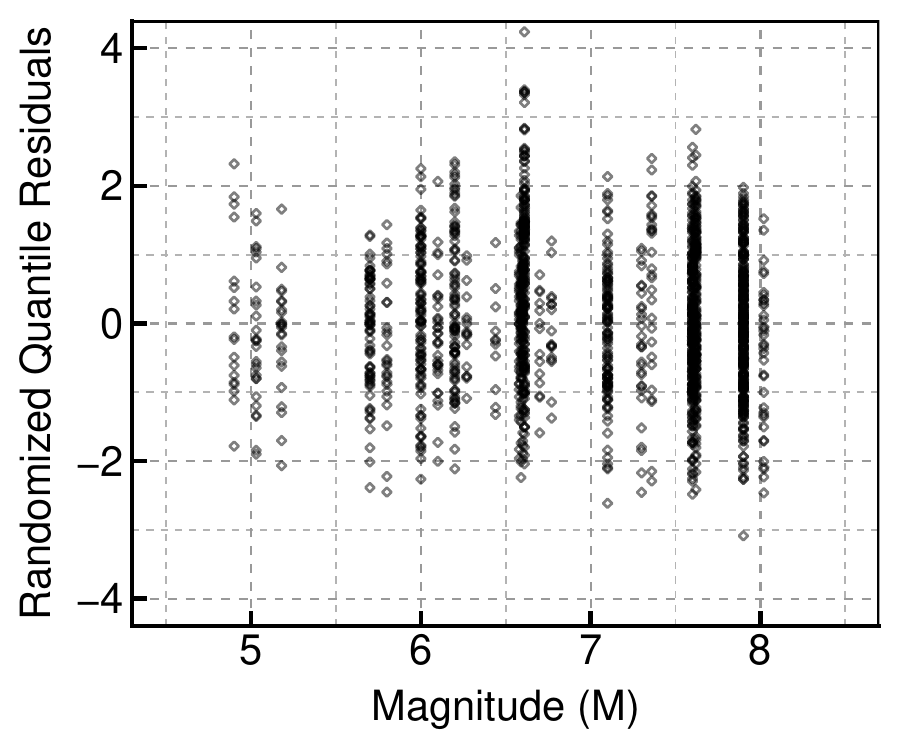}
\end{subfigure}
\hfill
\begin{subfigure}[t]{0.31\linewidth}
    \centering
    \caption{}
    \includegraphics[width=\linewidth]{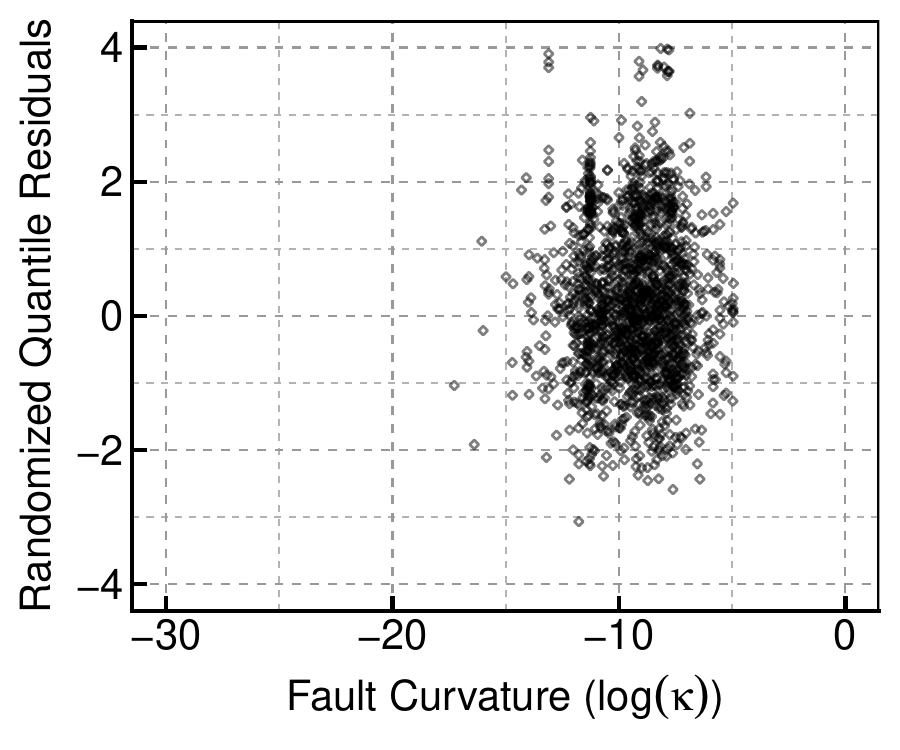}
\end{subfigure}

\caption{Assessment of residual trends in the fitted hurdle model for the number of overlapping principal ruptures using randomized quantile residuals, continued.}
\end{figure}

\begin{figure}[p]
\centering

\makebox[\textwidth][c]{%
\begin{subfigure}[t]{0.31\linewidth}
    \centering
    \caption{}
    \includegraphics[width=\linewidth]{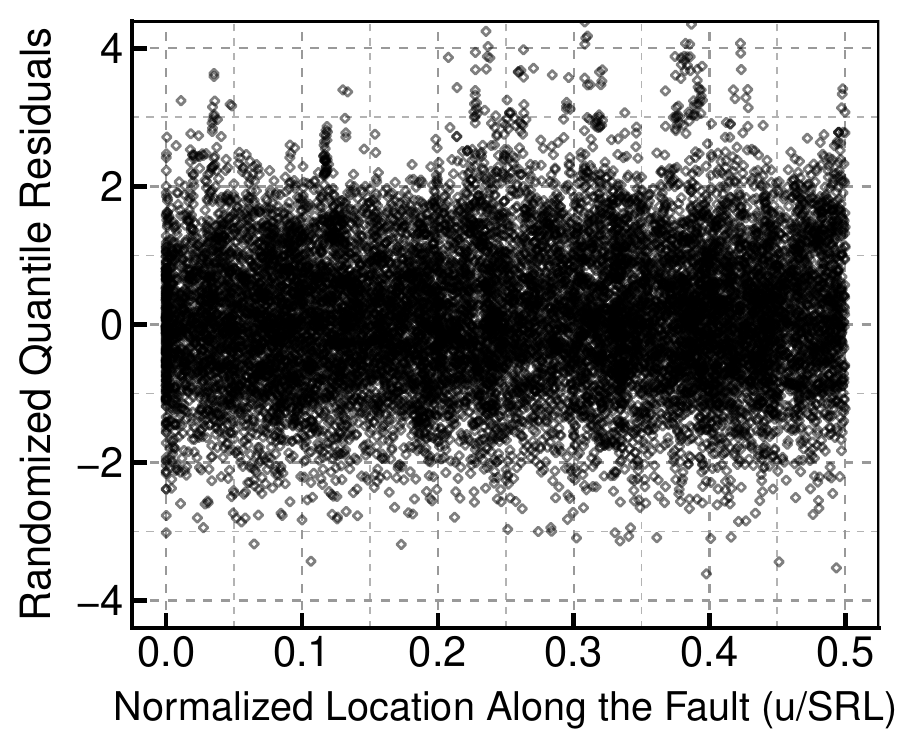}    
\end{subfigure}
\hspace{0.06\linewidth}
\begin{subfigure}[t]{0.31\linewidth}
    \centering
    \caption{}
    \includegraphics[width=\linewidth]{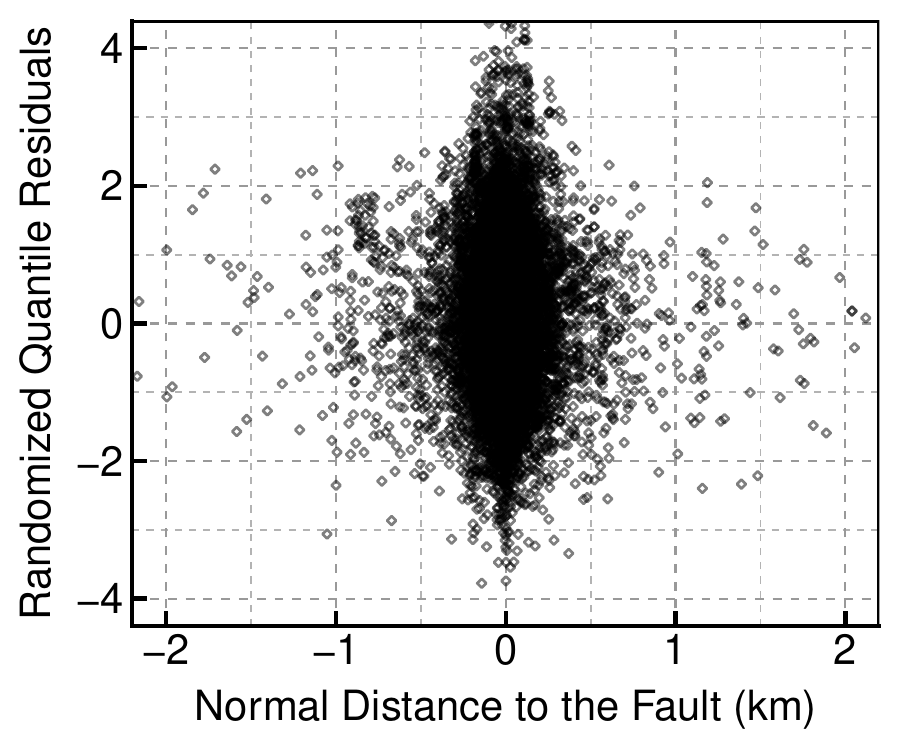}    
\end{subfigure}
}

\vspace{0.25cm}

\makebox[\textwidth][c]{%
\begin{subfigure}[t]{0.31\linewidth}
    \centering
    \caption{}
    \includegraphics[width=\linewidth]{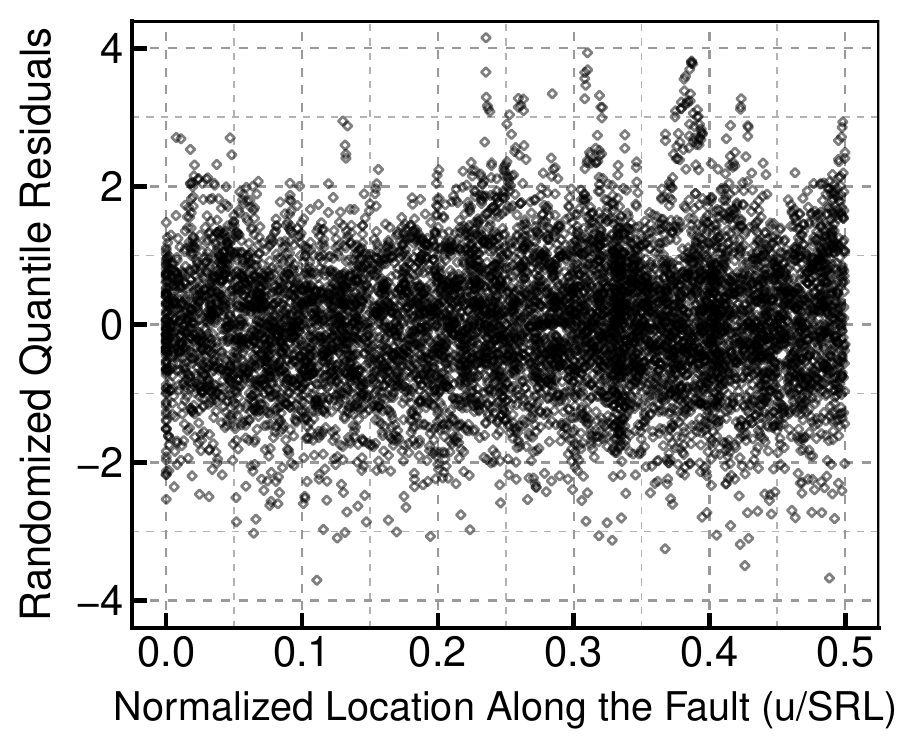}
\end{subfigure}
\hspace{0.06\linewidth}
\begin{subfigure}[t]{0.31\linewidth}
    \centering
    \caption{}
    \includegraphics[width=\linewidth]{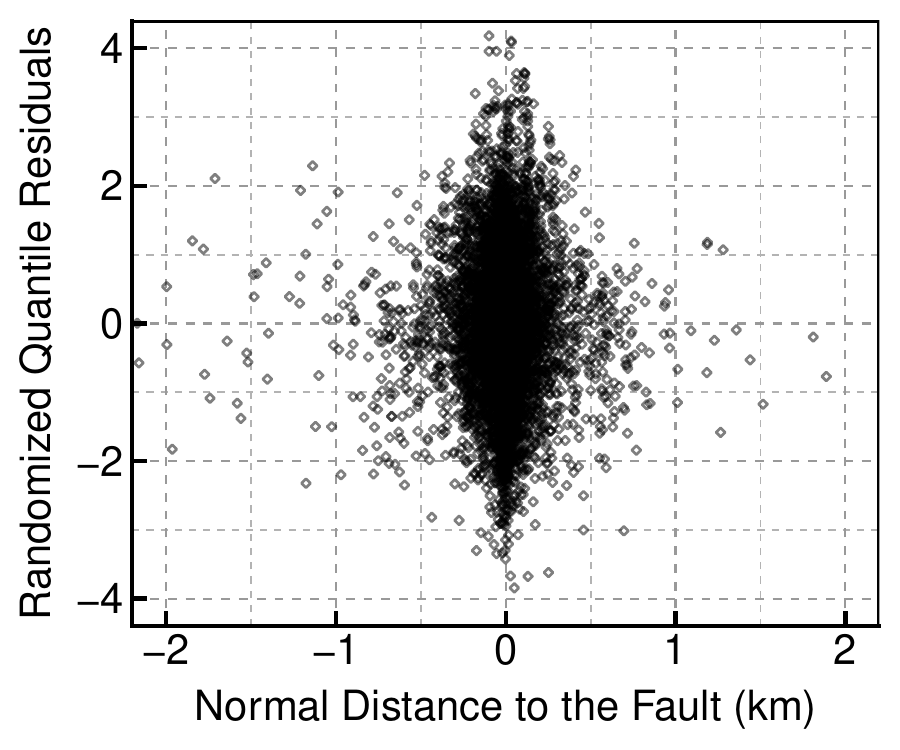}
\end{subfigure}
}

\vspace{0.25cm}

\makebox[\textwidth][c]{%
\begin{subfigure}[t]{0.31\linewidth}
    \centering
    \caption{}
    \includegraphics[width=\linewidth]{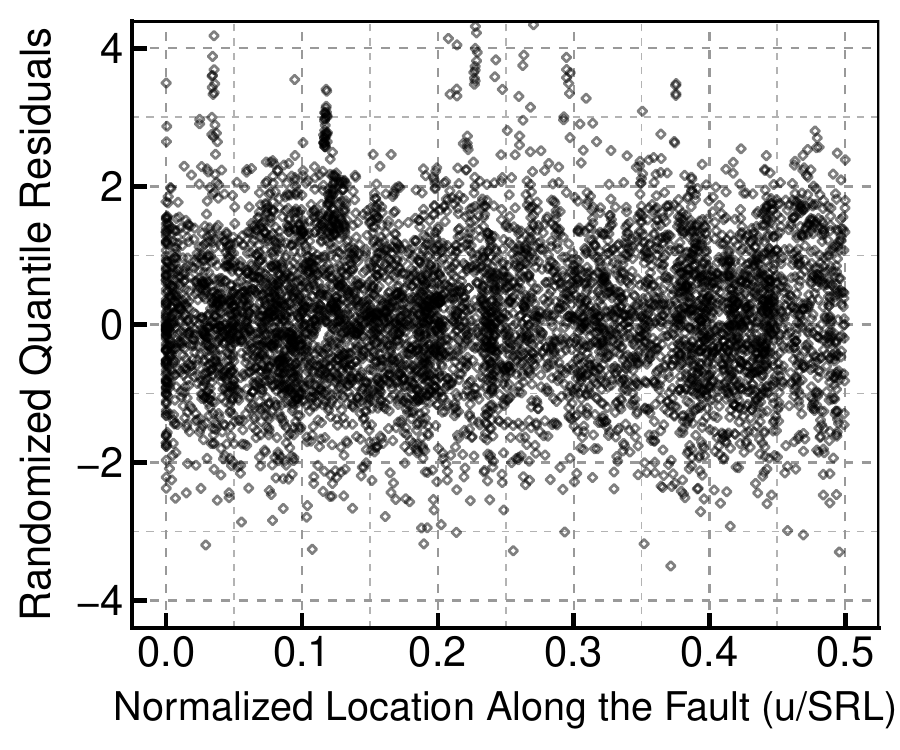}
\end{subfigure}
\hspace{0.06\linewidth}
\begin{subfigure}[t]{0.31\linewidth}
    \centering
    \caption{}
    \includegraphics[width=\linewidth]{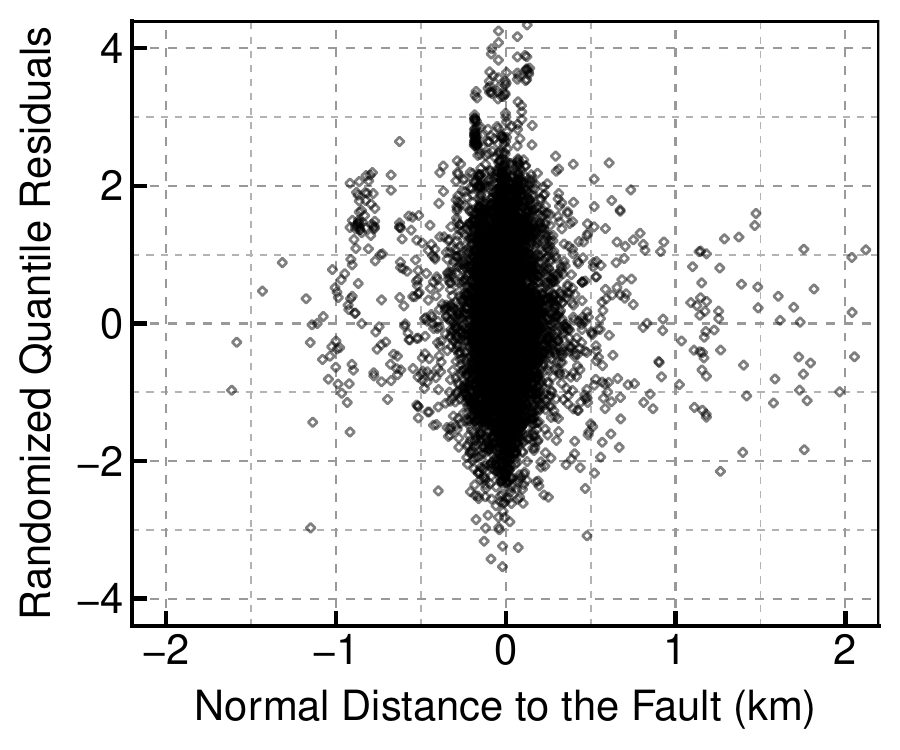}
\end{subfigure}
}

\caption{Assessment of residual trends in the fitted hurdle model for the number of overlapping principal ruptures using randomized quantile residuals. Rows correspond to faulting-style groups (all style of fault, strike-slip, normal and reverse, normal, and reverse), and columns show (left) residuals plotted against normalized location along the fault ($x/L$) and (right) residuals plotted against normal distance from the fault.}
\label{fig:z_Nr_ut}

\end{figure}

\begin{figure}[p]\ContinuedFloat
\centering

\makebox[\textwidth][c]{%
\begin{subfigure}[t]{0.31\linewidth}
    \centering
    \caption{}
    \includegraphics[width=\linewidth]{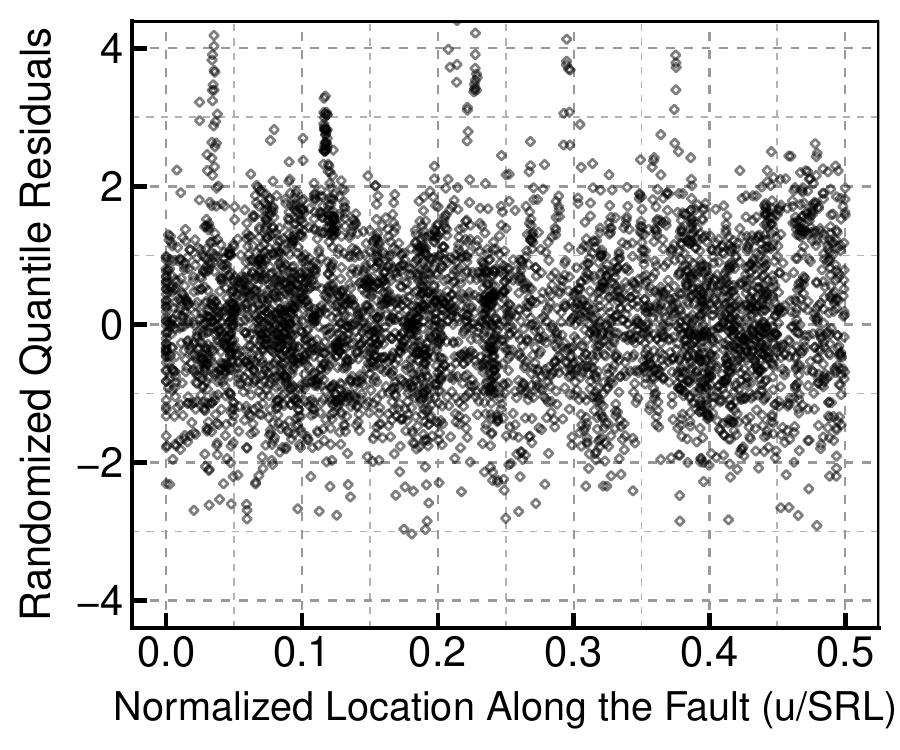}
\end{subfigure}
\hspace{0.06\linewidth}
\begin{subfigure}[t]{0.31\linewidth}
    \centering
    \caption{}
    \includegraphics[width=\linewidth]{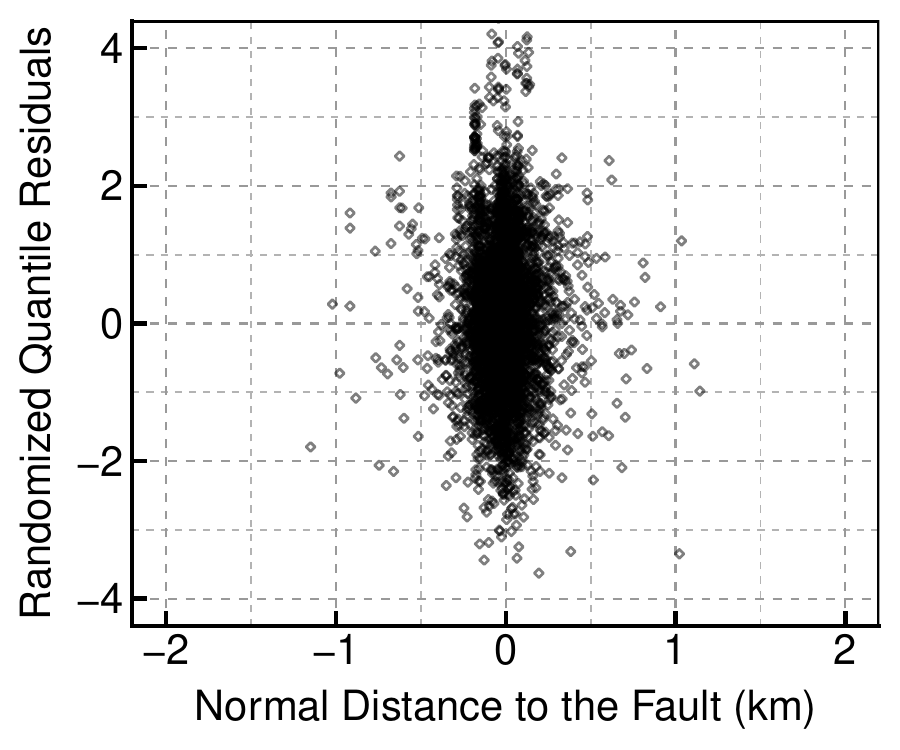}
\end{subfigure}
}

\vspace{0.25cm}

\makebox[\textwidth][c]{%
\begin{subfigure}[t]{0.31\linewidth}
    \centering
    \caption{}
    \includegraphics[width=\linewidth]{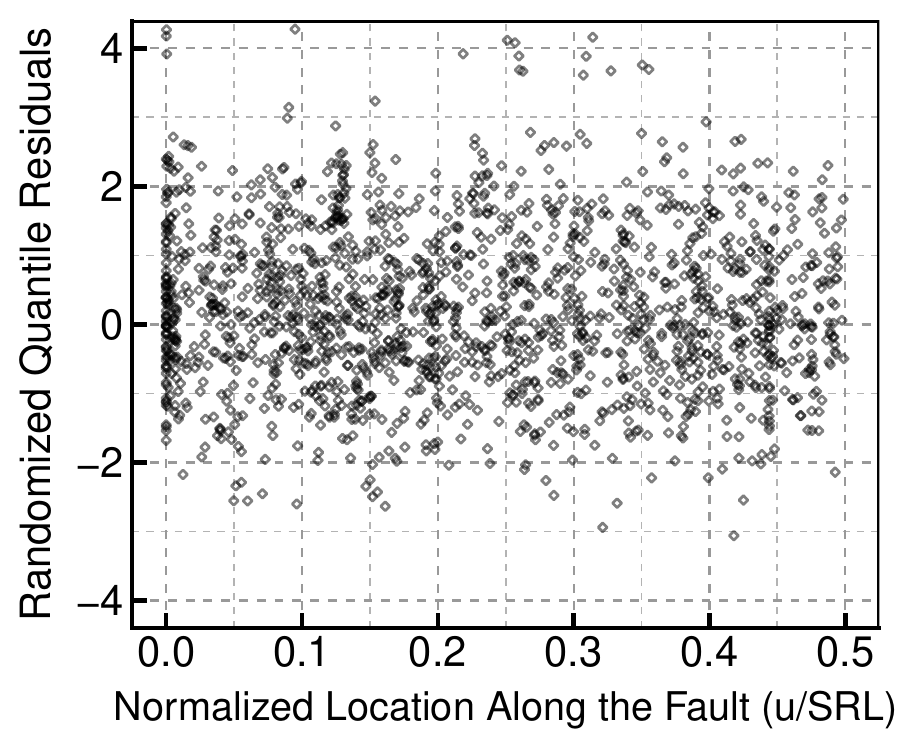}
\end{subfigure}
\hspace{0.06\linewidth}
\begin{subfigure}[t]{0.31\linewidth}
    \centering
    \caption{}
    \includegraphics[width=\linewidth]{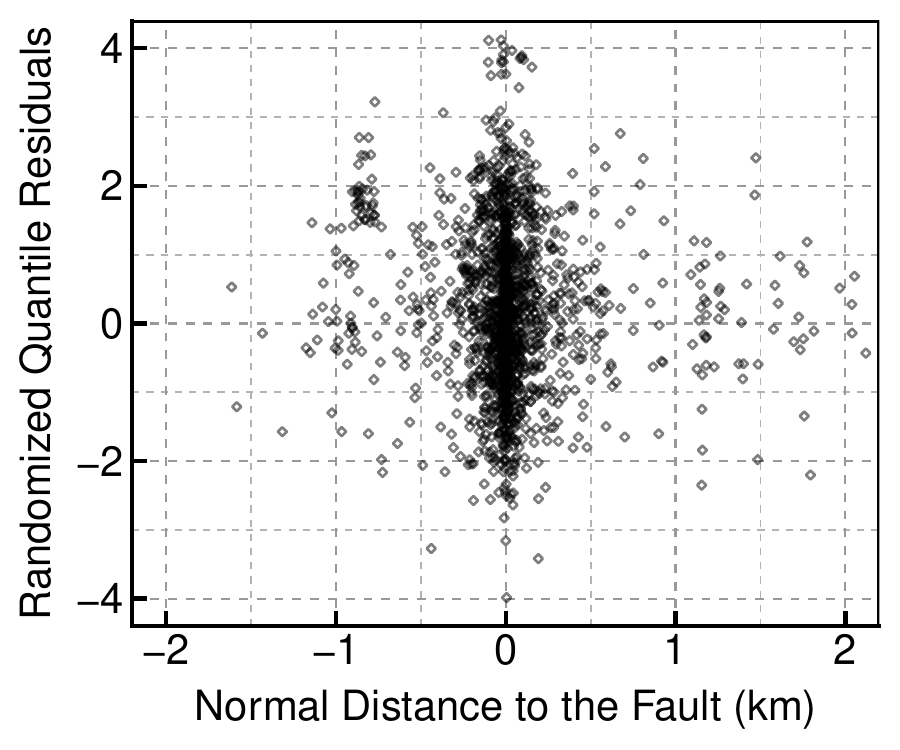}
\end{subfigure}
}

\caption{Assessment of residual trends in the fitted hurdle model for the number of overlapping principal ruptures using randomized quantile residuals, continued.}
\end{figure}

\begin{figure}[p]
\centering

\begin{subfigure}[t]{0.31\linewidth}
    \centering
    \caption{}
    \includegraphics[width=\linewidth]{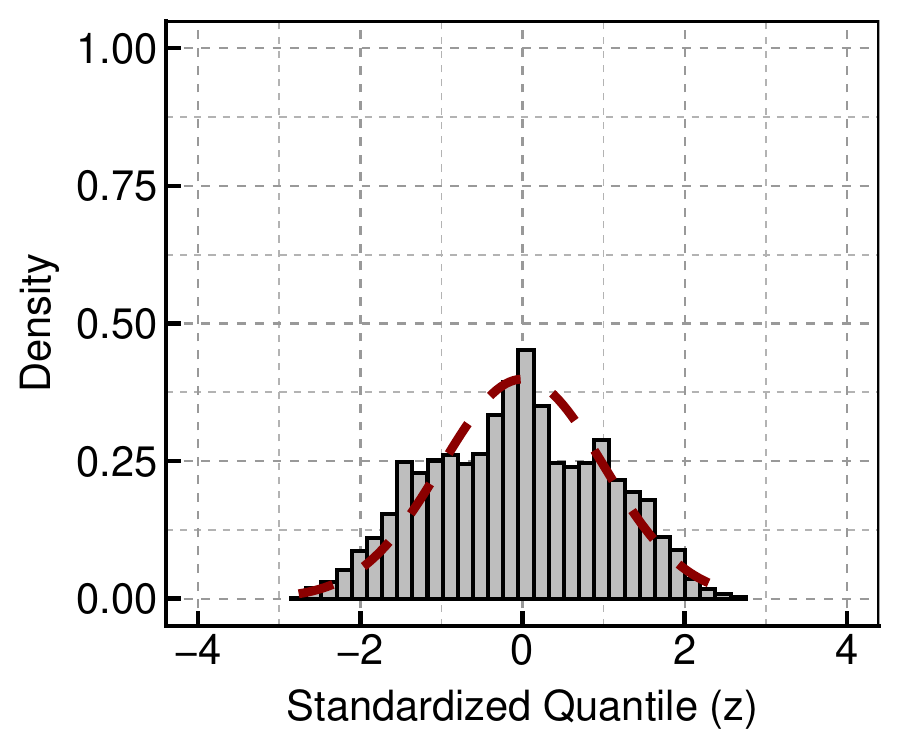}
\end{subfigure}
\hfill
\begin{subfigure}[t]{0.31\linewidth}
    \centering
    \caption{}
    \includegraphics[width=\linewidth]{figures/Distance/Distance_SCS/model_t_All_z_vs_M.pdf}
\end{subfigure}
\hfill
\begin{subfigure}[t]{0.31\linewidth}
    \centering
    \caption{}
    \includegraphics[width=\linewidth]{figures/Distance/Distance_SCS/model_t_All_z_vs_D.pdf}
\end{subfigure}

\vspace{0.25cm}

\begin{subfigure}[t]{0.31\linewidth}
    \centering
    \caption{}
    \includegraphics[width=\linewidth]{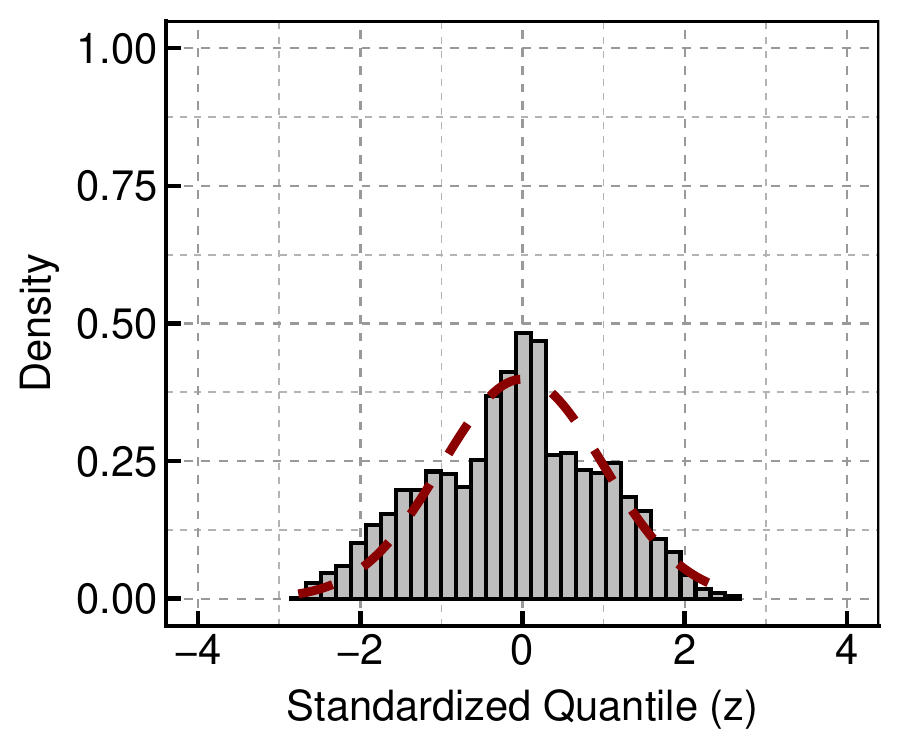}
\end{subfigure}
\hfill
\begin{subfigure}[t]{0.31\linewidth}
    \centering
    \caption{}
    \includegraphics[width=\linewidth]{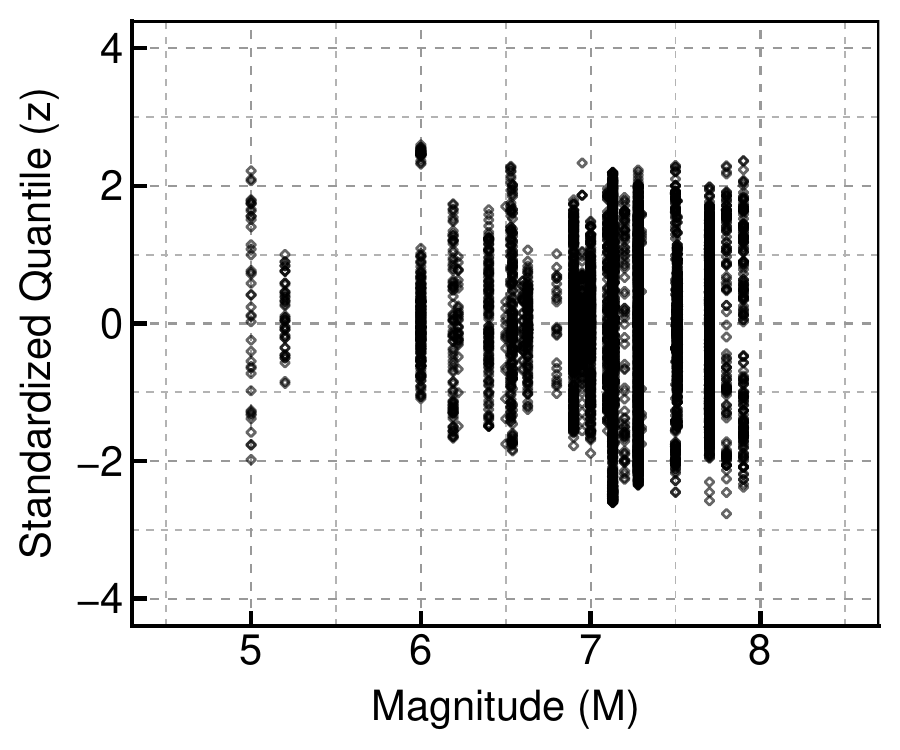}
\end{subfigure}
\hfill
\begin{subfigure}[t]{0.31\linewidth}
    \centering
    \caption{}
    \includegraphics[width=\linewidth]{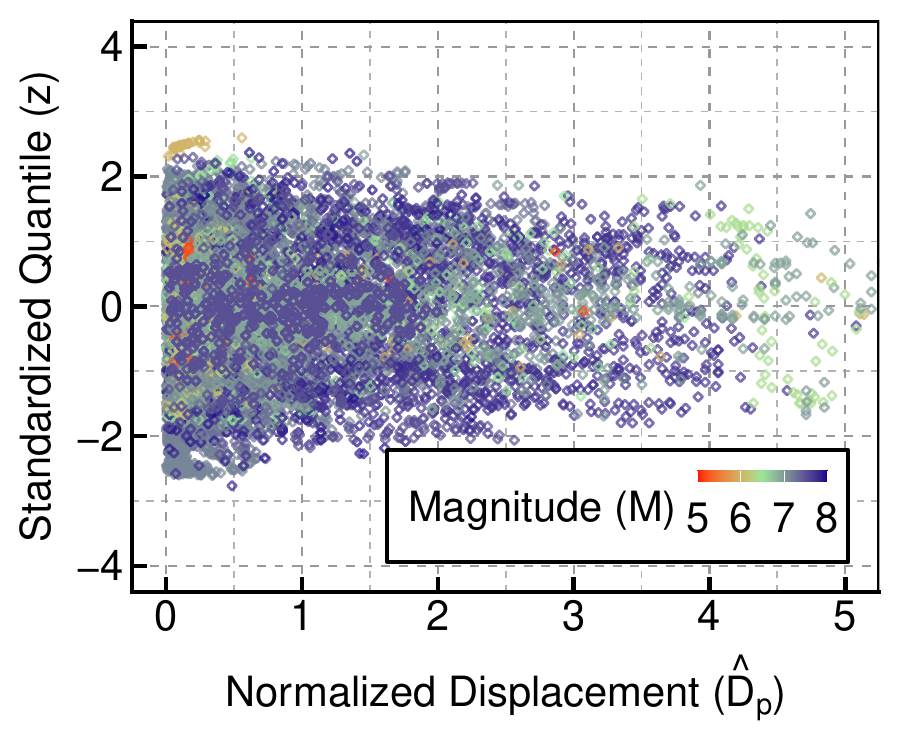}
\end{subfigure}

\vspace{0.25cm}

\begin{subfigure}[t]{0.31\linewidth}
    \centering
    \caption{}
    \includegraphics[width=\linewidth]{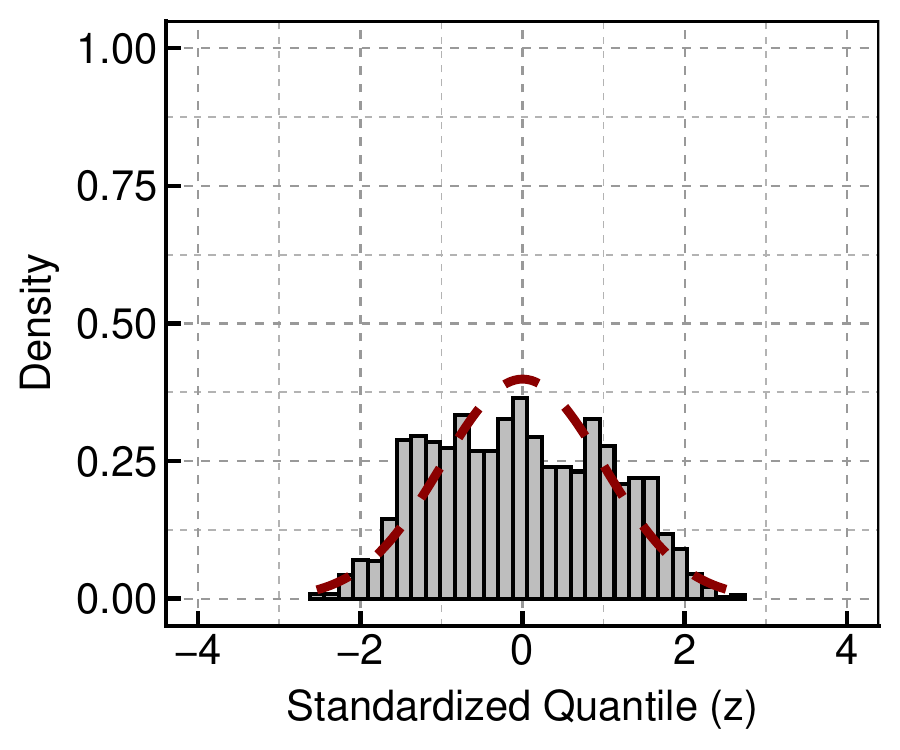}
\end{subfigure}
\hfill
\begin{subfigure}[t]{0.31\linewidth}
    \centering
    \caption{}
    \includegraphics[width=\linewidth]{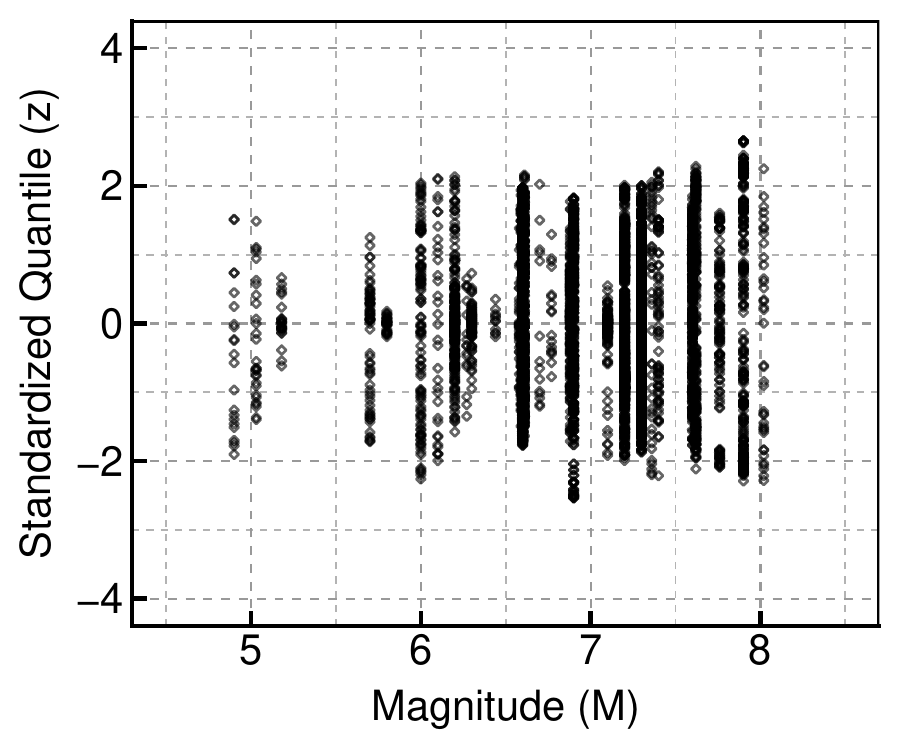}
\end{subfigure}
\hfill
\begin{subfigure}[t]{0.31\linewidth}
    \centering
    \caption{}
    \includegraphics[width=\linewidth]{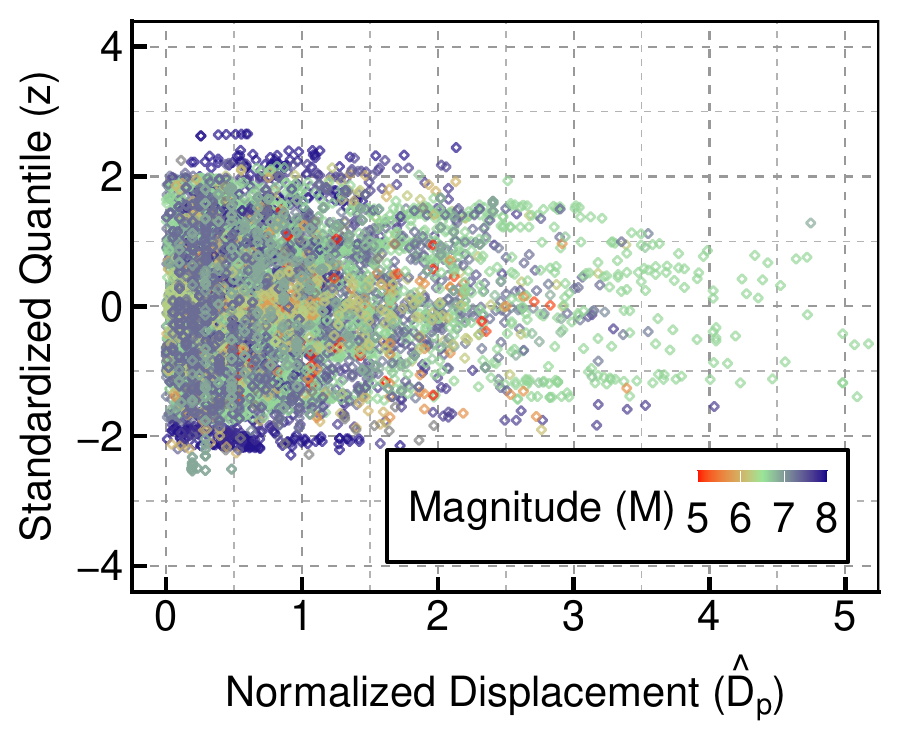}
\end{subfigure}

\caption{Assessment of residual trends in the fitted Student's t distribution for normal distance to the SCS fault trace using standardized quantiles. Rows correspond to faulting-style groups (all style of fault, strike-slip, normal and reverse, normal, and reverse), and columns show (left) histograms of residual quantiles, (middle) residuals plotted against earthquake magnitude ($M$) and (right) residuals plotted against normalized displacement, ($\hat D_p$).}
\label{fig:dist_z_scs}

\end{figure}

\begin{figure}[p]\ContinuedFloat
\centering

\begin{subfigure}[t]{0.31\linewidth}
    \centering
    \caption{}
    \includegraphics[width=\linewidth]{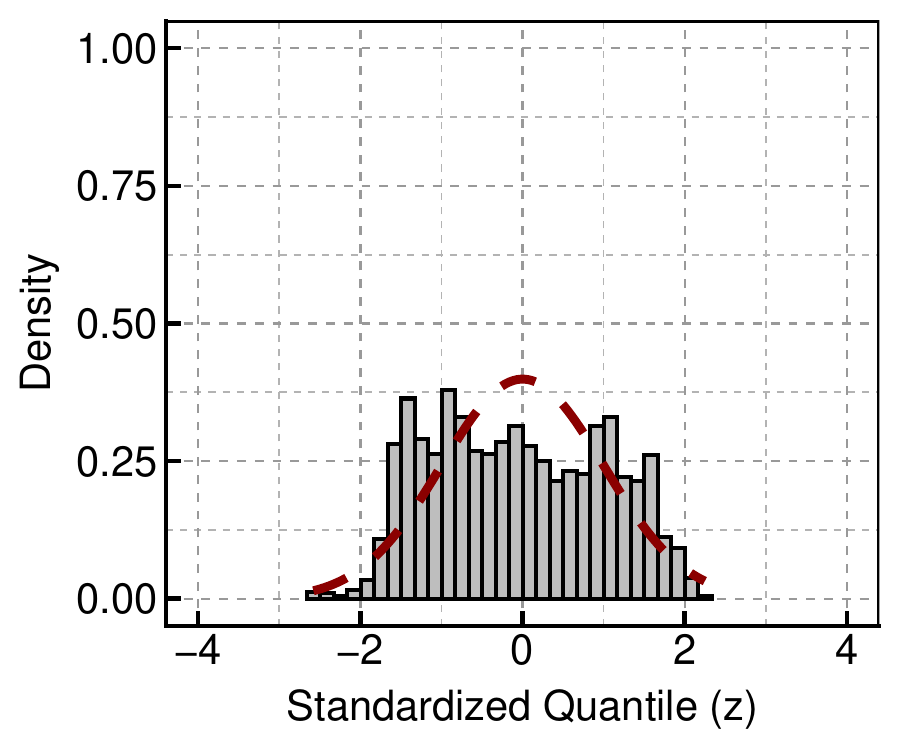}
\end{subfigure}
\hfill
\begin{subfigure}[t]{0.31\linewidth}
    \centering
    \caption{}
    \includegraphics[width=\linewidth]{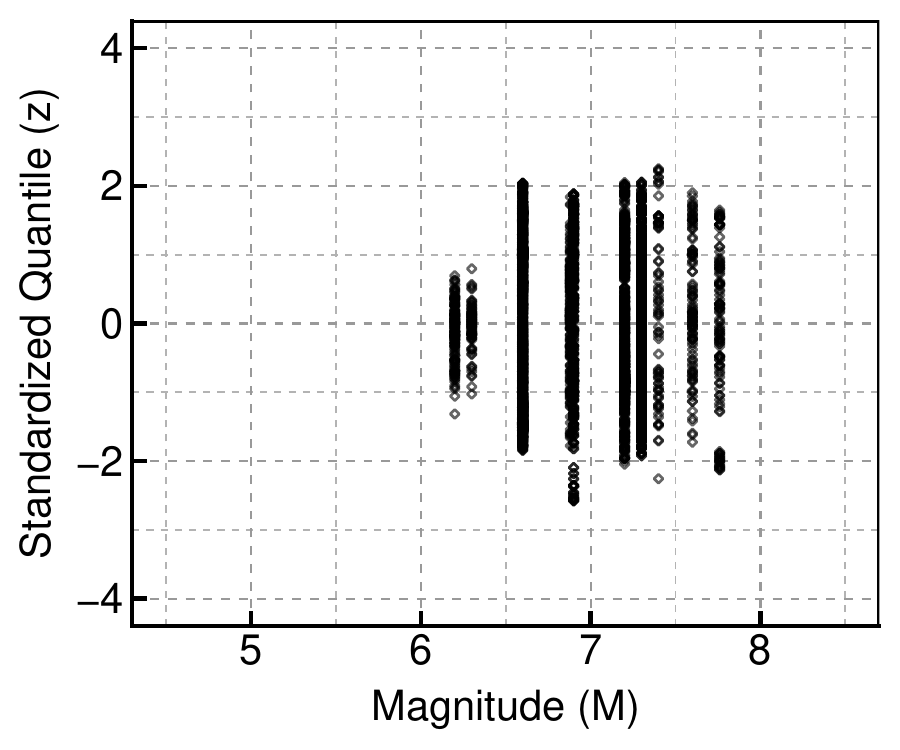}
\end{subfigure}
\hfill
\begin{subfigure}[t]{0.31\linewidth}
    \centering
    \caption{}
    \includegraphics[width=\linewidth]{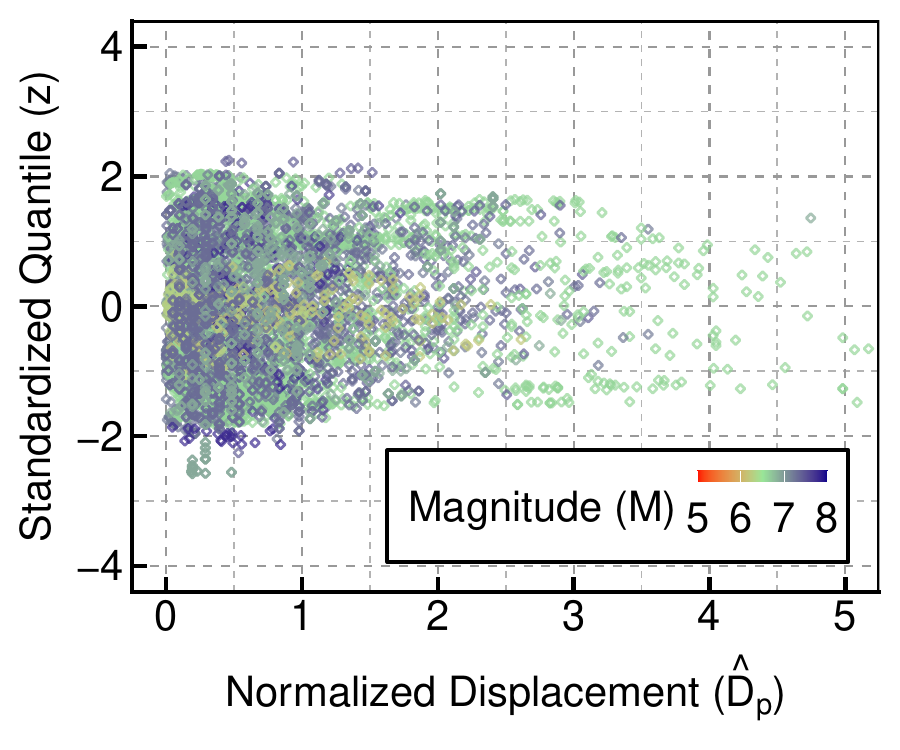}
\end{subfigure}

\vspace{0.25cm}

\begin{subfigure}[t]{0.31\linewidth}
    \centering
    \caption{}
    \includegraphics[width=\linewidth]{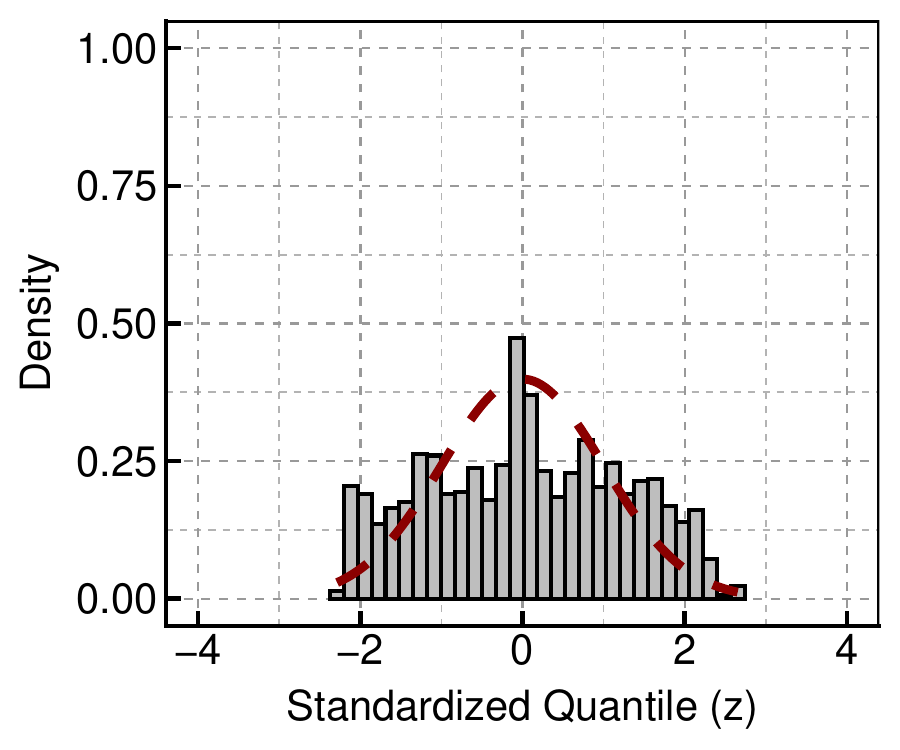}
\end{subfigure}
\hfill
\begin{subfigure}[t]{0.31\linewidth}
    \centering
    \caption{}
    \includegraphics[width=\linewidth]{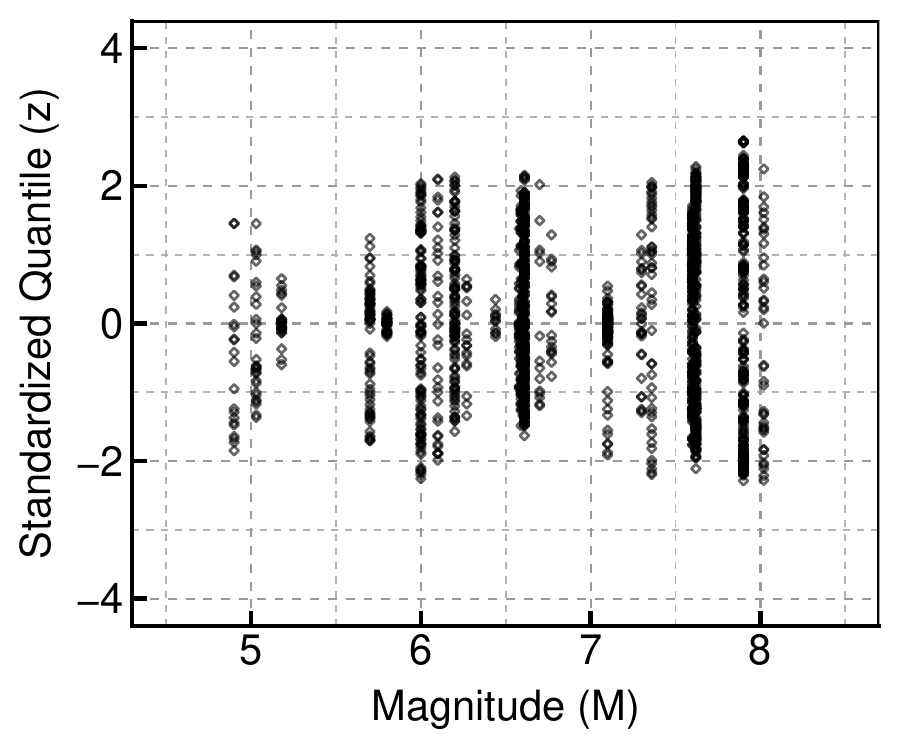}
\end{subfigure}
\hfill
\begin{subfigure}[t]{0.31\linewidth}
    \centering
    \caption{}
    \includegraphics[width=\linewidth]{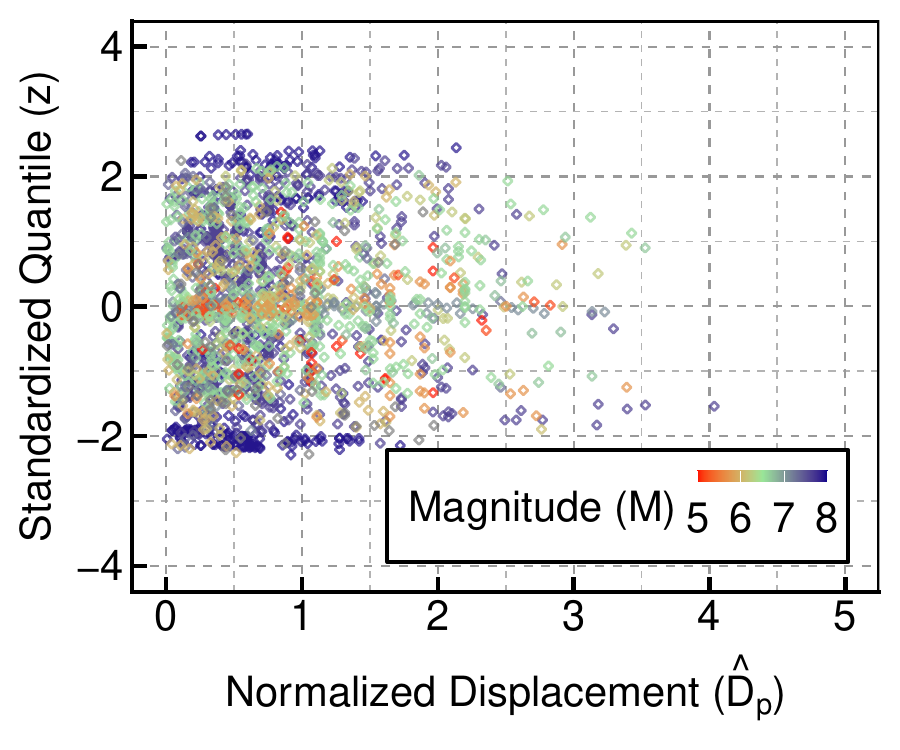}
\end{subfigure}

\caption[]{Assessment of residual trends in the fitted Student's t distribution for normal distance to the SCS fault trace using standardized quantiles, continued.}
\end{figure}

\begin{figure}[p]
\centering

\begin{subfigure}[t]{0.31\linewidth}
    \centering
    \caption{}
    \includegraphics[width=\linewidth]{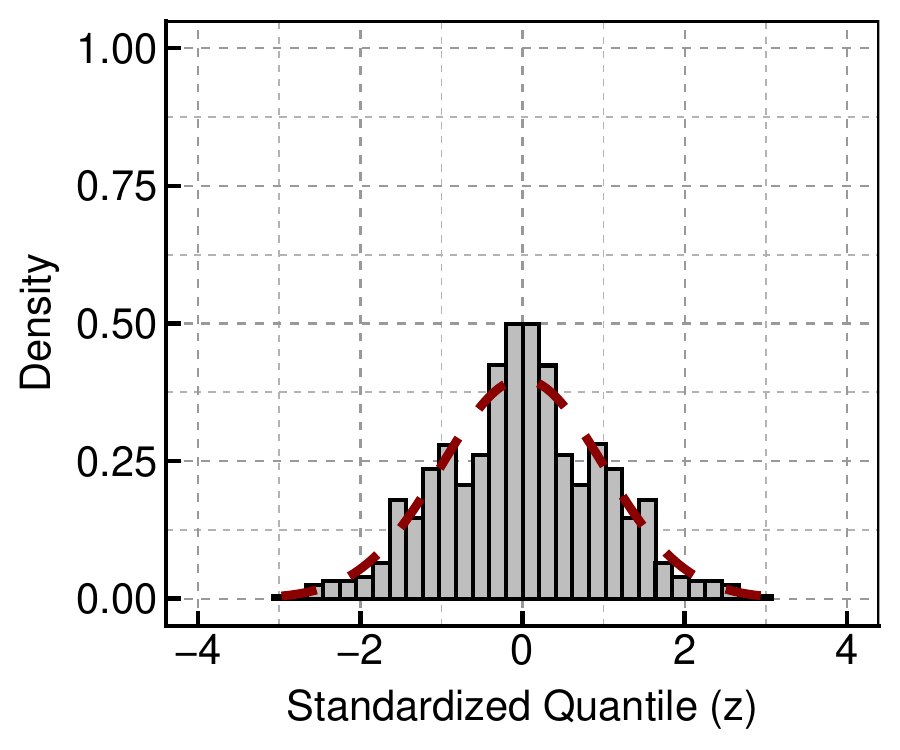}
\end{subfigure}
\hfill
\begin{subfigure}[t]{0.31\linewidth}
    \centering
    \caption{}
    \includegraphics[width=\linewidth]{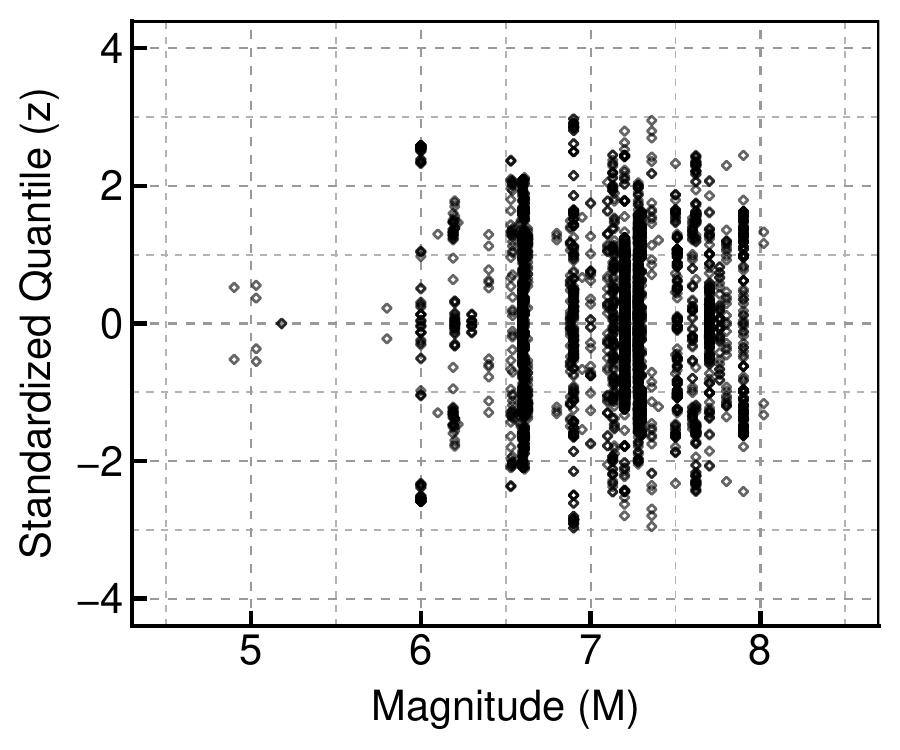}
\end{subfigure}
\hfill
\begin{subfigure}[t]{0.31\linewidth}
    \centering
    \caption{}
    \includegraphics[width=\linewidth]{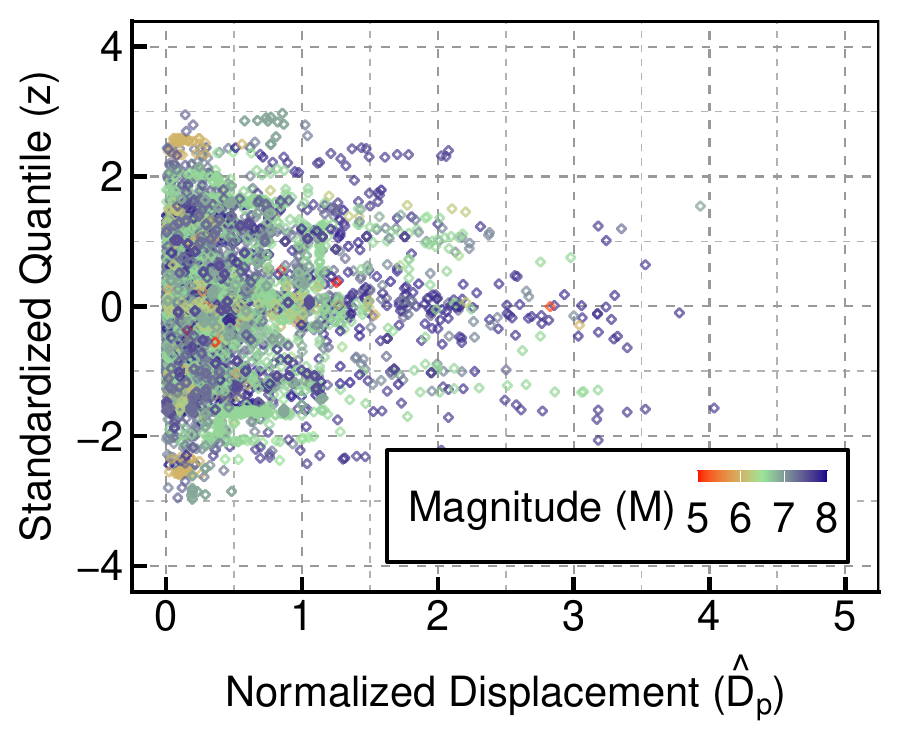}
\end{subfigure}

\vspace{0.25cm}

\begin{subfigure}[t]{0.31\linewidth}
    \centering
    \caption{}
    \includegraphics[width=\linewidth]{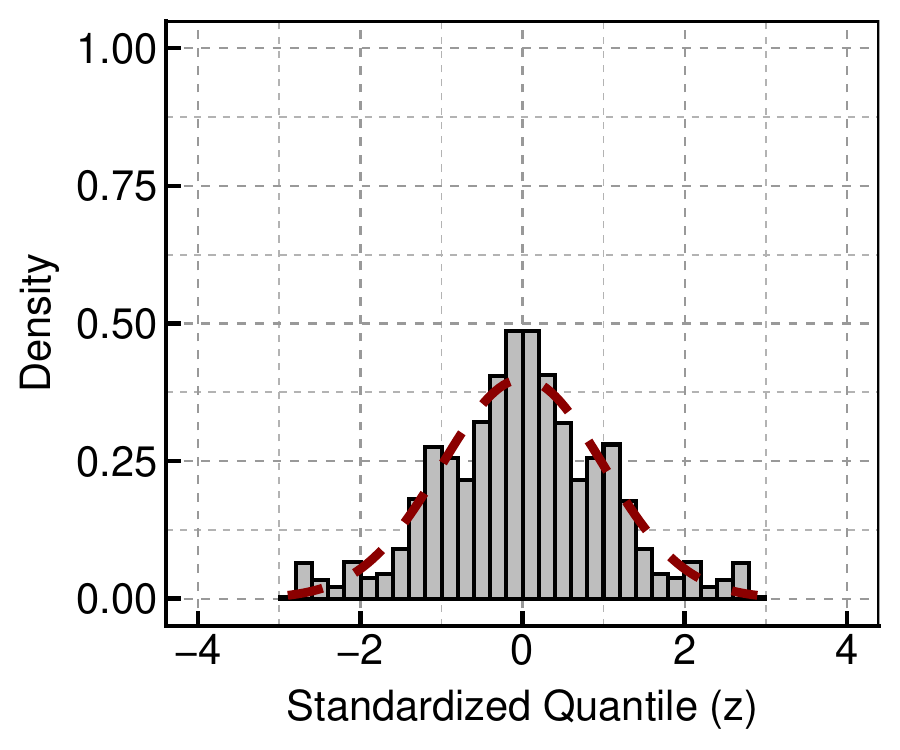}
\end{subfigure}
\hfill
\begin{subfigure}[t]{0.31\linewidth}
    \centering
    \caption{}
    \includegraphics[width=\linewidth]{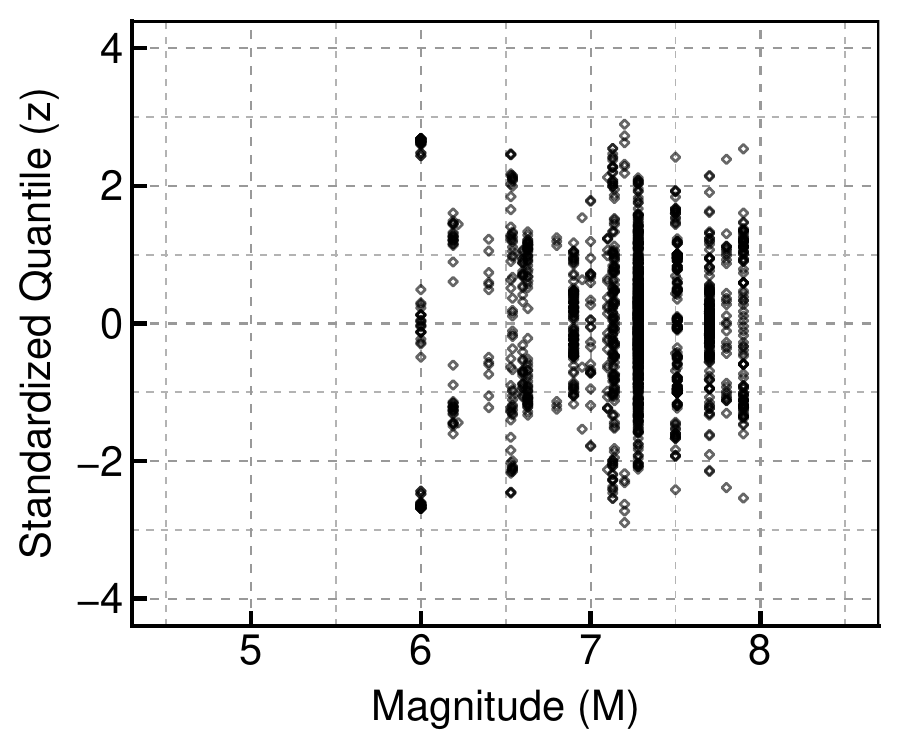}
\end{subfigure}
\hfill
\begin{subfigure}[t]{0.31\linewidth}
    \centering
    \caption{}
    \includegraphics[width=\linewidth]{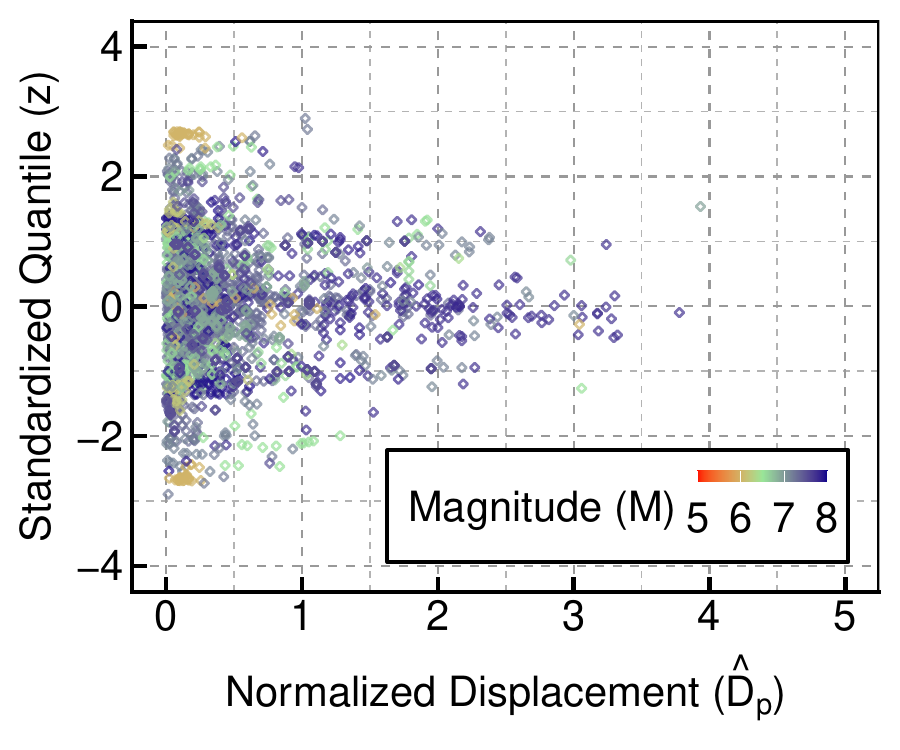}
\end{subfigure}

\vspace{0.25cm}

\begin{subfigure}[t]{0.31\linewidth}
    \centering
    \caption{}
    \includegraphics[width=\linewidth]{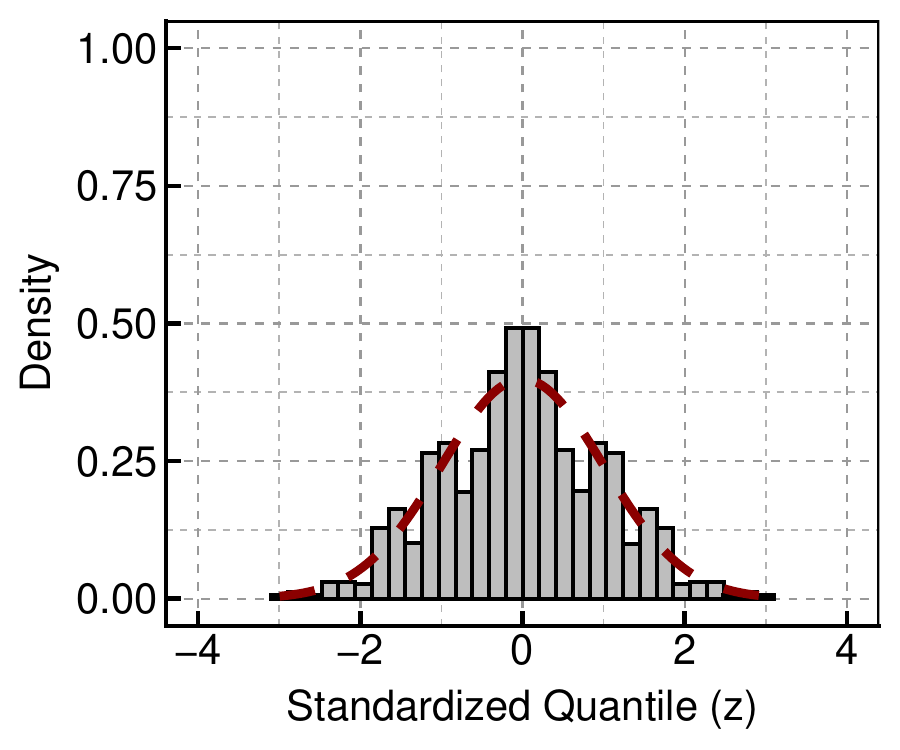}
\end{subfigure}
\hfill
\begin{subfigure}[t]{0.31\linewidth}
    \centering
    \caption{}
    \includegraphics[width=\linewidth]{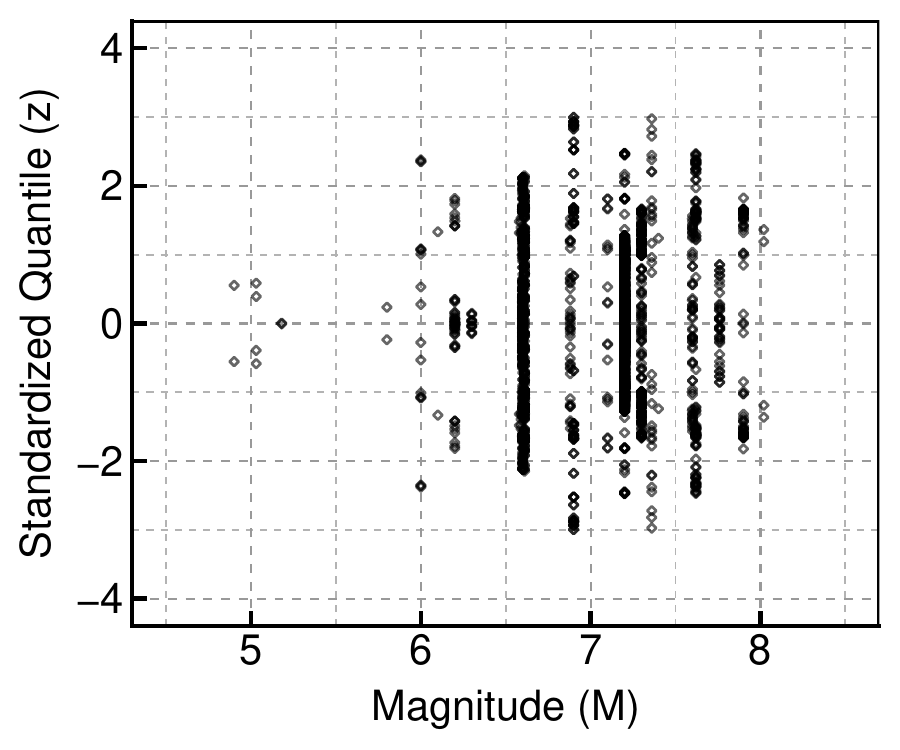}
\end{subfigure}
\hfill
\begin{subfigure}[t]{0.31\linewidth}
    \centering
    \caption{}
    \includegraphics[width=\linewidth]{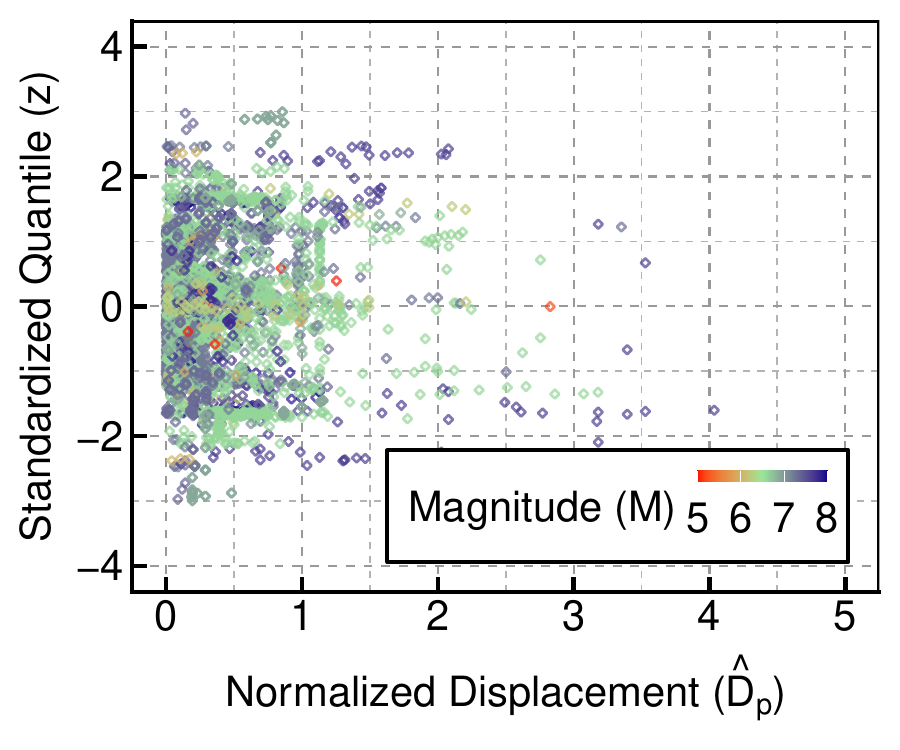}
\end{subfigure}

\caption{Assessment of residual trends in the fitted mix of two Student's t distributions for normal distance to the randomly selected fault trace using standardized quantiles. Rows correspond to faulting-style groups (all style of fault, strike-slip, normal and reverse, normal, and reverse), and columns show (left) histograms of residual quantiles, (middle) residuals plotted against earthquake magnitude ($M$) and (right) residuals plotted against normalized displacement, ($\hat D_p$).}
\label{fig:dist_z_random}

\end{figure}

\begin{figure}[p]\ContinuedFloat
\centering

\begin{subfigure}[t]{0.31\linewidth}
    \centering
    \caption{}
    \includegraphics[width=\linewidth]{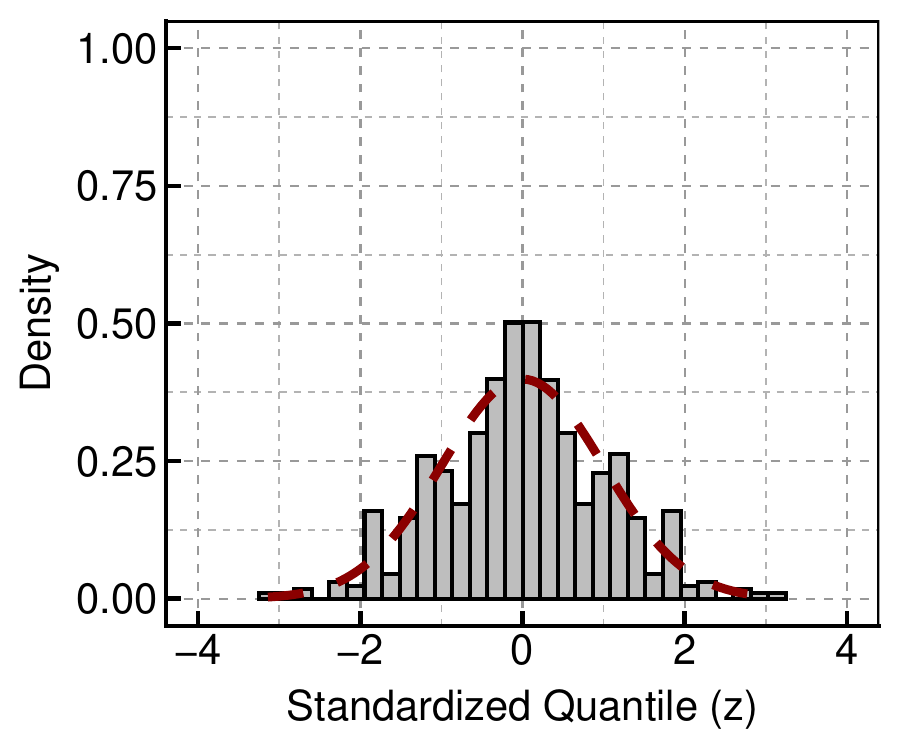}
\end{subfigure}
\hfill
\begin{subfigure}[t]{0.31\linewidth}
    \centering
    \caption{}
    \includegraphics[width=\linewidth]{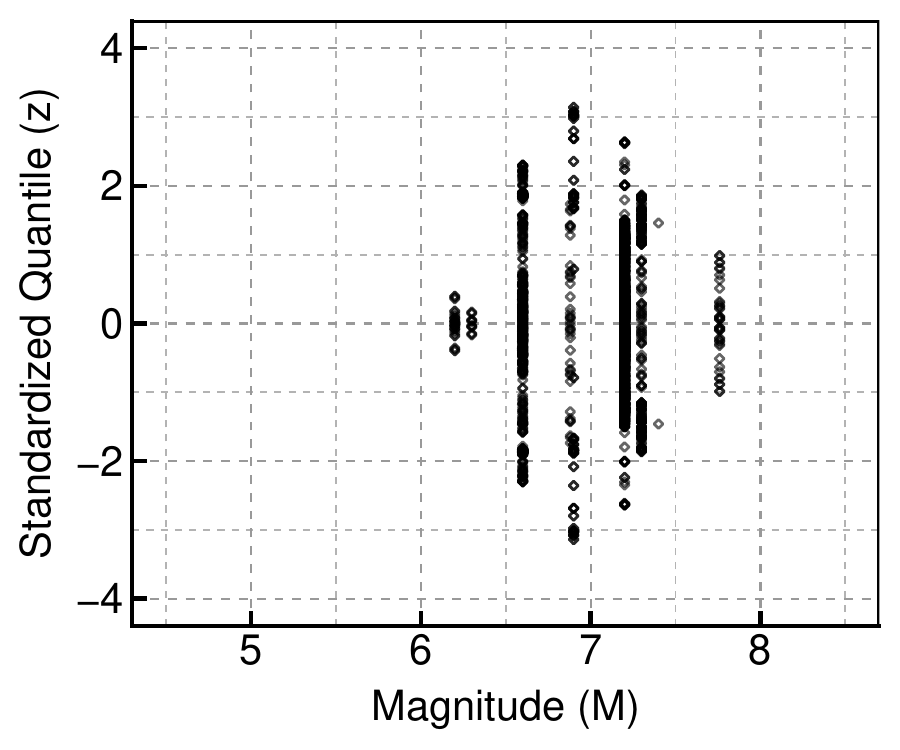}
\end{subfigure}
\hfill
\begin{subfigure}[t]{0.31\linewidth}
    \centering
    \caption{}
    \includegraphics[width=\linewidth]{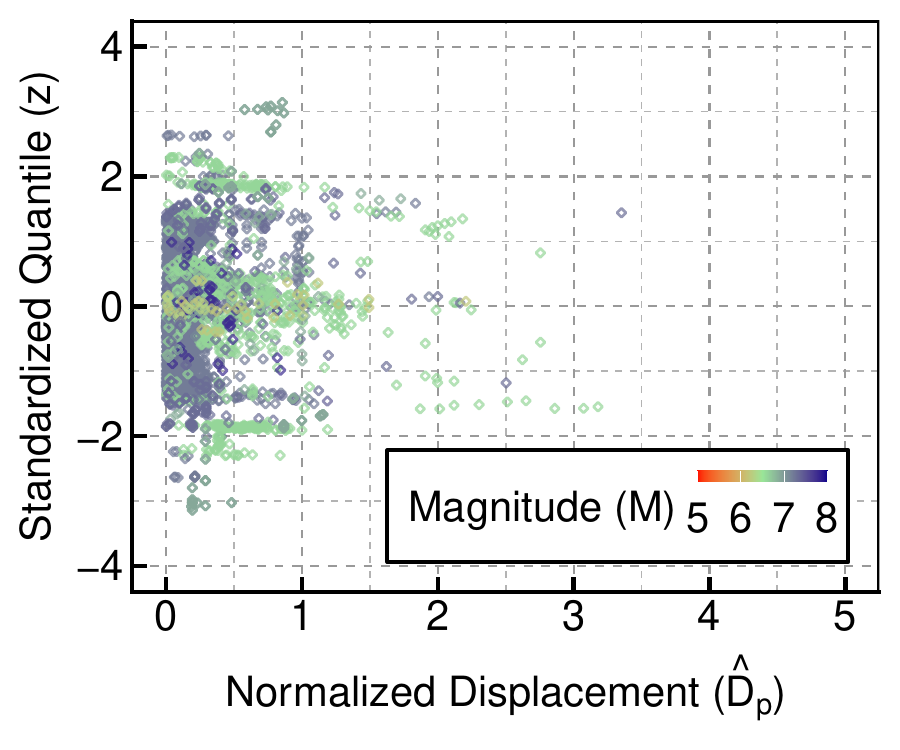}
\end{subfigure}

\vspace{0.25cm}

\begin{subfigure}[t]{0.31\linewidth}
    \centering
    \caption{}
    \includegraphics[width=\linewidth]{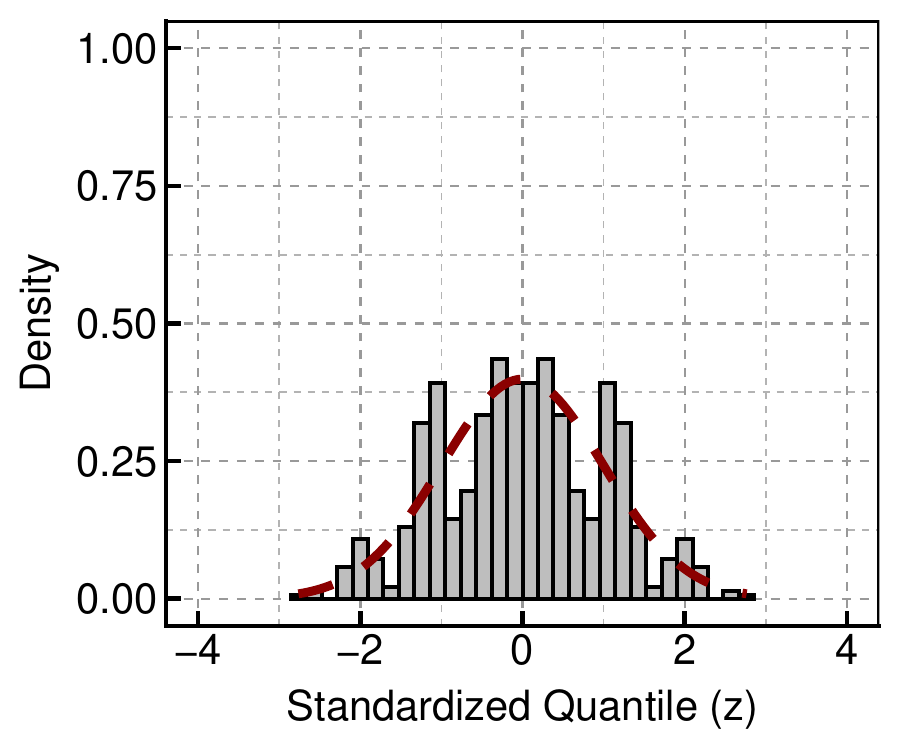}
\end{subfigure}
\hfill
\begin{subfigure}[t]{0.31\linewidth}
    \centering
    \caption{}
    \includegraphics[width=\linewidth]{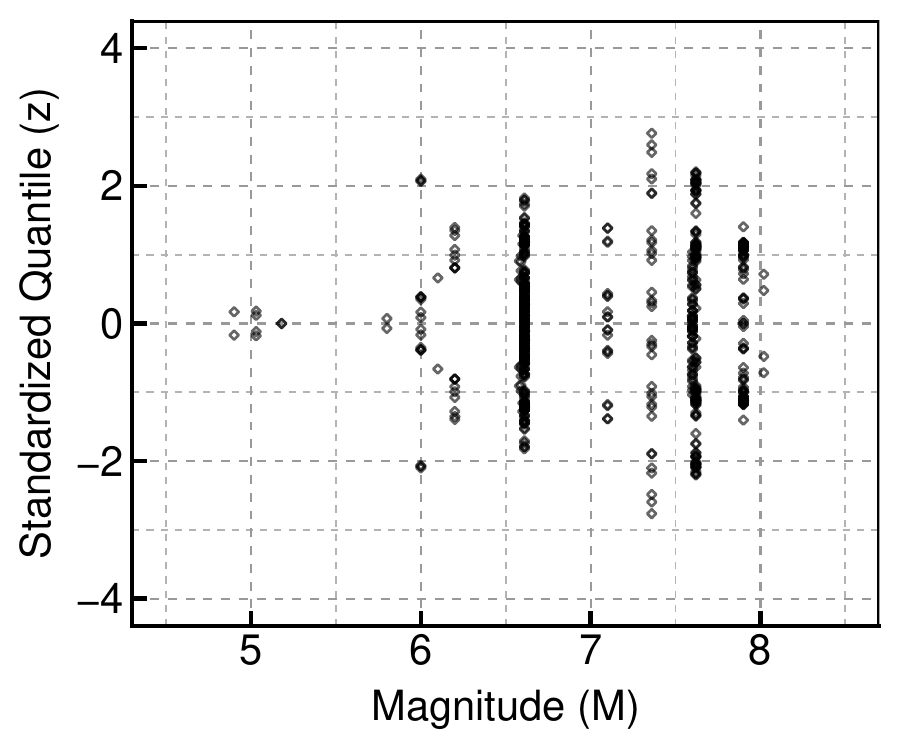}
\end{subfigure}
\hfill
\begin{subfigure}[t]{0.31\linewidth}
    \centering
    \caption{}
    \includegraphics[width=\linewidth]{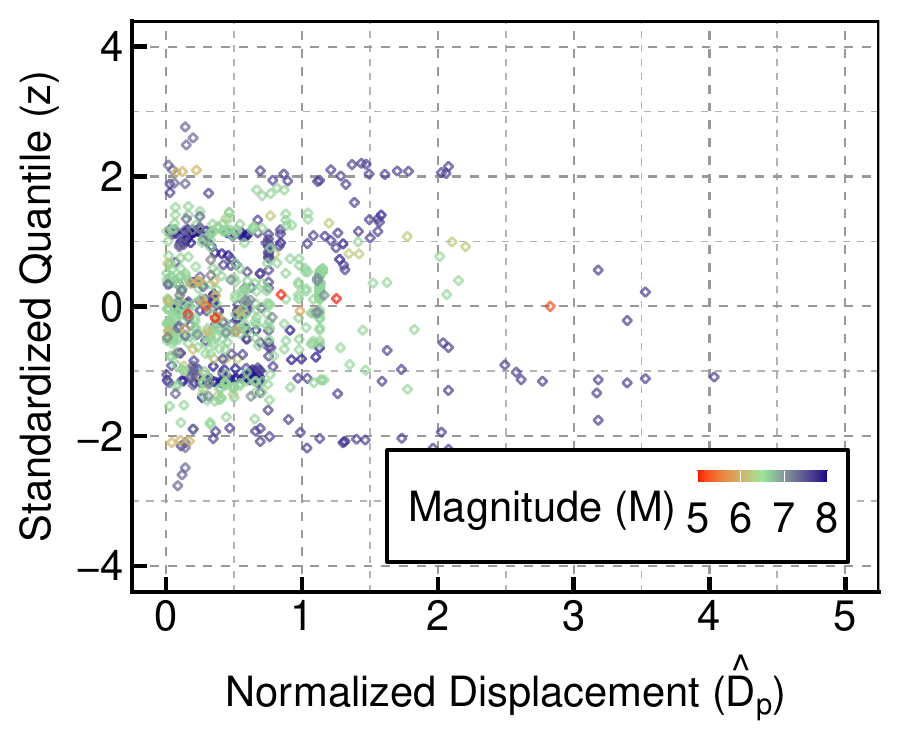}
\end{subfigure}

\caption[]{Assessment of residual trends in the fitted mix of two Student's t distributions for normal distance to the randomly selected fault trace using standardized quantiles, continued.}
\end{figure}

\begin{figure}[p]
\centering

\begin{subfigure}[t]{0.31\linewidth}
    \centering
    \caption{}
    \includegraphics[width=\linewidth]{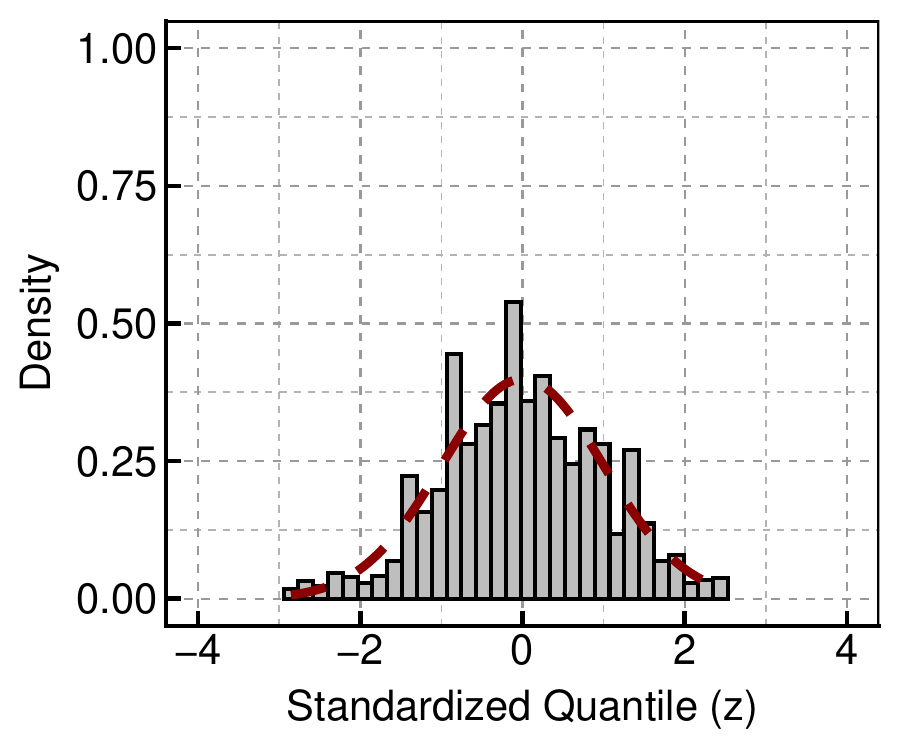}
\end{subfigure}
\hfill
\begin{subfigure}[t]{0.31\linewidth}
    \centering
    \caption{}
    \includegraphics[width=\linewidth]{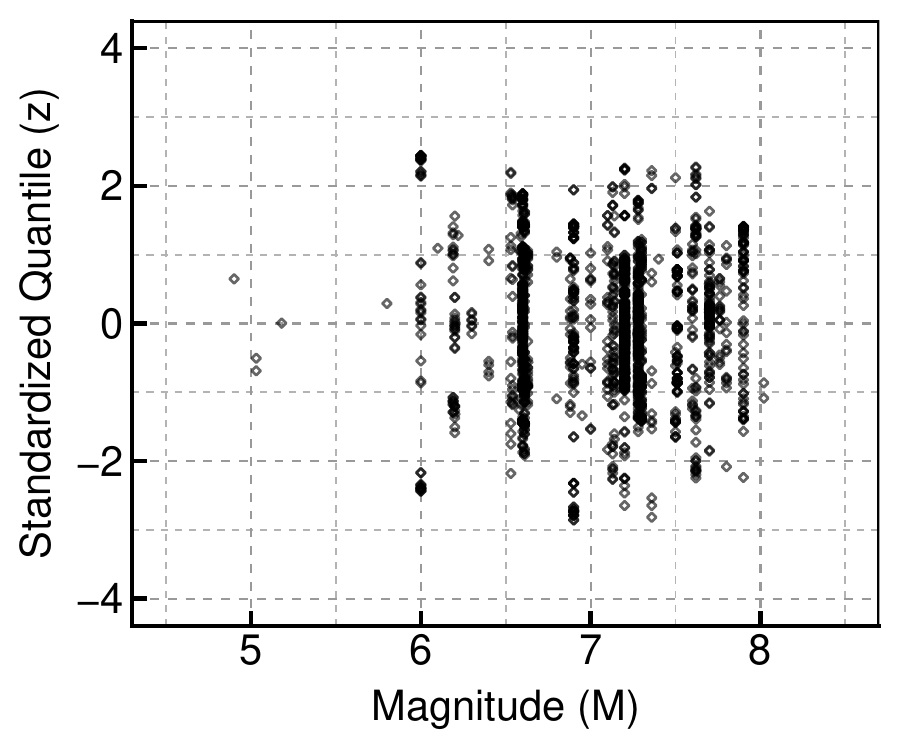}
\end{subfigure}
\hfill
\begin{subfigure}[t]{0.31\linewidth}
    \centering
    \caption{}
    \includegraphics[width=\linewidth]{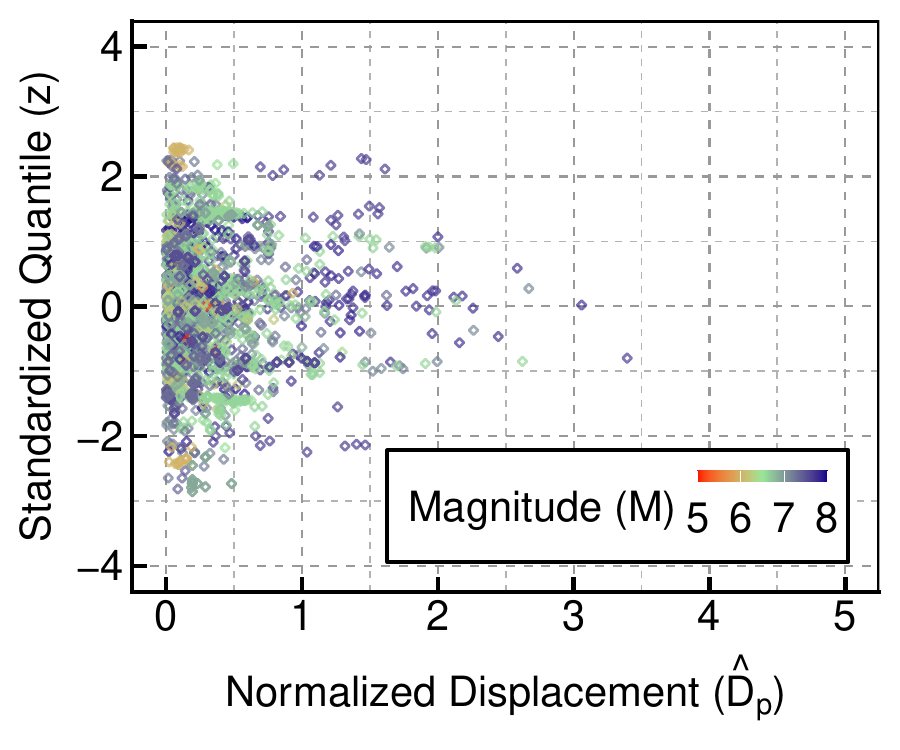}
\end{subfigure}

\vspace{0.25cm}

\begin{subfigure}[t]{0.31\linewidth}
    \centering
    \caption{}
    \includegraphics[width=\linewidth]{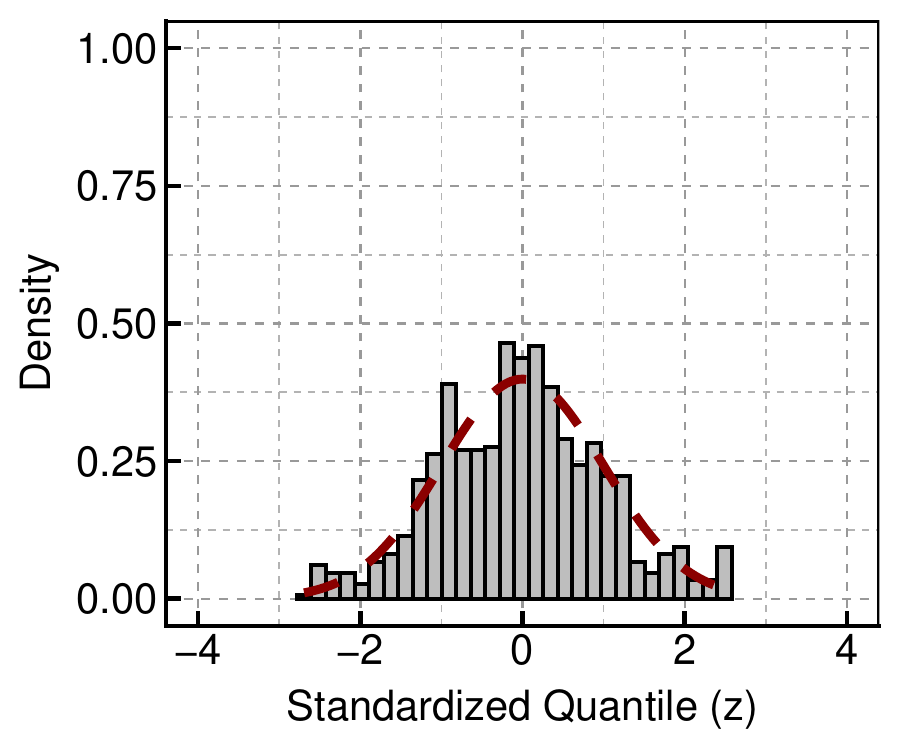}
\end{subfigure}
\hfill
\begin{subfigure}[t]{0.31\linewidth}
    \centering
    \caption{}
    \includegraphics[width=\linewidth]{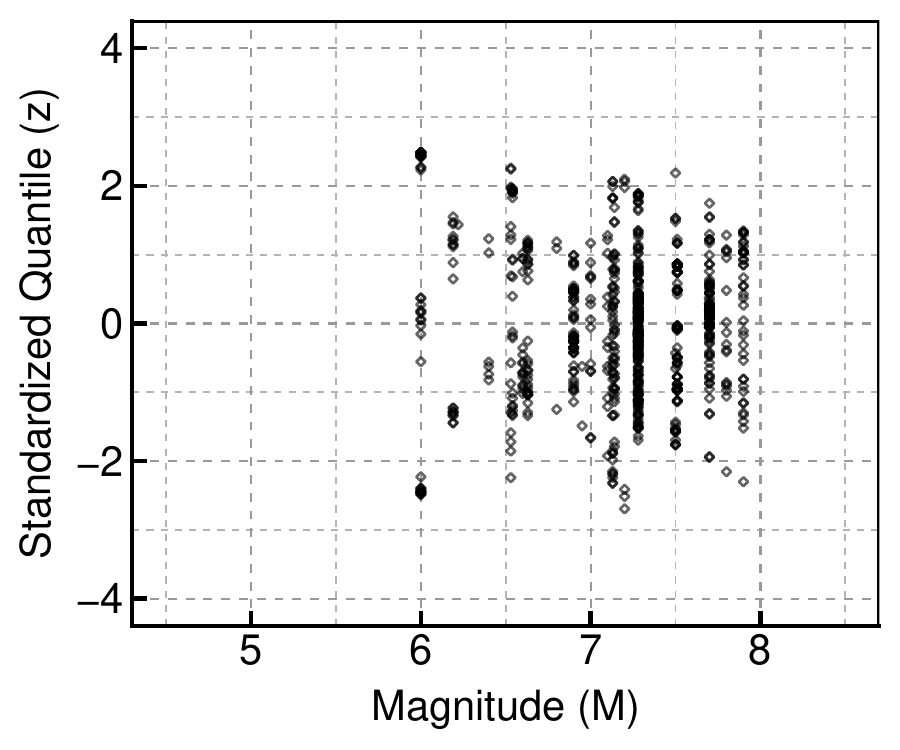}
\end{subfigure}
\hfill
\begin{subfigure}[t]{0.31\linewidth}
    \centering
    \caption{}
    \includegraphics[width=\linewidth]{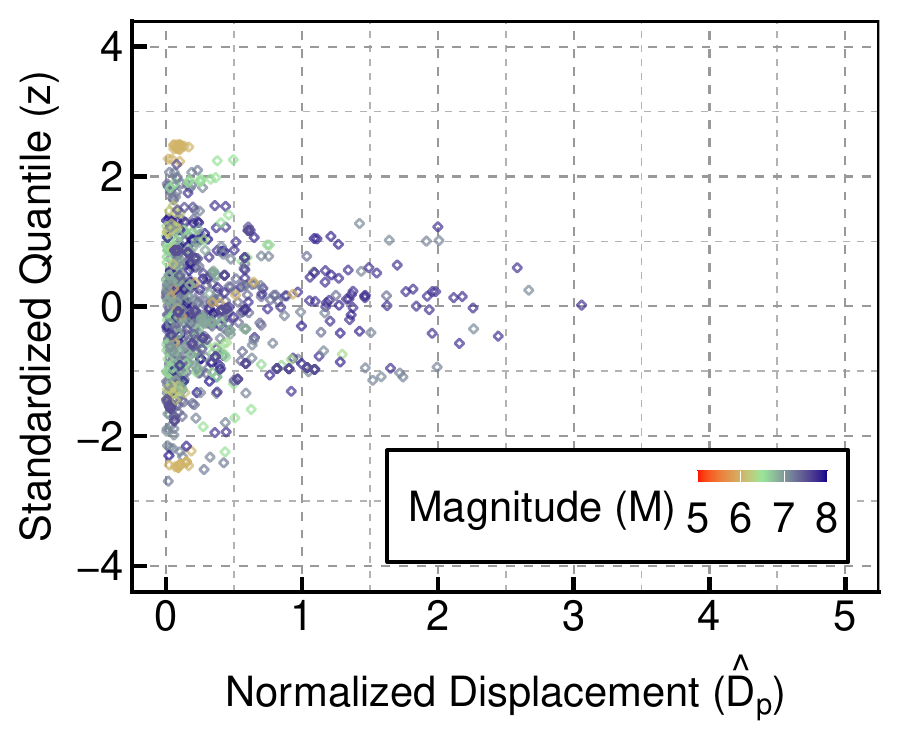}
\end{subfigure}

\vspace{0.25cm}

\begin{subfigure}[t]{0.31\linewidth}
    \centering
    \caption{}
    \includegraphics[width=\linewidth]{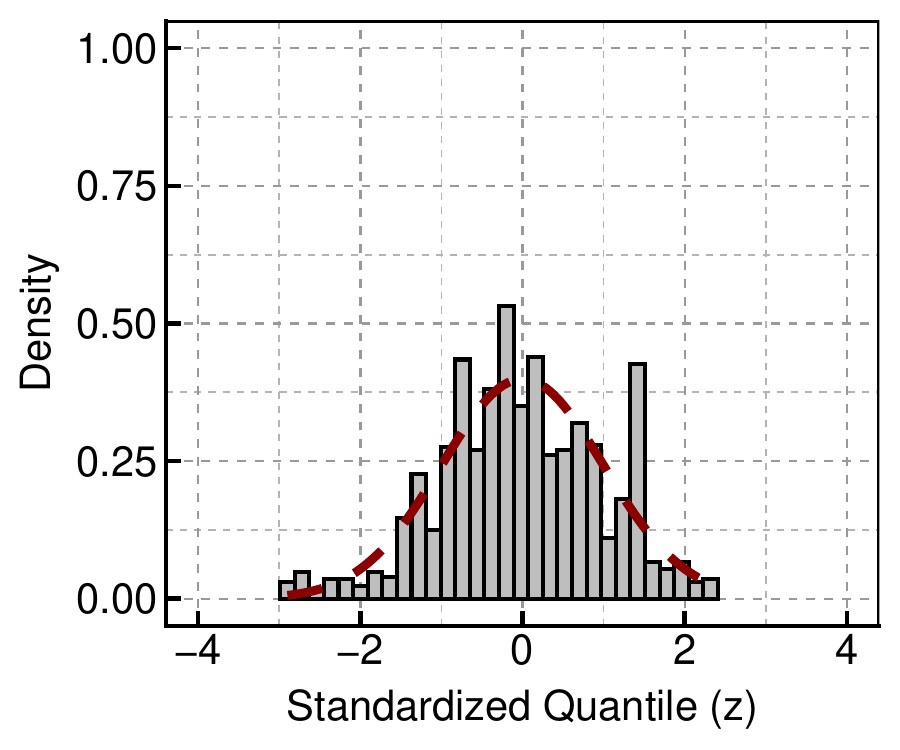}
\end{subfigure}
\hfill
\begin{subfigure}[t]{0.31\linewidth}
    \centering
    \caption{}
    \includegraphics[width=\linewidth]{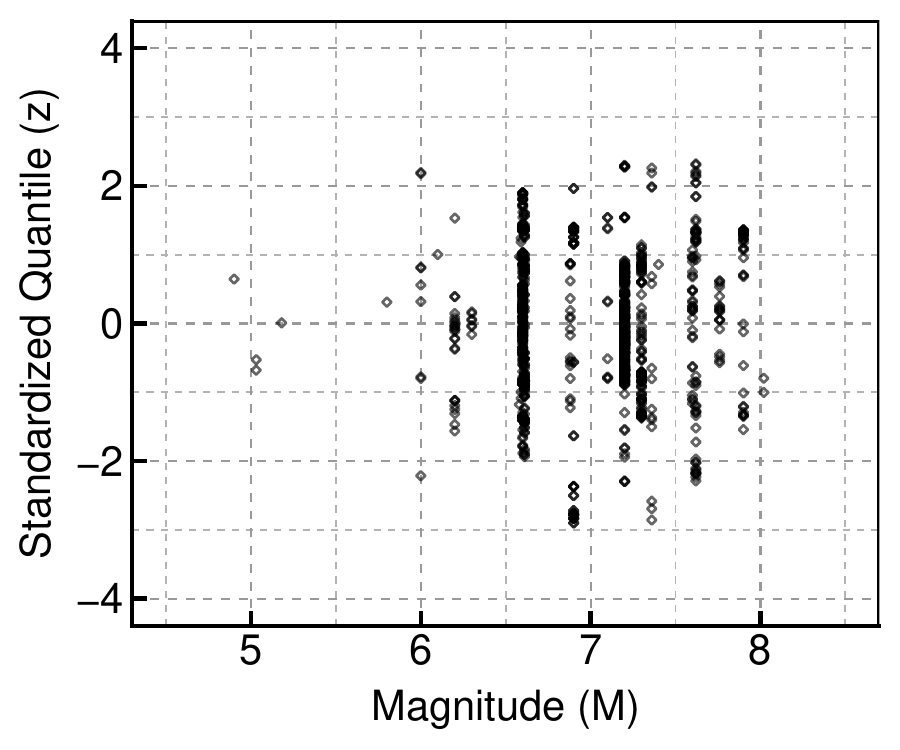}
\end{subfigure}
\hfill
\begin{subfigure}[t]{0.31\linewidth}
    \centering
    \caption{}
    \includegraphics[width=\linewidth]{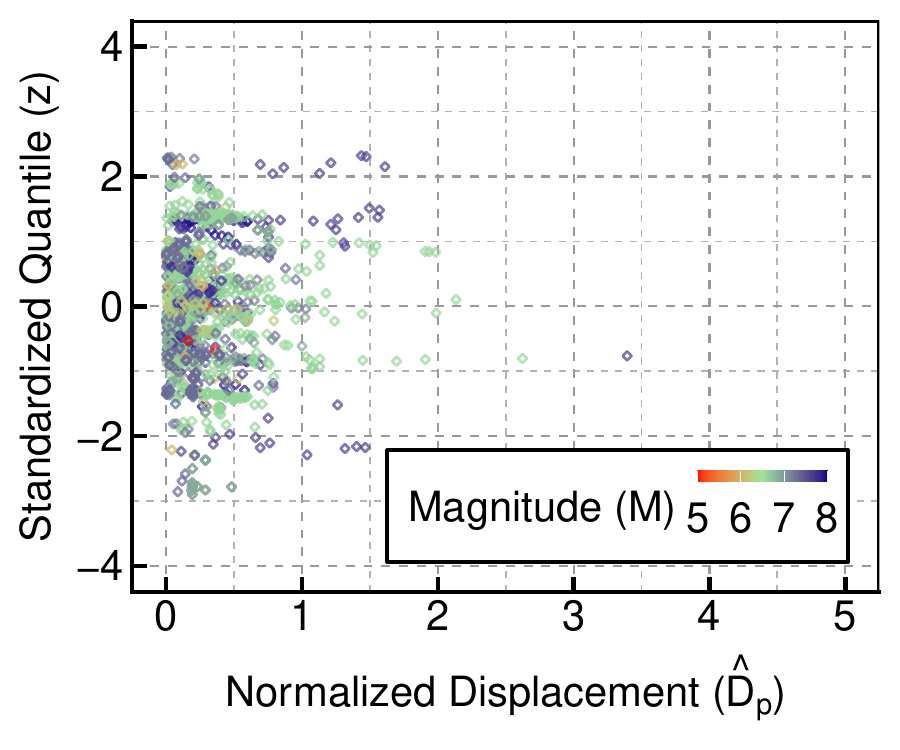}
\end{subfigure}

\caption{Assessment of residual trends in the fitted mix of two Student's t distributions for normal distance to the maximum principal displacement fault trace using standardized quantiles. Rows correspond to faulting-style groups (all style of fault, strike-slip, normal and reverse, normal, and reverse), and columns show (left) histograms of residual quantiles, (middle) residuals plotted against earthquake magnitude ($M$) and (right) residuals plotted against normalized displacement, ($\hat D_p$).}
\label{fig:dist_z_max}

\end{figure}

\begin{figure}[p]\ContinuedFloat
\centering

\begin{subfigure}[t]{0.31\linewidth}
    \centering
    \caption{}
    \includegraphics[width=\linewidth]{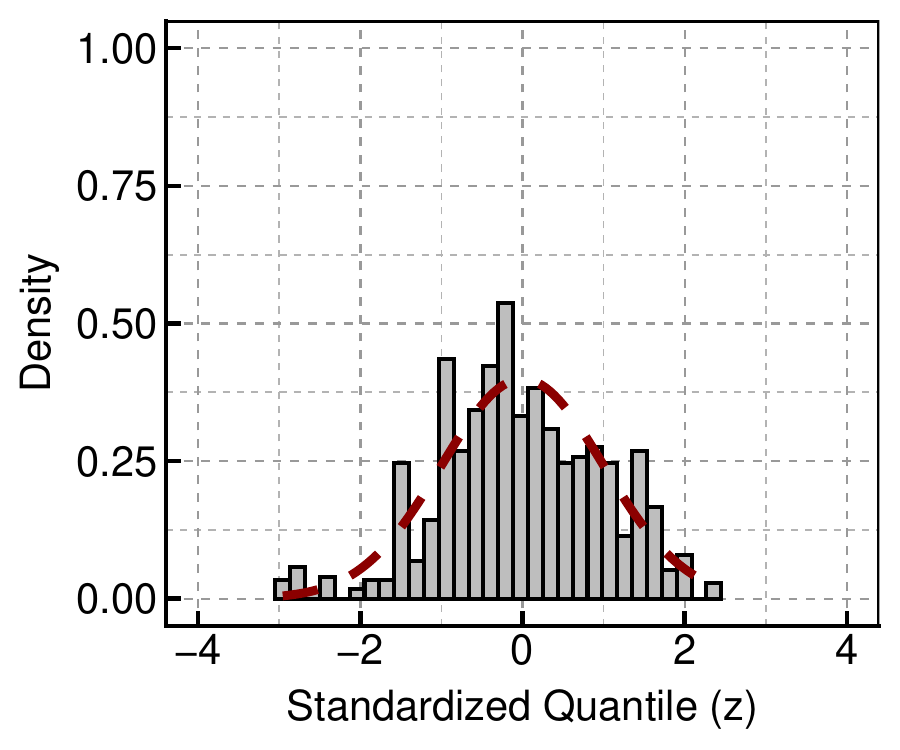}
\end{subfigure}
\hfill
\begin{subfigure}[t]{0.31\linewidth}
    \centering
    \caption{}
    \includegraphics[width=\linewidth]{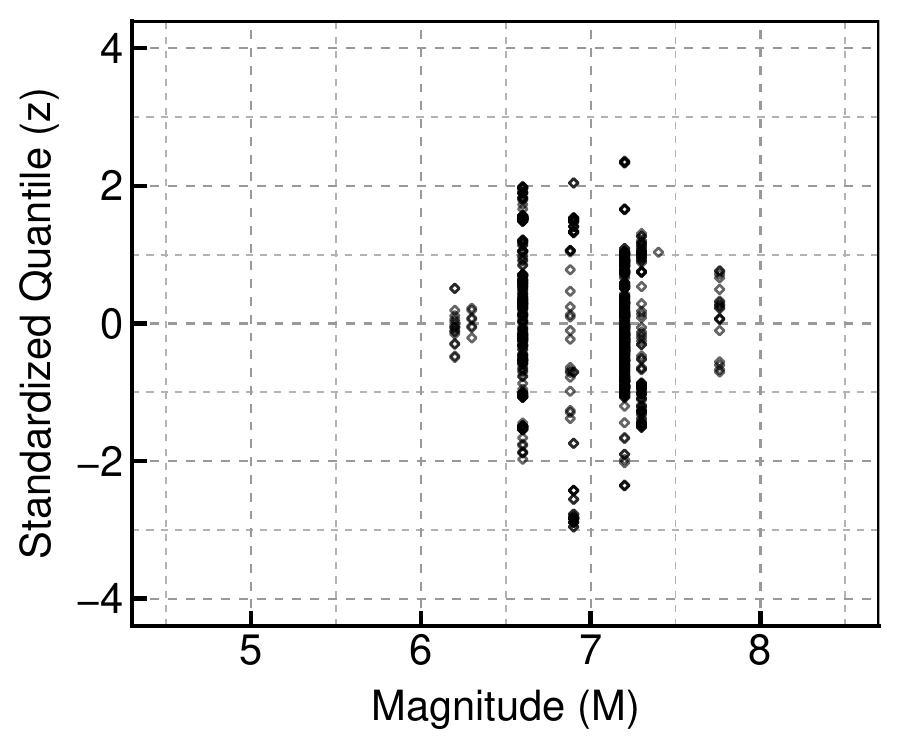}
\end{subfigure}
\hfill
\begin{subfigure}[t]{0.31\linewidth}
    \centering
    \caption{}
    \includegraphics[width=\linewidth]{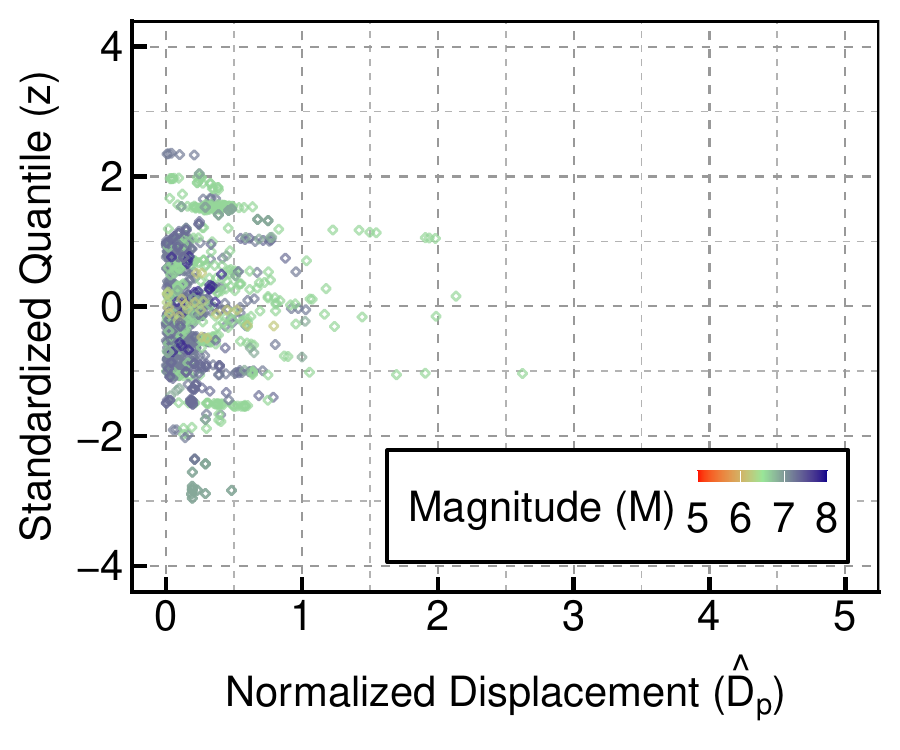}
\end{subfigure}

\vspace{0.25cm}

\begin{subfigure}[t]{0.31\linewidth}
    \centering
    \caption{}
    \includegraphics[width=\linewidth]{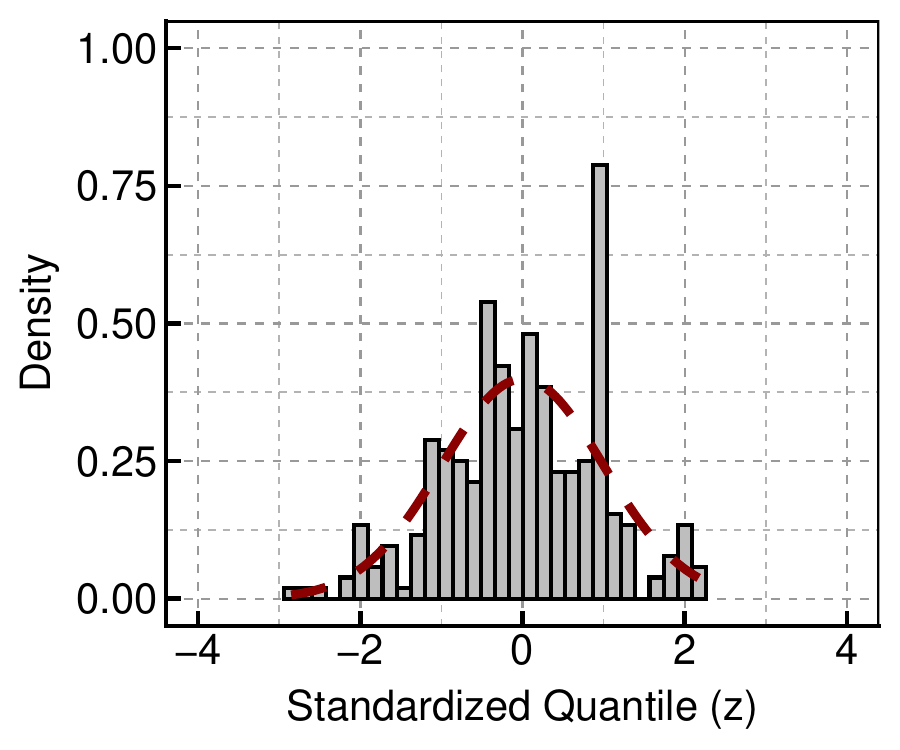}
\end{subfigure}
\hfill
\begin{subfigure}[t]{0.31\linewidth}
    \centering
    \caption{}
    \includegraphics[width=\linewidth]{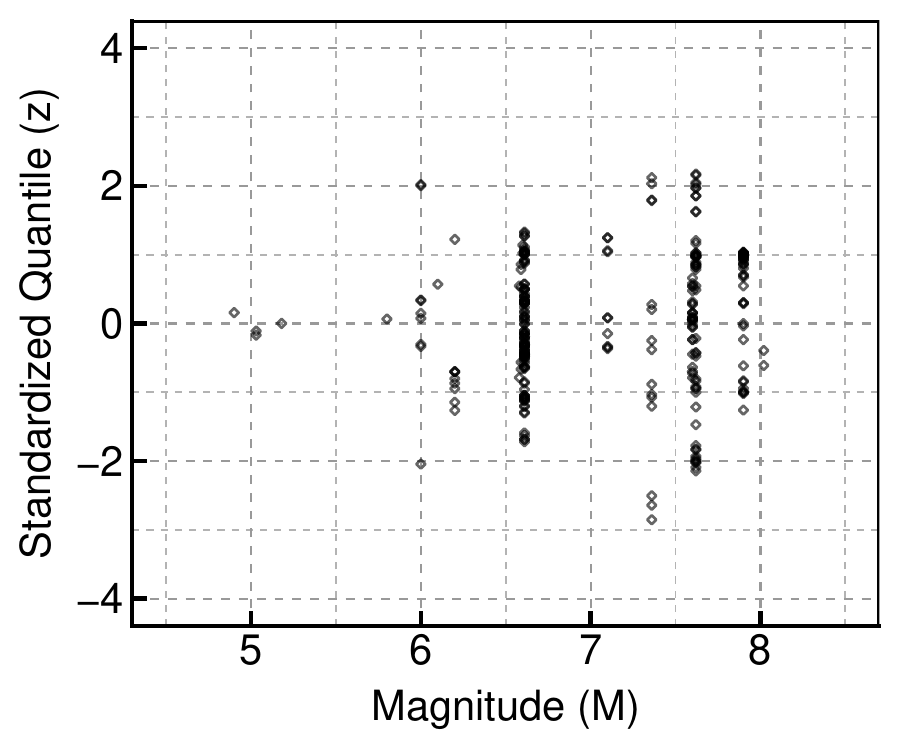}
\end{subfigure}
\hfill
\begin{subfigure}[t]{0.31\linewidth}
    \centering
    \caption{}
    \includegraphics[width=\linewidth]{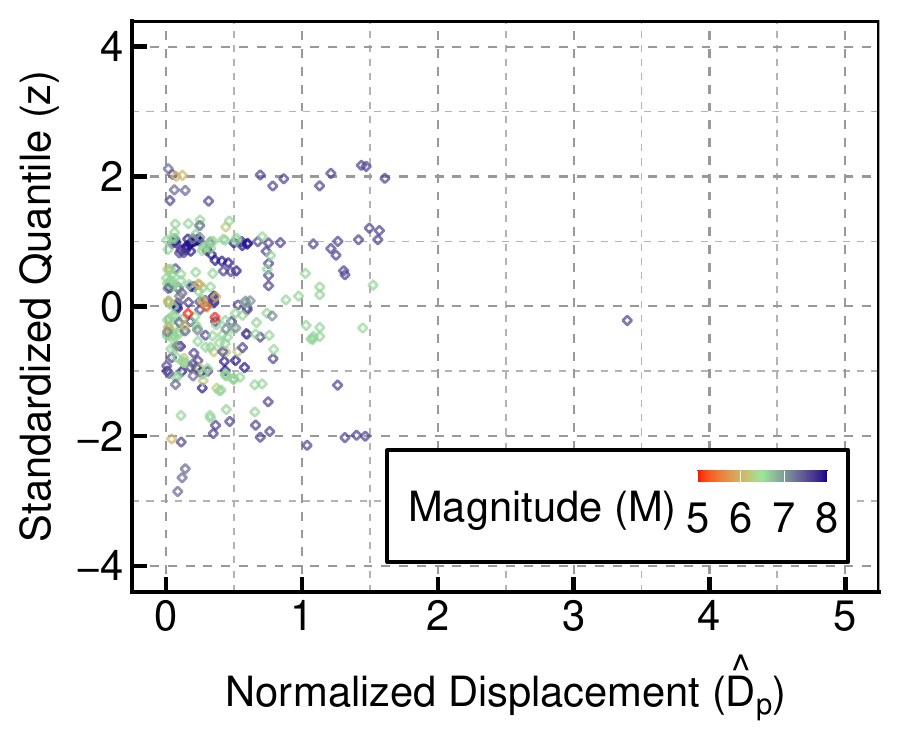}
\end{subfigure}

\caption[]{Assessment of residual trends in the fitted mix of two Student's t distributions for normal distance to the maximum principal displacement fault trace using standardized quantiles, continued.}
\end{figure}

\begin{figure}[p]
\centering

\begin{subfigure}[t]{0.31\linewidth}
    \centering
    \caption{}
    \includegraphics[width=\linewidth]{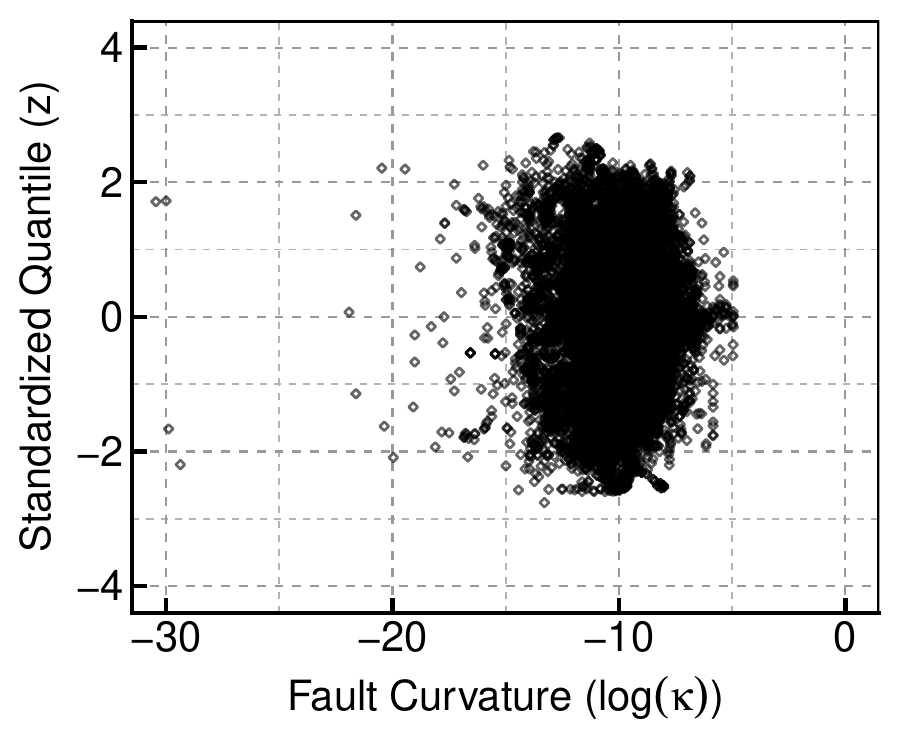}
\end{subfigure}
\hfill
\begin{subfigure}[t]{0.31\linewidth}
    \centering
    \caption{}
    \includegraphics[width=\linewidth]{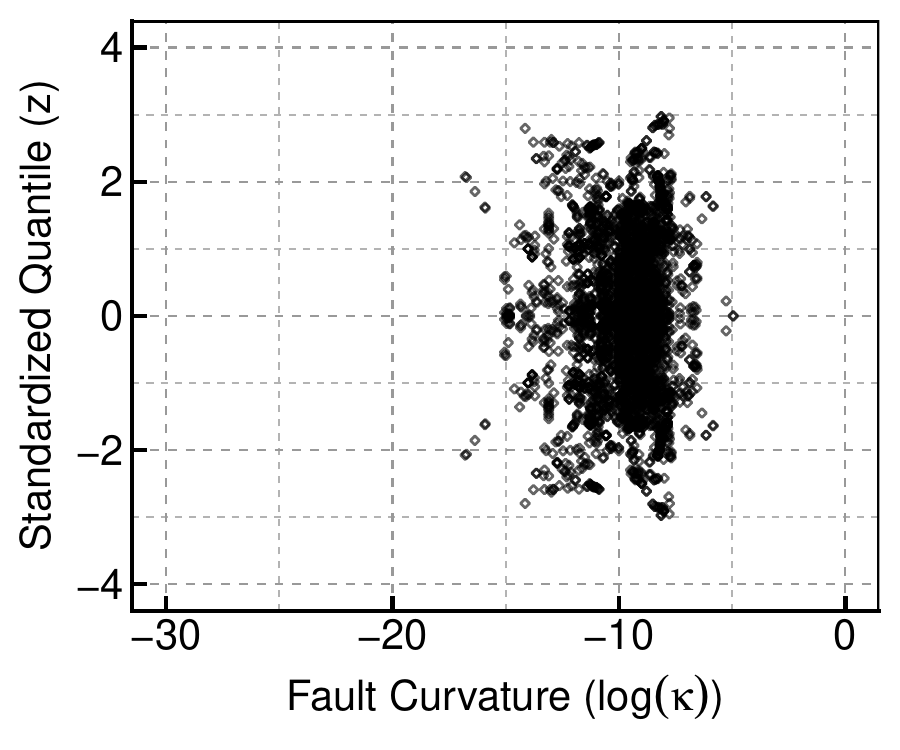}
\end{subfigure}
\hfill
\begin{subfigure}[t]{0.31\linewidth}
    \centering
    \caption{}
    \includegraphics[width=\linewidth]{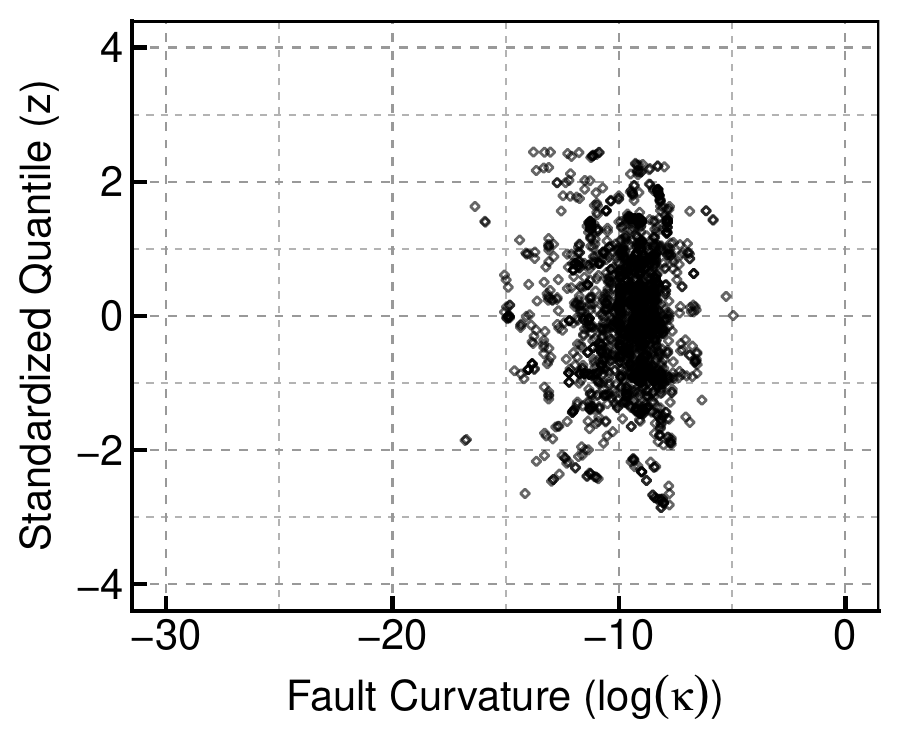}
\end{subfigure}

\vspace{0.25cm}

\begin{subfigure}[t]{0.31\linewidth}
    \centering
    \caption{}
    \includegraphics[width=\linewidth]{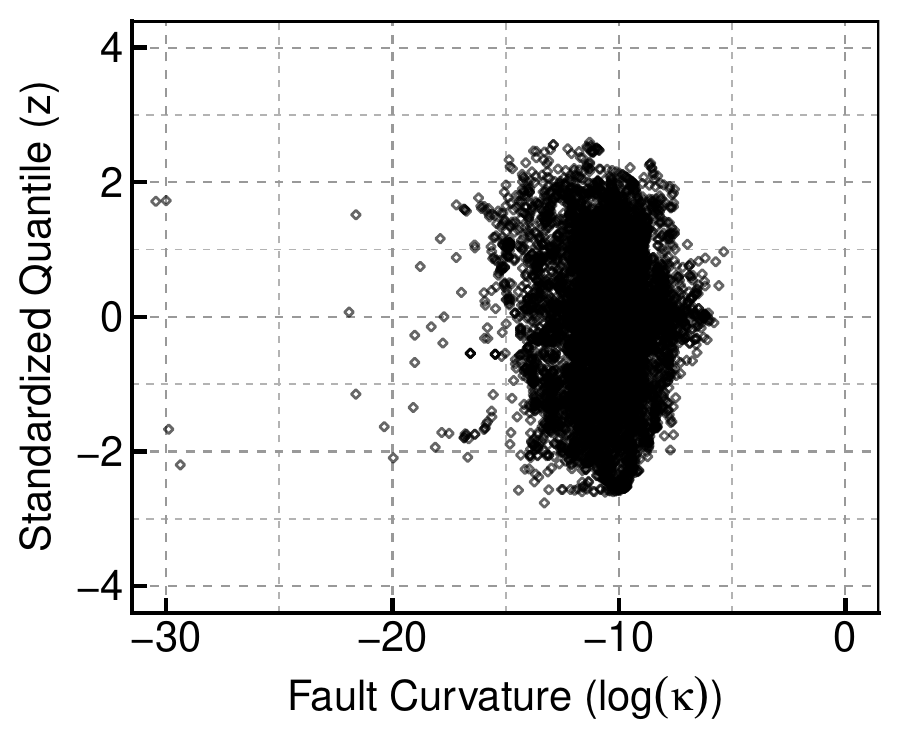}
\end{subfigure}
\hfill
\begin{subfigure}[t]{0.31\linewidth}
    \centering
    \caption{}
    \includegraphics[width=\linewidth]{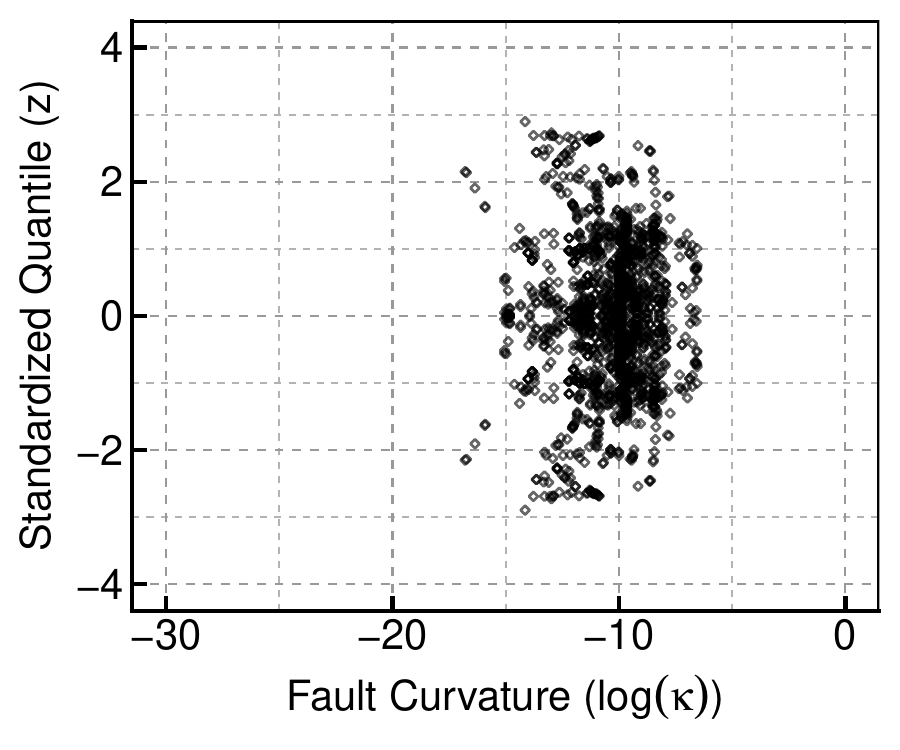}
\end{subfigure}
\hfill
\begin{subfigure}[t]{0.31\linewidth}
    \centering
    \caption{}
    \includegraphics[width=\linewidth]{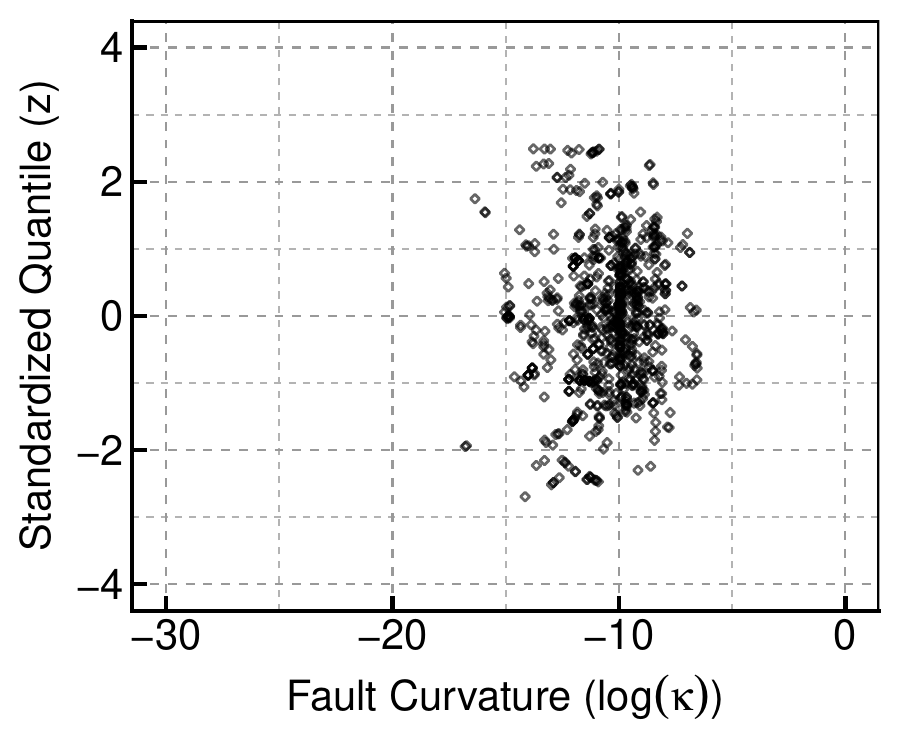}
\end{subfigure}

\vspace{0.25cm}

\begin{subfigure}[t]{0.31\linewidth}
    \centering
    \caption{}
    \includegraphics[width=\linewidth]{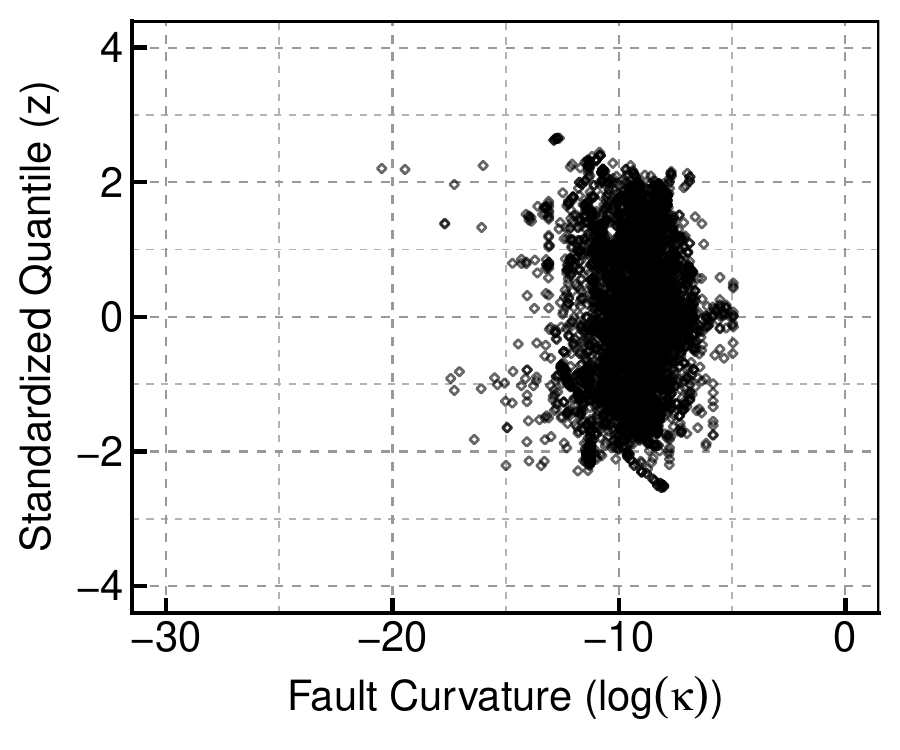}
\end{subfigure}
\hfill
\begin{subfigure}[t]{0.31\linewidth}
    \centering
    \caption{}
    \includegraphics[width=\linewidth]{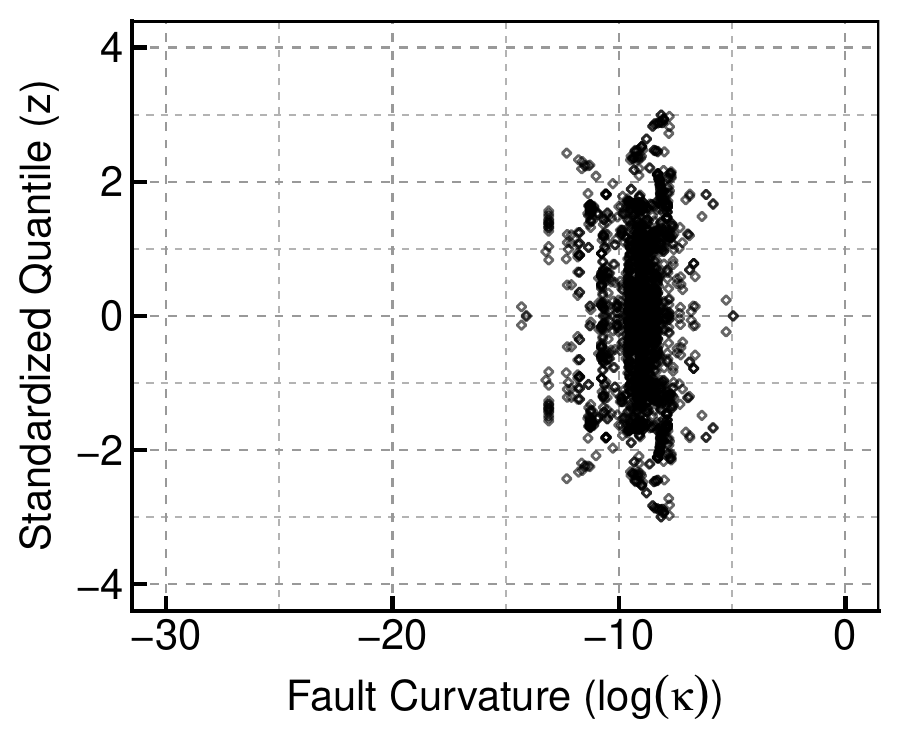}
\end{subfigure}
\hfill
\begin{subfigure}[t]{0.31\linewidth}
    \centering
    \caption{}
    \includegraphics[width=\linewidth]{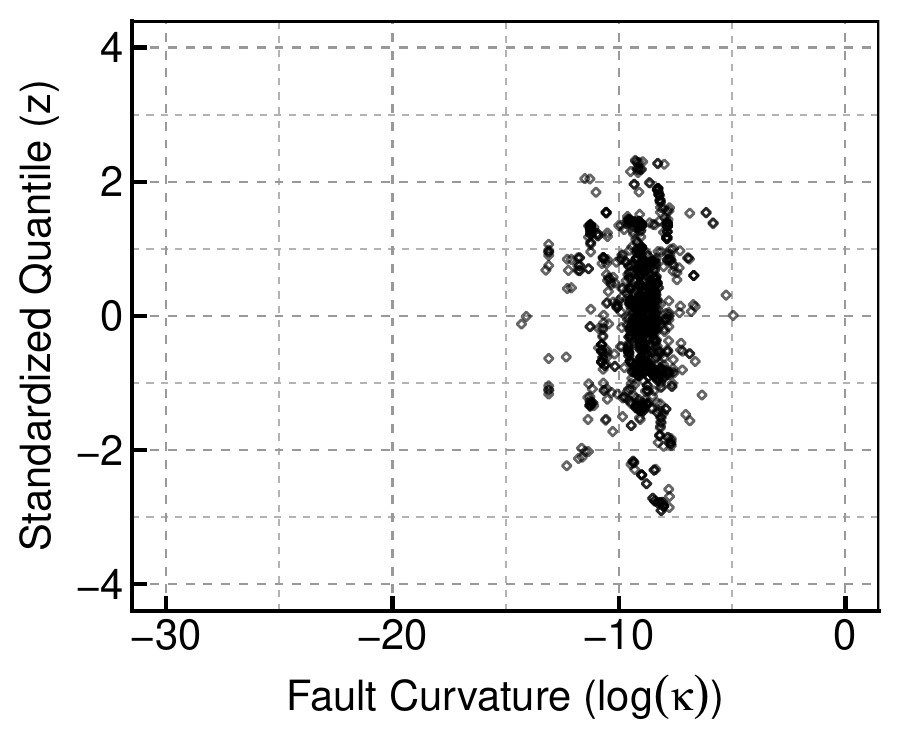}
\end{subfigure}

\caption{Assessment of residual trends with respect to fault curvature ($\log(\kappa)$) for the fitted normal-distance models using standardized quantiles. Rows correspond to faulting-style groups (all faulting styles, strike-slip, normal-reverse, normal, and reverse), and columns present residuals for normal distance to the (left) SCS trace, (middle) randomly selected trace, and (right) maximum principal displacement trace definitions.}
\label{fig:dist_z_curv}

\end{figure}

\begin{figure}[p]\ContinuedFloat
\centering

\begin{subfigure}[t]{0.31\linewidth}
    \centering
    \caption{}
    \includegraphics[width=\linewidth]{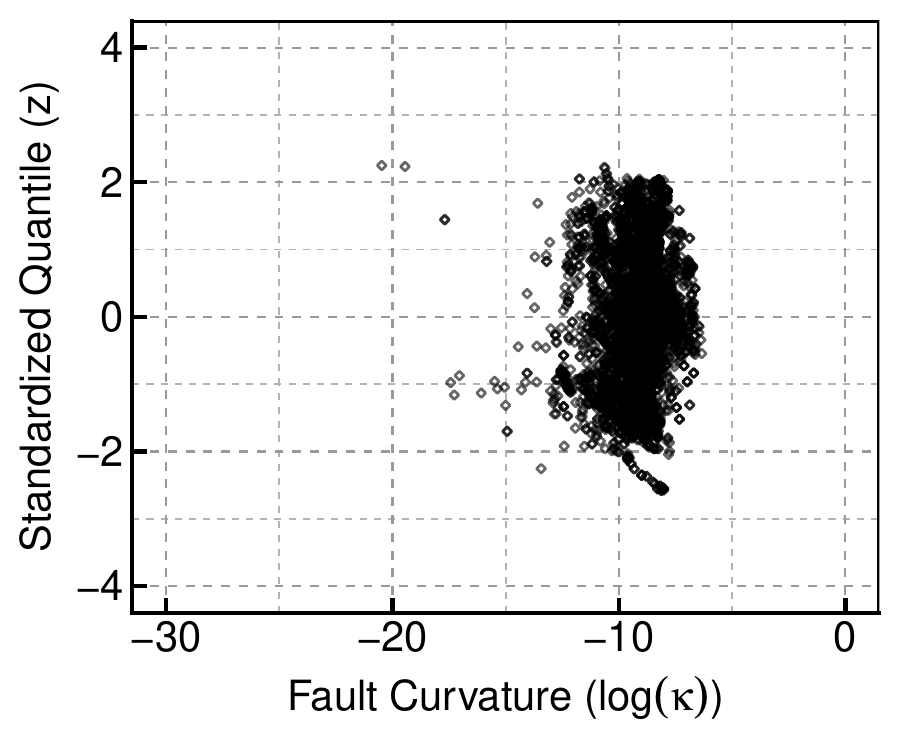}
\end{subfigure}
\hfill
\begin{subfigure}[t]{0.31\linewidth}
    \centering
    \caption{}
    \includegraphics[width=\linewidth]{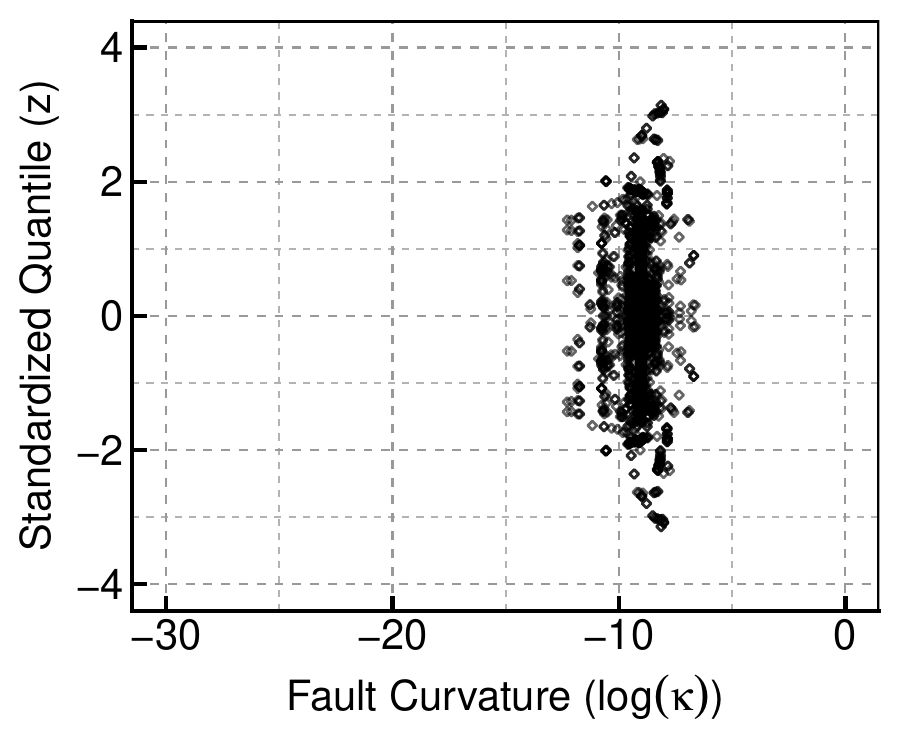}
\end{subfigure}
\hfill
\begin{subfigure}[t]{0.31\linewidth}
    \centering
    \caption{}
    \includegraphics[width=\linewidth]{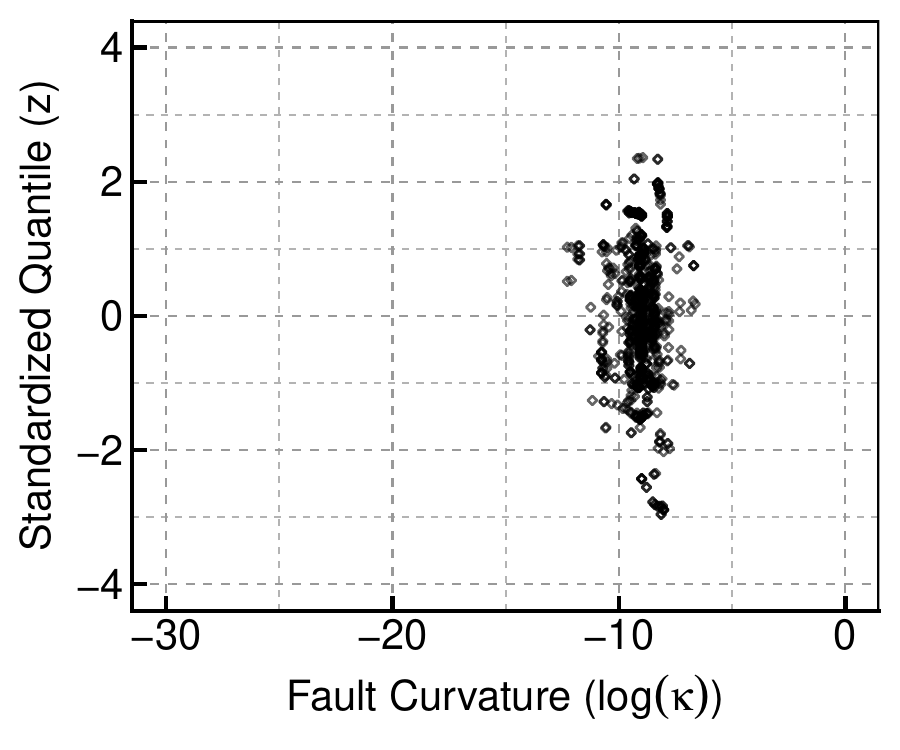}
\end{subfigure}

\vspace{0.25cm}

\begin{subfigure}[t]{0.31\linewidth}
    \centering
    \caption{}
    \includegraphics[width=\linewidth]{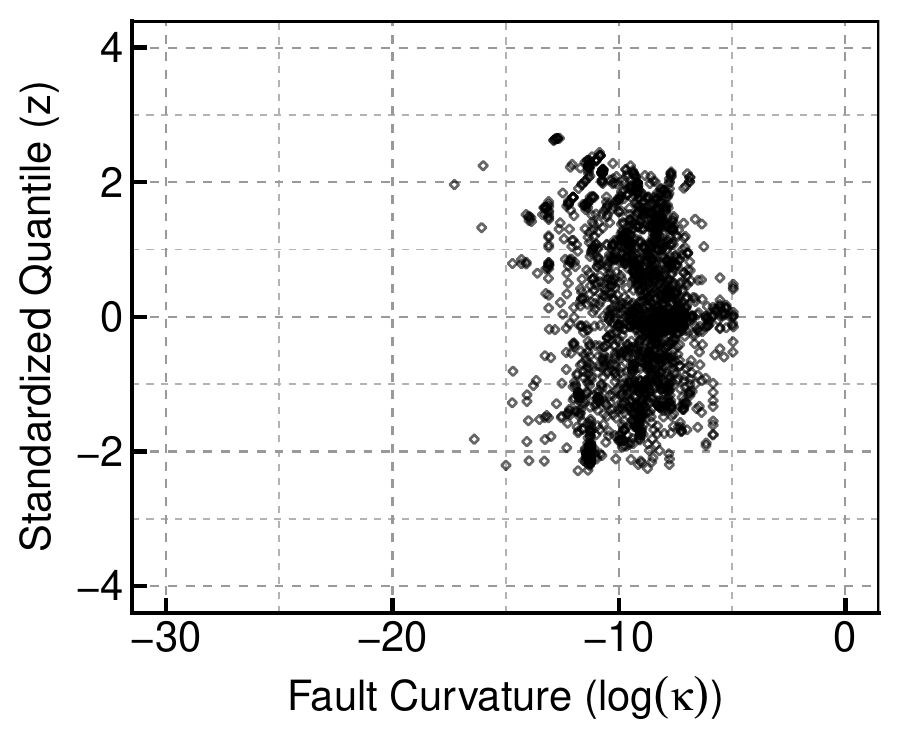}
\end{subfigure}
\hfill
\begin{subfigure}[t]{0.31\linewidth}
    \centering
    \caption{}
    \includegraphics[width=\linewidth]{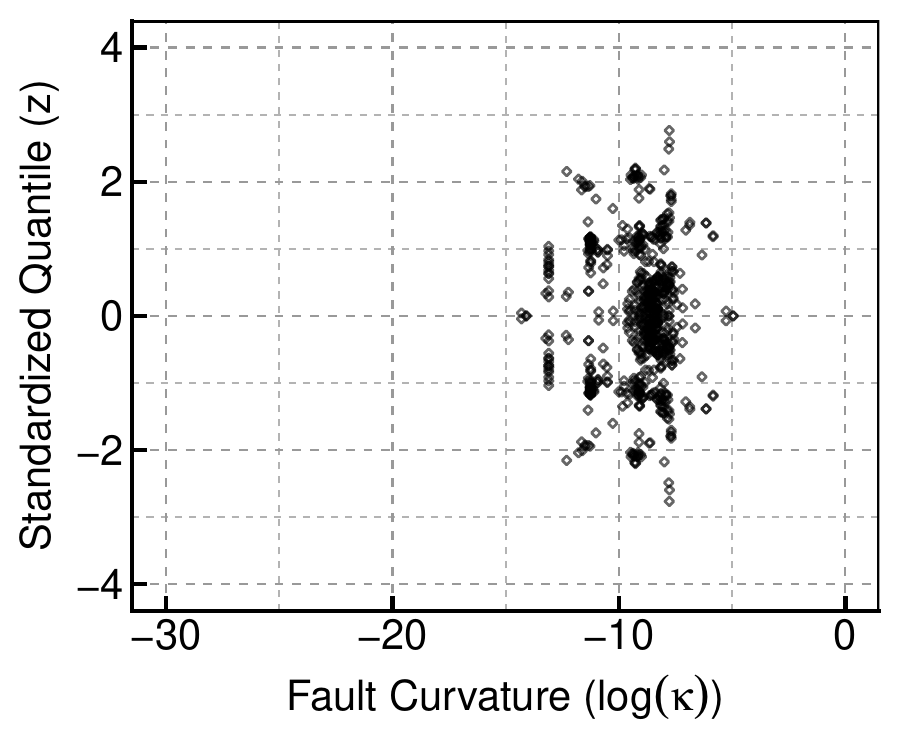}
\end{subfigure}
\hfill
\begin{subfigure}[t]{0.31\linewidth}
    \centering
    \caption{}
    \includegraphics[width=\linewidth]{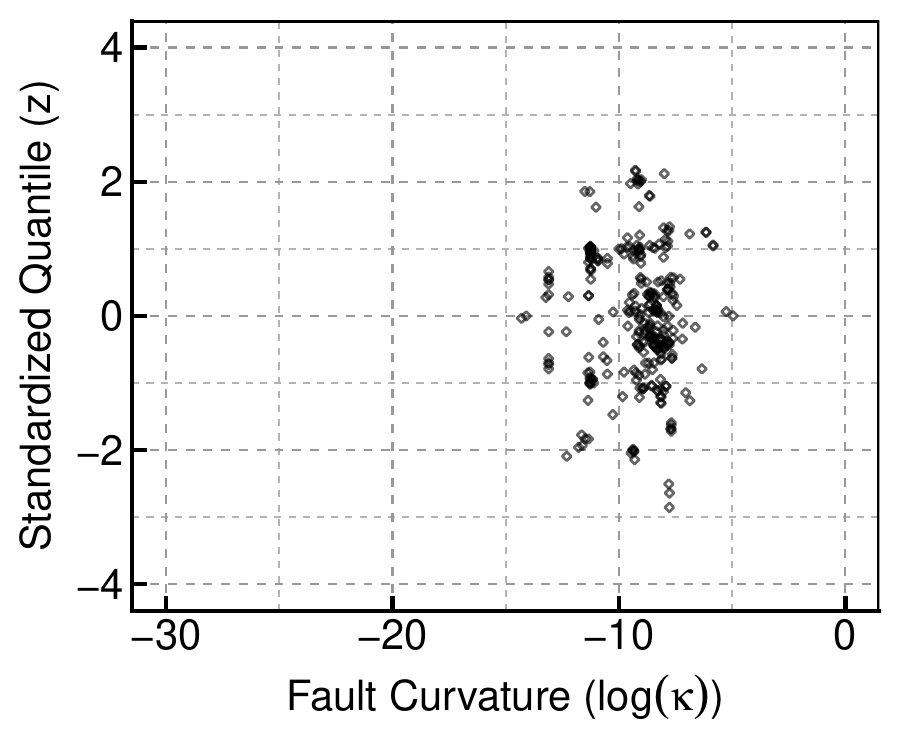}
\end{subfigure}

\caption[]{Assessment of residual trends with respect to fault curvature ($\log(\kappa)$) for the fitted normal-distance models using standardized quantiles, continued.}
\end{figure}

\clearpage
\phantomsection
\bibliography{references_PrincipalDisplacement}